\documentclass[rmp,twocolumn,showpacs,superscriptaddress]{revtex4}
\usepackage{amsmath}
\usepackage{amsfonts}
\usepackage{amssymb}
\usepackage{bm}
\usepackage{color}
\usepackage{multirow}
\usepackage{enumitem}
\usepackage{wrapfig}
\usepackage{graphicx}
\usepackage{epstopdf}
\usepackage{csquotes}
\usepackage{comment}
\usepackage{mathtools} 

\DeclareMathOperator{\Tr}{Tr}

\newcommand{\be}{\begin{equation}}
\newcommand{\ee}{\end{equation}}
\newcommand{\bea}{\begin{eqnarray}}
\newcommand{\eea}{\end{eqnarray}}

\newcommand{\br}{{\bf r}}
\newcommand{\bR}{{\bf R}}

\newcommand{\IV}{$IV$}

\newcommand{\eF}{\epsilon_\text{F}}  

\newcommand{\Gem}{G_\text{e$\mathcal{M}$}}
\newcommand{\Hem}{H_\text{e$\mathcal{M}$}}
\newcommand{\GR}{\Gamma_\mathcal{R}}
\newcommand{\GL}{\Gamma_\mathcal{L}}

\newcommand{\SL}{\Sigma_\mathcal{L}}
\newcommand{\SR}{\Sigma_\mathcal{R}}

\newcommand{\Hks}{H_\text{KS}}

\newcommand{\ci}{\mathfrak{i}}

\newcommand{\mfT}{\mathfrak{T}}

\newcommand*\Eval[3]{\left.#1\right\rvert_{#2}^{#3}}
\def\gowo{\,$G_0W_0$}
 
\def\stait{\,STAIT\ } 
\newcommand{\ea}{{\it et al.}\ } 
\newcommand{\go}{{\,$G_0$}} 
\newcommand{\qi}{{QI}} 
 
\newcommand{\mH}{H_\mathcal{M}}
\newcommand{\Do}{\Delta^{(0)}}
\newcommand{\tmH}{{\tilde H}_\mathcal{M}}
\newcommand{\mS}{\Sigma_\mathcal{M}}
\newcommand{\mff}{{\mathfrak f}}
\newcommand{\mfftwo}{\mff^{(2)}}
\newcommand{\ciss}{{CISS}} 
\newcommand{\soc}{{SOC}} 

\newcommand{\ket}[1]{| #1 \rangle}

\newcommand{\dsd}{\hat d_{\sigma }^\dagger}
\newcommand{\ds}{\hat d_{\sigma }^{\phantom\dagger}}

\newcommand{\ck}{\hat c_{k \sigma }^{\phantom\dagger}}
\newcommand{\ckd}[0]{\hat c_{k \sigma }^\dagger}
\newcommand{\did}{\hat d_{i \sigma }^\dagger}
\newcommand{\di}{\hat d_{i \sigma }^{\phantom\dagger}}
\newcommand{\dof}[1]{\hat d^{\phantom\dagger}_{#1}}
\newcommand{\dofd}[1]{\hat d^{\dagger}_{#1}}
\newcommand{\Hres}{\hat H_{\mathcal X}}

\newcommand{\temp}{T}
\newcommand{\transmission}{\mathcal T}
\newcommand{\Fermi}{\eF}
\newcommand{\Vg}{V_\text{g}}		
\newcommand{\Vb}{V_\text{b}}		

\begin{document}
\title{Advances and challenges in single-molecule electron transport}

\author{Ferdinand Evers}
\affiliation{Institut f{\" u}r Theoretische Physik, Universit{\" a}t Regensburg, D-93053 Regensburg, Germany}
\author{Richard Koryt{\' a}r}
\affiliation{Department of Condensed Matter Physics, Faculty of Mathematics and Physics,
Charles University, Ke Karlovu 5, 121 16 Praha 2, Czech Republic}
\author{Sumit Tewari}
\affiliation{Huygens-Kamerlingh Onnes Laboratory, Leiden University, Niels Bohrweg 2, 2333 CA Leiden, The Netherlands}
\affiliation{Department of Materials, University of Oxford, OX1 3PH, Oxford, United Kingdom}
\author{Jan M. van Ruitenbeek}
\email[Corresponding author:  ]{ruitenbeek@physics.leidenuniv.nl}
\affiliation{Huygens-Kamerlingh Onnes Laboratory, Leiden University, Niels Bohrweg 2, 2333 CA Leiden, The Netherlands}


\begin{abstract}
Electronic transport properties of single-molecule junctions have been widely measured by several techniques, including mechanically controllable break junctions, electromigration break junctions or by means of scanning tunneling microscopes. In parallel, many theoretical tools have been developed and refined for describing such transport properties and for obtaining numerical predictions. Most prominent among these theoretical tools are those based upon density functional theory. 
In this review, theory and experiment are critically compared and this confrontation leads to several important conclusions. The theoretically predicted trends nowadays reproduce the experimental findings quite well for series of molecules with a single well-defined control parameter, such as the length of the molecules. The quantitative agreement between theory and experiment usually is less convincing, however. 
 
{\color{black}
Two main sources for the quantitative discrepancies
can be identified: Experimentally, the atomic structure of the 
junction typically realized in the measurement is not well known, 
so that simulations rely on plausible scenarios. 
In theory, correlation effects can be included only in approximations
that are difficult to control for experimentally-relevant situations. 
Therefore, one typically expects a qualitative agreement 
with present modeling tools; encouragingly, in exceptional cases
also a quantitative agreement has already been achieved.}
For further progress, benchmark systems are required that are sufficiently well-defined by experiment to allow quantitative testing of the approximation schemes underlying the theoretical modeling. 
Several key experiments can be identified suggesting that the present description may even be qualitatively incomplete in some cases. Such key experimental observations and their current models are also discussed here, leading to several suggestions for extensions of the models towards including dynamic image charges, electron correlations, and polaron formation.

\end{abstract}

\maketitle

\tableofcontents

\section{Introduction}\label{s.Introduction}

Despite many experimental hurdles the understanding of electron transport of single-molecule junctions has seen impressive progress in recent years \cite{Cuevas2010}. It is fascinating to observe that it is now routinely possible to wire an organic molecule, an object as small as one nanometer, between two metallic leads and measure its electronic transport characteristics. Several approaches even allow bringing a third metal lead close enough to serve as a gate electrode, through which the conductance of the molecule can be adjusted electrostatically. 

Now that we control 
to some extent
the basic properties of molecular junctions the time is ripe to 
critically evaluate the question how well we understand 
electron transport in molecular junctions. 
Faithful modeling inevitably 
needs to take into account many details of the arrangements of the atoms and the molecule that make up the junction. 
{\color{black} Since molecular junctions are formed spontaneously under the influence of atomic and molecular interactions, which can be regarded as a form of self-assembly, and since }
imaging of the resulting 
structures has not been possible, experiment usually does not provide 
all of the atomistic information needed for comparison with 
theory. 

Theoretical approaches often employed for describing near-equilibrium electron transport 
are based on tight-binding methods, density functional theory (DFT), 
and sometimes also
{\color{black} rely on more 
advanced many-body techniques, such as the $GW$ approximation. Far from equilibrium, {\it i.e.} at high voltage bias, 
the non-equilibrium 
Green's function (NEGF) method has been widely used. DFT and $GW$  }
have been amply tested for bulk systems and gas-phase 
molecules, but molecular junctions pose new challenges. Moreover, suitable variants of the
NEGF formalism have been specially developed for this type of problems, which, regretfully, are difficult to benchmark 
for lack of reliably reference data. 

The question then arises: what is the predictive power of the theories? What are the critical experimental tests? DFT is 
used widely as a guide for interpreting experiments, but do we know how reliable it is, and how do we know this? How 
sensitive are the results to the choice of methods and to the assumptions? The problem lies partly in the computational 
methods themselves, where the level of approximation may be critical, the convergence needs to be controlled, and where 
it needs to be assessed whether the relevant physical mechanisms have been included in the description. On the other 
hand, when setting up a calculation many assumptions are made about the conditions of the experiments, while the 
validity of these assumptions in most cases cannot be directly verified from information obtainable  from the 
experiments. Without attempting to be exhaustive in the following we list a number of items that need to be considered 
in evaluating a specific molecular junction. 

{\it Molecule-metal binding motifs.} 
The binding sites of a molecule anchoring on a metal surface and its binding 
motifs may show a great variability. Indeed, the electron transport is sensitive to {\color{black}the atomic structure of the metal at the interface to the molecule \cite{Schull2011a}, to} the 
choice of binding sites ({\it e.g.}, top-, hollow- or bridge sites), and also to the orientation of the bond with respect to the surface, 
{\it c.f.} the review by \textcite{Hakkinen2012}. However, in considering the various possible binding configurations it is important to be aware 
that the experimental conditions are often such that more than just a single molecule is present at or near the specific 
junction site. Moreover, repeated contact making and breaking, which is widely employed in experiments, may lead to the 
formation of metal-molecule complexes, and produce molecule fragments. \textcite{Strange2010} have 
considered this much wider variability in binding motifs for benzenedithiol (HS--C$_6$H$_4$--SH) and Au electrodes, 
leading to a much larger range of computed conductance values than normally considered.
\begin{figure}
\includegraphics[width=0.8\columnwidth]{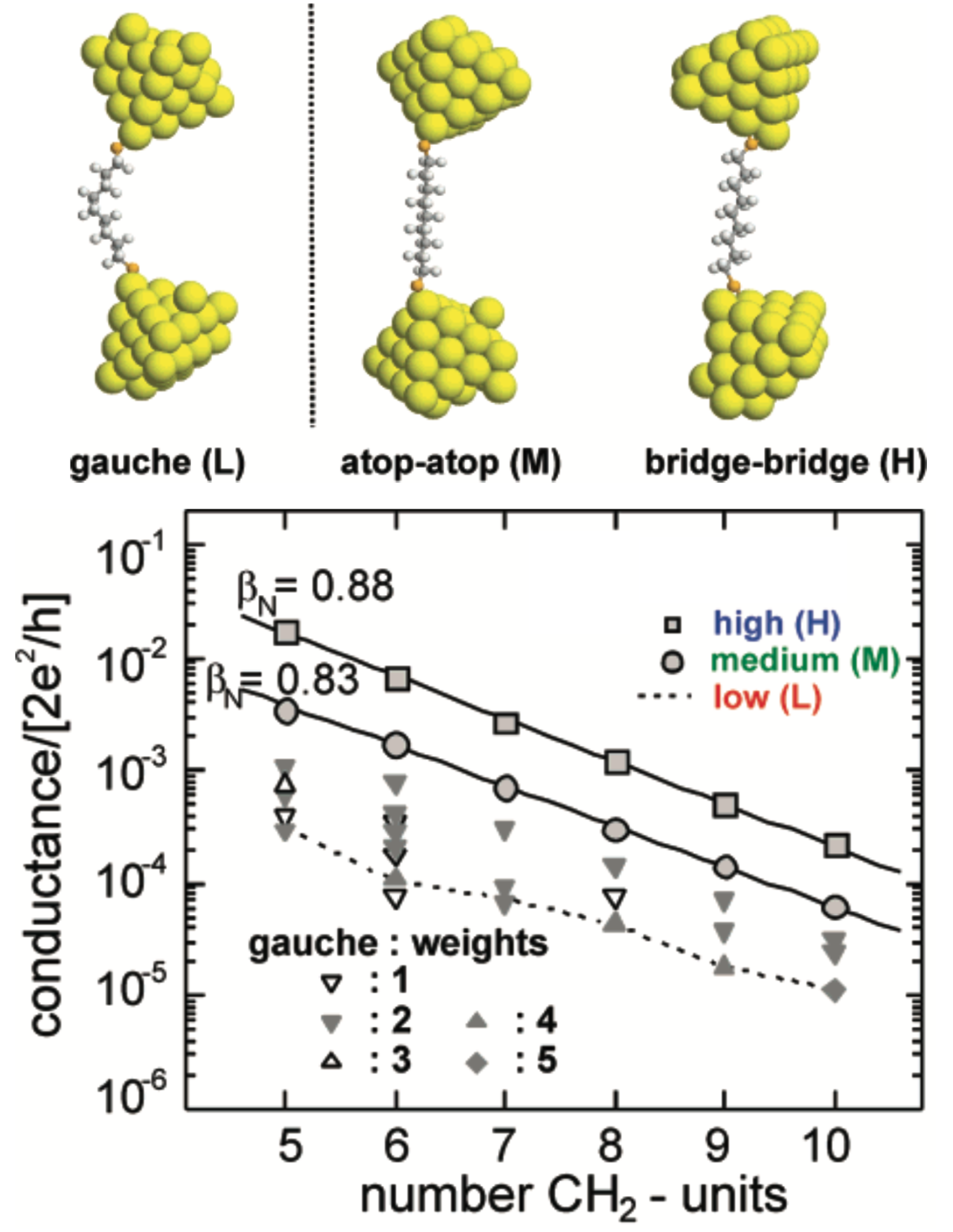}
\caption{\label{Fig:Li2008} Relaxed geometries representing
three typical arrangements of 
alkanedithiol molecule bridged between Au electrodes and calculated
length dependence of the conductance for these arrangements.
The results illustrate the spreading of conductances that can occur due to structural modifications. 
Reprinted with permission from Li {\it et al.}, J. Am. Chem. Soc. {\bf 130}, 318. Copyright (2008) American Chemical Society.}
\end{figure}

{\it Fluctuating geometries.} 
Longer molecules, such as the widely studied alkanedithiols (chemical formula 
HS--(CH$_2$)$_n$--SH), permit even wider variability, see Fig.~\ref{Fig:Li2008}. During the breaking of a junction the anchoring of the molecule 
may slide along the surfaces of the two electrodes, and the resulting conductance may vary during this process by more 
than an order of magnitude \cite{Paulsson2009}. Moreover, 
 the configuration of the molecule has a significant influence on the conductance, depending on the number of gauche 
defects in the molecular chain {\color{black} \cite{Jones2007,Li2008}}. 
At room temperature, such defects may form spontaneously and the 
conductance as measured will be an incoherent time average over the accessible configurations. 
Dramatic effects of such thermal averaging were shown in calculations \cite{Maul2009} 
for molecular wires containing up to four benzene rings coupled together (oligophenylenedithiol).

{\it Uncertainties of surface chemistry and level alignments.}  
The nature of the chemical bond between the molecule and the metal 
electrodes is another source of ambiguity. 
Notably the widely exploited Au-S-R
 anchoring, 
where R is the molecular group under study,
 is often obtained by adding thiol (SH) end-groups to the molecule. In the process of binding to Au one usually assumes 
that the hydrogen atom is split off and removed, but recent evidence suggests otherwise \cite{Stokbro2003,Inkpen2018}. 
Just as hydrogen remaining at or near the anchoring group, 
also the presence of other residuals or entire 
molecules on the surface has further consequences. Such surface coverage modifies the metal work function, and 
thus modifies the profile of the electrical potential drop along the junction axis. A very dramatic demonstration of 
this effect was given in the experiments by \textcite{Capozzi2015}. When working in solution the ions in the electrolyte 
dynamically adjust to the applied bias voltage, producing an asymmetric diode-like current-voltage (\IV) characteristic. 
Size and shape of the
electrodes on the nanometer scale also affect the details of the electron transport \cite{Hakkinen2012} 
and the profile 
of the electrical potential drop \cite{Brandbyge1999}. Information on such nanoscale details is not readily obtained 
from the experiment. One reason for the sensitivity of electron transport to the nanoscale shape of the electrodes is 
the effect of image charges 
 \cite{Perrin2013}.

Electron transport 
 {\color{black} for metal--molecule--metal junctions is typically off-resonant, } 
 which makes the conductance
highly sensitive to the {\color{black} energy} of a delocalized molecular orbital nearest to the Fermi level of the electrodes. 
This position is influenced by many of the factors listed above, and in addition this position self-adjusts by partial 
charge transfer between the metal and the molecule.

{\it Is our description complete?} 
Given these many poorly known factors one should conclude, as we will see below, that the agreement between experiment 
and computations is surprisingly good. To be more precise, 
conductance values for the same metal-molecule combinations, and 
most calculations find an agreement within an order of magnitude from the experiment 
 (although there are important exceptions, as we will see below).  
This raises three interesting questions: 
(i) Given the many unknowns, why is the agreement so close? 
(ii) If we 
could improve our knowledge of the experimental system to be described, how strong would the predictive power of theory 
be? 
(iii) 
Are we possibly missing some interesting physics in the description
?

The last question is the most important, in our view.
For example, the interplay between the bias-voltage, electrode screening and 
Coulomb-blockade can introduce nontrivial correlation effects, such as a negative-differential 
conductance \cite{Kaasbjerg2011}.
This regime escapes the single-particle doctrines and has hardly been explored. 
Electrons also interact with the ion cores by means of vibrations, leading to inelastic 
scattering signals that can be exploited for characterizing the molecular junction \cite{Smit2002}. 
The associated limit of strong electron-lattice interactions
has been briefly reviewed by \textcite{Thoss2018}.  
It is expected to lead to polaron formation \cite{Su1980},
which should have a strong impact on the current-voltage 
characteristics \cite{Galperin2005,Thoss2018}.  
Recently, this mechanism has been shown explicitly in experiments, although this result was not obtained for
a typical molecular junction, but rather a molecule in a STM tunneling configuration \cite{Fatayer2018}.


{\it Structure of this review.} 
Single-molecule transport is an extremely active 
and broad research field 
with a corresponding body of literature. A single review cannot hope to do full justice to all  
developments even when focusing on a few relevant aspects.
It is our aim in this paper to summarize and discuss 
the most significant experimental and theoretical results in 
the light of the set of specific questions raised above.
In particular, we critically evaluate 
the level of agreement between theory and experiment. 
We will further elaborate on selected experiments and 
calculations which indicate that the description 
of the systems may not be complete, and which suggest interesting 
physics beyond the standard approaches. 
For comprehensive reviews focusing on complementary aspects of molecular-scale 
transport we refer the reader to \textcite{Cuevas2010}, \textcite{Thoss2018}, \textcite{Su2016}, and \textcite{Jeong2017}.
While our focus is on single-molecule junctions, 
we occasionally also quote results obtained for 
self-assembled monolayers. 


\section{Experimental techniques}\label{s.Experimental}

In this section we briefly present various techniques used for studying electronic transport through single molecules, in order to make the reader acquainted with the methods that we will encounter in discussing the results. For a more detailed presentation of single-molecule techniques and their integration into various advanced measurement schemes we refer to previous reviews \cite{Agrait2003,Xiang2013,Xiang2016,Aradhya2013}. 

Since molecules have a typical size of 1~nm all existing top-down microfabrication techniques lack the required resolution for controlled wiring of molecules. Therefore, the methods employed rely on a combination of electromechanical fine-tuning of the nanometer-size gap between the contact electrodes and self-assembly of the molecules inside this gap. The three most frequently employed techniques are the mechanically controlled break junction (MCBJ) technique, the electromigration break junction (EBJ) technique and  methods using scanning tunneling microscopes (STM).

\subsection{Mechanically controllable break junctions}
\begin{figure}
\includegraphics[scale=0.079]{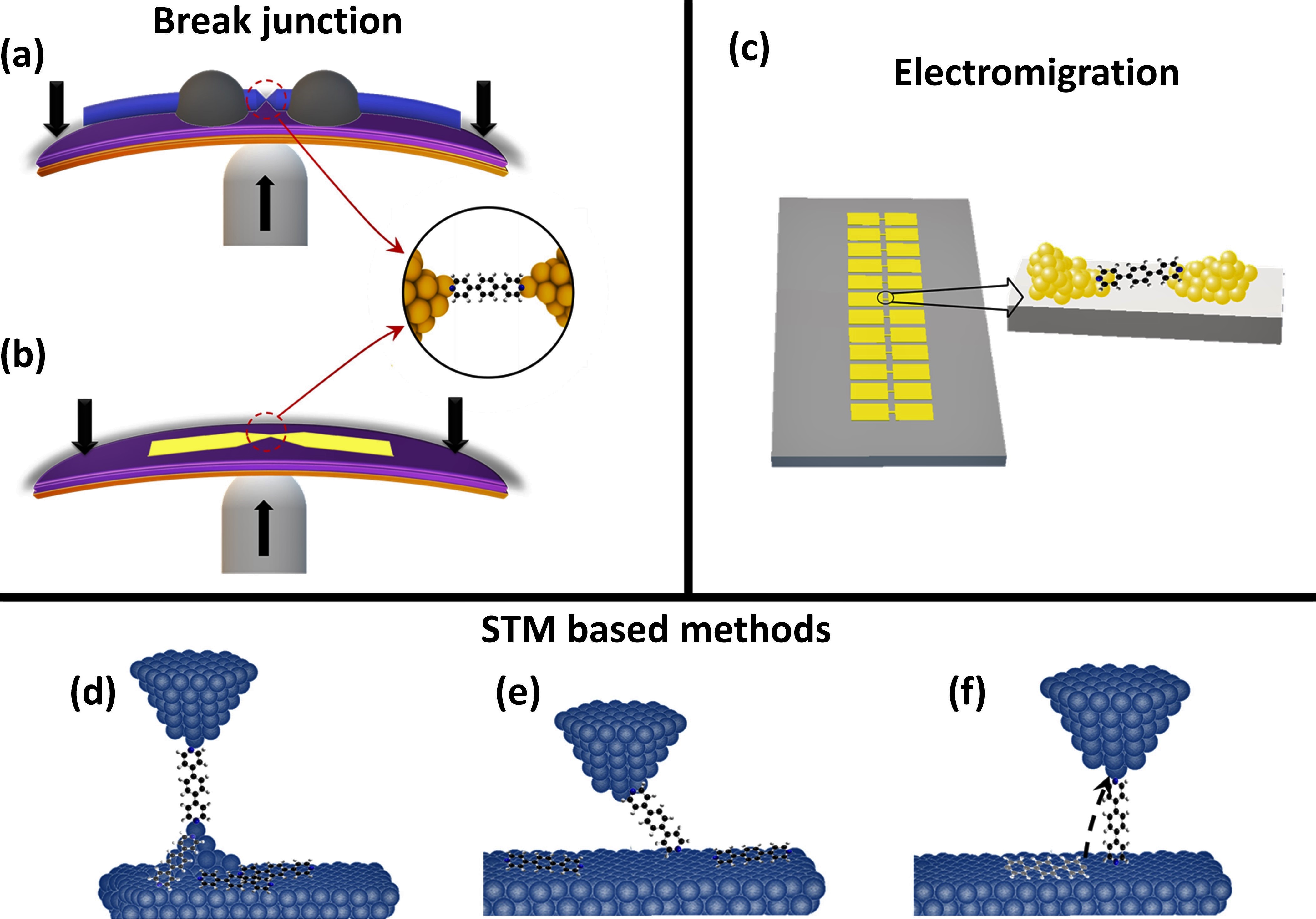}
\caption{Overview of experimental techniques aimed at measuring single-molecule transport. (a) Notched-wire mechanically controllable break junction. (b) Lithographically fabricated mechanically controllable break junction. (c) Electromigration break junctions (d) STM break junction by repeated indentation (e) $I(t)$ or $I(s)$ operation of STM (f) Low-temperature UHV STM manipulation of individual molecules.}
\label{fig:techniques}
\end{figure}

The MCBJ technique was developed for the study of atomic and molecular junctions \cite{Muller1992} based on an earlier method aimed at studying vacuum tunneling between superconductors \cite{Moreland1983}. We distinguish two fabrication methods: the notched-wire MCBJ, and the lithographically fabricated MCBJ. The first is the simplest and has the advantage that it can be easily adapted to nearly all metal electrodes. It is made starting from a macroscopic metal wire into which a weak spot is created by cutting a notch. The notched metal wire is placed on top of a flexible substrate (which is commonly stainless steel or phosphorous bronze) covered by an insulating sheet, usually Kapton. The wire is fixed by epoxy onto the substrate at either side and very close to the notch. This is then mounted in a three-point bending mechanism as shown in Fig.~\ref{fig:techniques}(a). 
Bending the substrate increases strain in the wire, 
which is concentrated at the weak spot created by the notch, until the wire breaks. The junction is first broken with a coarse mechanical drive, thereby exposing two fresh electrode surfaces. By relaxing the bending, and using fine control of the 
gap by means of a piezo-electric actuator, atomic-size contacts can be reformed and broken many times.

The lithographically fabricated MCBJ \cite{Ruitenbeek1996} shares the same principle as the  notched-wire MCBJ except that the pre-notched metal wire is replaced by a freely-suspended bridge in a thin metal film produced by electron-beam lithography. This metal film is electrically isolated from the flexible substrate using a $3-5\,\mu$m polyimide layer. The unsupported section of the bridge is reduced by about two orders of magnitude compared to the notched-wire MCBJ, to about $2\,\mu$m, or less. This has the effect that the mechanical displacement ratio, {\it i.e.} the ratio between the change of the gap size and the actuator motion, is reduced to about $10^{-5}$. The gain of using the lithographic technique is that the junctions are very insensitive to external mechanical perturbations as a result of the small displacement ratio. The added complications of clean-room preparation are offset by the possibility of producing multiple MCBJ samples on a single wafer \cite{Martin2008}. A drawback is the fact 
that 
by the very small displacement ratio the maximum extension of a typical piezo actuator produces less than 0.01~nm change in the distance between the electrodes. Therefore, the control of this distance is achieved by an electro-motor driven gear. Since such electromechanical control is much slower than piezo-electrical control it is much more time consuming to obtain enough statistics for a large number of contact breaking events (see below).

For most types of metal electrodes one can only take full advantage of the MCBJ method by performing the first breaking at cryogenic temperatures or under ultra-high vacuum (UHV). Otherwise, the surfaces will be contaminated with oxides and adsorbents will cover the surface within a fraction of a second, so that the atomic-size contact characteristics of the pure metal are lost. The main exception is Au, for which even under ambient conditions most of the intrinsic quantum conductance properties survive as a result of the low reactivity of the Au surface \cite{Pascual1993}. 

For the same reason Au stands out as the preferred electrode material for all other single-molecule transport 
experiments. Specific binding to target molecules can be achieved by selecting suitable anchor groups for the 
molecules, see also Section~\ref{ss:VC}. Typically, such molecules
having suitable anchor groups are deposited onto the
bridge of the MCBJ from solution, under ambient conditions.
This strategy has been first explored for lithographic MCBJ systems 
\cite{Reed1997,Reichert2002}, and this continues to be the most commonly employed approach, but recently it has also 
been demonstrated for the notched-wire MCBJ technique \cite{Bopp2017}.

The intrinsic cleanliness of the broken metal surfaces can  be more fully exploited by working under UHV and/or under cryogenic conditions. The deposition of molecules in these experiments proceeds by deposition onto the broken junction from the gas phase, either using an external vapor source \cite{Smit2002,Kiguchi2008} or employing a local cell for sublimation \cite{Kaneko2013,Rakhmilevitch2014}. By working under cryogenic or under UHV conditions it is possible to explore other metal electrodes and other forms of metal-molecule bonding. For example, hydrogen, H$_2$, binds to clean Pt electrodes without the need for anchoring groups \cite{Smit2002}, and this applies more widely also for many organic molecules such as benzene \cite{Kiguchi2008}, oligoacenes \cite{Yelin2016} and pyrazene \cite{Kaneko2013}.

\subsection{Electromigration break junctions}
Electromigration in metals \cite{Ho1989} results from an atom diffusion process driven by the `electron wind' force \cite{Huntington1961} exerted by the conducting electrons on the atoms in the system, under large current bias. This effect can be used to create nanogaps in metallic leads \cite{Park1999,Zant2006}, small enough for a single molecule to bridge. Such systems are prepared by first pre-patterning a narrow metal wire of about $100$~nm in a thin metallic film on an insulating substrate (usually SiO$_2$ on a Si wafer) using electron-beam lithography. Passing a large current through such narrow metallic leads gives rise to displacement of atoms, which is observed as an increasing resistance due to the gradual 
thinning of the wire. Initially, the reliability of the method was compromised by the fact that the strong local Joule heating leads to the formation of metallic nanoparticles in almost 30$\%$  of the junctions \cite{Houck2005,Zant2006}, which give rise to current-voltage (\IV) characteristics resembling 
those of molecules. However, by using a feedback circuit the electromigration process can be more precisely controlled, and further improvements are obtained by relying on self-breaking in the last stages of gap formation  \cite{Zant2006}. 

Molecules are deposited onto the nanowire before electromigration, and one relies on a molecule finding its way into the gap during the electromigration process. Alternatively, molecules can be allowed to self-assemble into the gap from solution after the electromigration process has been completed \cite{Osorio2007}.  In contrast to the other break junction techniques, junctions formed by electromigration can only be broken once and cannot be reformed. The gap distance depends on the details of the feedback-controlled breaking process, but it cannot be targeted very precisely. One cannot obtain a very precise value for the size of the gap, but a fair estimate can be obtained from fitting the \IV characteristics to the Simmons model \cite{Simmons1963,Vilan2007}. 

For imaging techniques the gap is better accessible than for any of the other techniques discussed here. High-resolution transmission electron microscopy imaging using transparent SiN$_{x}$ membranes was performed for gold electromigration junctions \cite{Strachan2008,
Gao2009} in order to study the breaking process and detect the nanogap size. 
The imaging resolution of transmission electron microscopy has not yet proven sufficient for detecting the position of an organic molecule.
  
The search for junctions bridged by a molecule is based on producing many (of order  several hundred) electromigration break junctions on a wafer, breaking each of them separately, and probing the resulting junctions for interesting \IV characteristics at room temperature, which may point at the presence of a molecule in the bridge. Such junctions, which are a minority of the order of a few percent, are then further studied, usually by more elaborate techniques. Although the method intrinsically allows obtaining only limited statistics over molecular junction configurations, and every junction formed has its particular characteristics, the more elaborate experiments permit very interesting case studies. Moreover, the rigid attachment of the electrodes to the substrate allows temperature and field cycling, it allows the fabrication of a metallic gate at close proximity to the junction \cite{Park1999,Zant2006}, and it permits easy optical access for Raman scattering \cite{Ward2008}. 

\subsection{Methods based on scanning probe microscopy}
The break junction methods described above do not permit imaging of the molecule in the junction. In contrast, scanning 
tunneling microscopy (STM) or atomic force microscopy allow imaging molecules on a surface before contacting them. This 
is possible only for very stable systems under UHV \cite{Joachim1995,Langlais1999} especially at cryogenic temperatures \cite{Temirov2008,Neel2007}. By imaging 
and manipulating single molecules on an atomically flat and clean metal surface it is possible to verify that the STM 
tip interacts with a single target molecule, and the shape of the bottom electrode contacting the molecule (the metal 
surface) is known. However, information on the shape of the tip cannot be easily obtained from experiments.\footnote{Progress 
in this direction has been made recently in atomic force microscopy (AFM) 
with CO-molecules on  Cu-tips. The symmetry of the AFM-data reveals the structure of the second layer of Cu-atoms 
that the apex atom couples to \cite{Welker2012}.}
Moreover, when approaching the tip for contacting the molecule and lifting it up from the surface the molecule and the 
metal atoms contacting it rearrange in ways that cannot be seen by the instrument.

\begin{figure}
\includegraphics[scale=0.37]{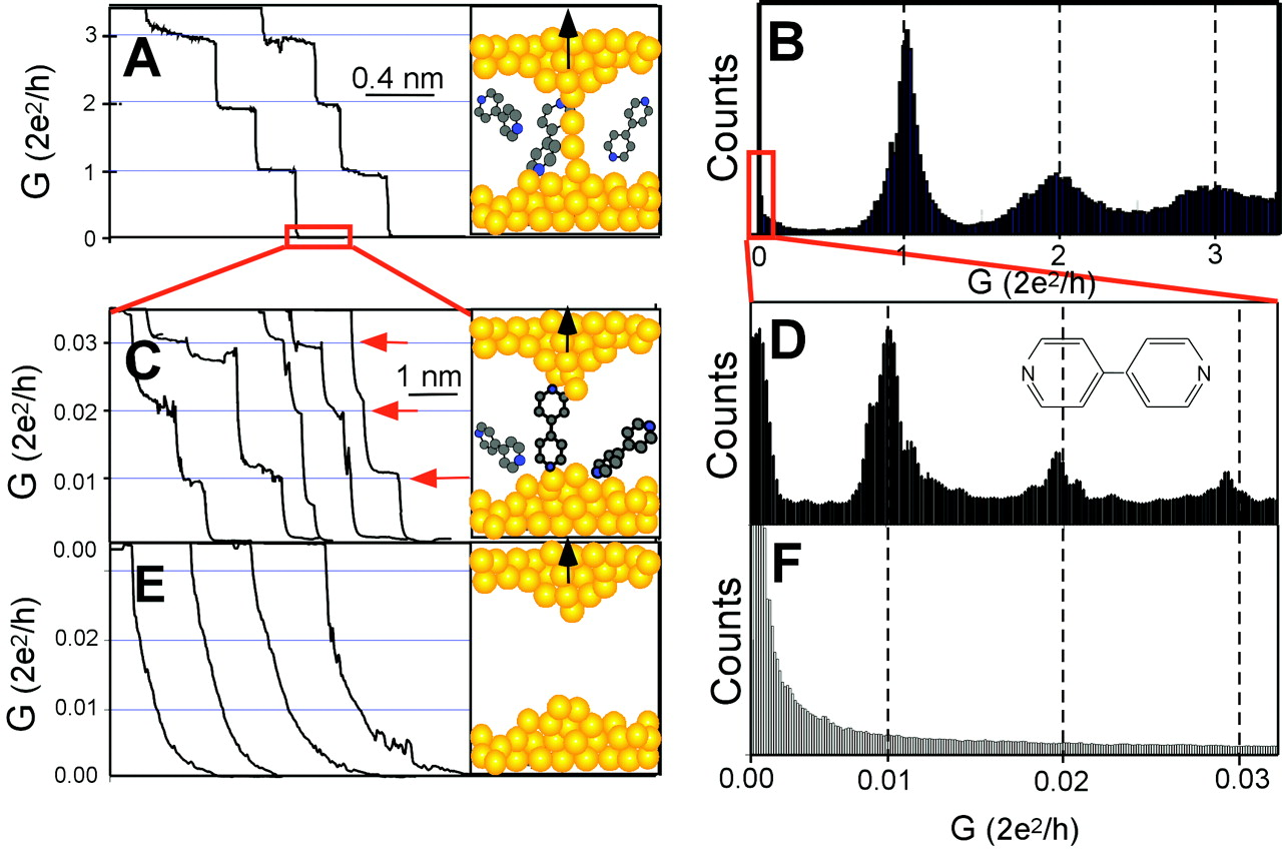}
\caption{Experiment probing the conductance of a single molecule by repeated indentation of a Au STM tip into the Au metal surface, in solution of 4,4-bipyridine. The breaking of the metal-metal contact is observed as steps in the conductance near multiples of \go\  (A), giving rise to peaks in the conductance histogram (B). Zooming in to lower conductance additional steps are resolved (C), and for many repeats of breaking this produces a new series of peaks in the conductance histograms at small fractions of \go\  (D). Tests with pure solvent show only tunneling characteristics (E,F). 
From Xu and Tao, Science {\bf 301}, 1221 (2003). Reprinted with permission from AAAS.} 
\label{fig:XuTao}
\end{figure}

While cryogenic UHV STM holds great promise, it is also a very demanding technique. A  versatile method for investigating the conductance of single molecules by STM at room temperature and in solution was introduced by \textcite{Xu2003}, which has inspired many other researchers, see Fig.~\ref{fig:XuTao}. 
Ignoring the scanning capability of STM, the instrument is used for approaching the tip to the surface and repeatedly 
indenting the tip into the surface and retracting. 
In this mode of operation the atomic structure of the 
junction is subject to fluctuations, so that the information obtained by this technique is statistical in nature, 
{\it i.e.} ensemble-based, and thus close in spirit to MCBJ-experiments.
The indentation of 
the (Au) tip into the (Au) metal surface to a depth corresponding to a conductance of 10--40 times the conductance 
quantum  ($G_0 = 2e^2/h$) restructures the shape of the electrodes with every indentation. Upon retraction a neck is 
formed that thins down until it snaps. The resulting gap is then frequently bridged by a molecule equipped with 
suitable anchoring groups through a self-organization process, which is observed as a plateau in the conductance during 
retraction. These plateaus usually have a lot of structure and appear at different levels for each retraction event. 
Therefore, the indentation and retraction cycles are repeated many times and the resulting conductance traces are  
combined in the form of conductance histograms, as had been previously introduced for MCBJ experiments 
\cite{Krans1993,Smit2002}. 
 
These room temperature experiments have the great advantage that they permit 
evaluating single-molecule junctions much faster than the other available techniques, and thereby allow exploring trends 
as a function of molecular composition. On the other hand, the information obtained is limited mostly to 
{\color{black} statistical properties, such as average 
and typical values of the conductance, the breaking 
length \cite{Chen2006}, the force holding the 
junction together \cite{Xu2003,Aradhya2013},}  and the thermopower \cite{Reddy2007}.
{\color{black} 

\subsection{Data analysis and conductance histograms \label{ss:IID}}

Most of the MCBJ and STM experiments have in common
that, as a result of the 
{\color{black} self-arranging} 
process
involved in the formation of the junction, little is
known about the atomic-scale shape and structure of the
electrodes, the configuration
of the molecule in the junction, and about its bonds to
the metal surfaces. 
%
As a result, the conductance can fluctuate from one contact-breaking trace to the next  by an order of magnitude or more. 
Notice that even for a given trace the current at fixed dc-voltage is usually not time-independent,
 due to thermal or bias-induced fluctuations in the junction geometry. 
 {\color{black} For example, during the process of breaking of a molecular junction in MCBJ or STM-BJ experiments, 
 which can take place on time scales between about 1 ms and several seconds, one often observes jumps around the 
 typical conductance value for the molecule. Between these jumps, that can have an amplitude of an order of magnitude or more, one observes 
 rapid fluctuations. The bandwidth of the experiment is usually limited to about 1 MHz or less, so that even the most rapidly observed fluctuations 
 already represent an incoherent average over different junction configurations due to thermally accessible vibrations. }

A widely adopted practice to deal with fluctuating observables is to study the fluctuation statistics, 
{\it i.e.}, the conductance distribution taken over an ensemble of junctions realized in a series of experimental 
measurements. In practice, one repeatedly forms and breaks many junctions, records the 
digitized conductance during the contact breaking process, 
and collects all data in a histogram, as illustrated in 
Fig.~\ref{fig:XuTao}.
It seems reasonable to expect that sufficiently deep 
indentation between recording traces restructures the metal leads 
and the molecular junction, 
so that correlations between subsequent recordings 
are negligible. By combining the displacement length, measured from the point of metal-metal contact breaking, with the 
evolution of the conductance one can also build two-dimensional histograms \cite{Martin2008b}, which are helpful for 
detecting multiple stable configurations and for obtaining a measure of the molecular bridge length.

The precise statistical properties of the ensembles generated 
in this way are hardly known and very difficult to predict. 
At this stage an important simplification should arise, 
because very often the experimental recording cycles can be assumed to 
be very slow as compared to the atomistic relaxation rates. 
Due to the resulting separation of time scales, one expects that 
there is time enough for the junction to relax into a set of 
particularly stable, `optimal' junction geometries. 
Presumably, at a slow-enough recording rate this set can be considered very small.
This is the justification for the histogram technique 
to operate with concepts 
like `typical' junction geometries. It explains, in particular, 
why the corresponding atomistic shape used in theoretical simulations 
may possibly be derived from a variational principle, rather 
than simulating the junction geneses as they occur in the actual measurement.\footnote{
A further justification for the general practice may be found in the following 
argument. For the junction not to break in the presence of thermal fluctuations 
or bias-induced forces, there should be a notion of stability. 
This suggests that there is an optimization principle, which should become identical, 
at zero bias, with the optimization of the free energy under the boundary condition 
that the contact exists.}
In the simplest case, the typical junction is identified as 
the most stable one, {\it i.e.}, the one with the maximum binding energy.\footnote{
To develop a statistical theory of the histogram technique 
would be rewarding, but also goes beyond the scope of this review.
Two closely related issues should be briefly mentioned, nevertheless.
(a) Molecular-dynamics (MD) investigations, {\it e.g.} as presented by \textcite{French2013}, 
have been put forward as an attempt to simulate the breaking of a molecular junction.
For such studies simulating the very long experimental time scales, which  
are associated with {\em plastic} deformation of the molecular junction under pulling, 
is very challenging. Large system sizes and simulation times up to microseconds might be required. 
If these are affordable at all, then the approximations underlying the solution of the 
equations of motion required are extremely difficult to control.
Examples of slow relaxation processes that should be properly described to be realistic 
include the temperature driven diffusion of ad-atoms or multi-atom 
exchange processes. Both processes can optimize or 
destabilize the junction geometry.
(b) For similar reasons, also the breaking of a molecular junction in the absence 
of a pulling force may be ill-described by MD-simulations. This can happen in situations
where breaking occurs due to rare, temperature driven fluctuations. 
-- A careful discussion of the consequences of 
computational limitations for the interpretation of simulation results
is not standard. For the reasons outlined above 
the interpretation of MD-type studies for the statistical properties 
of molecular junctions needs to be done with precaution. 
}~ 
Adopting this logic, the peaks in the histogram are usually interpreted as representing the energetically 
favorable junction configurations, and these are the most relevant parameters used for comparison with model 
calculations. 

In the breaking process the last-atom metal-to-metal contact is usually clearly visible as a plateau near 1~\go, and this produces a sharp peak in the conductance histogram. Breaking of this last metal contact is followed by a jump out of contact \cite{Agrait1993} to a conductance that is one or two orders of magnitude lower. In many cases, after this jump the current exponentially decreases with increasing separation of the electrodes, as expected for  vacuum tunneling. Only for a fraction of the breaking events one or more plateaus appear, signaling the successful bridging of the junction by a molecule.  The large number of traces without a molecular signal results in a large background in the histograms. Initially, curves without a clear molecular signature were manually removed from the data set. This practice has some danger of introducing experimenter-bias in the data selection, and this practice has now been abandoned. The background problem can be reduced by the use of automated routines, for example 
routines that detect the last step in the conductance \cite{Jang2006}. 
A widely adopted solution to the background problem is the use of  
histograms of the logarithm of conductance, rather than the linear conductance \cite{Gonzalez2006}. In this case the 
background tunneling contribution reduces to a nearly constant contribution and the relevant features related to the 
molecule will be more clearly visible in a data set, that now comprises all breaking traces. 

{\it $I(t)$ and $I(s)$ techniques.} 
The appearance of the shape of the histograms and the positions of the peaks for the same metal-molecule system do not 
reproduce perfectly between experimental groups, and even from one experimental run to the next. This implies that the 
underlying assumption that the repeated indentation effectively averages over all configurations is not fully justified. 
For example, one may anticipate that the results will be sensitive to parameters such as the voltage or current bias 
applied, and 
the depth of indentation. This has motivated Haiss {\it et al.} to avoid indenting the surface, in order to maintain a common surface and tip structure. They developed the so-called $I(t)$ and $I(s)$ techniques \cite{Haiss2003,Haiss2004}. These techniques operate near room temperature and rely on bringing the STM tip close to the surface by the usual current feedback control. For low surface coverage, molecules with suitable anchoring groups are expected to jump stochastically into and out of contact with the tip. The difference 
between $I(s)$ and $I(t)$ is that the tip is moved in and out of close distance to the surface repeatedly for the former, 
while in the latter case, the tip is held at a stable tunneling distance and the events are recorded as a function of 
time. The conductance values measured by $I(t)$ or $I(s)$ are typically found to be up to an order of magnitude smaller than the ones 
obtained from histograms produced by MCBJ or STM techniques.

\section{Computational techniques}\label{s.Computational}
\subsection{\color{black}A guided tour through quantum transport theories}

The transport of charge, spin and heat through a single molecule 
is a prime example of quantum-transport through a mesoscopic device, 
where quantum coherence and 
correlations dominate the measured observables. For this reason 
the standard mesoscopic transport technologies 
apply also in the case of single molecules. 

An important line of research focuses on model studies, 
{\it e.g.} the single-impurity Anderson model, 
the Hubbard model, the Holstein model, etc.; 
for a recent review see \onlinecite{Thoss2018}.
{\color{black} Models relevant for molecular transport will be discussed in Section~\ref{s.models}.

In contrast to most mesoscopic systems, single-molecule junctions consist of relatively few atoms, 
typically only a few hundred; moreover, their arrangement within the molecule is well known.
This begs for {\it ab-initio} electronic structure calculations. }
Concerning {\it ab-initio} transport computations, 
we identify three archetypical approaches as most prevalent:

(i) The non-equilibrium Green's function formalism 
(NEGF, Kadanoff-Baym formalism) is a very general approach. 
It applies to linear- and non-linear responses of interacting systems, 
in quasi-static and also in time-dependent situations.
An additional attractive feature is that the coupling of electrons to 
vibrations is straightforward to implement \cite{Pecchia2004,Paulsson2005,Paulsson2008}. 

This generality comes in situations where simplifications arise 
at the price of being somewhat inconvenient to use as compared to 
competing methods. 
\textcite{Meir1992} have worked out the most popular application of NEGF 
in mesoscopic transport. They have derived explicit expressions for the 
\IV-curve that apply to generic quantum dots under the assumption of 
non-interacting electrodes.

(ii) When interested only in linear responses, the Kubo-formula offers 
a viable alternative to NEGF. This formulation is advantageous because 
it involves only advanced and retarded Green's 
functions and therefore takes as an input only `equilibrium' 
(usually ground state)
electronic structure information. Moreover, these Green's functions 
are available, at least in principle, already in standard electronic 
structure codes. 
The reason is that advanced electronic structure 
methods, such as the $GW-$theory, already operate with these 
objects.\footnote{{\color{black} The $GW-$theory has been developed 
as a self-consistent leading-order approximation that 
emerges from a diagrammatically exact representation of 
the many-body Green's function 
\cite{Hedin1999,Aryasetiawan1998,Aulbur1999,Bechstedt2015}.
Intuitively, it is understood as improving 
over Hartree-Fock theory by computing the 
Hartree-potential with a screened interaction that is calculated 
on the level of the random phase approximation.
}}

(iii) To the extent that interaction effects can be treated on a mean-field 
level, the Landauer-B\"uttiker formalism is efficient. It derives in a 
straight-forward manner from NEGF, see \textcite{Meir1992} and applies 
also to the non-linear regime. This formulation underlies 
the standard {\it ab-initio} based transport theory described below. 

We emphasize that the list of methods here mentioned is not exhaustive. 
For example, a formalism based on the density-matrix theory
as described in \textcite{Bruus2004} 
has also been used with success \cite{Donarini2010,Niklas2017}.

\subsection{Brief overview of electronic structure calculations for 
molecular junctions \label{Brief-electr.struct.calc}}

No matter what transport formalism is used, 
an input concerning the electronic structure of the device is needed.
Indeed, molecular junctions  pose one of the most difficult challenges of
electronic structure theory. 

To see why this is so we recall that even
an isolated molecule requires advanced many-body techniques, {\it e.g.}, 
for calculating ionization potentials (IP) and electron affinities (EA), 
see \textcite{Setten2015} for a recent review.
This observation is of relevance also here, because uncertainties in IPs (EAs)
translate, in general, into errors in the position of transport
resonances related to the highest occupied (HOMO) and lowest 
unoccupied (LUMO) molecular level. 
Summarizing, estimates of IPs for small molecules based on H\"uckel
studies or Kohn-Sham (KS) energies of density functional theory (DFT)
typically deviate from higher level methods by 1eV or more 				
\cite{Setten2015}.\footnote{
In the case of KS-theory the IP can be calculated in two 
ways that are equivalent for exact DFT: One retrieves the IP 
either from the HOMO-energy or, alternatively, from the
difference in ground-state energies of the charged and charge-neutral 
molecular species (Self-Consistent Field, SCF-method). While the SCF-method is known to give 
much more accurate results for the IP (`error cancellation'), 
it is the HOMO-energy that actually enters the transport calculations. 
} 
For larger molecules or metallic wires, 
the absolute error in IPs sometimes decreases with the system size. 
This happens, {\it e.g.}, when the workfunction is dominated by a subsystem, 
such as a large metallic segment, for which the DFT-functional applied 
is working well. 
This observation can be deceptive, however, because the most 
interesting molecular junctions display weakly connected subsystems  
(`molecular quantum dots') for which the errors in the computed 
level alignments remain large, even though the error 
in the overall workfunction could be relatively minor. 

One, therefore, might  have the impression that higher level methods, 
such as perturbative, Green's-function-based approaches 
(\gowo) or wavefunction-based methods 
({\color{black} {\it e.g.}, configuration-interaction methods or coupled-cluster theory})  
should provide the next generation 
standard tools of {\it ab-initio} transport
calculations. However, there is an extra challenge, so the situation is not as clear. 
Despite of its well documented shortcomings, molecular transport studies 
still mostly rely on KS-based scattering theory. 
The basic reason for the popularity of KS-based transport studies  
is that KS-calculations, dealing essentially with a single particle picture, 
digest large enough systems. `Large enough', here, means that an approximation for the electronic 
structure can be found for the {\it extended molecule}, which 
comprises the molecule itself plus a part of the leads, Fig.~\ref{fig.extended-molecule}.\footnote{
It would perhaps be preferable to speak of `molecule' vs `extended molecule', or `junction' vs. `extended 
junction.' The name `extended molecule' follows the established nomenclature, 
and we prefer to adhere to it, here. The word `junction', on the other hand, 
was chosen to indicate the part of the system that 
connects the metallic electrodes. 
This is not sharply defined, 
but it need not be because only the extended molecule 
plays a role in the calculations.
} 
Dealing with the extended molecule is important because 
transport phenomena are 
sensitive to how the molecular orbitals hybridize with the 
electrodes.  This hybridization can be described 
consistently within KS-simulations of extended molecules, 
but usually not so at affordable cost with higher level methods. 

\begin{figure}
\includegraphics[scale=0.38]{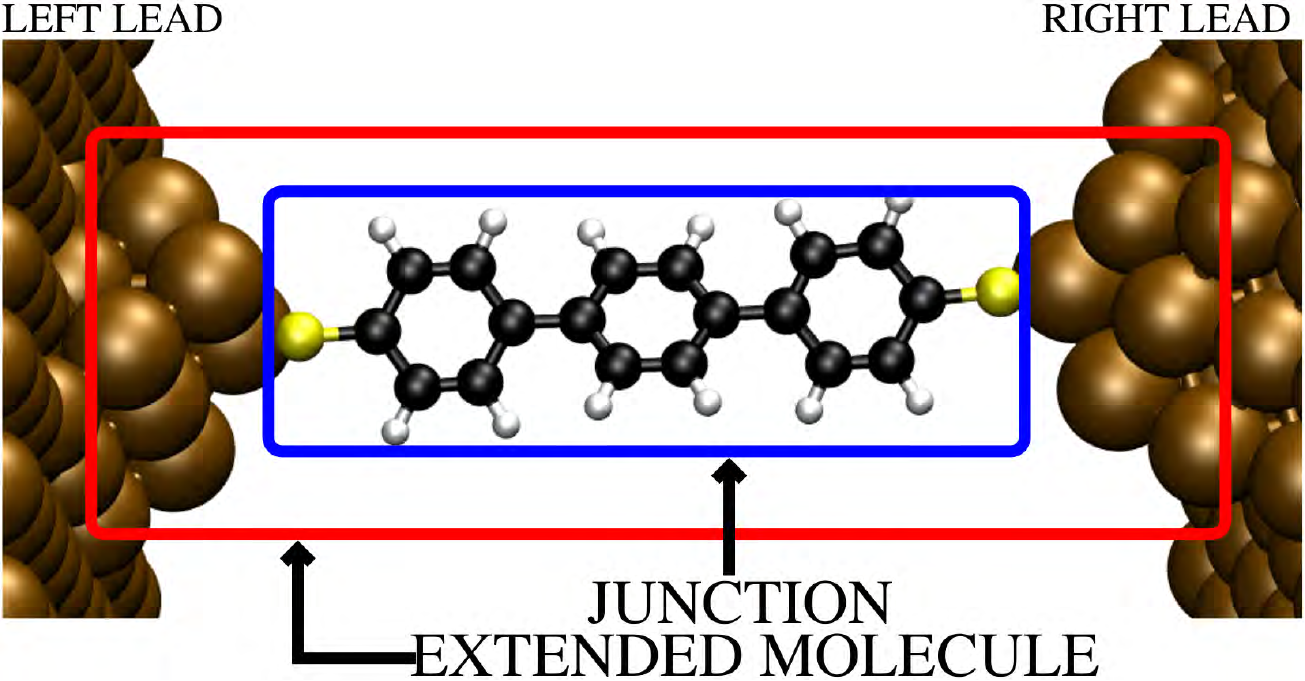}
\caption{Illustration of partitioning in model calculations: Molecule, `extended  molecule', and 
semi-infinite leads. Reproduced with permission from L. Delle Site, `{Simulation of many-electron systems that exchange matter with the environment}', Adv. Theory Simul. {\bf 1}, 1800056 (2018).
Copyright Wiley-VCH Verlag GmbH \& Co. KGaA.  }
\label{fig.extended-molecule}
\end{figure}

\subsection{Verification and validation of transport computations}

The geometry of a given molecular junction can be fluctuating in time
driven, {\it e.g.}, by thermal effects or the current flow. As we have argued 
in Sec. \ref{ss:IID}, the concept of a typical junction 
configuration should be well defined, nevertheless, for a great many 
experimentally relevant situations. 
Notice, that the statement is not completely obvious, perhaps, 
because many well investigated molecular
junctions work with highly flexible molecules, such as alkanes, 
that do not by themselves ({\it i.e.}, in the gas phase) provide a stable geometry.

The instance that the molecular geometry or the ensemble of geometries
is not usually well known in experiments
provides a major challenge for {\it ab-initio} simulations.
Since in such computations the geometry usually is taken as given input, 
simulations mostly work with a plausible scenario for the geometry.
Often, they provide a consistent and plausible description, 
sometimes even quantitative, 
but hardly ever are scenarios microscopically validated by experiment. 

It is rather straightforward to perform an internal 
consistency check on the simulation results: 
One determines to what extent the conclusions of the simulation 
are sensitive to variations of the geometry and the approximation level of 
the transport calculation; thus a certain verification is possible.  
Nevertheless, the atomistic geometry remains a 
degree of uncertainty to keep in mind when comparing computations with experimental data.
It superimposes the inherent theory uncertainty of electronic 
structure calculations that results from (parametrically) uncontrolled 
approximations.

\subsection{The standard theory of {\it ab-initio} transport (\stait)}

The standard theory of {\it ab-initio} transport has been reviewed 
in several textbooks \cite{Cuevas2010,DiVentra2008,Haug2008}. 
Efficient formulations of \stait have been devised so that it can be
implemented conveniently into many electronic structure 
codes. The sheer number of implementations that have been reported over 
the years gives an impressive illustration of how important 
\stait has grown; an incomplete list includes 
McDCal \cite{Taylor2001}, 
TranSIESTA \cite{Brandbyge2002,Papior2017}, 
SMEAGOL/Gollum \cite{Rocha2006,Ferrer2014}, 
two Turbomole-based codes \cite{Pauly2008} and AITRANSS \cite{Evers2004,Arnold2007},  
GPAW \cite{Enkovaara2010}, 
OpenMX \cite{Ozaki2010}, 
Atomistic NanoTransport \cite{Jacob2011},  
ASE \cite{Larsen2017},  
ATK \cite{Smidstrup2019}.  

In the following we briefly recapitulate 
\stait focusing on the conceptual underpinnings.

\subsubsection{Single-particle aspect, scattering theory and partioning} 
 \stait is a single-particle theory; 
 it is effectively assumed that the many-body
states of the molecular junction  
(at least in the low-energy sector) are reasonably well approximated 
by single Slater determinants. 
Equivalently, one assumes that the salient physics of the junction can 
be described in terms of an effective, single-particle Hamiltonian
$\Hem$ for the extended molecule. 
By now, an almost universally met practice is to adopt the 
Kohn-Sham Hamiltonian, $\Hks$, for $\Hem$.  

For isolated molecules the assumption that a single-Slater-determinant
dominates is almost certainly doomed to fail, because the interaction 
energy between valence electrons, $U$, tends to exceed the typical 
level spacing. 
If this latter observation were to be true also for molecules within the
junction, the phenomenon of Coulomb-blockade would
preempt the domain of validity of \stait.				

However, the Coulomb-interaction within the molecular junction is 
screened, reducing $U$ to a screened $U_\text{scr}$, 
so that the overall situation can be very complicated to analyze.
As it turns out, there is a 
significant number 		
of experimental situations 
where an effective single-particle theory provides a useful basis 
for data analysis. \stait is the standard tool for evaluating 
what such a single-particle description would typically predict. 

Depending on the emphasis, the transport formalism 
has been cast into different languages, 	
including the non-equilibrium  Greens function formalism 
(NEGF) 
\cite{DiVentra2008,Cuevas2010,Stefanucci2013,Haug2008}	
or the Landauer-B\"uttiker approach
\cite{Brandbyge2002,Evers2004}.
In either one the current is expressed as, 
\be
\label{et1} 
I = \frac{e}{h} \int_{-\infty}^\infty dE \ \transmission(E) \left( f_{\mathcal
    L}(E)-f_{\mathcal R}(E)\right) ,
\ee
where $f_{\mathcal L,R}$ denote the Fermi-distributions in the
left and right  contacts. The transmission function $\transmission(E)$ 
has the interpretation of a probability weight 
for a particle to be transmitted when it approaches the junction 
with energy close to $E$. 

The most widely spread way for calculating $\transmission(E)$ 
is the partitioning approach. 
It distinguishes three regions: left lead ($\mathcal{L}$), 
right lead ($\mathcal{R}$) and the device
region, that  should be thought of as an extended molecule
(e$\mathcal{M}$), see Fig. \ref{fig.extended-molecule}. 
Thus, partitioning amounts to 
separating the Hilbert space of the full system into three 
sectors. In this formalism one has, 
\be
\label{et2} 
\transmission(E) = \Tr\left[ \GL(E)\Gem(E)\GR(E)\Gem^\dagger(E) \right],
\ee
where the trace is to be taken over the device sector
of the Hilbert space. 
%
The  formula has been derived first for non-interacting 
particles \cite{Caroli1971}; 
it remains valid at zero temperature also for {\color{black} systems with electron-electron interactions
under the condition that the interaction with charge-carriers in the leads (beyond mean-field) 
can be neglected \cite{Meir1992}.
When applied to electrons in the tunneling regime, Eq. \eqref{et2} can be viewed 
as a generalization of Bardeen's theory of tunneling transport, 
going beyond the leading order in the tunneling amplitudes \cite{Bardeen1961}. 

The advantage of partitioning becomes apparent in the 
definition of the Greens-function that describes charge propagation on
the extended molecule in the presence of the reservoirs,
\be
\label{et3} 
\Gem(E) = \frac{1}{E -H_{e\mathcal{M}}-\SL(E)-\SR(E)};   
\ee
it features a single-particle Hamiltonian $\Hem$ that feeds into the transport 
formalism the electronic structure of the extended molecule,  
as it is provided, {\it e.g.}, by KS-based DFT calculations. 

The matrices $\GL,\GR$ are electrode specific and do not carry
information about the molecule; they denote the anti-hermitian parts of the 
self energies $\SL,\SR$ that describe the coupling of the extended 
molecule to the reservoirs: $\GL = \ci(\SL - \SL^\dagger)$, 
and similarly for $\mathcal{R}$. 
They can be calculated exactly, in principle, {\it e.g.} employing standard recursion
methods \cite{Walz2015,Groth2014}. 			
Alternatively, also simple approximative expressions can be used that become accurate when
sufficiently many contact atoms are included in the extended molecule \cite{Arnold2007}.

{\it A typical case.} In the most common scenario $\transmission(E)$  shows
a single peak near the Fermi energy of the reservoirs, 
$\Fermi$, due to either the HOMO
or the LUMO. As an example, we discuss Fig.~\ref{fig:Adak2015};  
at low temperatures the LUMO is the only transport 
active molecular orbital.
The transmission peak is characterized
by its position and width. Although the width is much smaller than the
energy distance to the nearby levels,
the shape of the peak is not Lorentzian in 
the tails due to quantum interference (\qi). 
We elaborate more on the \qi-effects
in Sections~\ref{sIV.A1}, and \ref{ss.QI}.
The paradigm Fig.~\ref{fig:Adak2015} also shows 
that the conductance is strongly
sensitive to the peak position, {\it i.e.}, alignment of the LUMO with
respect to $\Fermi$.
\begin{figure}
\includegraphics[width=\columnwidth]{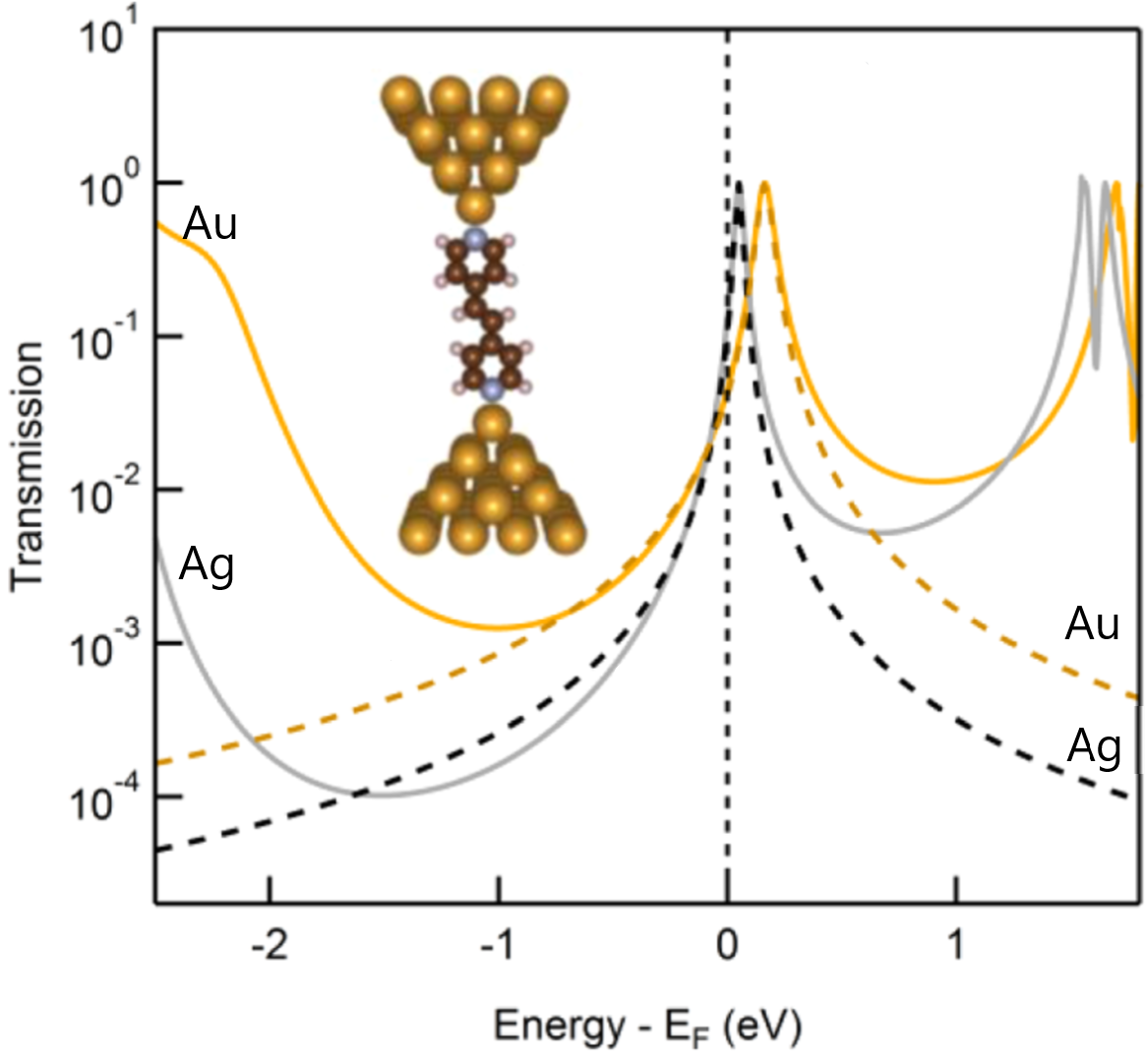}
\caption{\label{fig:Adak2015}\color{black}
Transmission function of 4,4'-vinylenedipyridine junction with Au  
and Ag electrodes calculated using STAIT.
The vertical dashed line indicates the Fermi energy $\Fermi$. 
Dashed curves are Lorentzian fits. The inset shows the relaxed geometry with
gold leads. 
Reprinted with permission from Adak {\it et al.}, Nano Lett. {\bf 15}, 3716. Copyright (2015) American Chemical Society.
}
\end{figure}

}

\subsubsection{Discussion of Kohn-Sham transport calculations} 

A theoretical perspective on  
\stait has recently been given by 
\textcite{Thoss2018}.
We briefly summarize the situation with 
a focus on KS-transport calculations.  
The main issue for us is to what extent 
the KS-Green's function, $G_\text{KS}$, 
can be a useful approximation to the 
real Green's function of the physical system.

(a) As is well known, in equilibrium, the KS-Green's function 
$
G_\text{KS}=1/(E-H_\text{KS}-\SL-\SR)
$
of the extended molecule relates to the local 
electron density 
$ 
n({\bf r})= 2\int_{-\infty}^{\Fermi} dE \ A_\text{KS}(E,{\bf x}) 
$
with a local spectral function 
$
A_\text{KS}(E,{\bf x})= - (1/\pi) \Im 
\langle {\bf x}|G_\text{KS}(E)|{\bf x}\rangle. 
$
When employing exact exchange-correlation 
(XC) functionals, the KS-Green's function 
reproduces the exact density $n({\bf r})$. 
This does not imply that $A_\text{KS}$ also 
is a good approximation to the physical spectral 
function $ A(E, {\bf x})$; in general, it is not.
For example, in the Coulomb-blockade regime the 
physical spectral function $A$ exhibits 
pronounced Hubbard side-bands, which are absent in $A_\text{KS}$.

(b) The relation between $A_\text{KS}(E)$ and 
the true spectral function $A(E)$ has been discussed 
since the 1980s, when band-structure calculations 
started using KS-eigenvalues as approximations 
for quasi-particle energies  \cite{Sham1983,Perdew1982,Perdew1983,Yang2012}.
It is clear that there is no rigorous argument supporting 
this wide-spread practice; 
even with exact XC-functionals, 
there is no known theorem guaranteeing that 
$G_\text{KS}(E)$ will provide an accurate approximation for the 
exact Green's function, $G(E)$. 

Indeed, in the presence of strong Coulomb-correlations, 
this is certainly not the case. As has been pointed out by 
\textcite{Burke2006}, when evaluating the Kubo-formula 
for non-interacting electrons with $G_\text{KS}$ 
the resulting KS-conductance reproduces the 
true conductance only up to a factor that
accounts for an XC-contribution to the voltage 
seen by KS-particles. 

(c) In the special case of very well separated 
transport resonances there may only be a 
single transport-active level, HOMO$^*$ or LUMO$^*${\color{black}, see Fig.~\ref{fig:Adak2015}}. 
In this situation the single-impurity Anderson model (SIAM) 
applies; it features the Friedel-sum rule, 
which allows to express the conductance as a functional 
of the occupation of the frontier orbital, ${\mathcal G}[n]$.
Since the functional ${\mathcal G}[n]$ happens to be the same for 
interacting and non-interacting particles, 
the KS-conductance can be quantitative, 
even though the spectral function is not physical 
\cite{Stefanucci2011,Bergfield2012b,Troester2012}. 
While the argument reproduced here is rigorous, 
it actually assumes symmetric coupling,
$\Gamma_\mathcal{L}{=}\Gamma_\mathcal{R}$.
A generalization to the experimentally 
much more important case of asymmetric couplings 
has also been found \cite{Evers2013}.	
It hinges on the (perhaps surprising) observation 
that the specific ratio of rates 
$\Gamma_\mathcal{L}\Gamma_\mathcal{R}/(\Gamma_\mathcal{R}+\Gamma_\mathcal{L})^2$ 
can be represented as a
parameter-free density functional.  

Summarizing, these considerations lead to an interesting situation: 
the conductance functional ${\mathcal G}[n]$ can reproduce correctly 
the Kondo-effect in the transmission function, Eq. \eqref{et2}, 
at the Fermi-energy despite the KS-Green's 
function $G_\text{KS}$ failing exhibit the Abrikosov-Suhl resonance. 

(d) While in many experimentally relevant cases 
the assumption of a single transport-active level 
may indeed apply, nevertheless, the corresponding 
KS-conductance, $G_\text{KS}$, 
may not be quantitative.
Two important factors intervene. First, 
the arguments employing Friedel's sum rule apply 
at temperatures below the Kondo-temperature, 
$\temp_\text{K}$, only. Experiments often are performed 
at elevated temperatures, $\temp>\temp_\text{K}$, where 
the Coulomb-blockade prevails. In this regime, 
the unphysical nature of $G_\text{KS}$ renders 
the transport nearly resonant, while in reality 
the transmission is strongly suppressed 
\cite{Stefanucci2011}. 
Second, explicit calculations operate 
with approximate XC-functionals. As a consequence, 
the density profile $n({\bf r})$ and, therefore, the input
into ${\mathcal G}[n]$ are not sufficiently realistic
for delivering quantitative conductances near $\temp{=}0$. 

(e) In the majority of cases, the current is carried by more 
than one resonance, so the SIAM is not a fair description
and extra quantum-interference effects can intervene. 
As a consequence, the connection between transport 
and Friedel's sum rule 
breaks down \cite{Hackenbroich2001}, 					
and the protective mechanism that it provides for KS-transport 
calculations (presumably) is not active. 
Hence, one is back to the lowest order expectation 
based on Eq. \eqref{et2}, namely that $G_\text{KS}$ is limited in accuracy 	
by the mismatch between $G_\text{KS}$ and the exact 
Green's function. In other words, KS-transport calculations 
are only as good as is the KS-estimate of the electronic structure, 
which is embedded, {\it e.g.}, in $A_\text{KS}(E,{\bf x})$.


\subsubsection{Proposed improvements over GGA-based Kohn-Sham calculations}\label{sss.Beyond-GGA} 

{\color{black} 
In the previous paragraph, the {\em principle} applicability of 
KS-theory for transport calculations has been discussed. 
In {\em practice}, additional difficulties arise, because actual 
computations always rely on approximate 
XC-functionals, mostly local and semi-local ones, 
such as the local density approximation (LDA), 
generalized gradient approximations (GGA)
or the PBE functional 
(for an overview of functionals 
see \textcite{Fiolhais2003}).
All these approximations neglect the `derivative discontinuity'
\cite{Perdew1983,Sham1983,Yang2012}. 
This implies, roughly speaking, that Coulomb-blockade and 
related phenomena, {\it e.g.} partial charge transfer, 
are treated incorrectly, namely on mean-field level 
\cite{Evers2013}.
There are numerous consequences, 
which have been investigated over the past three decades 
in quantum chemistry and computational materials sciences 
that we cannot cover here. For a first orientation  
see, {\it e.g.}, \textcite{Onida2002} and \textcite{Evers2007}. 
We briefly mention a few selected developments 
representative for the impact of the missing
derivative discontinuity on {\it ab-initio} transport simulations:

(a) Charge transfer can be a process that is critical for 
 the properties of molecules on substrates including their
 transmission properties. In their seminal work 
 \textcite{Neaton2006} have developed an understanding of 
 the relevant microscopic processes and analyzed to what extent 
 they are captured by semi-local XC-functionals.

 (b) In KS-theory charge transfer is controlled by the alignment 
 of energy levels of weakly-coupled subsystems. Therefore, 
 the charge-transfer problem goes along with an incorrect alignment of 
 energy levels of weakly coupled subsystems. 
 \textcite{Ke2007} have investigated the consequences of incorrect level 
 alignments for the transmission function. 
 
 (c) A problem of approximated XC-functionals that derives 
 from the fact that Hartree- and exchange-interaction 
 are not being treated on the same footing is the so called 
 `self-interaction' error. Its impact on the conductance 
 has been discussed by \textcite{Toher2005}.

In order to improve upon the Green's functions, $G_\text{GGA}$, 
thus obtained several procedures have been devised; 
an overview is given by \textcite{Thoss2018}. 
Three main themes can be identified: 

(i) One stays within the realm of KS-theory, 
but one improves upon known artifacts of the GGA-functionals. 
Specifically, optimized long-range separated functionals are introduced 
that provide a significantly better description of 
the partial charge transfer between molecule 
and substrate \cite{Liu2017}.

(ii) 
Alternatively, one leaves the realm of KS-theory and 
computes a Green's function employing conventional 
many-body techniques, {\it e.g.}, the \gowo-method \cite{Bechstedt2015}. 
Indeed, implementations of powerful \gowo-solvers for 
molecular matter are under way \cite{Faber2016,Wilhelm2016,Holzer2017,Wilhelm2018}. 
They open prospects for treating extended molecules 
with thousands of atoms and large enough basis sets, so that controlled 
simulations can be performed with size-converged 
computational parameters \cite{Setten2015}.   

An early attempt in this direction 
has been made by \textcite{Thygessen2007,Strange2011}. 
Due to computational limitations, the system sizes available 
at the time have not been sufficiently large to demonstrate 
convergence with respect to the simulation volume. 
Therefore, the results are not fully conclusive. 
However, relevant fundamental 
questions have been formulated that certainly need to be clarified 
in future research, for instance 
concerning the importance of self-consistency \cite{Thygesen2008} 
and dynamical image charge effects \cite{Jin2014}.

(iii) Rather than systematically computing a Green's function 
within a closed formalism (as in (ii) above) 
one modifies the bare $G_\text{KS}$ following a
physically motivated recipe (`{scissors operators}'
and `image-charge corrections') \cite{Mowbray2008,Quek2009,Quek2007}. 		
The procedure carries a manifestly {\it ad hoc} 
character and therefore its validity is difficult to  
evaluate systematically.

In this realm, a significant advancement has been made in recent work by 
\textcite{CelisGil2017}. 		
These authors determine the shift-parameters for the 
scissors operators in a self-consistent procedure 
by (computationally) gating the molecule 
inside the junction and monitoring the evolution of charge 
with the gate voltage, $Q(\Vg)$. 
As is well known, approximate 
DFT-functionals, such as generalized gradient-corrected 
functionals (GGA), do not properly predict the shape 
of the charge evolution:  
as in typical mean-field approximations, 
$Q^\text{GGA}(\Vg)$ fails to exhibit 
a plateau at integer filling (`Coulomb-blockade') 
in closed-shell calculations. 
Nevertheless, $Q^\text{GGA}(\Vg)$
is a useful object to study, because the 
gate-values that it takes to (de-)populate 
the LUMO (HOMO) allow to reconstruct $U$, 
which is the key scissors parameter.

\subsubsection{Discussion of Nonlinearities in the \IV-characteristics} \label{ss.comp.non-linear}

Generically, the current-voltage (\IV) characteristics  
exhibits a non-linear shape, that for many molecules reveals 
on a scale well above $10$~meV.
As is seen in Eq. \eqref{et1}, nonlinearities can be due to 
the transmission function, $\transmission(E)$, 
varying with energy $E$.\footnote{\color{black} In this section we do not consider inelastic (vibronic) interactions. 
They also introduce nonlinearities in the \IV curve, but these are not captured 
by $\transmission(E)$.} 
Because these terms are still linear in the difference of the 
Fermi-functions $f_\mathcal{L}{-}f_\mathcal{R}$, 
we refer to them as nonlinearities of order zero. 

Higher-order non-linearities arise, {\it e.g.}, because
the bias voltage, $\Vb$, can polarize the molecules 
and therefore affect the scattering potential, 
as illustrated in Fig.~\ref{fig.high-bias-calculation}. 
Within the framework of \stait such non-linearities  
are conveniently included 
by allowing for a bias voltage dependent transmission function,
$\transmission(E,\Vb)=\transmission_0(E)+\transmission_1(E)\Vb+\cdots$, 
in Eq.~\eqref{et1}.
\begin{figure}
\includegraphics[scale=0.58]{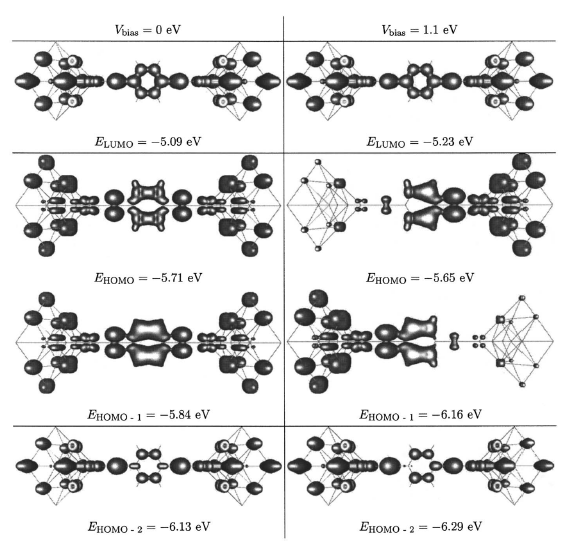}
\caption{Computational results for Au-benzenedithiol-Au 
junctions under high applied bias, $\Vb$. 
Atomic structure is indicated together with the electronic orbitals 
(density clouds) nearest to $\Fermi$. At $\Vb$=1.1 V 
(right) the orbitals shift in energy, but are 
also heavily distorted as compared to 0 V (left). 
Reprinted from Arnold {\it et al.}, J. Chem. Phys. {\bf 126}, 174101 (2007) with the permission of AIP Publishing.
}
\label{fig.high-bias-calculation}
\end{figure}
\begin{figure}
\includegraphics[width=\columnwidth]{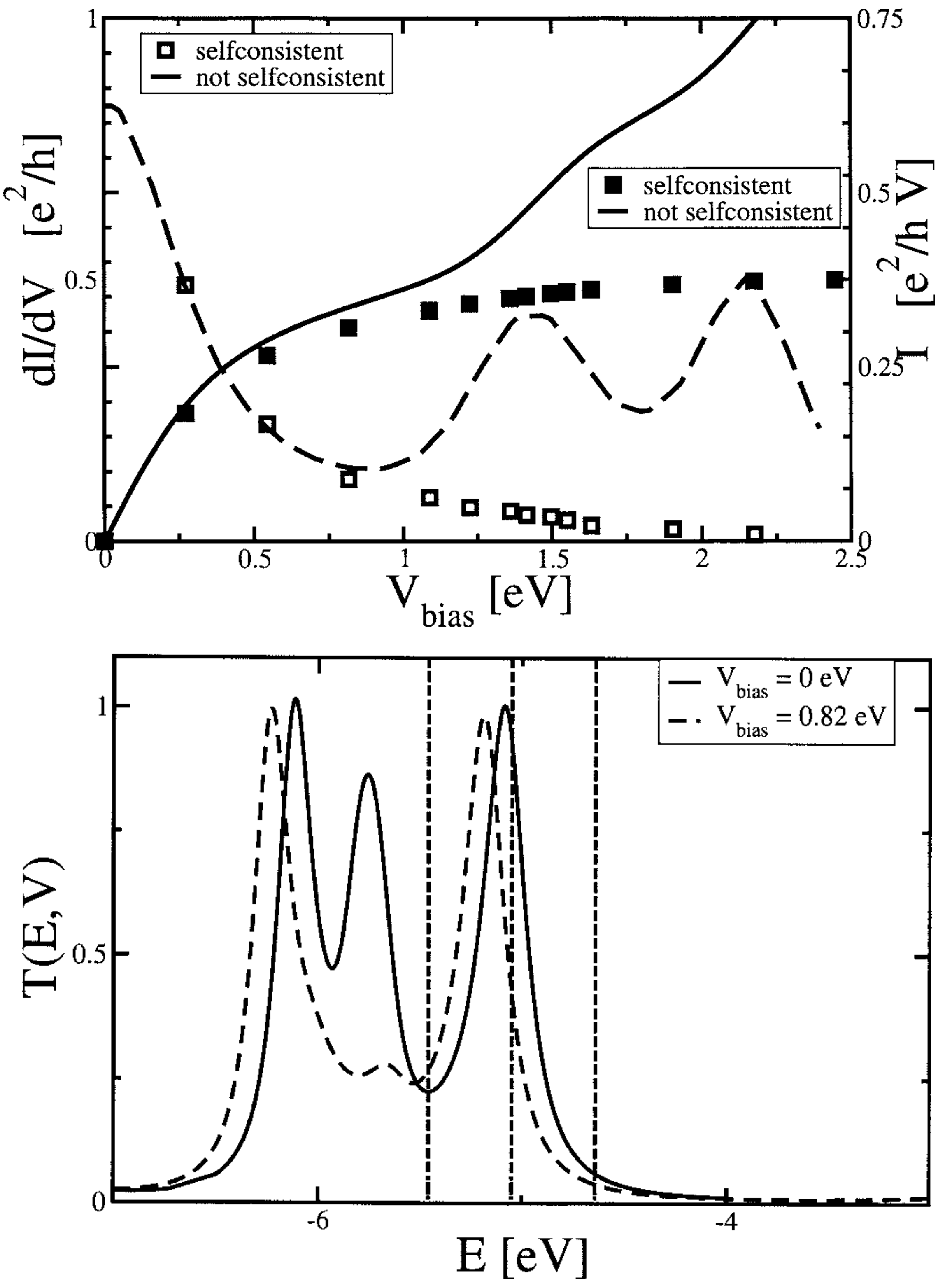}
\caption{
\label{Fig:Arnold2007b}\color{black}
Impact of self-consistency achieved under bias
in transport computations for the Au-BDT-Au 
junction of Fig.~\ref{fig.high-bias-calculation}.
Top: 
\IV-curve (right axis) and differential conductance $dI/dV$
(left axis). 
Bottom: Comparison of the transmission at zero bias
and at $\Vb=0.82$~V.
The vertical dashed lines
are placed at $\Fermi- \Vb, \Fermi, \Fermi + \Vb$.
The three peaks visible at zero bias correspond to 
LUMO, the pair HOMO/HOMO-1, and HOMO-2.
The central peak is suppressed at the finite bias, see the text
for explanation.
Reprinted from Arnold {\it et al.}, J. Chem. Phys. {\bf 126}, 174101 (2007) with the permission of AIP Publishing.
}
\end{figure}
The proper calculation of
$\transmission(E,\Vb)$ requires care. We include a corresponding 
discussion because it reveals, apart from technicalities, also 
aspects of the basic (mean-field type) physics of non-linear {\IV}s. 

{\it  Self-consistent calculations at finite bias.} 
Consider an extended molecule, consisting of 
the molecule plus segments of left and right electrodes. 
In mean-field
theories the effective single-particle potential, $v_\text{s}(\br)$, 	
that defines $\Hem$ has to be constructed self-consistently 
from its eigenstates and eigenvalues. 
The calculation of the potential requires the density matrix, 
$D(\br,\br')$, 
so that the potential can be expressed as a functional of the density matrix 
$v_\text{s}[D]$. In matrix notation (including spin)	
we can write, 
\be
\label{et4a} 
D = \frac{-1}{\pi} \int_{-\infty}^{\infty} \! dE\ 
\Gem\left( \GR f_\mathcal{L} + \GR f_\mathcal{R} \right) \Gem^\dagger,   
\ee
implying for the particle density $n(\br) = D(\br,\br)$. 
When focusing on zero-order nonlinearities,
{\it i.e.}, ignoring the feedback of the bias voltage on the transmission,
one replaces the Fermi-functions $f_\mathcal{L,R}$ by the equilibrium 
distribution $f_\text{eq}$; this usually is also the first iteration 
step in a self-consistent non-equilibrium calculation.
At the fixed-point of the self-consistency loop			
the full form, \eqref{et4a}, is used, for calculating 
$v_\text{s}[D]$ and the Hamiltonian $\Hem$, respectively. 
As long as $\Vb$ is not too large, 
one expects the fixed-point to be unique. 

Starting from equilibrium the self-consistent field 
cycle reshuffles electrons from 
one lead to the other, always keeping the net number of  
electrons of the extended molecule invariant 
(charge-neutrality condition).\footnote{
In practical terms, particle number ($N$) conservation 
can be enforced within the iteration cycle in the 
following way: In each step one keeps fixed the difference  
$\Delta \mu = \mu_\mathcal{L}-\mu_\mathcal{R}$, 
but varies the average 
$\bar \mu = (\mu_\mathcal{L}+\mu_\mathcal{R})/2$ 
so as to conserve $N$ \cite{Arnold2007}. 
}
At the fixed point an amount of charge $Q$ 
has been moved from one side to the other. 
For large enough electrodes
taking the shape of a plate capacitor, $Q$ is 
proportional to the face area giving rise to a 
finite surface charge density $\sigma$. 
The bias-induced charge surplus feeds back into the 
single-particle energies of the electrode state and thus 
enters $D$. 
Thereby, the corresponding electric fields
(surface dipole and capacitor field) are properly included in    
$v_\text{s}$ and so become part of the mean-field solution
\cite{Arnold2007}.	
Finally, the self-consistently
calculated KS-system yields the transmission function.
The effect of the bias is shown in Fig.~\ref{Fig:Arnold2007b}
for Au-BDT-Au.  
At voltages $\Vb<1$~eV the transport is dominated by the
LUMO. The corresponding transmission resonance experiences
a weak shift induced by the bias and its real-space
structure is largely unchanged (see Fig.~\ref{fig.high-bias-calculation}).
The effect of self-consistency on the \IV characteristics
is therefore weak at low bias. 
At bias $\Vb > 1$~V the orbital
pair HOMO/HOMO-1 plays an important role. These nearly-degenerate
states mix strongly under the effect of bias, as shown by
the wave functions in Fig.~\ref{fig.high-bias-calculation}.
The resulting states are each asymmetric, leading
to the suppression of the corresponding transmission resonances
(around -5.5~eV in Fig.~\ref{Fig:Arnold2007b}). 
This non-equilibrium Stark effect renders the molecular orbital
pair `dark'.
The mechanism described above leads to additional non-linearity
of the \IV-curve, suppressing the resulting current at higher bias.						

{\it  Voltage drop.} 
At the fixed-point of the self-consistency iteration cycle, the  
orbitals of the leads (metal clusters) 
are shifted in energy away from their equilibrium position, 
up-shifted in one electrode (by $\mu_\mathcal{L}{-}\eF$),  
and down-shifted in the other (by $\eF{-}\mu_\mathcal{R}$). 
The relative shift defines the bias voltage, $\Vb$. 
Like in experiments, $\Vb$ can therefore be `measured' also in 
computational simulations by evaluating the relative energy shift
in the raw data \cite{Arnold2007}. 									

We mention that even if the molecular junction exhibits an 
inversion or mirror symmetry along the axis of charge transport, 
the voltage drop cannot, in general, be expected to reflect this 
symmetric behavior as  {\color{black}$\mu_\mathcal{L}{-}\eF{=}\eF {-} \mu_\mathcal{R}{=} \tfrac{1}{2}e\Vb$}.
Namely, the chemical potential of 
a lead, {\it i.e.}, its workfunction, is sensitive to the surplus density $\sigma$, 
because the excess charge modifies the surface dipole. 
The detailed response depends on the atomistic 
structure of the electrode surface and is difficult to predict
quantitatively, even with {\it ab-initio} calculations. 
Generally speaking, metal surfaces cannot be expected
to exhibit a kind of particle-hole symmetry. Hence, 
one would expect that adding and subtracting charge will 
not usually have the same quantitative
effect (up to the sign) on the workfunction.\footnote{\color{black} In this respect, the case of Au 
could potentially be exceptional, because 
the (bulk) density of states is relatively flat near
$E_\text{Fermi}$.} 

{\it  Potential profile.} 
The profile of the voltage drop, $\phi_\text{b}(\br)$, 
can be read-off at the self-consistent fixed-point. 
It is essentially given by the contribution to the 
single-particle potential, $\Delta^{\text{\tiny $Q$}} v_\text{s}(\br)$, 
that arises due to the charge $Q$ being 
transferred within the self-consistency loop from one electrode 
to the other: 
$\Delta^{\text{\tiny $Q$}} v_\text{s}(\br){=}e\phi_\text{b}(\br)$.
In practical calculations the potential profile 
depends on the contact geometry, the shape 
of the electrode clusters and, in particular, on their size.
Since the Coulomb-interaction is long ranged, special care has to be taken 
with respect to the convergence of the transport simulation with system size;
correspondingly, finite-size converged computations can be demanding 
\cite{Arnold2007}. 	

{\it  Beyond zero-order nonlinearities.}  
We consider the Green's function of the 
(real) molecule, $G_\mathcal{M}$, that emerges if we shrink the extended molecule by eliminating the metal clusters; 
it exhibits a structure analogous to \eqref{et3}. 
At the self-consistent fixed point, 
the molecular Hamiltonian $H_\mathcal{M}$ 
and the corresponding self-energies
develop shifts away from their equilibrium values, 
$\Delta H_\mathcal{M}=H_\mathcal{M}(\Vb) - H_\mathcal{M}$. 
The bias-induced shift 
$\Delta H_\mathcal{M}$ will, in general, move energy levels
with respect to the electrode chemical potentials; 
also, it will deform molecular wavefunctions, so that the 
charge-distribution on the molecule changes. 

For example, as a consequence of the level shifts
the molecule can charge or discharge.  
Also the dipole moment can change, {\it e.g.}, due to the action 
of $\phi_\text{b}(\br)$. 
It enters 
$\Delta H_\mathcal{M}$ as an external potential
and summarizes 
the effects of the surface charges, $\sigma$, 
accumulated on both electrodes. 
Under its action the molecule polarizes and a 
Stark-shift of the molecular energy levels appears,
both feeding into $\transmission_1(E)$. 

{\it  Bias and current induced forces.}  
Since the charge distribution in the molecular 
junction reacts to the applied bias, 
electrostatic forces should appear. 
The molecule will move under their action 
from its equilibrium position. This, 
in turn, modifies the molecular orbitals affecting  
higher-order nonlinearities in the \IV\  and, 
potentially, also leads to switching behavior.

Such 
{\it bias}-induced forces exist even in the 	
absence of a current flowing, and therefore 
should be distinguished from
{\it current}-induced forces	
\cite{Todorov2001,DiVentra2002}. 
While theoretical studies of the former are still scarce
\cite{thesisSchnaebele14}, 
the latter have received 		
considerable attention, see, e.g., 
\textcite{Dundas2009}, \textcite{Bode2012}, \textcite{Todorov2014}, and \textcite{lue2015}  
as recent examples. 
The physical mechanisms behind  current induced 
forces are reciprocal: they are also felt by ion cores that 
move through an electronic bath. 
Therefore, the same mechanisms driving 
current induced forces also have implications for molecular dynamics 
simulations; the corresponding generalized Langevin theory is 
reviewed in \textcite{Lu2019}.
While experiments capable of resolving current-induced forces on the 
molecular scale are challenging, first indications of the effects have been reported 
\cite{Sabater2015}.

The origin of current-induced forces has been 
discussed in a particularly illuminating way 
in \textcite{Lu2010}. Our presentation is 
inspired by this work.
Consider a kinetic equation for 
the vector $\bR$ comprising the coordinates of all atoms
measured with respect to their equilibrium positions. 
The equation takes the following form,
\begin{equation}
\label{e6a} 
 M\ddot\bR +\eta \dot \bR + D \bR = {\bf F}_\text{f}(V_\text{b}),  
\end{equation}
 where the usual assumptions underlying such kinetic equations 
have been made. Most notably, a separation of time scales is assumed,
so that a Markovian ansatz is justified.
The left-hand side (lhs) of Eq. \eqref{e6a} is merely the statement that the relaxation dynamics 
of $\bR$ can be modeled by a collection of damped oscillators 
with mass tensor $M$. The matrix $D$ accounts for the 
restoring forces and is symmetric, reflecting 
Newton's third law. The matrix $\eta$ incorporates dissipation 
and also is symmetric, as can be seen, {\it e.g.}, from the fluctuation-dissipation theorem. 
In addition, it is positive semi-definite to guarantee 
the second law of thermodynamics. 

The right-hand side includes the fluctuating
forces typical of Langevin-type descriptions. 
The equilibrium part of these forces, 
${\bf F}_\text{f}^\text{eq}\coloneqq
{\bf F}_\text{f}(0)$, 
is trend-less by construction of $\eta$ 
and $D$. 
Out of equilibrium, for $V_\text{b}{\neq} 0$, trends exist, which are quite naturally 
cast into a form analogous to the lhs of \eqref{e6a}, 
\begin{equation}
\label{e7a} 
 {\bf F}_\text{f}(V_\text{b}) = {\bf F}_\text{f}^\text{eq} + {\mathcal B} \dot \bR + {\mathcal A}\bR
+ \ldots 
 \end{equation}
Formally, the matrices ${\mathcal A}$ and ${\mathcal B}$ can be decomposed into symmetric and anti-symmetric constituents. 
Symmetric pieces, if they exist, combine with $\eta$ and $D$ 
and do not give rise to qualitatively new phenomena 
- at least at small enough $V_\text{b}$. Therefore, the symmetric pieces will be 
ignored and ${\mathcal A}$ and ${\mathcal B}$ are considered as antisymmetric.

The matrix ${\mathcal A}$, being antisymmetric, cannot be understood 
as a second derivative of some energy-functional with respect to a coordinate.
It therefore represents a non-conservative force. 
It's effect on the dynamics is best illustrated by recalling that 
antisymmetry allows for 
rewriting the matrix-vector products 
appearing in {\color{black}\eqref{e7a}} as vector products
\begin{equation}
\label{e8a} 
 {\bf F}_\text{f}(V_\text{b}) = {\bf F}_\text{f}^\text{eq} + 
 {\bf B}\times \dot \bR + 
 {\bf A}\times \bR
+ \ldots 
 \end{equation}
where  ${\bf B}=(-{\mathcal B}_{yz}, {\mathcal B}_{xz}, -{\mathcal B}_{xy})$, 
and analogously for ${\bf A}$. 
Hence, the third term of \eqref{e8a} represents a 
force that tends to rotate the direction of 
displacement, $\bR$. Since a rotation requires the definition 
of an axis to rotate about, the term arises   
because in non-equilibrium the currents flowing break isotropy. 
The effect of this term has been observed by 
\textcite{Dundas2009} as `water-wheel' forces.

The second term in \eqref{e8a} rotates 
the direction of the velocity $\dot \bR$; it represents an 
effective `Lorentz force', where quotation marks remind us that 
the entries of ${\bf B}$ are matrix-valued. 
Effective Lorentz-forces are symmetry allowed because 
away from equilibrium with currents flowing, 
time-reversal invariance and isotropy are broken.
Since Lorentz-forces are energy conserving, 
they actually allow for periodic orbits.
In quantum-models such orbits are closely associated 
with geometric phases (also known as Berry-phases). For the present context, 
Berry-phases have been discussed further by \textcite{Lu2010}.  

The motion of the ion cores that results from the current-induced forces
feeds back into the electronic current. The effect has been considered 
by \textcite{Kershaw2017} by including corrections to the adiabatic 
response that are small in the ion velocities $\dot {\bf R}^2$. 
{\color{black} In extreme cases, the current-induced forces can lead to 
bond rupture. Progress towards a better understanding of this 
important phenomenon has been made in recent work by 
\textcite{Erpenbeck2018}}.

\subsection{Transport Viewed as Relaxation and Incoherent Processes} 
So far charge transport has been considered from the point of 
view of scattering theory. 
Here, we will slightly change our viewpoint and		
consider charge transmission as a relaxation problem.
This alternative perspective 
allows for a relatively simple extension of the single particle model 
including also inelastic effects. 
While the extension here presented is qualitative,
a more formal relation has also been worked out recently by
\textcite{Sowa2018}. 

\subsubsection{Alternative Derivation of the Trace Formula} 
We illustrate the strength  
of the relaxation perspective by using it to derive
the key equation \eqref{et1} in just very few lines. 
The transmission process is viewed as  
a decay of an electronic state of the left reservoir (source) 
into another one in right reservoir (drain). 
This perspective is very close in spirit to electron transfer theory,
a connection that has been made before \cite{Nitzan2001,Solomon2008a}\footnote{One may also note that the structure of the equations below resembles those for Bardeen's theory of electron tunneling, as often applied for STM \cite{Bardeen1961}. However, there are several important differences. E.g., in the Bardeen approximation the electronic structure of the states on the molecule is not taken into account, and states in the leads are assumed to remain unaffected by the formation of a junction.}	

We introduce nomenclature: 
the wavefunctions of the left electrode with
energy $\epsilon_{n}(k)$ we label by $|n,k\rangle$ for 
incoming and $|n,-k\rangle$ for outgoing states with $n$ 
denoting the channel index and $k>0$ the wavenumber. Similarly, for the 
right lead $\epsilon_{n'}(k')$ and  $|n',-k'\rangle$ for the incoming and $|n',k'\rangle$ for the outgoing states. 
The current flowing from the right to the left can then be written as, 
\be
\label{et11}
I = \frac{e}{\hbar} \sum_{n,n'} \iint \! dk \ dk'\ \Gamma_{n'n}(k',k)
[f_\mathcal{L}(\epsilon_{n}(k))-f_\mathcal{R}(\epsilon_{n'}(k'))] 
\ee
very much in the spirit of a rate equation: 
the current through the molecule 
is due to the decay of the states in the left lead 
that have energies $E$ within the voltage window. 
The associated decay rate, $\Gamma_{n'n}(k'k)$, has an exact 
representation in terms of the $\mfT$-matrix  (see below), 				
\be
\Gamma_{n'n}(k',k) = 2\pi \delta(\epsilon_{n}(k){-}\epsilon_{n'}(k')) 
|\langle n'k'|\mfT(E)|nk\rangle|^2, 
\ee
which is readily understood as a generalization of Fermi's Golden Rule. 
Employing this relation and matching \eqref{et11} to \eqref{et1}, we 
obtain for the transmission function, 
\bea
\transmission(E) &=& (2\pi)^2 \sum_{nn'}\iint \! dk\ dk' 
\delta(E-\epsilon_{n}(k))\delta(E-\epsilon_{n'}(k')) \nonumber \\
&& 
\label{et13} 
\times |\langle n'k'|\mfT(E)|nk\rangle|^2  \\
&=& 2\pi \text{Tr}_\mathcal{R} \delta(E-H_\mathcal{R}) \mfT(E) \delta(E-H_\mathcal{L}) \mfT^\dagger(E);   
\eea
$H_\mathcal{L,R}$ denote the Hamiltonians of the left/right leads and the trace is over 
the degrees of freedom of the right lead only. 								
We arrive at Eqs. \eqref{et1} and \eqref{et2} by recalling that, 
\be
\label{et10} 
\mfT(E)  = v \Gem(E) u^\dagger ,
\ee
and defining $\GL=2\pi u\delta(E-H_\mathcal{L})u^\dagger$ and 
$\GR=2\pi v\delta(E-H_\mathcal{R})v^\dagger$. 
The matrices $u$ ($v$) denote the couplings of the extended molecule 
to the left/right reservoir. They connect states of the 
Hilbert space of the leads $\mathcal{R}$ and $\mathcal{L}$ to the 
Hilbert space of the extended molecule. 

\subsubsection{Eigenchannel decomposition} 
We briefly comment on a misconception frequently met in connection 
with the trace formula
\be
\transmission(E) = \Tr\left[ \GL\Gem\GR\Gem^\dagger \right] \nonumber. 
\ee
The original version of the Landauer formula employs a representation 
of the transmission function 
\be
\label{et14}
\transmission(E) =  \Tr_\text{src}\ {\bf t}{\bf t}^\dagger
\ee
where ${\bf t}$ denotes the matrix of transmission coefficients that describe 
the transfer of charge from a channel incoming from the source into a channel 
leaving into the drain \cite{Imry2000}.
They constitute the off-diagonal elements of the scattering 
matrix and can be written as 
\be
t_{n,n'} = \frac{-2\pi\ci}{\sqrt{v_n v_{n'}}} \langle n'k'|\mfT(E)|nk\rangle
\ee
where $v_n{=} d\epsilon_n(k)/dk$ and 
it is understood that $E{=}\epsilon_n(k){=}\epsilon_{n'}(k')$.
Correspondingly, the trace in Eq. \eqref{et14} is to be taken over the transverse degrees of freedom 
(`channels') of  the source as indicated 
by our nomenclature $\Tr_\text{src}$. 
The eigenvalues of ${\bf t}{\bf t}^\dagger$ 
are the transmission coefficients, which are proper observables. 

A tradition has been widely established that effectively identifies the object 
\be
{\bf \tilde t} = \GL^{1/2}\Gem\GR^{1/2}  
\ee
with ${\bf t}$, see {\it e.g}, {\color{black} the well cited paper by \textcite{Brandbyge2002}, or} \textcite{Cuevas2010}, chapter 8.1.   
Unfortunately, this identification is misleading, because ${\bf t}$ and ${\bf \tilde t}$ 
are conceptually very different: ${\bf t}$
{\color{black} carries indices that correspond to channel numbers, so it } 
is acting on the (transverse part) of the Hilbert 
space of the leads. In contrast, ${\bf \tilde t}$ is acting on the Hilbert space 
of the extended molecules. The former is physically uniquely defined, while the latter is subject to the partitioning 
scheme and therefore is of arbitrary size. This implies, in particular, that the number of eigenvalues of 
${\bf \tilde t}{\bf \tilde t}^\dagger$ depends on the partitioning scheme, 
so that these eigenvalues are not, in general, observables. 

Despite of the basic conceptual problem, eigenvalues of ${\bf \tilde t}{\bf \tilde t}^\dagger$ have been 
used successfully in the past in order to interpret experiments and one may wonder how this is possible. 
Presumably, the answer is that the dominating eigenvalues of ${\bf \tilde t}{\bf \tilde t}^\dagger$ 
approach the ones of ${\bf tt^\dagger}$ reasonably quickly 
once the Hilbert space of the extended molecule allows for enough transverse degrees of freedom. 
A careful analysis of the conditions of convergence has not been performed up to date.  
In this context we note that a decomposition of the transmission alternative to ${\bf \tilde t}{\bf \tilde t}^\dagger$  into a product of $q$ and $p$ matrices was investigated by \textcite{Krstic2002}.

\subsubsection{Limit of Sequential Transport and relation to the Marcus Theory of Charge Transfer} 

We now briefly turn to the strongly incoherent limit: 
the electron after flowing from an electrode onto the molecule dwells  
there for a very long time. `Very long' means that the electron 
loses all phase coherences due to its interactions 
with many molecular and environmental degrees of freedom. 
In this situation, transport can be considered sequential  
and the transmission probability takes a product form.  
Actually, the source-drain picture of transport that we have 
embarked upon so far then is very closely related to 
the donor-acceptor concept familiar from electron transfer 
theory \cite{Nitzan2006}.
The observation is useful, because the latter theory 
suggests a phenomenological formulation of transport theory in the spirit of 
Marcus Theory.  The generalization captures incoherent 
and even inelastic aspects in the case of very weak coupling, 
where the dwell time of the charge carriers on the molecule is 
long enough for (a segment of) the molecule and/or its environment 
to restructure and thus destroy phase coherences. 
In this incoherent (sequential) limit 
charge is transferred in a sequence 
of two hopping processes. 

Along these lines concepts from electron transfer 
theory have been adopted for transport on molecular 
junctions \cite{Nitzan2001}.  	
Recently, applications to heat transfer across molecular
interfaces  and also to  charge-transfer networks have been
worked out \cite{Craven2017a,Craven2017}. 	
To the extent that conduction in the latter system class is diffusive, 
the connection to the macroscopic transport theories 
of material sciences, such as phonon-assisted hopping, 
has thus been made. 

\section{Model based analytical results} \label{s.models}

Models are an indispensable tool of understanding.
In molecular transport, 
they serve to illucidate the physical principles involved, 
for deriving explicit formul{\ae}, 
for estimating the relevant parameters, and for analyzing trends in the data. 
In addition, 
they are also needed to set up, analyze, interpret and motivate further  
elaborate numerical computations. Therefore, in this section 
we give a brief overview of the models 
most relevant for understanding molecular junctions.

\subsection{Qualitative discussion of few-level models}

In the vast majority of cases only very few orbitals, typically only one 		
or two, appear to be involved in molecular transport. 
These orbitals are usually weakly coupled in the sense that the contact mediated 
lifetime broadenings, $\GL,\GR$, are much smaller than the relevant 
molecular energy scales, which would be, {\it e.g.}, the 
HOMO-LUMO gap. 
In this situation, impurity models can provide a reliable description. 
Correspondingly, they are often employed for fitting and interpreting 
experimental data. We recapitulate the most basic facts.

\subsubsection{\label{sIV.A1} Two-level model without interactions} 
\newcommand{\tgammal}[1]{\widetilde\Gamma_{\mathcal L #1}}
\newcommand{\tgammar}[1]{\widetilde\Gamma_{\mathcal R #1}}

A situation with only two transport active orbitals is captured
by a two-level model (TLM),  
\be
\hat H = \sum_{\sigma=\uparrow,\downarrow} \sum_{i=0,1} \varepsilon_i \hat n_{i\sigma} ,
\ee
as long as interactions can be ignored. 
The corresponding transmission function is 
straightforward to derive. The (retarded) resolvent operator
takes the form $\hat G = (E-\hat H - \hat \Sigma(E))^{-1}$, 
where as usual the self energy facilitates the coupling to the reservoirs. 
Owing to the two-level structure (and ignoring the spin), 
the resolvent can be represented by a $2\times 2$-matrix,
$G(E)$, 
whose explicit structure depends on the choice of the basis. 
The corresponding matrix elements define the Green's function. 
Irrespective of the basis choice, 
{\color{black} 
a `rotation' $Q$ can be found so that 
$G(E)$ takes a form, 
\begin{equation}
G(E) = Q
\begin{pmatrix} E-z_0 & 0 \\ 0 & E-z_1\end{pmatrix}^{-1}Q^{-1},
\end{equation}
with $z_{i}{=}\varepsilon_i {+}\Sigma_{i}, i{=}0,1$;  
here, $\Sigma_{i}(E)$ is denoting the lead-induced shift of the 
pole positions into the complex plane.
The columns of $Q$, $\psi^\text{r}_i$, are given by the solutions of 
the eigenvalue problem 
\begin{equation}
\nonumber
[\hat H + \hat \Sigma(E)] \ \psi^\text{r}_{i}(E) = z_i(E) \ \psi^\text{r}_i(E), \quad i=0,1
\end{equation}
while the rows of $Q^{-1}$, ${\psi_i^\text{l}}^*$, solve 
\begin{equation}
\nonumber
\ {\psi^\text{l}_{i}}^*(E) [\hat H + \hat \Sigma(E)]  = z_i(E) \ {\psi^\text{l}_i}^*(E)
\end{equation}
For a detailed mathematical discussion see 
\textcite{Farid1999}.

Motivated by the trace formula \eqref{et2}, we introduce the abbreviations 
$\widetilde\Gamma_{\mathcal L} {=} Q^\dagger \Gamma_{\mathcal L} Q$
and $\widetilde \Gamma_{\mathcal R} {=}Q^{-1} \Gamma_{\mathcal R} Q^{\dagger-1}$.}
Then, employing \eqref{et2}, the transmission can be written as a sum of three terms,
\begin{equation} 
\transmission_{\text{TLM}}(E)= \transmission_0 + \transmission_1 + \transmission_{01} ,
\end{equation}
with two direct terms $i{=}0,1$ and an interference term,  
\begin{equation} 
\transmission_i = 2 \frac{\tgammal{ii}\tgammar{ii}}{|E{-}z_i|^2}, \nonumber
\transmission_{01} = {\Re}\frac{4\tgammal{01}\tgammar{10}}{(E{-}z_1)(E{-}z_0)^*}.
\nonumber
\end{equation}
In these exact expressions  
the pole positions as well as the residues 
are functions of energy $E$. 
If we assume that the energy variation of the 
self-energy $\Sigma(E)$ due to coupling with the leads is sufficiently 
weak, a simple two-pole structure is recovered,
 \be
\label{et9} 
\transmission_\text{TLM}(E) = 2 \left| 
\frac{\sqrt{{\GL}_0{\GR}_0}}{E-z_0} 
+ e^{\ci\Psi_c}\frac{\sqrt{{\GL}_1{\GR}_1}}{E-z_1} 
\right|^2 .
\ee
Here, 
$\Gamma_{\mathcal{L}i},\Gamma_{\mathcal{R}i}$
denote (twice) the imaginary parts of $\Sigma_{i}$
taken at the pole positions and resolved per left/right 
lead contribution. 
The overall prefactor of two accounts for the spin and the phase factor,
$e^{\ci\Psi_c(E)}$, parameterizes
interference effects \cite{Geranton2012}.									

Experimental \IV-traces of molecular junctions can often be modeled, 
phenomenologically, in terms of formul{\ae} like \eqref{et9}. 
We stress that this observation does not 
necessarily imply that the corresponding fitting parameters 
have meaningful interpretations in terms of a picture of 
non-interacting particles. As will be pointed out in the following section, 
a two-pole structure in the Green's function can also arise as a 
consequence of strong Coulomb interactions. 
In this case, fitting a two-resonance transmission similar to 
\eqref{et9} can be successful, while the interpretation 
of the resulting fitting parameters will be fundamentally different. 			

Eq.~\eqref{et9} also provides the basic concepts for discussion of 
quantum interference, a topic that will be elaborated in Section~\ref{ss.QI}, below.
To make contact to the conventional representation 
of a Fano line-shape,\footnote{We use the term Fano-resonance in a 
loose sense. Traditionally, it refers to the scattering 
of a free particle (continuous spectrum) 
off a potential with a bound state (discrete spectrum) 
and includes a two-path \qi\ contribution \cite{Fano1961}.
We include cases where the dominating interfering paths 
all run through (bound) molecular orbitals (e.g. through HOMO and LUMO).
Our motivation is that the characteristic 
line-shape, Eq. \eqref{eq.fano}, does not 
distinguish both situations, emphasizing that they are conceptually 
the same. }
we introduce  two real-valued, dimensionless parameters,  
\newcommand{\mfA}{A}
\newcommand{\mfB}{B}							
\begin{eqnarray}
\mfA &=&  4{\Im}\left [
\frac{\tgammar{01}\tgammal{10}}{(E-z_0)\Gamma_{1}}\right ] +
4\frac{\tgammar{11}\tgammal{11}}{\Gamma_{1}^2} ,\\
\mfB &=& 4{\Re} \left [\frac{\tgammar{01}\tgammal{10}}{\Gamma_{1}(E-z_0)}
\right ] ,
\end{eqnarray}
and a dimensionless energy 
$\epsilon = (E - E_{1})/(\Gamma_{1}/2)$ defined by the real and imaginary parts
of the pole $z_1 = E_{1} + \mathfrak i \Gamma_{1}/2$;  
we thus obtain,
\begin{equation}
\label{eq.fano}
\transmission_\text{TLM}(E) = 2\left[ \frac{\tgammal{00}\tgammar{00}}{|E-z_0|^2} +
\frac{\mfA + \mfB\epsilon}{1 + \epsilon^2}\right].
\end{equation}
The second term in Eq.~(\ref{eq.fano}) displays the 
typical Fano shape under the assumption 
that $\Gamma_{1}/2$ sets
the smallest energy scale, 
so that $E_{1}$, $\mfA$, and $\mfB$ are approximately 
constant on this scale. 
Then $\Gamma_{1}$ defines the width of the
asymmetric line-shape, 
the sign of $\mfA$ determines its resonant 
versus anti-resonant 
character, and $\mfB$ controls the degree of asymmetry.
Under the assumptions here made, 
the first term in Eq.~(\ref{eq.fano})
is weakly varying on the resonance scale $\Gamma_{1}$; 
correspondingly, it plays the role of a 
background transmission.

We mention that in the (artificial) case in which 
the three matrices $G(E)$ and $\Gamma_{\mathcal{L,R}}$ 
commute, we have $\mfB=0$ and
$\transmission_\text{TLM}(E)$ decomposes into 
independent resonances with two orthogonal transmission 
channels (i.e. without quantum interference).

\subsubsection{\label{sIV.A2} Basics of SIAM, Coulomb blockade, Kondo effect}

The single-impurity Anderson model (SIAM) considers 
a single transport-active orbital 
with interaction $U$, 
\begin{eqnarray}
\hat H &=& \hat H_\mathcal M + \hat H_\mathcal T + \Hres , \label{eq.siam} \\
\Hres &=& \sum_{k} \sum_\sigma \varepsilon_{k} \ckd \ck , \nonumber \\
\hat H_\mathcal T &=& \sum_{k} \sum_\sigma V_{k} \ckd \ds + \text{h.c.} , \nonumber \\
\hat H_\mathcal M &=& \sum_{\sigma} \varepsilon_0 \hat n_{0 \sigma} + 
U \hat n_{0\uparrow}\hat n_{0\downarrow}. \nonumber
\end{eqnarray}
 It has proven useful 
 in diverse physical contexts, such as transition-metal
 impurities in metals, or semiconductor quantum dots
\cite{Kouwenhoven2001}.
Similar to STAIT, also the SIAM 
employs partitioning featuring the canonical 
creation and annihilation operators of
the electrons in the leads, $\ck,\ckd$, and of the 
molecular `quantum dot,' $\ds,\dsd$.
The molecule is represented by the dot 
Hamiltonian $\hat H_\mathcal M$: 
the on-site energy, $\varepsilon_0$, is defined with respect to the 		
chemical potential in the leads; $\hat n_{0 \sigma} {=} \dsd\ds$
denotes the number operator for electrons occupying the single 	
level with either spin up or down, 
$\sigma={\uparrow,\downarrow}$; 
$U$ represents the charging energy of the level.
The Hamiltonian $\Hres$ implements the 
left and right reservoirs, for ${\mathcal X}={\mathcal L,R}$. 
For the sake of compactness of notation 
we denote all degrees of freedom of the electrons in the reservoir	
by $k$ and $\sigma$ ($k$ encodes band index, lead index, wavenumbers, etc.).
Finally, $\hat H_\mathcal T$ is the tunneling Hamiltonian, with the matrix elements $V_k$ corresponding  
to the matrices $u$ and $v$ in Eq.~\eqref{et10}.

{\it  Coulomb blockade.} 
In the limit $U=0$ the SIAM reduces to a single non-interacting
resonant level
with the width $\Gamma = \Gamma_\mathcal L + \Gamma_\mathcal R
= 2\pi \sum_k |V_k|^2 \delta(\varepsilon_k)$.
The limit other limit $U\ne 0$ and small $\Gamma$, 
the level is decoupled from the leads and 
the molecular Hilbert space comprises four states, which describe 
the empty, the singly occupied (spin up/down) or the doubly occupied level. 
The salient point of SIAM is 
that it can describe a configuration with partially filled orbitals (`open shells')
and integer filling, 			
$N_0 =\sum_\sigma\langle\hat n_{0\sigma}\rangle \approx 1$,
that does not break spin-rotational invariance, i.e., 
does not display magnetism.
This defines the regime of 
Coulomb-blockade, which prevails 
at non-vanishing $\Gamma$ and at large 
enough charging energy $U$ \cite{Anderson1961}, 
such that,\footnote{We reiterate:  
With non-interacting particles orbital fillings are either zero or two,  
except when the orbital energy, $\varepsilon_0$, is resonant with 
the chemical potential: 
$ |\varepsilon_0 - \Fermi|\lesssim \Gamma$.
At large $U$ the filling can be odd-integer
in a much larger regime $\Fermi-\varepsilon_0 \lesssim U$.
Also mean-field theories, such as Hartree-Fock, 
can realize odd-integer fillings, but at the price of 
invoking magnetism, {\it i.e.}, by 
spontaneously breaking spin-rotational invariance.
This mean-field artifact is known as the `symmetry dilemma.' 
}
\begin{equation}
\label{eq.atomiclimit}
\varepsilon_0 < 0 < \varepsilon_0 + U,
\quad \Gamma/|\varepsilon_0| \ll 1,
\quad \Gamma /|\varepsilon_0 + U| \ll 1. \end{equation}
The (retarded) Green's function of the isolated 
level is a $2\times2$ matrix in spin space; 
it is spin-diagonal with entries, 
\be
G_{0,\uparrow}(E) = \frac{1-\langle \hat n_{\text{0}\downarrow}\rangle}{E{-}\varepsilon_\text{0} {+} \ci0} +
\frac{\langle \hat n_{\text{0}\downarrow}\rangle}{E{-}\varepsilon_\text{0} {-} U {+} \ci0} ,
\ee
and similar for the spin-down component, in the limit $\Gamma=0^+$ 								
\cite{Bruus2004,Haug2008}.
Introducing the lead coupling,
the two peaks of the $G_{0,\uparrow}(E)$ shift and acquire some
broadening. In addition, a third peak emerges at the Fermi-energy, 
the Abrikosov-Suhl (also: Kondo) resonance, 
which signals the onset of the Kondo-effect.

{\it  Kondo effect.} 
In the regime given by the conditions (\ref{eq.atomiclimit}), the
electronic spin of the molecule represents a degenerate two-level
system. The associated quantum fluctuations 
become increasingly important as the temperature is
lowered, leading to the Kondo effect. A manifestation of the latter 
is the screening of the local magnetic moment of the molecule 
that goes together with the emergence of the Kondo-resonance. The characteristic
energy scale is \cite{Hewson1993},
\begin{equation}
\label{eq.tk}
k_\mathrm B \temp_\mathrm K =
 E_0 \exp \left [ - \frac {\pi} \Gamma 
\left ( \frac 1 {\varepsilon_0 + U} - \frac 1{\varepsilon_0}\right )^{-1}
\right],
\end{equation}
where the factor $E_0\simeq\text{max}(D,U)$
and $D$ denotes the half-width of the lead's conduction 
band.\footnote{Since $\temp_\mathrm K$ is the temperature
of a cross-over, some ambiguity exists in the definition of $E_0$. 
In any case, the
exponential implies a strong suppression of $\temp_\mathrm K$ in the 
limits (\ref{eq.atomiclimit}).}

Since the Kondo temperature $\temp_\text{K}$
is exponentially small in the ratio $U/\Gamma$, 
the Kondo effect is not commonly observed in molecular junctions, 
where the coupling $\Gamma$ is typically small compared to the charging energy $U$. 
Focusing on temperatures above $\temp_\text{K}$, where we can ignore the Kondo-peak, we 
arrive at the following simplified representation, 
\be
\label{et6}
G_{\uparrow} = 
\frac{1-\langle \hat n_{\text{0}\downarrow}\rangle}{E{-}\varepsilon_\text{0} {-} \Sigma} +
\frac{\langle \hat n_{\text{0}\downarrow}\rangle}{E{-}\varepsilon_\text{0} {-} U {-}\Sigma}
 \ee
 \cite{Haug2008},
 with $\Sigma(E) = \SL(E)+\SR(E)$. 
 It features only two resonances, 
the upper and lower Hubbard peak, 
for a half-filled level, 
$\langle \hat n_{\text{0},\uparrow}\rangle
{=}\langle \hat n_{\text{0},\downarrow}\rangle {=} 1/2$.
In more realistic theoretical treatments (still keeping $\temp {\gg} \temp_\text{K}$), 
the width of the peaks can differ \cite{Pruschke1989,
Koenemann2006}.

{\it  Discussion of transmission and interference features.}  
After feeding \eqref{et6} into the trace  
formula \eqref{et2}, we obtain,
\be
\label{et7}
\transmission_\text{SIAM}(E) = 2 \GL\GR
\left| \frac{1-\langle \hat n_{\text{0}\downarrow}\rangle}{E{-}\varepsilon_\text{0}{-} \Sigma} +
\frac{\langle \hat n_{\text{0}\downarrow}\rangle}{E{-}\varepsilon_\text{0} {-} U {-}\Sigma}
\right|^2.
\ee
Due to the charging effects, two Hubbard-peaks emerge 
that are well separated from each other under the conditions 
(\ref{eq.atomiclimit}).
While the transmission functions, Eqs. \eqref{et9} and \eqref{et7}, 
describe two very different physical situations
-- with and without strong correlations --
they share the same analytical structure. 	
Therefore, interpreting measurements   
by fitting a single-particle theory (such as STAIT, Eq. \eqref{et9}) 
can be misleading.
The difficulty exists, in particular, in situations where 
the {\it ab-initio} theory for the ground-state 
predicts half-filled levels.  
\begin{figure}
\includegraphics[scale=0.44]{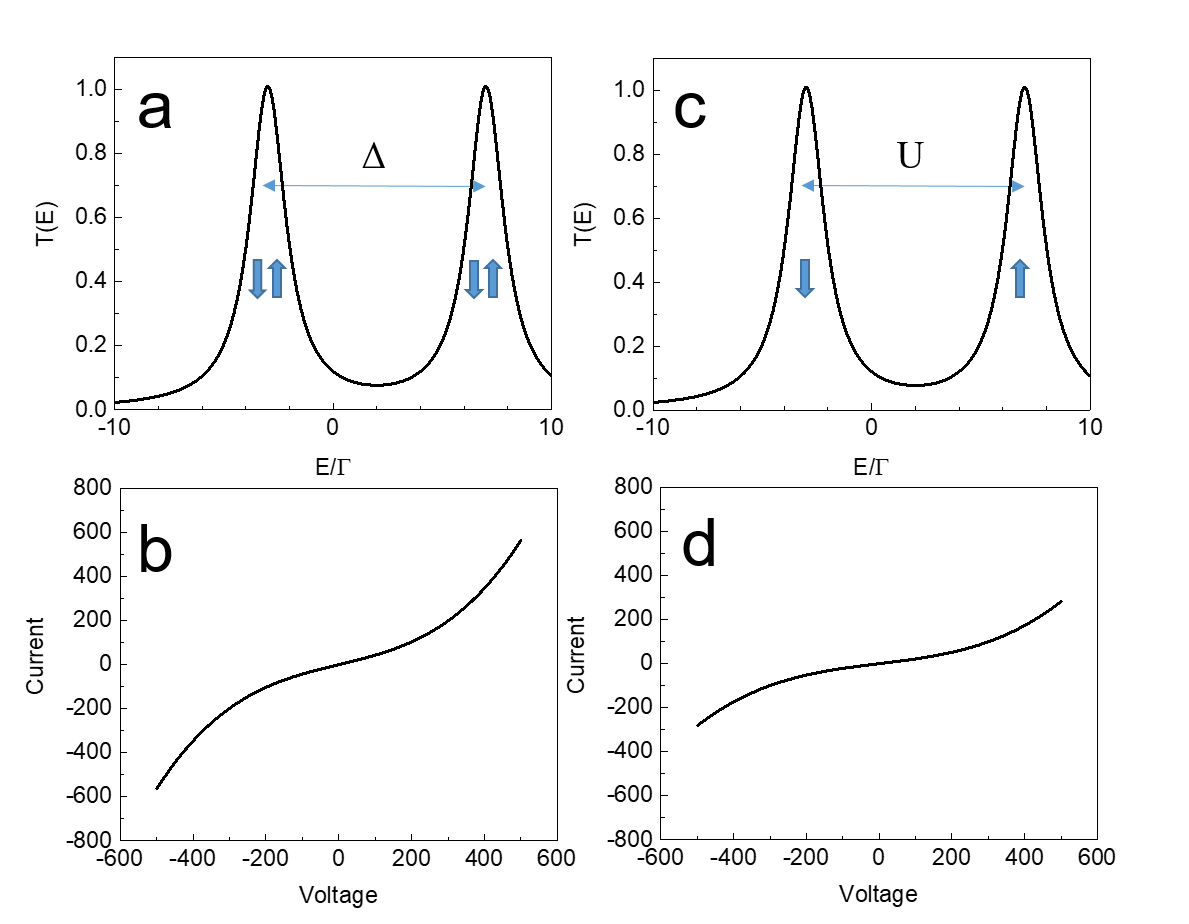}
	\caption{Schematic illustration of the similarity between two models at two extreme limits. 
	(a) When the single-particle level separation $\Delta=\varepsilon_1 -\varepsilon_0$ and the
        level broadening $\Gamma$ are much larger than
	the on-site Coulomb repulsion $U$ we may neglect the latter. The transmission has two spin degenerate
resonances 
	and the \IV-characteristics (b) are expected to show a generic S-shape determined predominantly by the nearest
level, in this example the HOMO.
	(c) In the other limit, when $U \gg \Gamma $, we obtain again two resonances, but here they are
single-spin resonances
	separated by $U$. 
	However, when measuring an \IV curve the shape (d) will be indistinguishable 
	from the non-interacting two-level model, 
	although the amplitudes may differ. Here, we 
	assume that 
	$\temp_\text{K}\ll \temp$. }
\label{fig:Delta-U}
\end{figure}

In metallic or semiconducting 
quantum dots one usually encounters a 
spacing of single-particle levels that is
much smaller than the interaction strength, 
$|\varepsilon_1-\varepsilon_0|\ll U$. 
Such a clear separation of energy scales is less common in molecular junctions, 
where the HOMO-LUMO gap is typically of the order of $U$. 
This often makes it challenging to clearly identify correlation effects in 
molecular junctions from the measured \IV-characteristics alone, 
as illustrated in Fig.~\ref{fig:Delta-U}.


{\it  Remark on Fano line-shapes in the Kondo regime.}  
In the Kondo regime
the elementary excitations are of Fermi-liquid type \cite{Nozieres1974}.
Correspondingly, the Kondo-resonance is simply understood as an extra pole 
in the Green's function, 
\begin{equation}
	G_{\sigma} (E)\ \approx \frac{Z_\mathrm K}{E + \mathfrak i E_\mathrm K} + \widetilde g
\label{et32xxx}
	\end{equation}
which naturally appears at zero energy; its width is given by 
$E_\mathrm K\approx k_\mathrm B \temp_\mathrm K$.
The smooth background $\widetilde g$ is small compared to the resonant term.

By virtue of the Fermi-liquid nature of the Kondo ground state, 
Eq. \eqref{et32xxx} can be used in combination with
Eq. \eqref{et2}; the resulting transmission has the structure
of a single resonant level (neglecting the small $\widetilde g$). 
The residue
 $Z_\mathrm K \le E_\mathrm K/\Gamma $
guarantees that the maximum conductance 
is limited to 1~\go.

When another transport orbital is present, or when there is direct lead-to-lead tunneling, 
this can be represented by a `background' scattering
amplitude $\tilde t$, and 
the $\mfT$-matrix \eqref{et10} can be written in the form,
\begin{equation}
  \mfT(E) = vG_{\sigma} (E)u^\dagger + \tilde t(E).
\end{equation}
Typically, the energy dependence of $\tilde t(E)$ is smooth on the scale
of $E_\mathrm K$ and the transmission probability attains the same
structure as for quantum interference, Eq.~(\ref{eq.fano}),
with a Fano line-shape \cite{Ujsaghy2000,Plihal2001}.

{\it  Kondo blockade.}  
\textcite{Mitchell2017} made an interesting prediction
that the Kondo effect can lead to a suppression of
the conductance, termed \emph{Kondo blockade}. The crucial
condition for its realization is that the molecular
spin is coupled to {\em two} independent conduction channels. 
The conductance is suppressed because of an intricate combination of many-body 
effects that may be viewed as a specific hallmark 
of Kondo-physics in molecular systems.

Let us consider a molecule 
with a spin-half ground state coupled to a pair of reservoirs.
Due to spin-rotational invariance, the low-energy description has
the form of a Kondo Hamiltonian \cite{Nozieres1980},
\newcommand{\ua}{\underline{\alpha}}
\newcommand{\us}{\underline{\sigma}}
\begin{equation}
\label{e31} 
\hat H_\text{2CK} = \sum_{{\bf k}}\sum_{\alpha\ua\sigma\us}
\left(\frac 12 J_{\alpha\ua}\mathbf{\hat S}\cdot 
\bm \tau_{\sigma\us} + W_{\alpha\ua}\delta_{\sigma\us}
\right )
\hat c^\dagger_{\alpha\sigma}({\bf k}) 
\hat c^{\phantom\dagger}_{\ua\us}({\bf k}), 
\end{equation}
where $\alpha,\underline{\alpha}$ 
each sum over the states in 
right and left leads, $\alpha,\ua\in \{\mathcal L,R\}$; 
 $\mathbf{\hat S}$ is the spin operator of the molecule and
$\bm\tau$ are Pauli matrices, with $\sigma$ and $\us$ summing 
over both spin directions. 

The second term in \eqref{e31} represents a description of the
molecule in terms of a scattering potential $W$. 
The first term has the form of exchange; it is the one that 
gives rise to spin-flip scattering, which
ultimately leads to the Kondo effect. 
In the presence of two electrodes, the coupling $J$ and also $W$ 
take the form of $2\times 2$-matrices. The conventional treatment 
proceeds with a rotation of the basis, so 
$J$ is brought into a diagonal form. Thus two superpositions of electrode
states arise, $\hat \psi_\text{e,o}$, that are referred to as even and odd channels. 

It turns out that due its single level nature, the coupling $J$ 
has a zero eigenvalue in the case of the SIAM, $J_\text{o}=0$
\cite{Glazman1988}.
In more generic situations, both eigenvalues are non-vanishing. 
If $J$ and $W$ can be simultaneously diagonalized ($J$ and $W$
commute), then we arrive at the conventional two-channel Kondo model
\cite{Nozieres1980}.
With molecules, this is not typically the case, however. 
Several new energy scales enter with the implication that the 
conductance is no longer universal ({\it i.e.}, a function of $\temp/\temp_\mathrm K$
only). One of the new phenomena arising is a possible interference 
between the first and the second channel that is mediated via the 
potential-scattering term. 
As \textcite{Mitchell2017} have shown, this interference phenomenon can lead 
to a suppression of conductance, which is the essence of 
the \emph{Kondo blockade}. 

\subsubsection{Two-impurity Anderson model} \label{sss.models.TIAM}

In this section we discuss correlation effects
involving a second transport-active molecular orbital, as it 
arises, {\it e.g.}, in open d-shells of transition metal complexes
or organic polyradicals. 
Qualitatively new phenomena occur, {\it e.g.}, 
the \emph{underscreened}
Kondo effect. It was predicted for impurity systems in
a seminal paper by \textcite{Nozieres1980} and has finally 
been observed for the first time in a molecular junction \cite{Roch2009}.

A very general way to include two transport-active orbitals is the 
two impurity  Anderson model (TIAM) \cite{Alexander1964}. It 
reveals a rich phenomenology. We review the corresponding experiments 
that have been performed in the context of molecular junctions 
in  Section~\ref{s.Achievements}.  In this section we provide the reader with 
the theoretical background that will be helpful for the interpretation
of  these experiments. 

In full analogy with the SIAM, the TIAM-Hamiltonian reads,
\begin{eqnarray}
\hat H &{=}& \hat H_\mathcal M + \hat H_\mathcal T + \Hres , \label{Eq.TIAM.1} \\
\Hres &{=}& \sum_{k} \sum_{\sigma=\uparrow,\downarrow}
\varepsilon_{k} \ckd \ck ,  \label{Eq.TIAM.2}\\
\hat H_\mathcal T &{=}& \sum_{k} \sum_\sigma \sum_{i{=}1}^2
V_{ki} \ckd \di + \text{ h.c.},   \label{Eq.TIAM.3}\\
\hat H_\mathcal M &{=}& \sum_{i\sigma} \varepsilon_i \did\di 
- t \sum_\sigma \left (\hat d_{1 \sigma }^\dagger \hat d_{2 \sigma }^{\phantom{\dagger}}
+ \text{h.c.}\right) + \hat H_2 , \label{Eq.TIAM.4}\\
\hat H_2 &{=}& \frac 12\sum_{\left\{ i_j\sigma_j \right \} }
U_{i_1\sigma_1i_2\sigma_2i_3\sigma_3i_4\sigma_4}
\dofd{i_1\sigma_1}\dofd{i_2\sigma_2}\dof{i_3\sigma_3} \dof{i_4\sigma_4}.
\label{Eq:Hd}
\end{eqnarray}
The molecular (`double dot') Hamiltonian $\hat H_\mathcal M$ 
now contains bilinear terms: the on-site energies in 
$\varepsilon_i$ and a hybridization term in $t$. 
The two-particle interaction term $\hat H_2$
is written here in the most general form. It can represent
the Coulomb repulsion or an effectively  induced interaction 
({\it e.g.} kinetic exchange).

The kind of physical phenomena appearing in this model
depend on the (average) charge state 
of the molecule.
We focus on two situations with nearly one or two electrons.
They correspond to 
filling fraction
$\nu{=}1/4$ and $\nu{=}1/2$, 
where $\nu:= \frac{1}{4}\langle \sum_{i\sigma} \did\di\rangle$.
A comprehensive overview of the TIAM at arbitrary filling
can be found in \textcite{Logan2009}.

{\it  Quarter filling.}  
At quarter filling, the correlated subspace hosts a single electron 
that can take four different states. A second electron is not allowed 
to enter due to repulsive interactions (`Coulomb blockade'). 
A particularly interesting situation arises in the degenerate
case, where all four states are energetically identical. 
Thus motivated we consider, 
\begin{equation}
t=0,\ \varepsilon_1=\varepsilon_2
\qquad \text{and}\qquad 
\Sigma_{ii'}(\omega) = \delta_{ii'}\Sigma(\omega).
\label{eq:et33} 
\end{equation}
The last expression implies that the particle leaving the molecule 
for excursions in the leads later returns into the same orbital. 
The condition ensures that the lead coupling does not lift 
the four-fold degeneracy.\footnote{The charge exchange between molecule and reservoirs can be represented by the
single-particle (hybridization) self-energy,
\begin{equation}
 \Sigma_{ii'}(\omega) = \sum_k V^*_{ki}
\frac{1}{\omega - \varepsilon_k + i0+}V_{ki'}^{\phantom *}.\nonumber
\end{equation}
}
When charge fluctuations are suppressed due to Coulomb blockade, 
the low-energy physics is described by the Kondo Hamiltonian, 
however now with an SU(4) hyperspin accounting for four-fold 
degeneracy. The associated SU(4) Kondo temperature 
$\temp_\mathrm K^{(4)} \propto e^{-1/2J\rho}$ carries an extra factor of 
$1/2$ and therefore is exponentially
enhanced compared to an SU(2) Kondo scale 
$\temp_\mathrm K^{(2)} \propto e^{-1/J\rho}$ \cite{Bickers1987}.
Here, 
$J=2V^2/|\varepsilon| + 2V^2/|\varepsilon + U|$, and 
$V^2$ denotes $|V_{ki}|^2$ averaged over the Fermi
surface. 

More realistic descriptions will, in general, 
contain terms violating the symmetries (\ref{eq:et33}). 
There are two kinds of such terms.
The first kind is exemplified by
orbital splitting $(\varepsilon_2-\varepsilon_1)$ and
magnetic field.
The SU(4) physics is stable as long as these splittings
remain much smaller than $k_\mathrm B \temp_\mathrm K^{(4)}$ 
\cite{LeHur2003,Borda2003}.\footnote{These terms couple as
marginal operators in the renormalization-group sense.}
The second kind of operators are mixing terms
($-t + \Sigma_{12}(\omega)$); these terms are relevant:
when cooling below a temperature $\temp_\mathrm K^{(2)}$ exponentially lower than
$\temp_\mathrm K^{(4)}$, there is a crossover from the SU(4) to  
the SU(2) fixed point \cite{Lim2006}. We mention that often the $\temp_\mathrm K^{(2)}$
temperature is experimentally inaccessible and the actual observations
are still determined by the SU(4) fixed point. 

When the perturbations of both kinds
become large enough to 
overcome the $k_\mathrm B \temp_\mathrm K^{(4)}$ scale, the Kondo resonance
splits. 
The resulting peaks represent transitions within the low-energy
quartet and are a residual signature of the
SU(4) symmetry \cite{Choi2005}.

{\it  Half filling.} 
The TIAM at $\nu =\frac 12$ has been thoroughly investigated, because it
offers insights into a competition between Kondo screening and magnetic
ordering.  We give a brief overview
of certain regimes of particular relevance for molecules. For a 
detailed account of the general situation, we direct the reader
to the original research articles and reviews 
\cite{Varma2002,Vojta2006,Florens2011,Bulla2008}.

It is convenient to rewrite Eq.~\eqref{Eq:Hd} in
the following form: $\hat H_2 = 
\sum_i U_i \hat n_{i\uparrow}\hat n_{i\downarrow} + I_\mathrm d \hat{\mathbf S}_1
\cdot\hat{\mathbf S}_2 + \hat H_2'$. $U_i$ is the energy
of the on-site repulsion, $I_\mathrm d$ is the exchange
energy and the remaining interaction terms are lumped 
into $\hat H'_2$.

We consider a regime 
in which charge fluctuations are suppressed on each 
orbital. This regime is delineated by the conditions
 (\ref{eq.atomiclimit}), 
with $\epsilon_0, U, \Gamma$ substituted by 
$\varepsilon_i, U_i,\Gamma_i$, where $\Gamma_i
= 2\pi\sum_k |V_{ki}|^2 \delta(\varepsilon_k)$ and $i=1,2$.
Additionally, to suppress inter-orbital charge fluctuations,
we require $|t|, |\Sigma_{12}(E_\mathrm F)| \ll U$ and
neglect $\hat H_2'$ for simplicity.

In this limit, it can be shown that the physics at low energies
is governed by the two-impurity Kondo model (TIKM)
\cite{Jayaprakash1982, Zitko2006}. 
The TIKM has found applications
in diverse fields, describing interactions between ad-atoms
or double quantum dots, to mention only a few. The only remaining
degrees of freedom of the molecule are the spins, $\hat{\mathbf S}_i$.
An effective Hamiltonian for the molecule has the form 
$I \hat{\mathbf S}_1\cdot \hat{\mathbf S}_2$. Here, $I$ may contain
contributions from direct exchange $I_\mathrm d$, Heisenberg
exchange $4t^2/U$, or terms generated by the environment,
such as RKKY exchange 
\cite{Jayaprakash1981,Proetto1981} or super-exchange \cite{Lee2010}.

Upon connecting the leads, spin fluctuations are induced, as in
the single-impurity Kondo effect.
The resulting Kondo screening of the individual impurity, 
scale $\temp_\mathrm K$, competes with the mutual interaction of the two spins, represented by the
energy scale $I$, that is trying to pair them up. 

{\it  Phenomenology of the TIKM.}  
Based on the previous discussion, three regimes can be distinguished \cite{Varma2002}:

  (i)  At strong antiferromagnetic coupling, 
   $I\gg k_\mathrm B \temp_\mathrm K$, both spins are locked
   into a singlet state. The two impurities then act merely as a potential
   scatterer.
   
  (ii) In the opposite limit of a strong ferro-magnetic coupling,
   $I \ll -k_\mathrm B \temp_\mathrm K$, the two spins behave as a compound
   spin-one object. If only a single screening channel applies
   (we specify
   this condition for molecular junctions in the paragraph below), 
   the Kondo physics is of the
   \emph{underscreened} type. Full Kondo screening
   is achieved close to $\temp=0$ if a second channel is present.
   
  (iii) For intermediate coupling $|I| \approx k_\mathrm B \temp_\mathrm K$
   we have a transition regime. The nature of the transition
   depends on the number of applicable screening channels
   \cite{Vojta2006,Logan2009}.

The number of screening channels 
(i.e. number of available Fermi-surfaces) 
is an important parameter that discriminates between different classes of low-energy
behaviors \cite{Nozieres1980}.
In a generic molecular junction the molecule usually couples 
to a three-dimensional electrode rather than to single-channel wire. 
Therefore, one might suspect that in molecular junctions a description 
in terms of a two channel model is more generic. 
While this is true in principle, this does not imply that the 
underscreened Kondo effect is irrelevant in molecular junctions. 
Namely, the two channels will not be fully equivalent. Even small differences
in atomistic energy scales will in general lead to significant differences in
the Kondo-temperatures associated with each individual channel. 
The reason is the exponential sensitivity of $\temp_\mathrm K$ to atomistic 
energy scales. Therefore, with lowering temperature 
a wide pre-asymptotic regime exists exhibiting underscreened Kondo correlations 
even though the molecule is not coupled to a single-channel wire
\cite{Jayaprakash1981,Posazhennikova2005}.\footnote{A hallmark of the underscreened
Kondo state is the spin degeneracy of the ground state. In
electron transport setups, the degeneracy manifests by the
strong splitting of the Kondo peak in a weak magnetic field
\cite{Roch2009}.}

If only a single conduction channel needs to be taken into account,
a quantum phase transition (QPT) appears in the intermediate regime (iii).
The transition separates an underscreened Kondo state (doublet ground-state)
from the singlet, regime (i) \cite{Vojta2002}. A particular
manifestation of the QPT in the transport is the appearance
of a dip in the density of states on the singlet side due to
two-stage Kondo screening \cite{Hofstetter2001}. The width
of the dip is a dynamically generated energy scale
$\temp^*\propto \exp \left(-k_\mathrm B \temp_\mathrm K / |I-I_\mathrm c|\right )$,
where $I_\mathrm c$ is the value of $I$ at which the QPT occurs.

{\it  Inelastic transport signatures at half filling.}  
The non-equilibrium dynamics of the TIAM at half-filling 
in the Kondo regime was intensively investigated 
\cite{RouraBas2010,Florens2011}. The differential conductance
on the triplet side (regime (ii)) shows a Kondo peak with side peaks
located at energies $\pm I$ (triplet-singlet transitions). In the
regime (i) there are inelastic steps corresponding to singlet-triplet
excitations \cite{Paaske2006,Korytar2012}.  As $I$ approaches $I_\mathrm c$,
the latter merge into the dip ($k_\mathrm B \temp^*$) due to the
two-stage screening. An example of a 
theoretical differential conductance in these regimes is shown in
Fig.~\ref{Fig:Florens2011_fig11}.
\begin{figure}
\includegraphics[width=\columnwidth]{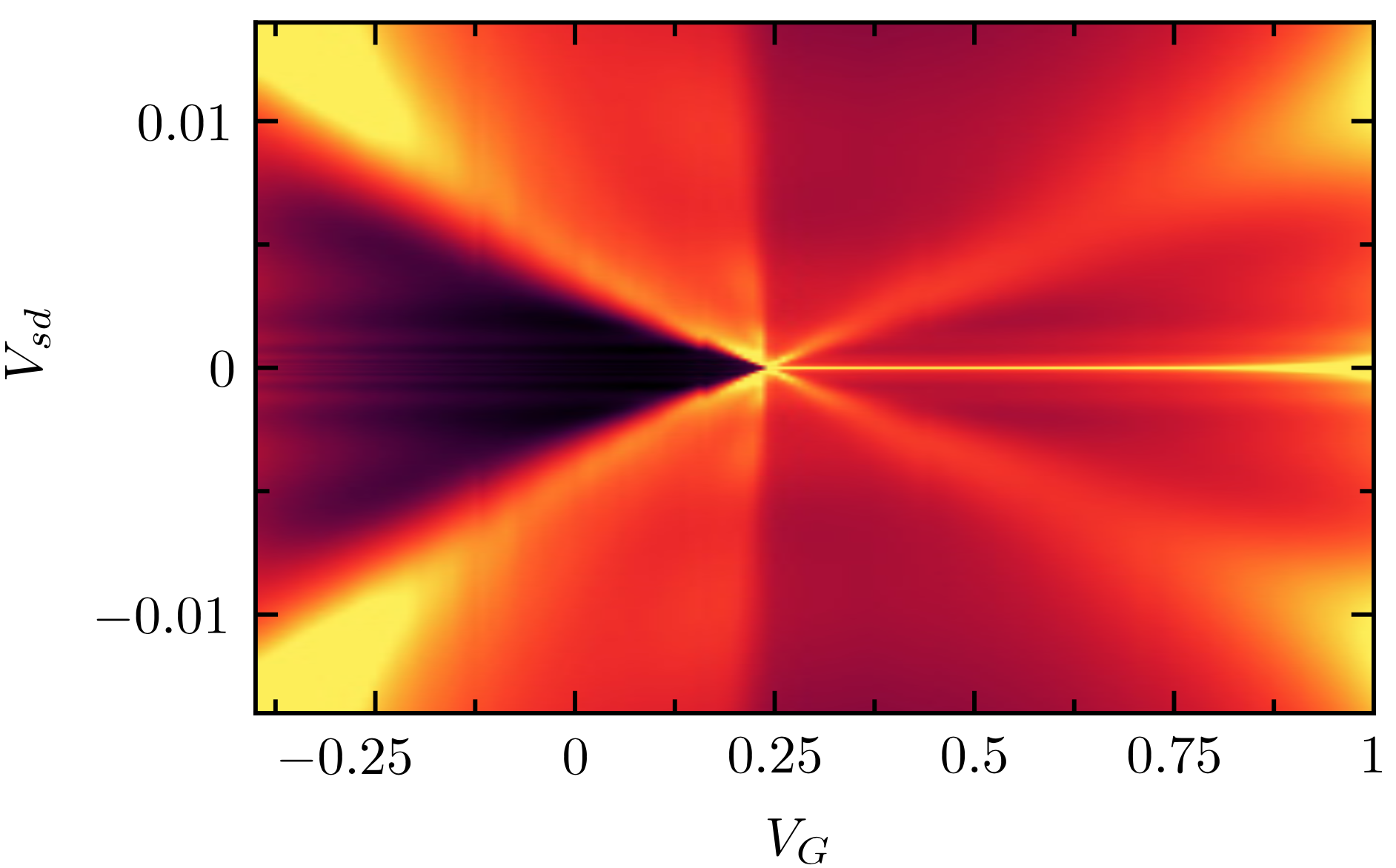}
\caption{\label{Fig:Florens2011_fig11} 
The differential conductance of
a TIAM near the singlet-triplet transition
displayed in the plane of gate/bias voltage
(units: $U_1{=}U_2$).
Effectively, the gate voltage fine-tunes the difference between singlet
and triplet energies with $V_G^*{=}0.25$ at criticality.
At $V_G>V_G^*$ a Kondo peak and triplet-singlet side-peaks emerge, 
while $V_G<V_G^*$ there is a singlet gap.
Color (gray scale) coding is such that the brightest level implies 
unitary conductance. Reproduced with permission from Florens {\it et al.}, `Universal transport signatures in two-electron molecular quantum dots: gate-tunable
   Hund's rule, underscreened Kondo effect and quantum phase transitions',  J. Phys.: Condens. Matt. {\bf 23}, 243202 (2011).  
$\copyright$ IOP Publishing. All rights reserved.
}
\end{figure}

\subsection{Quantum interference effects (\qi)} \label{ss.QI}

{\qi}-features tend to be strong and robust in molecular junctions. 
This is hardly surprising, because with a view on \eqref{et9} 
we see that usually only a few complex valued numbers 
need to be added for evaluating observables.
As a consequence, interfering probability amplitudes 
do not tend to cancel and interference effects are ubiquitous and significant. 
Therefore, they have received 
considerable attention, 
and proposals for applications have been made from early 
on \cite{Baer2002,Dijk2006,Maggio2014,Bergfield2015,Strange2015}. 
Correspondingly, the literature on {\qi} 
is sizable and we here focus our survey on 
theoretical concepts and mechanisms. 
For the discussion of important experimental tests see Section~\ref{ss.QI-exp}, 
for a basic pedagogical introduction 
see \textcite{Hansen2009} and \textcite{Lambert2015}; 
\color{black} for an overview taking 
a chemical perspective see \textcite{Su2016}
and for a recent comprehensive review \textcite{Tsuji2018}.
\normalcolor

\subsubsection{Symmetry considerations in orbital representation \label{sIVB1}}

Under certain conditions the amplitudes appearing 
in the two-level model \eqref{et9} cancel at some 
energy, $E$, so the transmission vanishes
and is strongly suppressed nearby; a {\em Fano 
(anti-)resonance} appears. It is this manifestation of {\em destructive} 
interference that is mostly discussed in the context of molecular transport. 
An intuition for the effect can be obtained 
by evaluating \eqref{et2} for a simple model system. We consider 
a tight-binding representation of a molecule with the source contacting a 
single site and the drain contacting (another) single site, only, 
see, e.g., Fig. \ref{ft6}. 
Then the matrices $\GL,\GR$ reduce to numbers and,  
\be
\label{et22} 
\transmission(E) = \GL\GR |\langle \text{d}|G(E)|\text{s}\rangle|^2 \equiv \GL\GR |G_\text{ds}(E)|^2 ,
\ee
where $\text{s, d}$ denote the contact sites for the 
source (left) and drain (right). 
The Green's function $G_\text{ds}(E)$ 
describes the probability amplitude for a particle with 
energy $E$ (measured with respect to $\Fermi$) 
to travel from source to drain, 
while the prefactor $\GL\GR$ defines the contact resistance.  
When comparing to experiments, the latter can be conveniently 
dealt with by investigating suitable conductance ratios 
in which the contact conductance 
cancels \cite{Geng2015, Manrique2015}.
\begin{figure}[b!]
\includegraphics[scale=0.25]{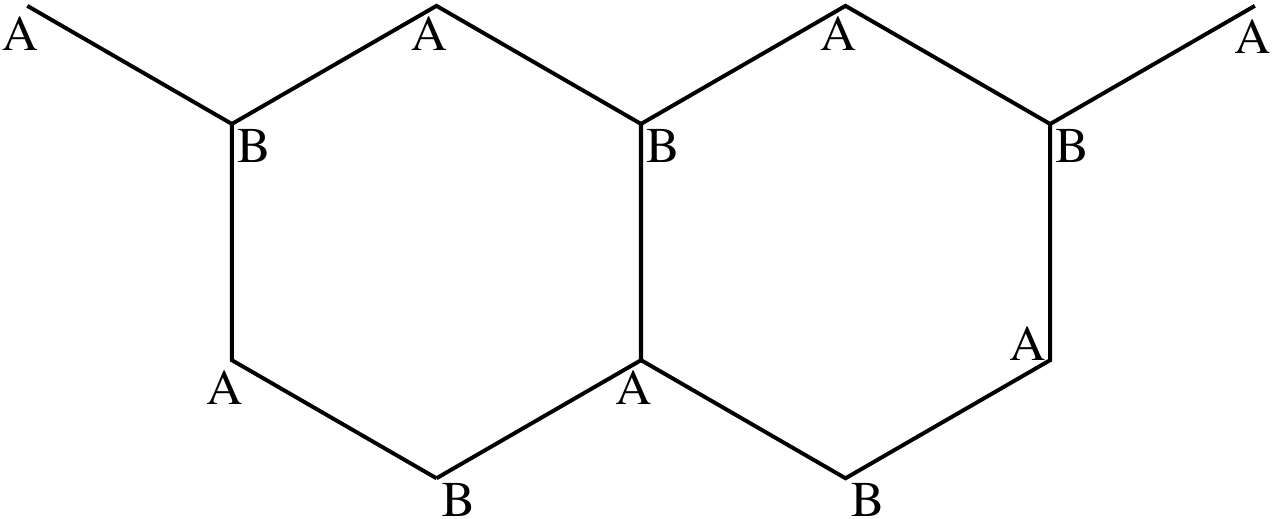}
	\caption{Schematic illustration of a molecule with two-point contacts
	and the sublattice structure of the honeycomb lattice.}
\label{ft6}
\end{figure}

We evaluate \eqref{et22} by using explicit 
representations of the matrix element $G_\text{ds}(E)$.
In the limit of weak coupling, we can neglect excursions into 
the leads, so that $G_\text{ds}$ has a simple representation in terms of 
the states of the (isolated) molecule, 
\be
\label{et40}
G^{(0)}_\text{ds}(E) = \sum_{n} \frac{\psi^*_{\text{d},n}\psi_{\text{s},n}}{E-\epsilon_n+\ci 0} . 
\ee
Here, $\psi_{\text{s},n},\psi_{\text{d},n}$ denote the amplitudes of 
orbitals with energy $\epsilon_n$ at the molecular sites that
contact to source and drain.
In the spirit of the two-level model, we now assume that only two
orbitals are relevant, HOMO and LUMO, 
\be
\label{et24} 
G^{(0)}_\text{ds}(E) \approx 
\frac{\psi^*_{\text{d,H}}\psi_{\text{s,H}}}{E-\epsilon_\text{H}+\ci 0} -  
\frac{\psi^*_{\text{d,L}}\psi_{\text{s,L}}}{\epsilon_\text{L}-E -\ci 0} .  
\ee
Let us, for simplicity, assume the wavefunctions to be real valued
in the preceding expression. 
One then realizes that tendency for cancellation exists if
$\epsilon_\text{H}<E<\epsilon_\text{L}$, 
provided that the products in the numerator have the same sign. 
The interesting aspect of this trivial observation is that 
for very generic and relevant classes of organic molecules this 
relative sign can be easily 
predicted \cite{Tada2002,Yoshizawa2008,Lovey2012,Tada2015}.

{\it Mirror plane, inversion symmetry.}  
Molecular junctions (molecule plus the source/drain contacts) can exhibit 
discrete spatial symmetries, {\it e.g.}, mirror planes and inversion centers. 
Even though, under typical experimental conditions, 
such symmetries are only approximate, 
they still can have a very important impact on the transmission. 
Specifically, we consider a parity symmetry that allows to sort 
wavefunctions into even and odd behavior under reversal of 
left and right. For broad molecular classes, in particular for 
molecular wires, the generic situation is that two wavefunctions 
neighboring in energy exhibit different parity.\footnote{Since in longer wires the longitudinal direction 
accommodates most of the phase-space, most states differ in the 
quantum-numbers associated with this direction.
`Generic' here accounts for the fact that for this reason 
most states neighboring in energy (but not all) differ in their longitudinal
(parity-based) quantum numbers.} 
This typically implies 
that, {\it e.g.}, the HOMO is symmetric (even) while the LUMO is anti-symmetric (odd). 
In this situation the numerators in \eqref{et24} exhibit opposite signs. 
{\qi} is constructive if also the denominators exhibit  the same signs, 
which is the case if 
$\epsilon_\text{H}<\Fermi<\epsilon_\text{L}$. 
 \begin{figure}[!t]
\includegraphics[scale=0.072]{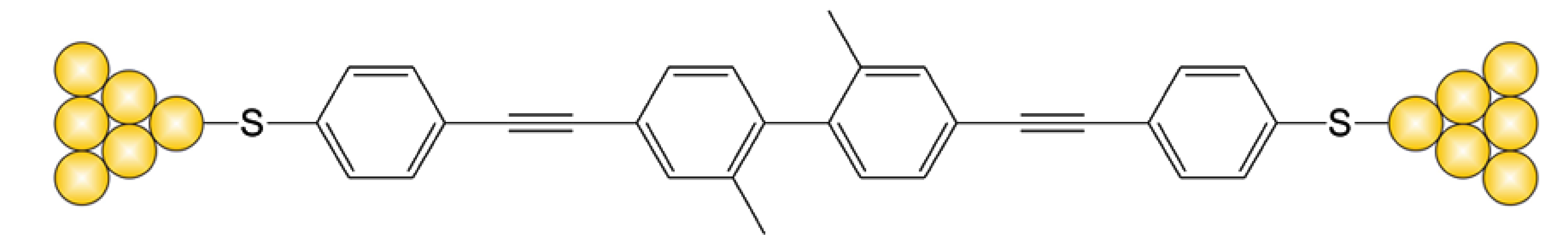}
\caption{Double-dot molecule used in the experiment
   \cite{Ballmann2012}. Due to the presence of the side groups, 
the center benzene rings are tilted against each other leading
to a partial decoupling of the left and right $\pi$-systems.}
 \label{ft4}
 \end{figure}
 
{\it Molecular double dots.} 
A special class of molecular junctions comprises molecules 
that consist of two identical, weakly coupled subunits (`double dots').
In such systems molecular orbitals tend to form parity-doublets
with respect to  mirroring the two dots. 				
For weakly coupled double dots arranged in series, the parity doublets 
hardly split in energy and therefore can exhibit very strong {\qi}-features.
Due to the smallness of the splitting, the doublet states tend to be either both 
occupied or both empty.
Under this condition interference typically is destructive. 
This case is realized in molecule {\bf 3} 
investigated by  \textcite{Ballmann2012}, 
Fig. \ref{ft4}, at least if screening is strong enough to 
suppress charging effects. 
The molecule exhibits a left and a right side with $\pi$-systems 
that have been partially decoupled by tilting the systems 
against each other, reducing the bond to a $\sigma$-bond.  

A situation with pronounced {\qi} can also arise
for two quantum dots arranged in parallel. 
If the dots are not 
too weakly coupled, then the (occupied) states near 
$\Fermi$ will exhibit even parity so that 
orbital contributions from each dot add constructively.
Hence, the transmission is seen to be enhanced 
up to a factor of two as compared to 
{\color{black} the parallel transport through two 
independent single dots.} 
The experimental demonstration of this effect by \textcite{Vazquez2012}
will be discussed in Section~\ref{ss.QI-exp}.

{\it Sublattice symmetry. \label{sec_IVD1}}  
We specialize to the important class 
of molecules with only one 
relevant orbital per atom, 
including graphene-type molecular matter, e.g.,
flakes, nano-ribbons and nano-tubes ($p_z$-orbital). 
Such systems can exhibit a bipartite symmetry, 
so that each atom (``site'') can be associated 
with one out of two sub-lattices, A and B, 
see Fig. \ref{ft6} 
(corresponding to starred and non-starred sites in chemistry nomenclature \cite{Tsuji2014,Tsuji2018}).
The associated tight-binding Hamiltonian 
(also {\em connectivity matrix} in graph theory), 
can be cast into the form, 
\be
\label{et25} 
H_\text{ch} = 
\left(
\begin{array}{cc}
 0 & t \\
 t^\dagger & 0
\end{array}
\right);  
\ee
the matrices $t,t^\dagger$ are 
$N_\text{A}\times N_\text{B}$ and $N_\text{B}\times N_\text{A}$, 
respectively. $t$ describes a 
hopping process $\text{B}\to\text{A}$ and 
$t^\dagger$ accounts for the time-reverse process.  
The on-site energies are taken to be all the same
and zero, assuming that all sites are equivalent; 
they act as the reference letting $\Fermi=0$ at half filling. 
As a consequence of the sub-lattice symmetry 
(also `chiral' symmetry), eigenvalues of $H_\text{ch}$ come in pairs
$\pm \epsilon$ (`Coulson-Rushbrooke pairs').
Moreover, if $\chi=(a,b)$ denotes an eigenstate of 
$H_\text{ch}$ with energy $\varepsilon$ and amplitudes $a,b$ on 
sub-lattices A,B, then $\chi=(a,-b)$ is the corresponding eigenstate for
$-\epsilon$.

With respect to {\qi} and Eq. \eqref{et24}, 
the sublattice symmetry has the following trivial implications:
(i) $\epsilon_\text{H}=-\epsilon_\text{L}$ 
at half filling ({\it i.e.} with one electron per site). 
(ii) The relative signs of the two numerators in \eqref{et24} 
depend on whether source and drain are located on the 
same sub-lattice or not: if source and drain sites belong to the same 
sub-lattice, the relative sign between HOMO and LUMO states is the same 
and the interference is destructive. In contrast, 
when two different sub-lattices are contacted the interference 
is constructive.  
For the transmission function, this argument has first been 
made by  \textcite{Solomon2008}. 
The sensitivity of the transmission to 
the relative choice of the contact lattice 
has been discussed, in particular in 
\textcite{Tsuji2014} and \textcite{Zhao2017}. Powerful selection rules have been worked 
out that we will re-derive and discuss in the following sections.

In more complete descriptions of organic molecules,     	
such as {\it ab-initio} calculations, 
the sub-lattice symmetry \eqref{et25} tends to be approximate, 
{\it e.g.}, due to the presence of next-nearest neighbor terms,
inhomogeneous on-site potentials and also due to additional bands, 
{\it e.g.} $\sigma$-bands. 
The latter give rise to superimposing transport channels that 
effectively mask the Fano-dip \cite{Ke2008}.
The leading effect of higher-order hopping terms
is to disturb the energy pairing, 
while the nodal structure of the $\pi$-electron orbitals is less affected. 
As a consequence, with higher-order hopping terms  
the Fano-dip is no longer 
situated symmetrically between the HOMO and the LUMO resonance. 

Without going into further detail, 
we mention an interesting 
development related to cases where a 
splitting of the Fano-resonance has been reported \cite{Solomon2011}.
A detailed investigation of this effect using a linear 
tight-binding chain as a paradigm has been performed  
by \textcite{Tsuji2014}. It has been found that
two kinds of Fano-resonances occur that differ, 
{\it e.g.}, by the way in which they react to perturbations; 
see also Sec. \ref{sIVB3}. 
While the first kind is easily understandable within 
a picture of molecular orbitals, the second kind requires a more 
elaborate analysis based, {\it e.g.}, on graphical methods, 
see the following section \cite{Zhao2017}.

\subsubsection{Sum-over-paths approach}

In Eq. \eqref{et40} the Green's function $G^{(0)}_\text{ds}$
has been represented as a sum over its poles 
(`Lehmann representation').
There is an alternative representation of 
$G^{(0)}_\text{ds}$ in terms of determinants, 
\be
\label{et26}
G^{(0)}_\text{ds}(E)=\frac{\det(E-\mH)_\text{d,s}}{\det(E -\mH)}, 
\ee
where $\mH$ denotes the tight-binding Hamiltonian of the 
isolated molecule.
The truncated matrix $(E-\mH)_\text{r,c}$, 
{\it i.e.} the `minor', 
derives from the parent matrix 
$(E-\mH)$ by eliminating the row `$\text{r}$'
and the column `$\text{c}$'. 
Writing \eqref{et26} we anticipate that $\det(E{-}\mH){\neq} 0$, 
so there is no spectral weight of $\mH$ at the energy $E$.
Eq. \eqref{et26} and variants thereof have 
been used as a starting point 
to derive graphical rules for predicting the presence of 
Fano-features \cite{Fowler2009,Pickup2008,Markussen2010nl,Markussen2011,Mayou2013,Stuyver2017}.

{\it The MST-rules.}  
A very general set of rules to exploit \eqref{et26} for deriving the 
transmission at zero energy, $E{=}0$, has been 
obtained by  \textcite{Markussen2010nl} (MST). 
We re-derive their result. 

A determinant of an arbitrary $N\times N$ matrix $H$ with elements $h_{ij}, 
(i,j{=}1,\ldots N)$
has an explicit representation,
\begin{equation}
 \det H = \sum_{\sigma} \text{sgn}(\sigma) \prod_{i=1}^{N} h_{i\sigma(i)} ,
 \label{et26a}
\end{equation}
with $\sigma$ abbreviating a permutation of the numbers $1,\ldots,N$ and 
$\text{sgn}(\sigma)=1$ for even permutations and $\text{sgn}(\sigma)=-1$ for odd ones. 
We view $H$ as the connectivity matrix of a graph with sites $i$ and $j$ and 
with connectivities $h_{ij}$ indicating a directed link from $j$ to $i$.
In this picture, individual terms appearing in \eqref{et26a} can be interpreted as 
paths on the graph. For instance in the case of $N{=}3$ sites, the term 
$h_{12}h_{23}h_{31}$ appears. It has the interpretation of a loop that 
starts from the first site, then visits the third, the second next 
and eventually returns to the starting point.
To stay within the picture of loops, we interpret a diagonal element, $h_{ii}$, 
as an undirected loop from a site into itself. 

Then, the entire determinant \eqref{et26a} has a transparent graphical representation: 
it is a summation over all directed paths on the graph that have the property that 
there is one incoming and one outgoing link per site.
It is thus clear that each path consists of one (or more) loops 
\cite{Harary1962}.

Following MST, we translate the graphical rules for determinants into a statement 
about the minor $\det(\mH)_\text{d,s}$. 
To this end, we will be using the familiar fact that determinants 
can be calculated by expanding into such minors:  
Suppose, the connectivity matrix $\mH$ features a direct link 
between source and drain, $h_\text{d,s}$. Then performing an expansion in terms of minors,
the determinant $\det \mH$ needs to contain a term 
$h_\text{d,s}\det (\mH)_\text{d,s}$. 
Recalling that only closed paths contribute, 
this term can be nonvanishing only if its associated 
graphical representation contains at least one
path with only closed loops. 
Correspondingly, we conclude that the minor can be non-vanishing only, 
$\det (\mH)_\text{d,s}\neq 0 $,
if the following two conditions are met by its representation in terms
of paths on the graph: 
(i) there is a path connecting source and drain (which is
closed eventually by $h_\text{d,s}$). 
(ii) All sites that do not belong to the path that connects source
and drain are bound in one or more closed loops. 

The graphical rules of MST incorporate these two conditions: 
consider the atom positions of the molecule as sites of a graph;  
check if a path can be found on that graph that satisfies (i) 
and (ii); if there is no such path, then $\det (\mH)_\text{d,s}= 0 $ 
and there is no transmission at zero energy  
\cite{Markussen2010nl}.

The original derivation has been given 
for the case where  $\mH$ features zero on-site energies, only. 
In this case, paths with isolated sites have zero weight 
and, hence, are not considered closed. 
We emphasize that the rules derived here are completely 
general and do not, in particular, require the connectivity 
matrix $\mH$ to be bipartite 
\cite{Xia2014,Stadler2015,Strange2015}. 

{\it  Zero-eigenvalues and radicals.} 
We extend the discussion of the sum-over-paths approach 
preparing relations that become important in the following section. 
The representation of determinants as sums over terms 
that represent closed loops can have interesting implications
for the evaluation of $\det(E{-}\mH)$. 
Namely, it may in fact not be possible to find a closed path: even 
with the `best' dressing of the graph with directed links, there may always
be certain sites left that cannot be made to participate in any loop. 
In that case, we safely conclude that $\det(\mH)=0$, so there are 
zero eigenvalues. 

The smallest number of such isolated sites (`radicals') that one can 
achieve we call $\zeta_\text{r}$. This number constitutes a lower bound for 
the multiplicity of the zero-energy root and, hence, also for the number of 
zero-energy eigenstates: $\zeta\geq \zeta_\text{r}$.

{\it  Bipartite symmetry: double bonds.} 
In the presence of bipartiteness, {\it e.g.}, 
for hydrocarbons with alternating single-double bonds, 
a lower bound for $\zeta_\text{r}$ can be derived. 
Suppose that the $\det \mH$ is non-vanishing. Then there must be
at least one closed path - potentially featuring disconnected loops -
that touches every graph site once and only once.  
Because of bipartiteness, each loop of this path visits every
sub-lattice  in an alternating fashion - and therefore always contains
an even number of sites. We conclude that $\det \mH$ can be
non-vanishing only, if the following necessary condition is met: 
the number of sites (atoms) is the same in each sub-lattice,  
$N_\text{A}{=}N_\text{B}$. 
Based on the analysis of the preceding
paragraph, a lower bound $\zeta_\text{r} \geq |N_\text{A}{-}N_\text{B}|$ is 
thus derived. 

We can give a practical guide to a better estimate, however. 
Consider one of the paths that features the minimum number of
radicals; it exhibits  one or more loops and a number of $\zeta_\text{r}$
isolated sites. Every one of the loops can be decorated by double-bonds following the rule 
that along the loop every site should participate in one and only one 
double-bond. In this way, paths on bipartite lattices are 
associated with a decoration of a number of $N_\text{db}$ 
double bonds. We derive for the number of 
radicals,\footnote{We mention that the number of radicals, $\zeta_r$, 
is closely related to the number of 
Kekul\'e structures, $K$, that are associated with a graph. 
In particular, if $K>0$, then $\zeta_r{=}0$.}
\begin{equation}
\zeta_\text{r} = N_\text{A}+N_\text{B}-2N_\text{db}\text{.} 
\end{equation} 

\subsubsection{Selection rules for destructive {\qi} \label{sIVB3}} 

The MST-rules are somewhat tedious to handle for larger molecules because of bookkeeping for
 a combination of two  geometrical objects - closed loops and an open
 path. This is a remnant of the fact that the minor \eqref{et26} 
is evaluated, directly. 
Working with minors can be avoided by exploiting  
determinant relations. They allow to express the ratio \eqref{et26} 
in terms determinants of proper connectivity matrices. 
Thus, a sum-over-path analysis of the transmission will involve only 
sums over closed paths. 

We define 
$
\Do_\text{r,c}(E){=}\det (E{-}\mH)_\text{r,c}$, so that 
$
G^{(0)}_\text{ds}(E)
{=} \Do_\text{d,s}(E)/\Do(E)
$. 
Using a general identity for determinants \cite{Fowler2009} we can 
relate the determinant of the minor, $\Do_\text{d,s}$,
to three new determinants in the following manner, 
$$
\Do_\text{ds,ds}{=}[\Do_\text{d,d}\Do_\text{s,s}{-}(\Do_\text{d,s})^2]/\Do. 
$$
This expression is attractive, because the new determinants have an appealing 
graphical interpretation: 
$\Do_\text{d,d},\Do_\text{s,s}$ represent determinants of the matrix 
$(E{-}\mH)$ with one site, $\text{d}$ or $\text{s}$, removed; similarly, $\Delta_\text{ds,ds}$ 
is a determinant  with source- and drain-sites removed from $(E{-}\mH)$.
Evaluating with the help of this expression \eqref{et22} then allows us to write,  
\be
\label{et29} 
\transmission(E) {=} \GL\GR [\Do_\text{d,d}\Do_\text{s,s}{-}\Do\Do_\text{ds,ds}]/{\Do}^2 + \ldots, 
\ee
valid to lowest order in $\GL,\GR$. 
Such a relation has been derived before by  \textcite{Stuyver2015} 
embarking on earlier work of \textcite{Pickup2008} and \textcite{Fowler2011}.

{\it  Bipartite symmetry.} 
We apply \eqref{et29} to molecules with bipartite symmetry, 
such as alternating hydrocarbons, 
and focus on the band center $E{=}0$. 
To meet the condition $\Do(0){\neq} 0$, 
we require that the number of sites 
in each sublattice is the same: $N_\text{A}{=}N_\text{B}$;
In other words, the $N_\text{A}{\times}N_\text{B}$-matrix 
$t$ must be square.
As we have seen before, 
otherwise $\mH$ exhibits {\em at least} 
$|N_\text{A}{-} N_\text{B}|$ 
zero-energy states. This statement has been derived first 
by \textcite{Longuet1950}.
By the same argument, we conclude that 
$\Do_\text{d,d},\Do_\text{s,s}{=}0$ at $E{=}0$: 
the removal of the drain or the source site 
implies a sub-lattice imbalance with $N_\text{A}{\neq}N_\text{B}$ 
and hence the existence of at least one zero eigenvalue. 

We thus arrive at the relation, 
\be
\label{et30} 
\transmission(0) = -\GL \GR \frac{\Do_\text{ds,ds}(0)}{\Do(0)}.  
\ee
Its consequences have been investigated, 
recently \cite{Fowler2009,Stuyver2017}. 
Towards deriving rules for {\qi}-induced transmission zeros, 
one arrives at the remarkable fact that Eq.~\eqref{et30} expresses 
the transmission as a ratio of determinants corresponding 
to matrices that can both be interpreted as Hamiltonians of a physical system. 
In particular, $\Delta_\text{ds,ds}$ corresponds to the original 
system with two vertices deleted (`vacancies' in the nomenclature of material sciences) 
at the original position of source and drain.  

For nearest-neighbor hopping, one can relate $\transmission(0)$ to known spectral properties
employing graphical rules that date back to the early work 
of \textcite{Longuet1950}; we have re-derived them using
the sum-over-path approach:  
the number of zero modes $\zeta$ associated with $\mH$ and $(\mH)_\text{ds,ds}$
is given by, 
\be
\label{et31} 
\zeta \geq N_\text{A}+N_\text{B} - 2 N_\text{db}, 
\ee
where $N_\text{db}$ denotes the maximum number of double bonds 
that can be placed on the graph.

One discriminates two kinds of zero modes.
The {\em predictable}  modes 
result from a sub-lattice imbalance: $\zeta^\text{pre}{=}|N_\text{A}{-}N_\text{B}|$. 
The remaining modes, called {\em supernumerary}, come in pairs, 
\be
\label{et32} 
\zeta^\text{sup}\geqslant 2 ( \text{min}(N_\text{A},N_\text{B}) - N_\text{db}),    
\ee
and $\zeta{=}\zeta^\text{pre}+\zeta^\text{sup}$. 
The result was sharpened later in benzoidal graph theory: 
the equal-sign holds for honeycomb lattices \cite{Fajtlowicz2005}.
We emphasize, that supernumerary modes are very far from being 
a mere curiosity. For instance, as has been demonstrated recently, they
play an important role for the thermodynamic
properties of graphene-flakes \cite{Haefner2014}. 
Such supernumerary modes are associated, {\it e.g.}, with incomplete 
parts of the honeycomb lattice at the edges, as will be illustrated below.

Thus prepared, we distinguish in the discussion of \eqref{et30} three cases: 

{\it  (i) The case $\Do(0){\neq} 0$ with source and drain located in the same sub-lattice.} 
Since the parent Hamiltonian $\mH$ exhibits balanced sub-lattices,
the truncated Hamiltonian, $(\mH)_\text{ds,ds}$, is imbalanced, 
$\zeta^\text{pre}{=}2$. 
There are at least two zero modes, so $\Do_\text{ds,ds}(0){=}0$
and $\transmission(0)$ vanishes. 
We thus confirm the qualitative findings based on wavefunction arguments
and make them rigorous within the H\"uckel model. 

{\it  (ii) The case $\Do(0){\neq} 0$ with source and drain located in different sublattices.} 
The truncated Hamiltonian $(\mH)_\text{ds,ds}$ 
is balanced; the presence of supernumerary 
zero-modes can be checked, {\it e.g.}, 
graphically employing \eqref{et32} for the 
case of benzenoids.
Certain substructures (`motifs') of alternating hydrocarbons, 
like dangling bonds, can contribute such a mode. 
By `dangling  bond' we refer to a situation where a lattice 
site couples to a single other site; 
a realization is found, {\it e.g.}, in cross-conjugated molecules
\cite{Solomon2008}. 
Further examples of such motifs have 
been listed by \textcite{Weik2016}.
However, cross-conjugation by itself is not a reliable 
indicator of destructive interference
\cite{Pedersen2015}. 
For instance, if dangling bonds come in pairs the respective zero 
modes can hybridize and split away from zero energy. 
In this case, a Fano-dip will survive only to the extent that 
hybridization can be considered very weak. 
We mention in this context that special classes of edge motifs can be
identified that are always accompanied with a zero mode
\cite{Weik2016}.
Examples from \textcite{Weik2016} have been reproduced in Fig. \ref{ft6c}. 
\begin{figure}[t]
\includegraphics[scale=0.45]{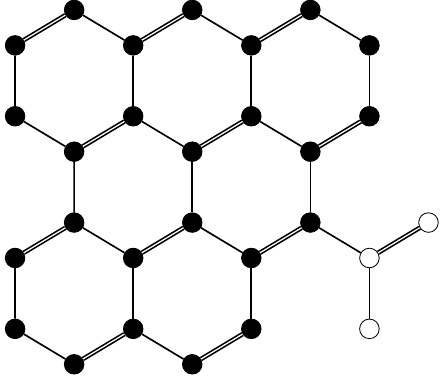} 
\includegraphics[scale=0.45]{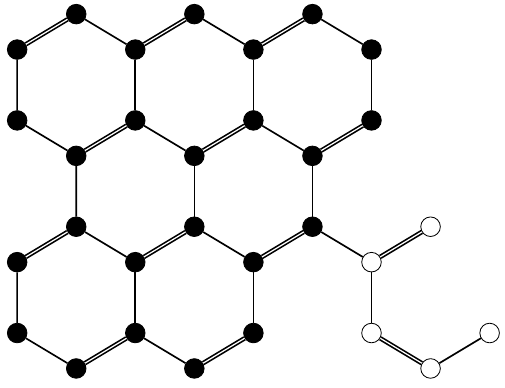}
 	\caption{Two examples of edge structures that are always 
 	associated with supernumerary zero modes. Left: The two singly 
        connecting sites share the same binding partner. Therefore
        only one of them can form a double bond; the other becomes a 
        radical. Right: Similar pattern. Reproduced from \protect\textcite{Weik2016}.} 
  \label{ft6c}
\end{figure}

{\it  (iii) The case $\Do(0){=}0$.}  
This situation has received much less attention, so far. 
A first discussion of the situation has been given by \textcite{Stuyver2017}.
We emphasize that a vanishing denominator in \eqref{et26} 
at $E{=}0$ is far from pathological: (i) a generic parent Hamiltonian $\mH$ will 
exhibit supernumerary zero modes also if it is balanced; 
(ii) \eqref{et26} has been written for the isolated molecule, 
because only a gapped spectrum can satisfy, in principle, 
the condition of zero spectral weight at a given energy $E$.  

We recall the exact relation, 
\bea
\transmission(E) = -\GL\GR \left| \Delta_\text{d,s}(E)/\Delta(E)\right|^2, 
\eea
where $\Delta{=}\det (E{-}\tmH)$ and $\Delta_\text{d,s}{=}\det(E{-}\tmH)_\text{d,s}$
with $\tmH=\mH{+}\mS$; $\mS$ denotes the self energy accounting 
for both electrodes. 
Since the molecule exhibits only a single contact orbital for source and drain
its transmission must be bounded, $\transmission(E) {\lesssim} 1$, for any energy
and electrode coupling. This implies that wherever the polynomial in
the denominator exhibits a root, there must be a corresponding root
with the same or higher multiplicity also in the numerator in the
limit of vanishing coupling, {\it i.e.}, in \eqref{et29}. 

We thus conclude that even in the case where $\mH$ exhibits zero
eigenvalues, the main ideas of the analyses outlined here remain
valid. Basically, the multiplicity of the roots of two polynomials has
to be determined with graphical rules. For bipartite systems this
implies counting radicals. If the number of radicals in the numerator 
determinants of \eqref{et29} exceeds the one in the denominator, 
then the transmission at zero energy is suppressed. 

The precise value of the transmission is going to depend on the cut-off mechanism. 
It can be determined, {\it e.g.}, by small violations of the symmetric form \eqref{et25}
as they are brought about by next-nearest neighbor terms or variations in on-site potentials. Alternatively, also the
tunnel coupling $\Gamma{=}\ci(\mS{-}\mS^\dagger){=}\GL{+}\GR$ 
can serve as a cut-off; for a first discussion see \textcite{Stuyver2017}.

\subsubsection{Applications} \label{sIVB4}
We {\color{black} illustrate the implications of} the rule \eqref{et30} by applying it to 
the three molecules shown in Figs.~\ref{ft6a} and \ref{ft6b}:
(a) the anthraquinone molecule, 
(b) the LC2-molecule of \textcite{Pedersen2015}, which is a combination of a carbon 6-site ring and a 4-ring,
(c) the azulene molecule (5-7 carbon double ring).\footnote{While anthraquinone has originally been proposed as a 
candidate for a redox-switch \cite{Dijk2006}, 
subsequent theoretical work revealed a strong variation of 
the transmission with energy near $\Fermi$, 
which could serve as gate-driven switch 
even in the absence of a genuine redox-reaction 
\cite{Markussen2010}.
When switching was later confirmed experimentally, 
the observed effect has then been 
attributed to electrochemically controlled {\qi} 
\cite{Darwish2012}.
}

\begin{figure}[b]
\includegraphics[scale=0.12]{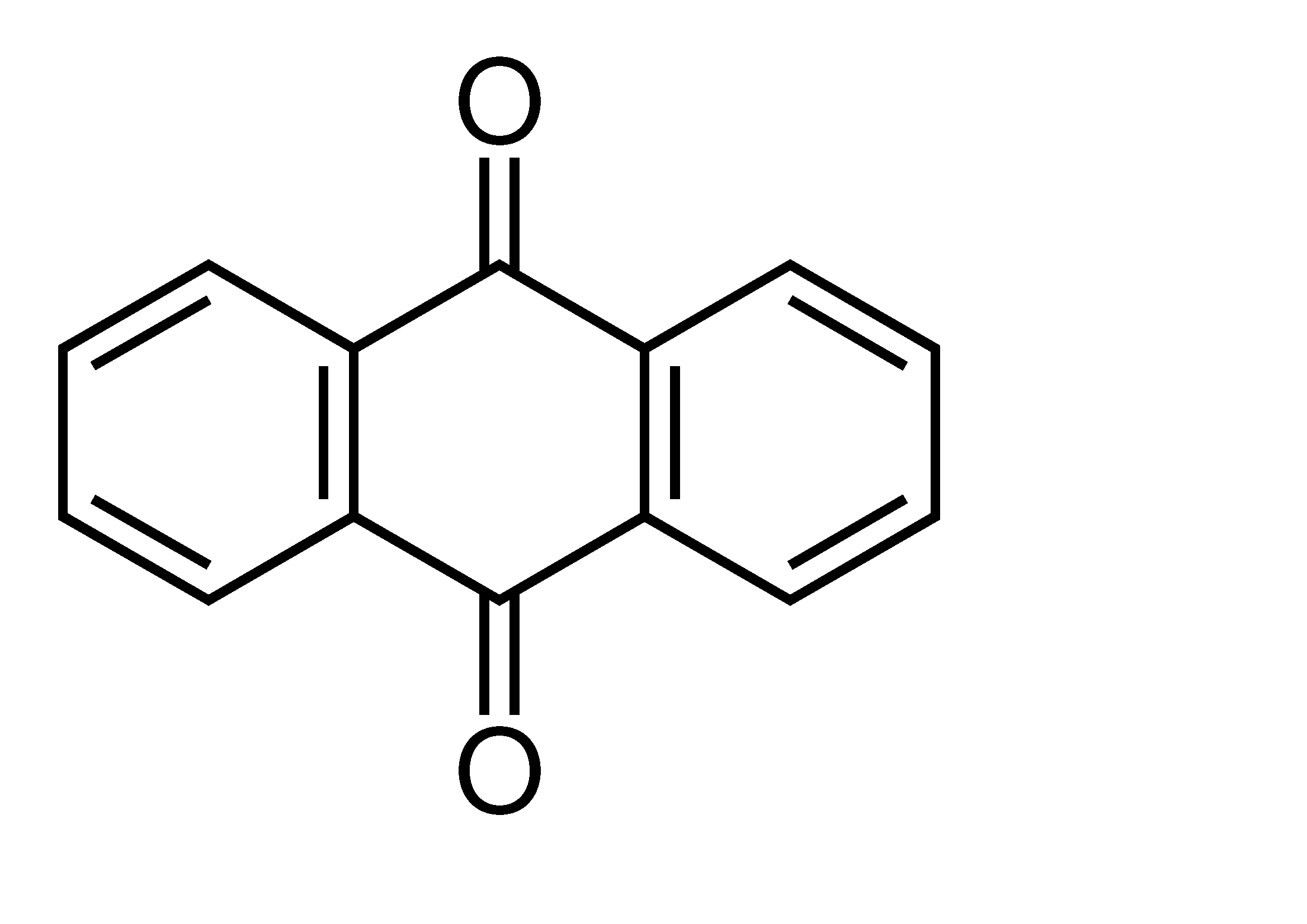}	
\includegraphics[scale=0.22]{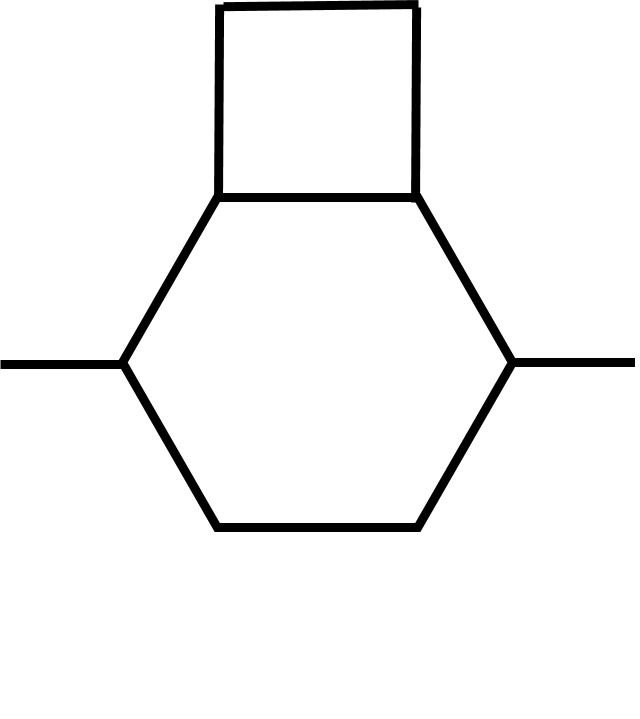}
	\caption{Structures of the molecules anthraquinone (left) and 
	the molecule LC2 from \textcite{Pedersen2015} (right).}
\label{ft6a}
\end{figure}

(a) We consider both oxygen sites of the anthraquinone 
as dangling and, following \textcite{Markussen2010} and \textcite{Guedon2012},
we attach source and drain to different sub-lattices.
The molecule exhibits a benzoid structure with sub-lattice 
symmetry. We have $\zeta^\text{sup}{=}0$ for the parent 
Hamiltonian, so $\Do(0)\neq 0$. 
However, after eliminating source and drain 
the truncated Hamiltonian has $\zeta^\text{sup}{=}2$, 
so $\Do_\text{ds,ds}(0){=}0$. 
Consistent with this result, a Fano-dip is seen in the 
model calculations. When the dangling bonds 
are removed, no supernumerary modes are found 
and Fano-dips are not expected, in agreement 
with \textcite{Markussen2010,Guedon2012,Stuyver2017}.
Since supernumerary zero modes come in pairs,  
a small perturbation added to $\mH$ will tend  
to lift this degeneracy and split the Fano-resonance. 
This prediction is in agreement with observations, 
see the related discussion in Section~\ref{sec_IVD1}.  

(b) LC2 carries contacts in para-position 
at the six-ring, Fig. \ref{ft6a}, and also exhibits a sub-lattice 
symmetry. Since it is balanced, there are no predictable zero modes.
However, the graph is not within the honeycomb class and 
 \eqref{et32} does not reduce to an equality;  
one needs to check for supernumerary zero modes, explicitly. 
While the parent graph turns out to have none, 
the truncated graph exhibits two supernumerary modes associated with the 
four ring. Hence, we predict destructive interference, consistent 
with \textcite{Pedersen2015}. 

(c) Following \textcite{Xia2014} 
  and  \textcite{Schwarz2016} 
we consider the azulene molecule, Fig. \ref{ft6b}. 
This molecule is not bipartite and, therefore, the full equation 
\eqref{et29} must be used. 
We first notice that the molecule is conjugated, i.e., 
there is a consistent covering of the graph with double-bonds. 
This implies the existence of a closed loop 
and therefore allows $\det{\mH}\neq 0$. 
For the transmission, we consider contacts at positions 
1 and 3 (Fig. \ref{ft6b}).  Then, after removing the source 1, 
and the drain, 3, the graph exhibits an isolated site, 2; 
hence, $\Do_\text{ds,ds}(0)=0$. 
Further, removing just one contact site, either 1 
or 3, the resulting graph can be covered with a loop for the 
seven-ring and a double bond between the two sites 
remaining from the five-ring. Hence, the determinants 
$\Do_\text{d,d}, \Do_\text{s,s}$ may be non-vanishing and
there is no prediction. As it turns out, there is indeed no 
Fano-dip at zero energy seen for this case 
\cite{Xia2014,Stadler2015,Schwarz2016}\color{black}, see 
Fig.~\ref{ft6b}.\normalcolor  

(d) Eq.  \eqref{et30} also applies to  
hexagonal graphene nanoflakes \textcite{Valli2018,Valli2019}. 
As one would expect, if source and drain couple to the same sublattice (`meta') a pronounced 
destructive \qi\ is observed, which is absent otherwise (`para' and `ortho').
The authors propose to use this effect 
for spin- and valley-filtering in electronic transport.

\begin{figure}
\includegraphics[width=\columnwidth]{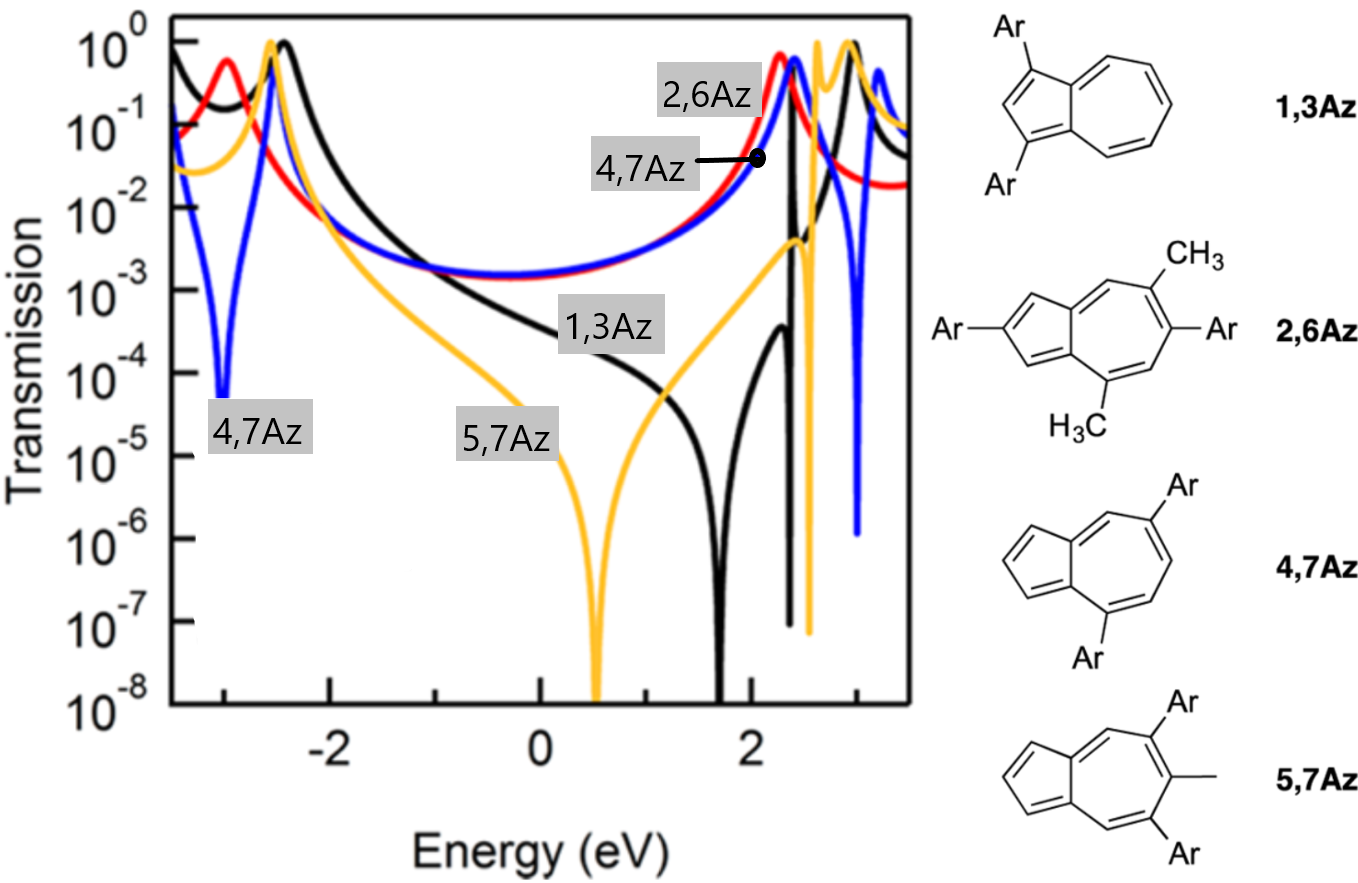}
\caption{\label{ft6b} 
\color{black}
Azulene with different positions of the linker 
group (denoted by Ar-) and corresponding
{\it ab-initio} transmissions. 
The molecule is not bipartite and 
so the (ab-initio) transmissions  
exhibit nodes shifted from the gap
center, in agreement with the rules discussed in the text.
Reprinted with permission from Xia {\it et al.}, Nano Lett. {\bf 14}, 2941. Copyright (2014) American Chemical Society.
}
\end{figure}


\subsubsection{{\qi} and ring currents}
The scattering states that enter the Landauer-B\"uttiker picture are not 
invariant under time reversal. For this reason, they 
generically support circulating (`ring' or transverse) currents,
unless these are suppressed by extra symmetries, 
such as mirror planes. 
They give rise to local, bias-induced magnetic 
fields and thus are physical observables that enjoy a unique 
definition \cite{Walz2014}.
A discussion is given by \textcite{Rai2010}. 
Their experimental detection has not yet been achieved, 
because (presumably) eddies are strongest on the atomic scale, 
where they are hard to resolve. 
However, a proposal has been made how the phenomenon could be 
studied experimentally in slightly larger, mesoscopic systems, 
where they actually might have a profound effect, for instance 
on spin-relaxation times \cite{Walz2014}.

Circulating currents are prevalent close to a Fano resonance, 
where the transport (longitudinal) current, $I_\text{tr}$, 
is suppressed.
Their qualitative behavior is 
discussed conveniently within a toy model
\cite{Walz2014}.  
At energies $E$ near resonance $\epsilon_\text{c}$ we have for the ratio of 
circulating to transport current 
$I_\text{circ}/I_\text{tr}\propto (E-\epsilon_\text{c})^{-1}$,  
which indicates that ring-currents can exceed 
the transport currents by orders of magnitude. 
In graphene samples with ad-atoms eddies with 
enhancements of two to three orders of magnitude have 
been computed \cite{Walz2014}. Note, that the ring-current switches 
sign when $E$ passes by the resonance, 
consistent with explicit model calculations \cite{Solomon2010,Rai2010}.

Due to its robust nature and strong signatures, 
ring currents have motivated a significant amount of 
theoretical research on transport in ring-shaped
molecular systems and, in particular, 
in Aharonov-Bohm type geometries. 
This includes the effect of magnetic fluxes on 
occupation numbers and the current-voltage 
characteristics \cite{Rai2011,Rai2012,Bedkihal2013}, 
and the interplay of spin-flip scattering and circular 
currents \cite{Rai2012a}. 
Transient phenomena have been investigated in detail 
\cite{Tu2012,Tu2016,Schoenauer2018}
and also effects of electronic correlations have 
been addressed \cite{Nuss2014}.

\subsubsection{Temperature and interaction effects}

The {\qi} effects discussed so far 
reflect the nodal structure of molecular wavefunctions. Changing this structure
costs energies that correspond to (purely) electronic excitations. 
Therefore, qualitatively {\qi} tends to be robust against
electron-electron interactions and thermal fluctuations
in small molecules 
\cite{Cardamone2006,Markussen2014}. 
A careful study of the Pariser-Parr-Pople (PPP) 
model, which simulates interactions in $\pi$-systems,  
came to a similar conclusion \cite{Pedersen2014}.

The most obvious effect of temperature on the transmission 
is smearing of the Fano-dip indicating  
an incoherent averaging over a thermal 
ensemble of molecular structures. 
Even though conceptually straightforward, 
the effect is somewhat tedious to describe 
from {\it ab-initio}, because many different vibrations 
are involved.

We continue describing additional, more subtle interference 
effects that appear {\em only} with interactions: 

{\it Vibrations.}  
A special situation can arise (near) degeneracies, 
where otherwise weak interactions can have significant effects. 
For instance, a coupling to vibrations, 
can enhance the inelastic scattering rate
so as to significantly weaken \qi, 
if interference is brought about 
by two nearly degenerate levels 
\cite{Haertle2011}. 
The temperature dependence of {\qi} in an anthraquinone has 
been attributed to this mechanism \cite{Rabache2013}.
Conversely, it has been reported that {\qi} 
can also selectively suppress signatures of vibrational 
modes in IETS spectra 
\cite{Lykkebo2014}.
The interplay of vibrations and destructive interference has recently 
been investigated also with graph-theoretical means \cite{Sykora2017}.

{\it Electron-electron interactions and many-body effects.}  
Much of the intuition that has been 
developed for {\qi} in molecules 
is based on tight-binding models. 
However, many-body effects have been identified  
that are not captured by effective single-particle descriptions. 

(a) Interaction effects lead to extra poles in 
the Green's function that indicate the existence 
many-body excitations. Such poles inevitably interfere 
with each other when being summed over in the construction 
of the many-body Green's function. 
An illustrative example is given by the Anderson model, where many-body 
excitations of a localized level emerge as lower and upper Hubbard peak, 
see \eqref{et6}. In this case, {\qi} manifests as an extra Fano-resonance, 
a `Mott node' in the terminology of \textcite{Bergfield2012}. 
In the context of many-body degeneracies, 
which appear, {\it e.g.}, in models of coupled quantum dots,
also more complicated interference scenarios 
can be realized \cite{Donarini2010,Niklas2017}. 
Of special interest is the Abrikosov-Suhl pole that is brought 
about by the Kondo effect. Like any other pole of the Green's
function,  it can give rise to interference phenomena. 
A most recent discovery in this context is the `Kondo blockade' 
discovered by \textcite{Mitchell2017} and discussed in the previous
Section~\ref{sIV.A2}.

(b) While the notion of molecular orbitals is robust 
against interaction effects \cite{Pedersen2014},
the energy ordering of orbitals can be modified as a consequence of 
strong Coulomb interactions. Evidently, this has a strong impact
on the relative weight and phase of the interfering poles in the 
many-body Green's function and therefore crucially enters \qi. 
Observation of this rather striking effect in STM-experiments 
has been reported recently \cite{Yu2017}.

\section{Key experimental results and their semi-quantitative understanding}\label{s.Achievements}

Before we turn to a quantitative comparison between theory and experiment 
 in the next chapter 
it will be useful to highlight a number of results that illustrate the great level of qualitative, or semi-quantitative, understanding that we have achieved. We will not attempt a full overview of the literature of single-molecule transport, but we will focus on results that uncover systematic trends and important physical effects in molecular junctions. By this overview we illustrate an important conclusion: despite the many unknown and poorly understood factors listed in chapters  \ref{s.Experimental} and \ref{s.Computational}
many of the qualitative features have been understood. This implies that such features are robust against  variations in electrode configurations and molecule-electrode bonding patterns and other poorly known factors, and that they are robust against the approximations made in developing the theory. It will be interesting to investigate why we find this robustness, but also to probe under which circumstances this breaks down. Roughly 
speaking, this chapter will be devoted to the robustness, and the next two chapters will explore the limits of validity and break-down of this robustness.

\subsection{Conductance as a function of length} \label{ss.length-dependence}  
The foremost systematic characteristics studied for molecular wires is the length dependence of the conductance. The 
case of alkanes has been 
investigated extensively and for us will serve as a paradigm
\cite{Akkerman2008}.
In addition to these carbon-based wires, also wires based on other elements, such as Si (silanes) 
and Ge (germanes), have been studied with 
qualitatively very similar conclusions \cite{Su2017};  
an overview may be found in \textcite{Su2016}, Table 1, 
and \textcite{Gunasekaran2018}.  

In the review by \textcite{Akkerman2008} the data obtained from many measurement
techniques was compared as a function of the number, $N$, of carbon atoms in the chain, 
ranging from $N= 2$ to $N=28$.
Such data will be discussed more quantitatively in the next chapter. 
Of relevance to us, here, is the observation that 
the conductance decreases typically exponentially with the number $N$, 
{\it i.e.}, as a function of the length of the chain, $G(N)\sim G_\text{c} \exp(-\beta N)$, (for more details, see Section~\ref{ss.ADT}). 
Here, the inverse of $G_\text{c}$ defines the contact 
resistance associated with left and right anchors while 
the exponent $\beta$ describes 
the attenuation coefficient of the transmission per 
wire unit. For alkanes the reported literature values range 
from $\beta = 0.8$ to 1.1 with only few exceptions \cite{Tewari2018}.\footnote{There 
are two established ways of expressing the exponential dependence: 1. By the number of monomers added, which is the most unambiguous, here indicated by the symbol $\beta$. 2. By the length in nm (or \AA) per monomer, for which we reserve the symbol $\beta_\ell$. The latter is useful when we are interested in the resistance dependence on length, but requires a conversion step. For alkane chains the C-C bond length is typically used in the conversion. However, the carbon wire backbone is not straight. In stead, one could use the C-C bond length projected along the wire axis, but this would be a property that depends on the state of stretching of the wire. When, adhering to widely adopted practice, we express $\beta_\ell$ as a function of length using a straight C-C bond length of 1.26 \AA, the decay constants are 0.63 -- 0.87 \AA$^{-1}$. We thank Latha Venkataraman for discussions on this topic.} 
For silanes and germanes smaller values have been found, $\beta = 0.75$ as compared to $\beta = 0.94$ for alkanes obtained under similar experimental conditions \cite{Su2015}. 

In this spirit, the length dependence of the conductance 
for many other molecular wires has been analyzed in terms of an effective exponent $\beta$. 
We will argue below that such exponential dependence is often not properly justified.

\subsubsection{Basic concepts}

For a convenient discussion of the experimental observations, 
we briefly recall the relevant theoretical concepts 
\cite{Gunasekaran2018}.
Long molecular wires that are built out of a single
repetitive unit can be categorized in terms of electronic 
band structure theory. In particular,
molecular orbitals take the form of Bloch-states 
with an associated crystal momentum $k$.
The $k$-state classification is highly useful
even for wires with a finite length, $N$. This is because
similar to the `particle-in-the-box' problem, 
the electronic properties of the molecule can be obtained  
from the properties of the crystalline wire by 
imposing selection rules on `allowed' $k$-space momenta. 
\newcommand{\NStar}{N^*} 
As a consequence, there will be a length $N{>}\NStar$
beyond which the molecular wire exhibits properties that 
fully reflect the insulating limit $N\rightarrow \infty$, as is applicable for alkane wires. 
In particular, in this asymptotic limit the HOMO-LUMO gap $\Delta_N$ 
approaches the bulk gap $\Delta_\text{bulk}$ and 
the attenuation,  
$$
\beta \coloneqq -\frac{d\log G(N)}{dN}
$$
takes a constant value, $\beta_\infty$. 
Depending on the molecule and its anchor groups 
the asymptotic regime, $N{>}\NStar$ may be 
very challenging to reach in experiments; 
even small but systematic deviations of $\beta$ 
from a constant may indicate that this regime is still 
very far. 

In order to rationalize how $\beta_\infty$ relates to 
the band structure of the infinite wire we recall
that 
$
\epsilon_\text{H} < \Fermi < \epsilon_\text{L}  
$
where $\epsilon_\text{H,L}$ approach the top of the 
valence band or bottom of the conduction band  
at $N{>}\NStar$, respectively. We recognize this as a tunneling problem
where the height of the barrier, $\delta$, is given (approximately) by 
the energy difference of $\Fermi$ to either 
$\epsilon_\text{H}$ or $ \epsilon_\text{L} $, depending on which is closer. 

To relate $\delta$ to $\beta_\infty$, we recall
a result for the exponent familiar 
from the one-dimensional tunneling problem,   
$\beta \sim \sqrt{2m(V_\text{barrier}-\Fermi)}/\hbar$. 
In the case of the molecular barrier, the effective mass, $m$, follows 
directly from the curvature of the band-structure around 
the band-edges. For instance, in the case where the LUMO is close
to $\Fermi$, we have an implicit definition 
$
\delta \approx -\varepsilon_\text{cond}(\mathfrak{i}\kappa) 
$
and $\beta_\infty=\kappa a$. The formula involves the 
band structure $\varepsilon_\text{cond}(k) $ of the conduction band 
(counted from bottom of the band)
and the crystalline lattice constant, $a$.
The expression is further motivated in 
complex band-structure theory \cite{Reuter2017}. 
{\color{black} A formula that interpolates between the two limiting 
cases where $\Fermi$ is close to either one 
of the frontier orbitals has been 
derived by \textcite{Joachim2002}.} 

We emphasize a basic consequence of these considerations that 
often is not fully appreciated: In the asymptotic limit, 
the exponent $\beta_\infty$ is a property of the molecule alone. 
It does not reflect any aspects of the molecular junction other than 
the location of the Fermi-level.
{\color{black} In particular, $\beta_\infty$ 
does not depend on the choice of the anchor 
groups.}\footnote{\color{black} \textcite{Stuyver2017a} consider 
the possibility that the anchor-groups affect 
$\beta_\infty$ by shifting the Fermi-energy $\Fermi$.
We would like to point out here that a single molecule attached 
to metallic substrates cannot modify $\Fermi$. 
Such an effect can take place only in the presence of a finite concentration 
of molecules, as they occur in self-assembled monolayers. 
In this case the surface-dipole of the substrate - and hence $\Fermi$ - 
can indeed be modified by an amount that scales with the concentration 
of molecules per surface area. {\it Ab-initio} calculations can illustrate this effect \cite{Obersteiner2017}.
}

With an eye on experiments we note that our discussion focuses 
on phase-coherent transport. If the wire length increases beyond 
the phase-coherence length, $N_\text{coh}$, the exponential decay of the conductance 
will give way to a weaker decay which reflects an incoherent and 
strongly temperature dependent dynamics, see Section~\ref{ss.incoherent}.

\subsubsection{Conjugation and metallicity}

As is well known in organic chemistry, the properties of conjugated molecular wires differ 
strongly from those of carbon chains with all saturated bonds 
as exemplified by the alkanes. 
Conjugated molecular wires are characterized by a 
path of alternating single and double C--C bonds, as a result of 
dangling p-orbitals on each of the carbons. 
Every unit cell contributes 
a single-electron to the conduction band. Therefore, 
conjugated wires generically exhibit a metallic behavior, 
{\it i.e.}, a HOMO-LUMO gap that vanishes like $1/N$ in the asymptotic limit.
Exceptions occur in the presence of strong interactions \cite{Schmitteckert2017},
or if instabilities interfere, such as the Peierls-transition
in polyacetylene \cite{Heeger1988}. 
In the absence of a band-gap, the conditions for a purely exponential 
length dependence of the conductance of a molecular 
wire are not fulfilled.  

Effects of breaking the conjugation have been discussed 
in Section~\ref{sIVB4}. In essence, breaking the conjugation, 
even at only a single point along the wire,
introduces a very strong scattering center and therefore 
leads to a reduction of the conductance, 
as illustrated, 
{\it e.g., } for 
single-molecule measurements on 
oligo(p-phenylene ethynylene) 
(OPE) derivatives \cite{Kaliginedi2012}. 

\begin{figure}
\begin{center}
\includegraphics[scale=0.065]{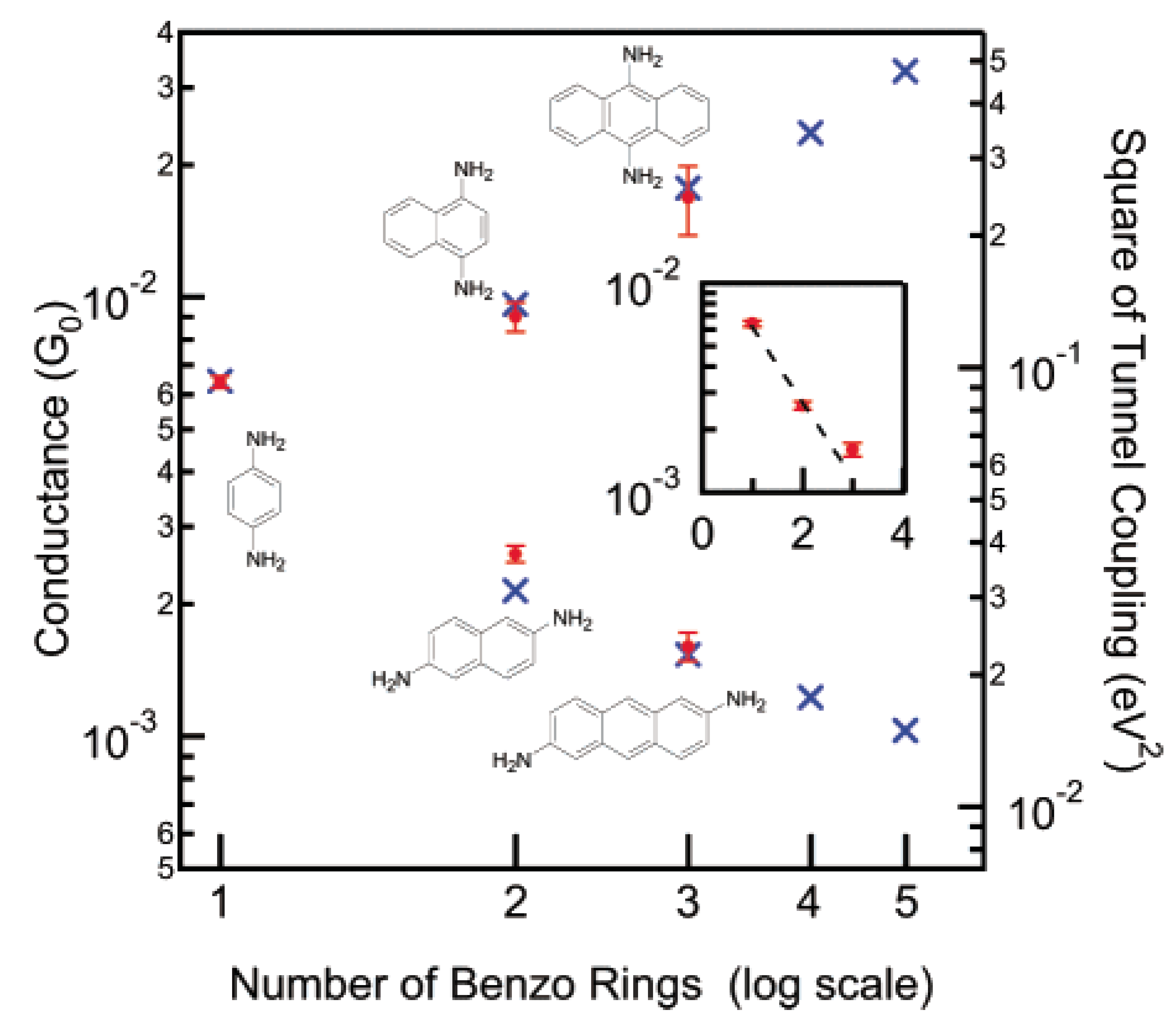}
\end{center}
\caption{Conductance measured for a series of diaminoacenes as a function of the number $N$ of benzo rings (dots with error bars, 
left axis). When measured in the transverse direction (upper data points) the conductance increases with $N$, while 
measured longitudinally the conductance decreases. The log-lin plot in the inset shows that the latter is not a simple 
exponential decrease. The two trends are reproduced by calculations for the square of the tunnel coupling (crosses, 
right scale). 
Reprinted with permission from Quinn {\it et al.}, J. Am. Chem. Soc. {\bf 129}, 6714. Copyright (2007) American Chemical Society.
}\label{fig.diaminoacenes}
\end{figure}

\subsubsection{Length dependence for conjugated wires}

In several series of experiments very small attenuation constants $\beta_\ell$ for molecular wires have been reported. 
For instance, for the OPE 
molecular wires just mentioned \cite{Kaliginedi2012,Liu2008} 
values for $\beta_\ell$ of 0.34 and  0.21~\AA$^{-1}$ were found.
Still smaller values have been obtained for 
oligothiophenes $\beta_\ell = 0.1$~\AA$^{-1}$ \cite{Yamada2008}, 
oligoyne $\beta_\ell = 0.06$~\AA$^{-1}$ \cite{Wang2009} 
and oligoporphyrins, $\beta_\ell = 0.04$~\AA$^{-1}$ \cite{Sedghi2011}. 
The small $\beta_\ell$ values have been used for arguing that these wires are 
in the metallic regime. We would like to add three warnings: 

First, we have argued above that one can equally well represent the decay constant 
$\beta$ in terms of the decay per monomer. 
The attenuation constants in units per monomer for OPE molecular wires are $\beta = 1.5$ \cite{Liu2008} and 2.35 \cite{Kaliginedi2012}. 
For oligothiophenes $\beta = 0.42$  \cite{Yamada2008}, for oligoyne $\beta = 0.18$  \cite{Wang2009} and for oligoporphyrins $\beta = 0.55$  \cite{Sedghi2011}, 
and for oligoacenes \cite{Quinn2007} in the longitudinal direction $\beta =  0.7$. This shows that most of these numbers are comparable to those typical of insulators. 

Second, the series of molecules considered in these works is small,
in nearly all cases covering only three points. From these three points one cannot
rigorously distinguish an exponential dependence from, {\it e.g.}, a dependence on inverse length. The exponential 
dependence would be consistent with a finite energy gap in the $N \rightarrow \infty$ limit, meaning that the wire is an insulator.
A decrease of conductance as $1/N$ would be consistent with a metallic wire in the hopping regime. 

Third, the widespread practice of extracting a simple exponent from the decay of the conductance 
is at variance with the expected behavior for phase coherent metallic wires, such as given 
in the theoretical considerations below.

Fig.~\ref{fig.diaminoacenes} reproduces experimental data 
obtained by \textcite{Quinn2007} 
showing how the transmission of an oligoacene evolves 
with increasing length, $N$, depending on the placement of the anchor groups.
The conductance decreases when measured along the wire length, 
but the trend is not purely exponential, as illustrated in the inset 
of the figure. In contrast, the conductance measured in the transverse 
direction increases with length, as shown for the upper series in the plot. 
We take this latter observation as a strong indication that the 
conductance in this type of wires can be thought of as phase-coherent.
For describing length-dependence of phase coherent transport 
in metallic molecular wires we will
invoke basic scaling arguments.

{\it Theoretical considerations.} We follow \textcite{Yelin2016} and 
focus on the situation of well-separated levels where the 
conductance is dominated by a single orbital, only. 
In this case the transmission $\transmission(E)$ can be approximated by three parameters, $\Gamma_\mathcal{L,R}(N)$ and the level position relative 
to $\Fermi$, $\epsilon(N)$,
\begin{equation}
\label{et53}
\transmission(E) = \frac{\Gamma_\mathcal{L}\Gamma_\mathcal{R}}{(E-\epsilon)^2 + 
(\Gamma_\mathcal{L}+\Gamma_\mathcal{R})^2/4}.  
\end{equation}
Since the band-structure typically is analytic near the Fermi-energy, 
we make an expansion  
\begin{equation}
\label{et54} 
\epsilon(N) = \epsilon_0 + \epsilon_1/N + \epsilon_2/N^2 + \ldots. 
\end{equation}
As written here, the expansion applies to metals and insulators.
In the latter case $\epsilon_0$ accounts for the offset between the Fermi-energy and 
the closest band-edge. For the case of metallic wires, which we consider here, 
$\epsilon_0 {=}0$.
Equation \eqref{et54} formalizes the idea 
that the spectrum of finite-length 
wires derives from the band-structure of the bulk 
by imposing $k$-space selection rules with 
$\delta k\sim 2\pi/N$ for neighboring $k$-values.
The slope $\epsilon_1$ is of the order of the 
Fermi energy, while the curvature $\epsilon_2$ 
corresponds to an (inverse) band mass. 

Notice that $k$-space selection rules are sensitive 
to boundary effects 
\cite{Dasgupta2012,Korytar2014}.
For example, electrophilic anchor groups
can shift a LUMO-based transport resonance
closer to $\Fermi$. Therefore, in general 
the expansion coefficients comprise information 
about the molecular wire {\em and} its anchoring. 

Concerning the level broadening, we observe that 
the wire's Bloch-states extend homogeneously 
over the wire. Hence, asymptotically, their overlap with the contact sites, 
which connect to the electrodes, 
is inversely proportional to the length of the wire, 
\begin{equation}
\label{et55} 
 \Gamma_\mathcal{L,R}\approx c_\mathcal{L,R}/N .
\end{equation}
Collecting formulae and inserting into \eqref{et53} we obtain,
\begin{equation}
\label{et56} 
 \transmission(\epsilon_0) \approx \frac{\transmission_\infty}{1 + 2(\epsilon_1/\epsilon_2) N^2_{c}/N + (N_{c}/N)^2} ,
\end{equation}
where $\transmission_\infty= 4c_\mathcal{L}c_\mathcal{R}/(4\epsilon_1^2 + (c_\mathcal{L}+c_\mathcal{R})^2)$
and $N^2_{c} = 4 \epsilon_2^2/(4\epsilon_1^2 + (c_\mathcal{L}+c_\mathcal{R})^2)$. 

As is readily seen from this result, 
if $\epsilon_1$ and $\epsilon_2$ have the same sign
the conductance increases monotonically 
approaching the asymptotic value, $\transmission_\infty$, 
from below.  
In the opposite situation, 
in which $\epsilon_1$ and $\epsilon_2$ have opposing signs, 
the evolution of the transmission can be 
non-monotonic. It will first move through a maximum of  
$\transmission_\text{max}=4 c_\mathcal{L}c_\mathcal{R}/(c_\mathcal{L} + c_\mathcal{R})^2$ at 
$N_\text{max}=|\epsilon_1/\epsilon_2|$ before approaching the asymptotic value 
from above. 

\newcommand{\eTrans}{\epsilon^\text{trans}}
\newcommand{\eLong}{\epsilon^\text{long}}
{\it Application to experiment.} 
The concepts developed above have been successfully applied 
for understanding the evolution of the conductance 
of oligoacene wires under conditions where these 
wires were attached directly to the metal leads (Ag or Pt), without 
employing anchor groups \cite{Yelin2016}. 
{\color{black}
\begin{figure} 
\includegraphics[width=\columnwidth]{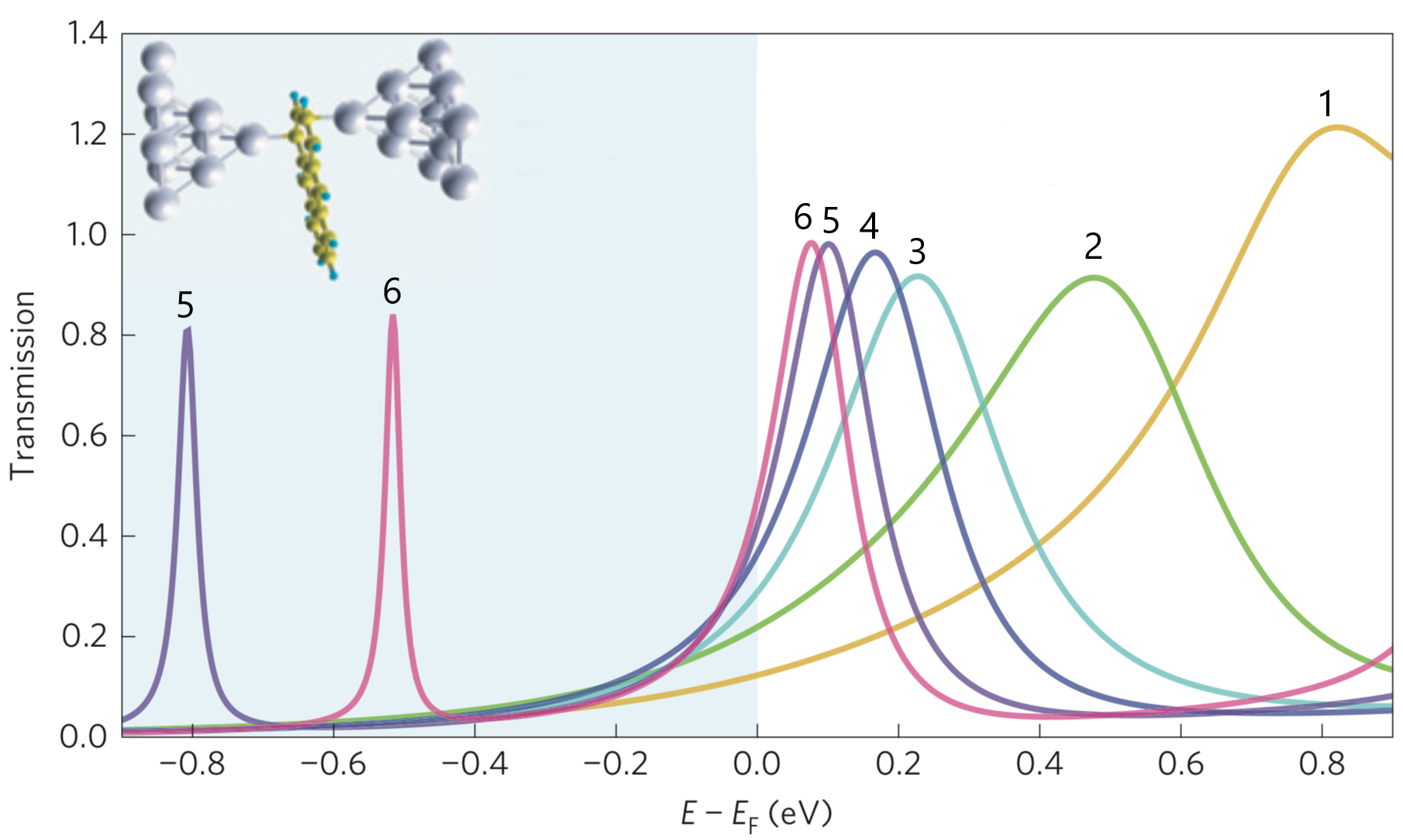}
\caption{\label{Fig:Yelin2015}\color{black}
Transmission resonances of oligoacenes directly bound to Ag contacts
for geometries as exemplified in the
inset for anthracene. The plot gives an example for the evolution
of the transmission resonances that exhibit growth of
the transmission with increasing molecular length. The numbers above the 
curves indicate the numbers of carbon rings: benzene (1), naphthalene (2), 
anthracene (3), tetracene (4), pentacene (5), and hexacene (6). 
Adapted with permission from Yelin {\it et al.}, `Conductance saturation in a series of highly transmitting molecular junctions', Nature Mater. {\bf 15}, 444. $\copyright$ Springer Nature (2016).
}
\end{figure}
In the case of Ag leads, the molecular level spacing is much
larger than the level broadening, and the transport is
entirely dominated by the LUMO's, as shown by 
{\it ab-initio} transmissions in Fig.~\ref{Fig:Yelin2015}.
The exception is benzene ($N=1$) with two transport-active
orbitals. For $N>1$, the resonances are approximately
Lorentzian, and the scaling arguments of Eqs.~(\ref{et54}-\ref{et56})
apply:
Both, the peak width and the peak position, decrease with $N$ and
the saturation is observed.

}

Returning to Fig.~\ref{fig.diaminoacenes}, 
where amine-anchors have been used in transport 
measurements for a series of acenes, we analyze the trends observed there
thus giving a fresh example to demonstrate how useful these 
concepts can be. 
In our analysis we will assume that the level-broadening is less 
sensitive to the position of the anchor groups, 
transverse or longitudinal, 
as compared to the level spacing, which we will refer to as 
$\eTrans(N)$ and $\eLong(N)$, respectively. 
Our assumption implies that the expansion coefficients 
$c_\mathcal{L,R}$ in \eqref{et55} 
for the transversal and longitudinal placing 
of the anchor groups are roughly the same.\footnote{
To motivate this statement, we recall that in tight-binding 
descriptions the level broadening is given by 
$|t|^2\rho_\text{Fermi}$ where $t$ denotes a hopping matrix element
between the molecule and the reservoir 
and $\rho_\text{Fermi}$ the density of states on the 
reservoir contact site. One would expect that neither $t$
nor $\rho_\text{Fermi}$ are very sensitive to 
the placement of the contacts, transverse or longitudinal. 
}
Furthermore, from  
\textcite{Yelin2016} we adopt the result that transport will be LUMO-based, 
so that $\epsilon_{1}$ is positive for both cases, 
longitudinal and transversal. Since the asymptotic values
of the conductance seen 
in Fig.~\ref{fig.diaminoacenes} are very different, 
$\transmission^\text{long}_{\infty}{\ll} \transmission^\text{trans}_\infty$,
we conclude $\eTrans_{1}{\ll} \eLong_1$.
This finding is understood as follows:  
Due to the electrophilic character 
of nitrogen, the amino-based anchor groups pull the LUMO-level closer to $\Fermi$. 
If the anchor groups are far apart (longitudinal) 
they compete when attracting molecule-based charge, 
while they cooperate if they are close (transverse). 
Therefore, the LUMO is expected to be 
closer to $\Fermi$ 
in the latter case and the transmission is 
enhanced.  The difference between $\eTrans_{1}$ and $\eLong_1$ is further amplified by
image charges in the leads. These have the effect of reducing the HOMO--LUMO gap, and this reduction
grows strongly when the electrodes are closer to the molecule.

We observe that in the limit of short wires, 
there is no pronounced difference between 
both cases, $\transmission^\text{long} = \transmission^\text{trans}$ for $N \downarrow 1$. This matching condition 
can be satisfied if the large contribution of the longitudinal 
case,  $\eLong_{1}$, is partially canceled by 
the second term $\eLong_{2}$ in \eqref{et54}. To 
facilitate this, the two coefficients
should have opposite signs, 
$\eLong_{1}{\approx} {-}\eLong_{2}$. 
Note that there is no 
such expectation in the transverse case. 
Correspondingly, 
Eq.~\eqref{et56} predicts an asymptotic decay of the 
transmission in the longitudinal 
case (crossover length $N_\text{max}{=}|\epsilon_1/\epsilon_2|{\approx} 1$) 
and an increase in the transverse case, qualitatively 
consistent with the experiment shown in Fig.~\ref{fig.diaminoacenes}.

\subsubsection{Incoherent transport limit \label{ss.incoherent}}

Several groups have reported a transition from exponential decay of the conductance with wire length, $N$, to a slower, nearly linear dependence above a certain value of $N$. 
Examples can be found in 
\textcite{Choi2008,Choi2010} for oligonaphthalenefluoreneimine up to $N=10$, 
and in \textcite{Hines2010} for conjugated molecular wires up to 9.4~nm in length. 
While the phenomenon is attributed to a crossover from coherent tunneling 
to thermally activated hopping, the deeply inelastic regime 
is not experimentally investigated in detail, yet.  
One expects that molecular-type Bloch states, which originally extend over the full length of the molecule, become localized. 
Several mechanisms are conceivable that can drive the process. 
One possibility is a spontaneous breaking of translation invariance due to the formation of a polaron. 
The process could be effective in wires, which have a soft molecular backbone or in 
strongly polarizable environments. 
Alternatively, a thermal activation of deformations of the molecule, notably ring 
rotations (see the next section) could be involved.  
This would suggest that the crossover is strongly temperature dependent, as is indeed 
observed \cite{Choi2010,Hines2010,Smith2015}.

\subsection{Conductance as a function of molecular conformation}
The breaking of conjugation in molecular wires has been studied systematically by designing a series of molecular wires for which the neighboring phenyl groups have a rotation fixed by suitable choice of side groups. For a series of biphenyl-based  molecules with varying degrees of sterically constrained rotation of the two phenyl rings \textcite{Venkataraman2006b} found that the conductance for this series decreases proportional to $\cos^2(\vartheta)$, with $\vartheta$ the angle between the two rings, see Fig.~\ref{fig.twist-angle}. This is the dependence expected as resulting from the overlap of the $\pi$-orbital systems on the two rings. This was confirmed by other methods of constraining the ring rotations in the study by \textcite{Mishchenko2010}.

The observed dependence $G=a\cos^2(\vartheta)$, agrees with detailed {\it ab-initio} computations \cite{Mishchenko2010,Hybertsen2008}. The twist angle dependence can be reproduced by a simple two-site model, where the two sites represent the two phenyl rings \cite{Mishchenko2010}. As long as the molecular levels are far removed from the Fermi energy, compared to the energy level broadening and the inter-site coupling, the angle dependence is purely given by $\cos^2(\vartheta)$. When the levels move closer to resonance, terms in $\cos^4(\vartheta)$ appear, but due to cancellations they remain small.   

Note that the slope $a$ obtained from fitting the DFT results is 3 orders of magnitude higher than the slope obtained from the experimental data, because all conductance values are so much lower in experiment. One of the possible explanations offered by \textcite{Mishchenko2010} is that the transport under the experimental conditions is not fully coherent. Even under such conditions, the $\cos^2(\vartheta)$-dependence would be robust and survive.

\begin{figure}[!h]
\includegraphics[scale=0.06]{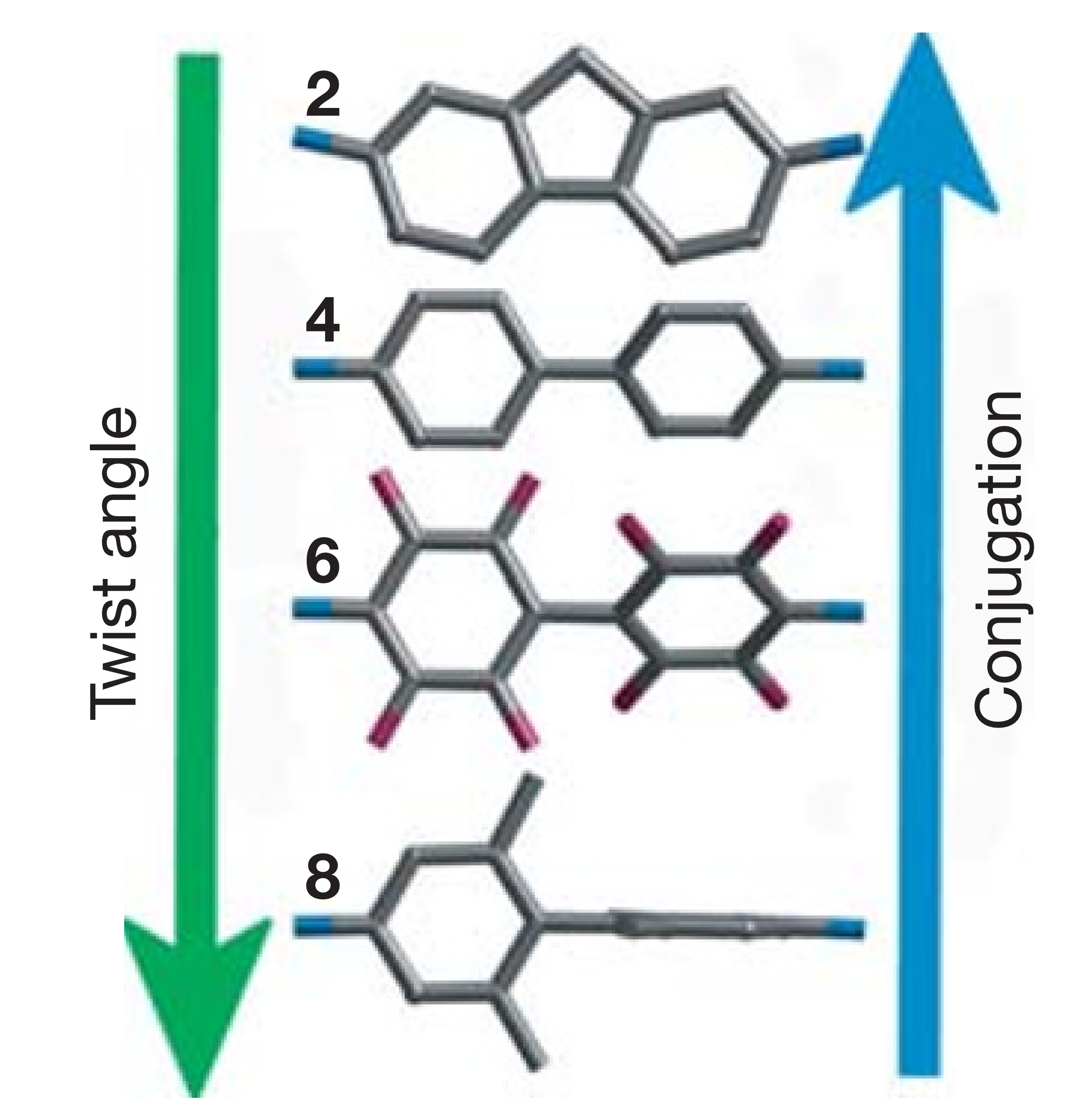}
\includegraphics[scale=0.06]{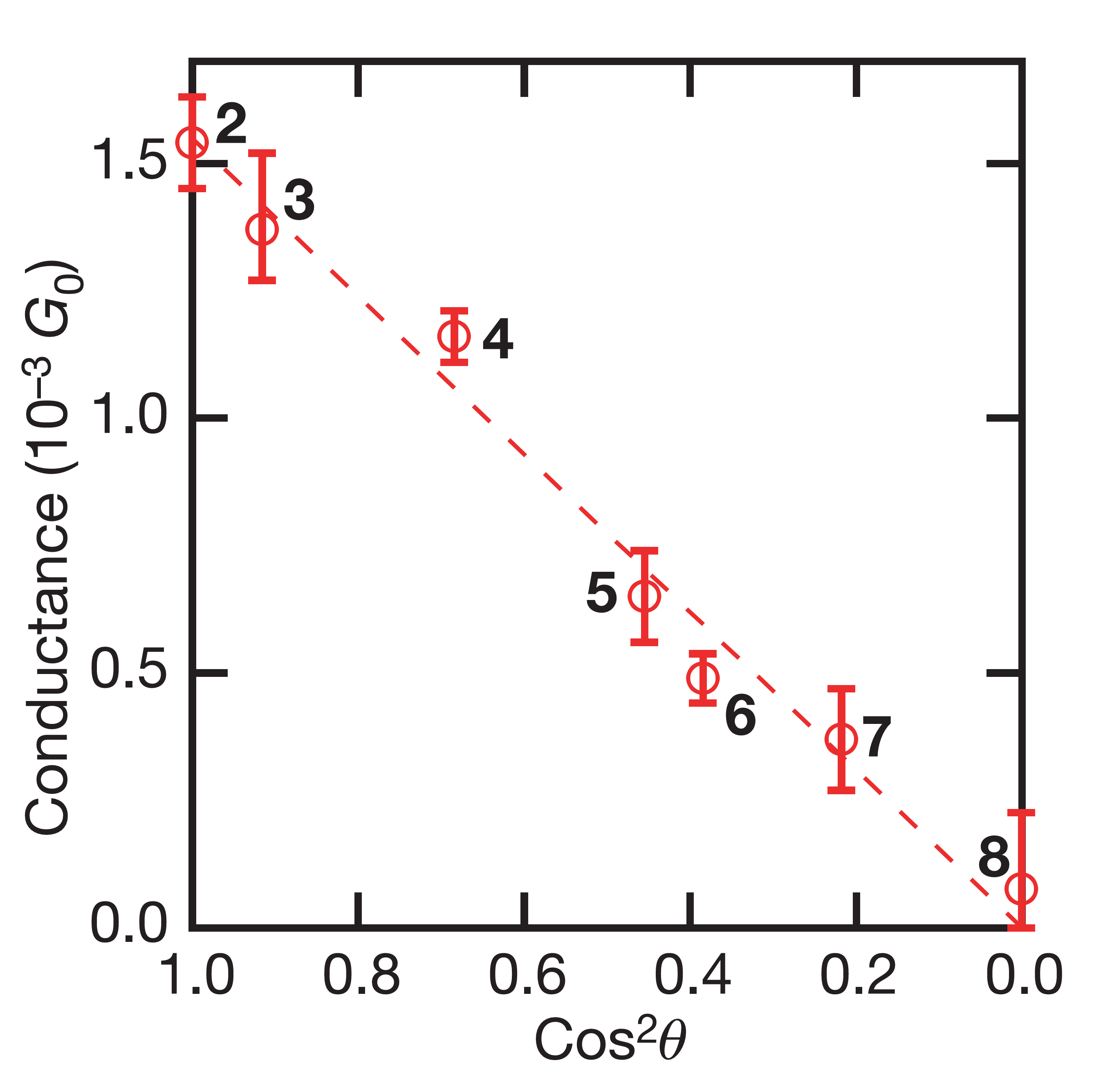}
\caption{Conductance as a function of the twist-angle between two phenyl rings. By design steric interactions constrain the twist angles between the two phenyl rings of a series of molecules (left). The smaller the angle, $\theta$, the larger is the overlap of the wavefunctions (the conjugation) between the two ring sections. The measured conductance follows the expected  $\cos^2(\theta)$-dependence (right). 
Adapted by permission from Venkataraman {\it et al.}, `Dependence of single-molecule junction conductance on molecular conformation', Nature {\bf 442}, 904. $\copyright$ Springer Nature (2006).
}
\label{fig.twist-angle}
\end{figure}


\subsection{Anchor groups \label{ss:VC}}

The physical properties of a molecular junction depend 
significantly on the atomic details of how the molecule 
binds to the electrodes. For this reason the optimal choice 
of the anchor group has been an important research 
topic in the field from the beginning. The role of the molecule-electrode interface has been recently covered 
in an excellent review by \textcite{Su2016}, 
highlighting the relevant chemical principles.
Further discussions, and a list of anchors that have been tested, 
{\color{black}
are given by \textcite{Jia2013} and \textcite{Hybertsen2016}.
}

One may identify three important roles that anchor groups have in determining the properties of single-molecule junctions. Obviously, the end groups need to provide mechanical anchoring of the molecule to the metal leads, in the sense of bonds that resist breaking by thermal or mechanical agitation. Second, they should provide electrical contact between the metal leads and the core of the molecule. In many cases, the goal has  been to achieve nearly unimpeded transmission of electrons from the metal to the core of the molecule such that the properties of the later dominate the junction properties. Finally, the anchor groups influence the alignment of the frontier orbitals of the core of the molecule with the Fermi level in the leads.

\subsubsection{Thiol-based anchoring groups}

In most experiments \cite{Cuevas2010} Au electrodes in conjunction with thiol anchors have been used. This prominent role deserves special mention here and, in addition, we will present arguments that the nature of the bonding is still a matter of debate.

Au surfaces are natural candidates because they exhibit a low tendency towards contamination. The choice for thiol-linkers is mostly motivated by the fact that these make strong bonds to Au, promoting a single point of contact between the molecule and each of the leads.
The mechanical coupling they provide is strong enough for producing frequently appearing plateaus in conductance breaking traces, and the electronic coupling is sufficient for the properties of the core of the molecule to be observable.
They have the drawback that thiols are easily oxidized, resulting in polymerization by the formation of $-$S-S$-$ bonds between the molecules. This can be avoided by replacing the H in the thiol group by a protection group, often chosen to be an acetyl group.  This protection group slows down the immobilization kinetics of the molecules \cite{Elbing2005}. The protection group can be removed during exposure of the Au surface to the molecular solution by adding NH$_4$OH \cite{Tour1995} or tetrabutylammonium hydroxide, TBAOH, \cite{Grunder2007} as a de-protection agent. Alternatively, the acetyl protected molecules can directly adsorb on the Au electrodes without the use of any de-protection agent. However, it has been shown \cite{Tour1995} that for the direct adsorption of the thioacetyl containing molecules and formation of self-assembled monolayers a larger concentration of molecules is required.

For all work on thiol anchors, an important issue was raised by \textcite{Stokbro2003}, namely the question whether in the interaction between the thiol group and the Au surface the hydrogen atom actually splits off. The computations suggest that the thiol bond (with the S-H bond intact) and the thiolate bond to the Au surface (with the H removed) are nearly equivalent in energy, and there is experimental evidence that both may occur \cite{Rzeznicka2005}. 
\color{black}
Recent experimental evidence based on STM break junctions for single molecules by \textcite{Inkpen2018} shows that the formation of thiol or thiolate bonds sensitively depends on the preparation conditions.
\normalcolor

\subsubsection{The role of the mechanical coupling}\label{sss.coupling}

One may be inclined to select anchor groups that provide the strongest mechanical coupling. On the other hand, arguments have been put forward that optimizing anchor groups towards strong mechanical coupling may not be favorable for producing clear signatures of molecular conductance in conductance histograms.  In STM break-junction experiments the conductance distribution (peak width) obtained with strongly binding thiol linkers was found to be much wider than that for amine anchors that have smaller binding strength \cite{Venkataraman2006a}. The interpretation offered is based on the flexibility of the Au-amine bond, which leaves the arrangement of the  Au surface atoms unaffected. In contrast, the Au-S bond is stronger than a Au-Au bond resulting in restructuring of the metal electrodes during stretching of the contacts. Consequently, conductance histograms will be based on many metal electrode surface configurations  (for a review, see \textcite{Li2007,Hybertsen2016}). 

Surprisingly, the opposite result was found for MCBJ break junctions by \textcite{Martin2008}: the thiol coupled molecules produced a stronger and sharper signature in the conductance histograms as compared to their amine-coupled counterparts. 
{\color{black} In a study comparing results for different anchoring groups \textcite{Chen2006} found no major difference between thiol-, amine- or carboxylic-acid anchoring of alkanes, except for minor shifts in the conductance peak position.
}
As suggested by \textcite{Martin2008} the outcome of the experiments may depend sensitively on the experimental conditions. The experiments by 
\textcite{Venkataraman2006a} were performed in solution, where bond breaking may be followed by spontaneous reforming of bonds. This may enhance the signature for a weak bond, such as the amine bond, in the conductance histograms. When performing experiments under vacuum with a sparse surface coverage, on the other hand, once a bond is broken it cannot be reformed spontaneously. In such experiments a stronger bond, such as a thiol bond, could be preferable.

Taking the last point one step further, multidentate bonds have been investigated. One may reason that multiple anchoring points at each anchoring site could lead to the molecule imposing the structure of the metal leads, which would suppress the variability in the conductance histograms. Such multidentate bonds have been explored in the forms of carbodithioate ($-$CS$_2$H) groups \cite{Tivanski2005,Xing2010}  and dithiocarbamate ($-$NCS$_2$H) groups \cite{Wrochem2010}. A systematic comparison of the conductance histograms with those for other anchoring groups has not yet been made.

\subsubsection{Anchor transparency and gateway states} 

The formation of electrical contact between the molecule and the electrodes  can be discussed 
in terms of the hybridization of the orbitals on the molecule
with the surface states of the electrodes, see Section~\ref{Brief-electr.struct.calc}. 
The degree of hybridization is determined by the amount of overlap that 
the anchor group orbitals have with the foremost electrode atom(s). 
Therefore, the classification of anchor groups follows largely 
the atomic orbital theory of the chemical bond. 
For instance, one distinguishes donor-acceptor type anchors from 
covalent anchors \cite{Su2016}.
The setting of the anchor and the associated hybridization of 
orbitals follows the local rules of optimizing atomic overlaps. 

In many cases anchoring orbitals do not strongly hybridize 
with the frontier orbitals of the molecular backbone, and lie at much lower energy. 
As a result, the effect of the anchor groups on the transmission of a 
molecular wire can be accounted for by a contact resistance. The picture is
that the anchor resembles a tunneling barrier for the charge carriers, that 
is characterized by only a single parameter, its transparency, which 
is assumed to be roughly independent of the energy of the incoming particle. 
This picture is applicable, {\it e.g.}, for 
some of 
the long insulating wires discussed above,
whose conductance 
is captured in the asymptotic expression $G=G_\text{c}e^{-\beta N}$, 
where $N$ denotes the length of the wire in units of its monomer and $G_\text{c}$ the limiting value due to the contacts, see Section~\ref{ss.length-dependence}. 

However,  
it has been noticed that more complicated  
situations can arise, when the atomic orbitals of the anchor groups
lie closer to the Fermi energy than the frontier orbitals of the molecular wire 
\cite{Li2008}. Specifically, consider the case of an alkane wire 
with a thiolate bond. The sulfur atom, when binding to a Au-electrode,  
exhibits a localized orbital with an energy situated in the 
band-gap of the alkane wire. This in-gap state 
(`contact' state \cite{Li2008} 
or `gateway' state \cite{Vazquez2012}) 
is associated with a very broad transport resonance; 
it can dominate the transmission of the shorter alkane-chains  
 \cite{Li2008}. 
Gateway states have been observed in various theoretical 
studies \cite{Brooke2015,Huser2015}, and need to be accounted for when quantitatively evaluating 
experiments on quantum interference \cite{Vazquez2012}.  

{\color{black} 
In the presence of gateway states, the asymptotic behavior
of the conductance, $G\sim e^{-\beta N}$, sets in only at large 
$N$ when, technically speaking, the passage through the 
insulating wire dominates the tunneling action.
Only in this limit the conductance is truly exponential in 
the length and the gap exponent, $\beta$, 
is a property of the band-structure of the long wire, 
{\it i.e.} independent of the contact arrangement.
In their recent experimental work \textcite{Sangtarash2018} 
observe in alkane wires with an extra aromatic center-unit 
an approximately exponential decay, $G(N)\approx e^{-\beta' N}$, 
with an effective exponent $\beta'$ that is considerably smaller than $\beta$.
The authors explain their observation invoking in-gap (gateway)
states. From our perspective one would expect that the effective exponent 
$\beta'$ characterizes a pre-asymptotic regime that crosses over into
a steeper decay at larger $N$. 

\begin{figure}
\includegraphics[width=0.8\columnwidth]{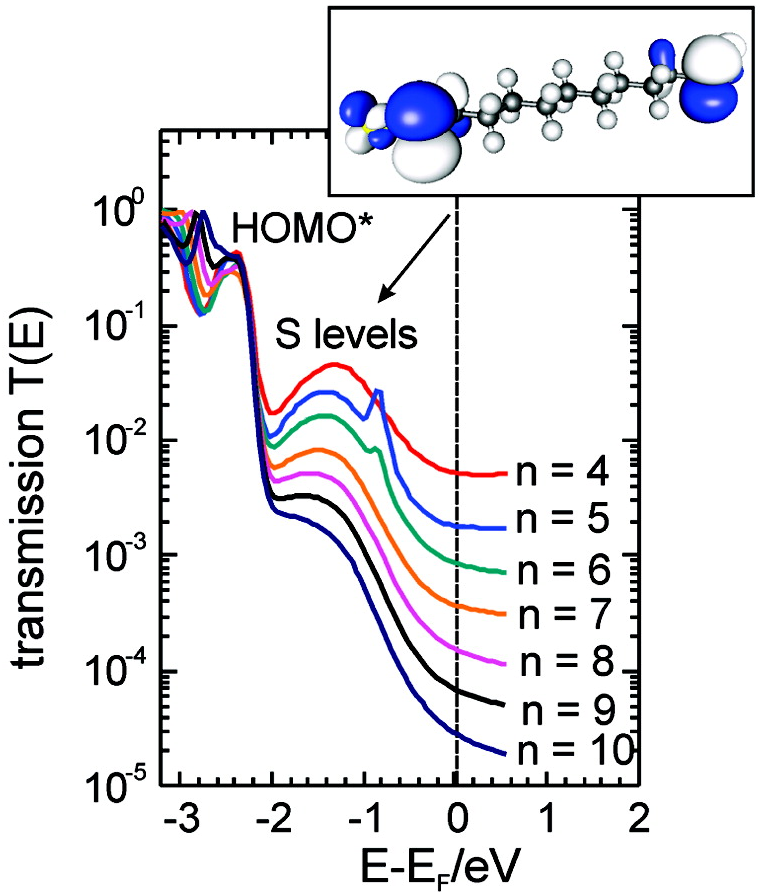}
\caption{\label{Fig:Li2008b}\color{black}
Transmission functions of Au-alkanedithiol-Au series,
where the number of carbons is denoted by $n$.
The length dependence of the conductance is determined by the
spatially delocalized HOMO$^*$ resonance and a broad resonance around $E=-1.5$ eV,
which corresponds to the gateway state (HOMO, `S level').
The inset shows the wavefunction isosurface of the
HOMO and the ball-and-stick model
with carbons (black), hydrogens (white) and sulphurs
at the very ends.
Reprinted with permission from Li {\it et al.}, J. Am. Chem. Soc. {\bf 130}, 318. Copyright (2008) American Chemical Society.
}
\end{figure}
The concepts discussed in this section are illustrated in
Fig.~\ref{Fig:Li2008b} for an alkanedithiol (ADT) gold
junction \cite{Li2008}. Two molecular orbitals
dominate the length-dependence of the conductance.
At $-2.4$ eV there is the HOMO$^*$ of ADT, which is the
HOMO of the alkanes and the backbone state. Its weight at the
Fermi energy decreases exponentially with length, consistent
with the tunneling picture. The resonance at $-1.5$ eV is the
HOMO of ADT, which is the gateway state, not present for bare 
alkanes. Its width is almost length-independent 
due to the orbital's location being close to one electrode, its
position approaches HOMO$^*$ and its weight drops exponentially
due to its localized nature. 
}

\subsubsection{Direct metal-molecule coupling} 

It is possible to form direct links between metal electrodes and the carbon backbone of molecules.
This can be achieved by interaction of organic molecules without anchoring groups 
with reactive metal electrodes such as Pt under cryogenic vacuum conditions \cite{Kiguchi2008,Yelin2016}. 
The coupling to Pt electrodes even results in conductance above 1~\go\   because multiple conductance channels are participating in transport.
Alternatively, coupling reactions have been exploited based on trimethyl-tin ($-$SnMe$_3$) terminations. 
Upon exposing the molecules to Au surfaces the Sn terminal groups are split off and are replaced by direct Au-C bonds \cite{Cheng2011}.  
Such direct Au-C coupled junctions have transmissions that exceed the ones for anchor-group coupled molecules{\color{black}, and can even produce nearly perfect transmission \cite{Chen2011}.  }

Further methods for direct coupling of Au to C exploit C$\equiv$C triple bonded end groups, as in the works by \textcite{Hong2012} and \textcite{Olavarria2016}. The 
Au--C$\equiv$C--coupling leads to sp-hybridization, which does not optimally couple the Au s states to the molecular backbone. Indeed, the conductance 
is lower than for analogous sp$^3$-hybridized Au--C bonds \cite{Olavarria2016}.  

Despite the absence of an explicit anchoring group, even for direct (sp$^3$ hybridized)  Au-C bonds gateway states appear, and can dominate the transmission \cite{Widawsky2013,Batra2013}.
In such cases the gateway state is formed by a Au-C $\sigma$ bonding orbital.
This insight is essential for rationalizing the observed combined data for conductance and thermopower for series of molecular wires \cite{Widawsky2013}.

\subsubsection{Level alignment} 

A non-local aspect of molecular junctions concerns the occupation of the 
junction states: Their filling is controlled by the alignment 
of their energy level with the electrode's work function. 
The occupation thus depends on certain global properties, 
like the surface orientation, 
and on the materials chosen. 
The (partial) filling of the frontier orbitals decides 
between particle (LUMO-based) versus hole (HOMO-based) transport. 
Since the filling depends on a combination of local and global 
aspects of the junction, precise rules for transport properties 
based on the nature of the anchor groups alone are difficult to establish. 
{\color{black} \textcite{Su2016} propose as rule of thumb that dative anchor groups 
tend to come with hole transport while electron-withdrawing groups favor 
particle transport. }

\subsection{Quantum interference} \label{ss.QI-exp}

After we have introduced the theory of quantum interference (\qi) 
in molecular junctions in Section~\ref{ss.QI}, 
we here review key measurements that  
demonstrate the experimental significance of these concepts. 
As a preparatory remark, we emphasize that 
experiments on {\qi}  in molecular junctions are
necessarily somewhat less direct 
as compared to those using 
microfabricated mesoscopic devices.
For the latter, it is relatively 
straightforward to manipulate {\qi}, {\it e.g.}, by application of magnetic fields. 
This route is not open for molecular junctions, because the field strength 
necessary to achieve a measurable effect is not practical. 
Therefore, for the systems of interest here one proceeds 
via a combination of measurement and theoretical analysis. 
This given, a remarkable number of experimental investigations 
has been performed over the last years that all support 
the existence of strong {\qi} effects in molecular electronics. 

An early piece of indirect evidence of {\qi}
has been reported in \textcite{Ballmann2012}, as mentioned in Section~\ref{ss.QI}. 
The authors explain the observed increase of the conductance
with temperature as a result of lifting of destructive {\qi} 
by molecular vibrations. 
\begin{figure}[b]
\includegraphics[scale=0.26]{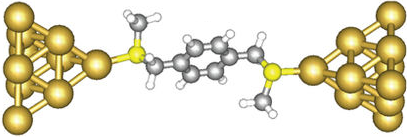}
\includegraphics[scale=0.26]{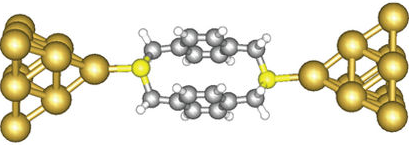}
\caption{One-path and two-path molecules used 
for demonstrating constructive quantum interference 
in the experiment by \textcite{Vazquez2012}. 
Adapted by permission from V\'azquez {\it et al.}, Nature Nanotechnol. {\bf 7}, 663. $\copyright$ $\copyright$ Springer Nature (2012). }
\label{ft8a}
\end{figure}
In another beautiful experiment, 
\textcite{Vazquez2012} have 
been able to perform a two-path experiment 
employing a special molecular design, as shown in Fig. \ref{ft8a}. 
Ideally, one expects the conductance to  increase by a factor of 
four when adding a second parallel channel. In the measurement, 
a factor of three has been observed, which indeed 
exceeds significantly the classical limit of two. 

Particularly strong effects of {\qi} occur in 
molecules that exhibit a Fano-type anti-resonance. 
Motivated by theoretical considerations, see Sec.~\ref{ss.QI}, 
molecules have been synthesized that exhibit the predicted  
conductance suppression.
For example, Gu\'edon {\it et al.} have observed that 
the conductance through an anthraquinone unit is strongly 
suppressed as compared to anthracene, see Fig.~\ref{ft8}. 
By combining 
DFT and tight-binding calculations the authors argue
that this effect results from destructive {\qi}; 
see Fig.~\ref{Fig:Guedon2012b} for computed transmissions.
\begin{figure}[!t]
\quad\includegraphics[scale=0.05]{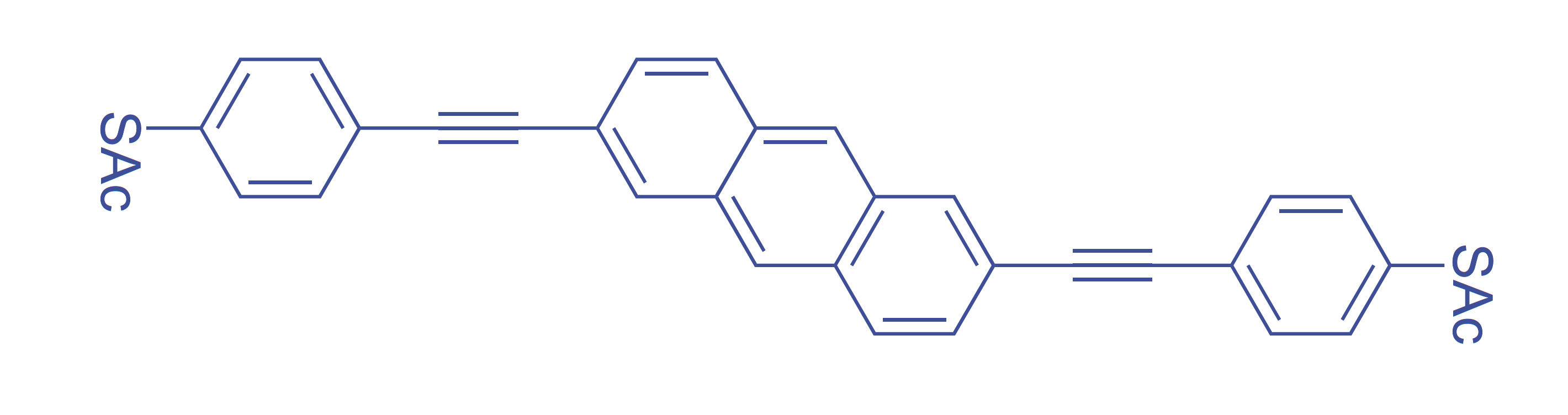}\quad
\includegraphics[scale=0.05]{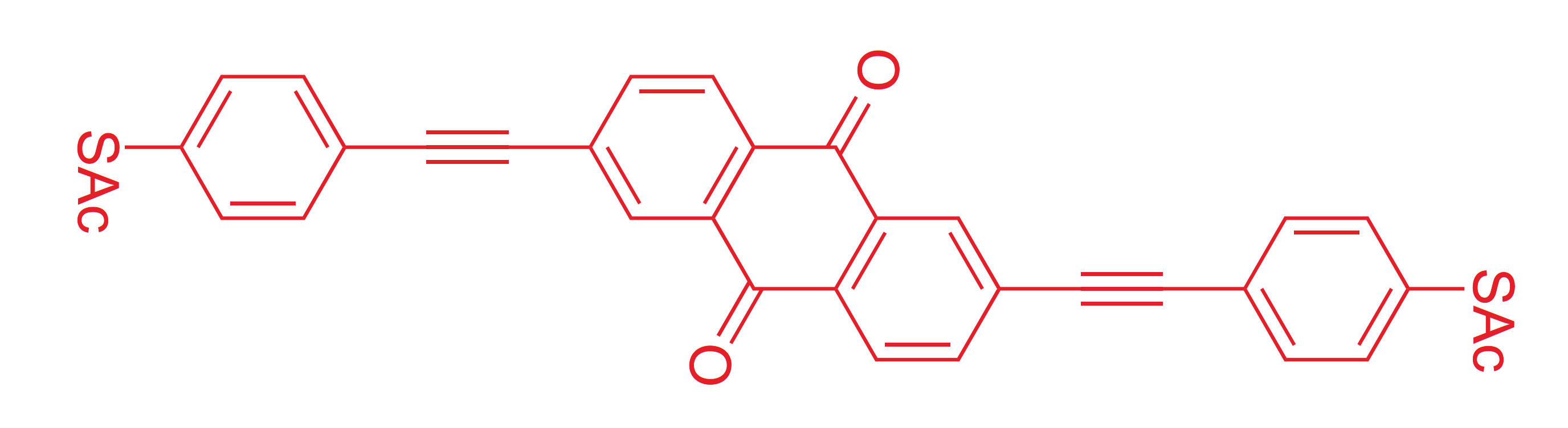}
\includegraphics[scale=0.245]{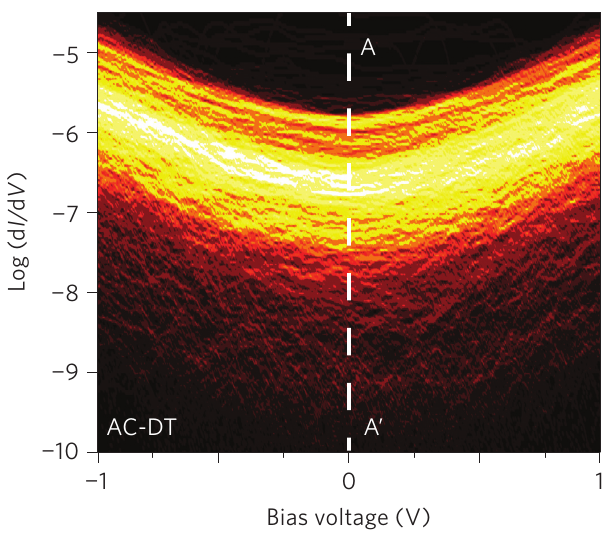}
\includegraphics[scale=0.255]{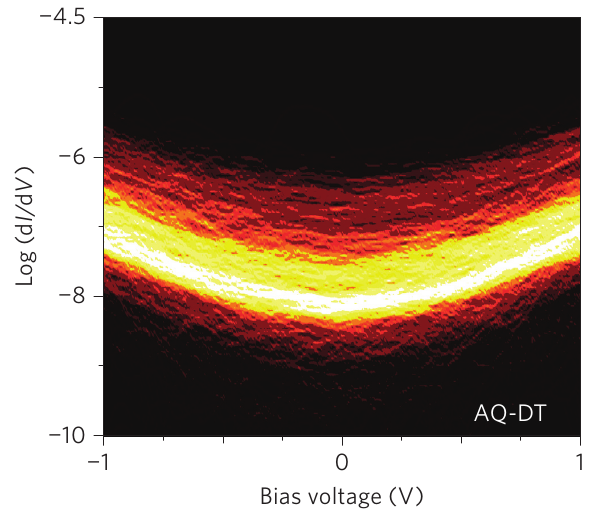}
	\caption{Conducting-tip AFM measurements of the conductance of linearly 
	(left, anthracene-based) and cross-conjugated (right, anthraquinone based) 
	molecules. Two-dimensional conductance histograms are
shown, constructed by logarithmic binning  
	of $dI/dV$ in units $\Omega^{-1}$ versus bias voltage. The color (gray) scale indicates the
number of counts, ranging from black (0 counts) to white ($> 40$ counts). 
Destructive \qi\ suppresses transport in cross-conjugated molecules (right).
Adapted by permission from Gu\'edon {\it et al.}, `Observation of quantum interference in molecular
charge transport', Nature Nanotechnol. {\bf 7}, 305. $\copyright$ Springer Nature (2012).  
}
\label{ft8}
\end{figure}

\begin{figure}[!b]
\includegraphics[width=0.9\columnwidth]{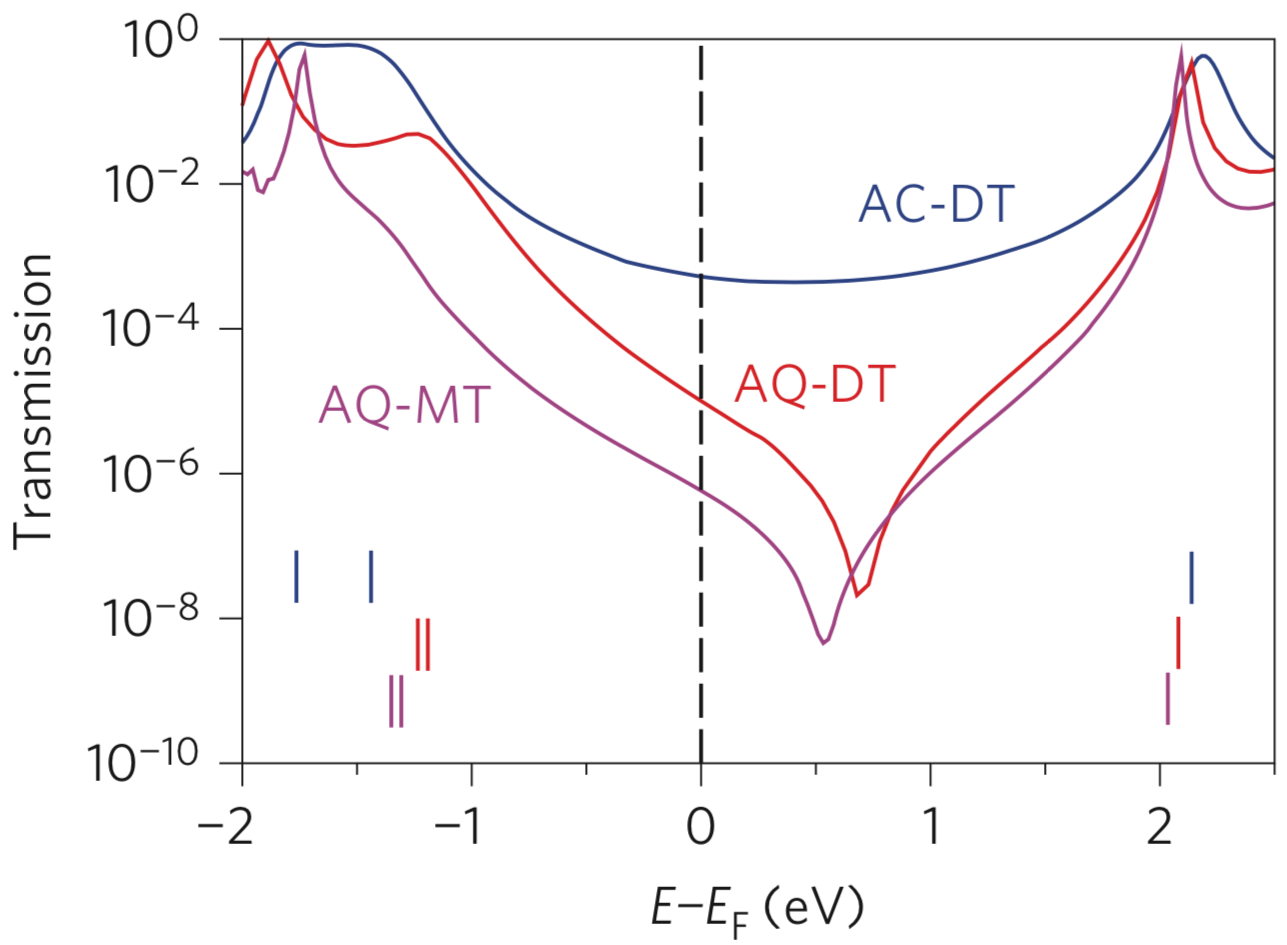}
\caption{\label{Fig:Guedon2012b}
\color{black}
Transmission curves from {\it ab-initio} for anthracene (AC) and anthraquinone (AQ) 
molecules shown in Fig.~\ref{ft8}; AQ-MT contains 
only one thiol end-group. Vertical bars at the bottom indicate resonance
centers of HOMO-1, HOMO and LUMO, for AC-DT (top), AQ-DT (middle), and AQ-MT (bottom).
The AC molecules show a pronounced dip between
HOMO and LUMO peaks, consistent with rules for bipartite lattices discussed in
Sec.~\ref{sIVB4}(a).
Adapted by permission from Gu\'edon {\it et al.}, `Observation of quantum interference in molecular
charge transport', Nature Nanotechnol. {\bf 7}, 305. $\copyright$ Springer Nature (2012).  
}
\end{figure}

In a further experiment by \textcite{Garner2018} a bicyclo[2.2.2]octasilane moiety has been
employed. The {\qi}-induced conduction suppression 
was so strong that the transmission fell below the vacuum value
associated with the gap deprived of its molecular bridge, which represents a
`single-molecule insulator.'

Signatures of {\qi} are very pronounced when 
$\pi-$type binding dominates the most transmitting states. 
Then, moving one of the contacts, {\it e.g.}, the drain, 
from one atom to a neighboring atom strongly affects the transmission.
In the context of alternating 
hydrocarbons such a contact displacement implies that in one situation 
the contacts couple to the same sub-lattice, while in the 
other they couple to different sub-lattices. The corresponding
conductance change is readily explained in terms of the concepts 
introduced in Section~\ref{sIVB3} 

The sensitivity of the transmission to shifts of the contact position has 
been investigated from early on. For instance, 
\textcite{Mayor2003} have observed that a benzene ring 
when used as a linker group with contacts 
in para-position (Fig. \ref{ft9}, {\bf 1}) 
carries a current much larger than when the 
contacts are in meta-position (Fig. \ref{ft9}, {\bf 2}). 
\begin{figure}[b]
\includegraphics[scale=0.075]{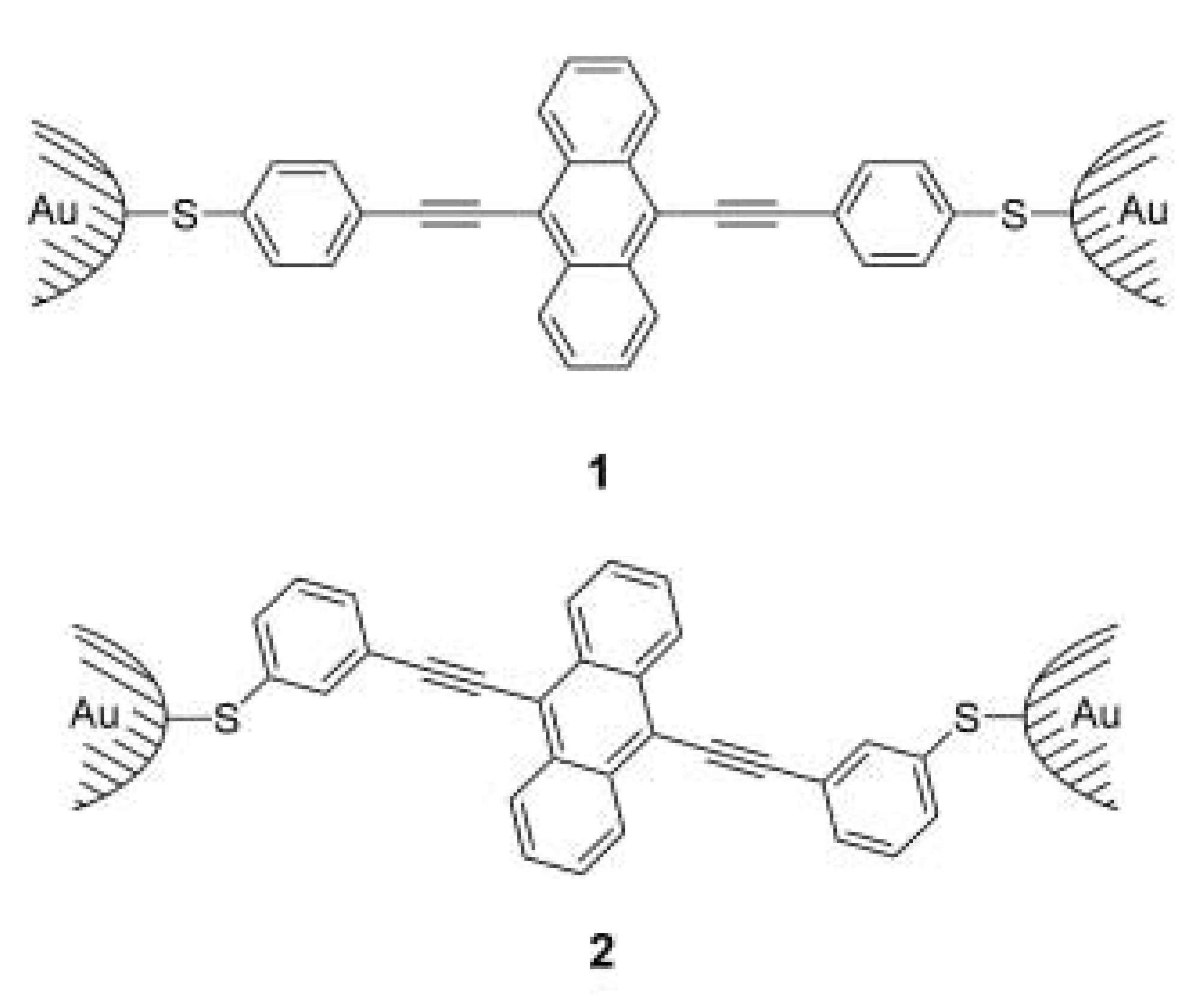}
 \hspace*{0.5cm}
\includegraphics[scale=0.14]{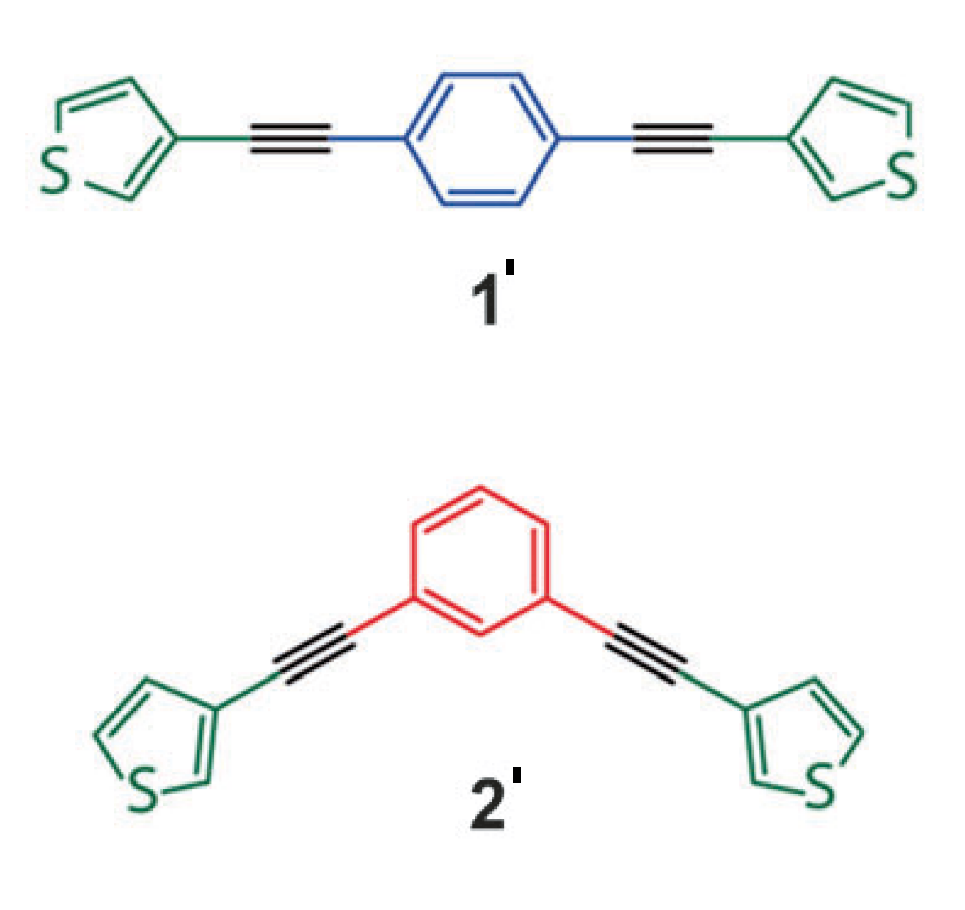}
	\caption{Left: Molecules investigated in the work by \textcite{Mayor2003}. Similar species 
	with mixed end groups have been investigated by \textcite{Ballmann2012}. 
	Right: Molecules investigated in the work by \textcite{Arroyo2013}.} 
\label{ft9}
\end{figure}
An intuitive picture put forward at the time to explain the effect 
was that a contact in meta-position couples to a node of the frontier orbitals while 
para-position couples to a maximum. The theoretical concepts 
presented above, connecting this 
observation to the sub-lattice structure{\color{black}, applied to each terminal phenyl group}, provide a broader scope
and, in particular, predictive counting rules.

The basic idea of investigating the change of transmission upon  
varying the contact positions has been followed in subsequent work. 
Arroyo {\it et al.} considered molecular wires having a 
benzene ring as a center unit, see Fig.~\ref{ft9}, {\bf 1'} and {\bf 2'} \cite{Arroyo2013}.  
These authors confirm experimentally the expectation that follows from 
counting rules: the transmission of a single para-coupled benzene ring 
largely exceeds the one with meta-coupling. 

Also \textcite{Manrique2015} confirm this conclusion. 
These authors use in their experiment pyridine rings for anchoring. Their explicit DFT-based calculations
show that the transmissions of pyridine rings in para-position and ortho-position are similar, while the transmission in 
meta-position is suppressed. This result, once more, illustrates that the 
transmission of benzene rings with contacts coupling to one 
sub-lattice only (meta-position) is suppressed as compared to contacts 
coupling to both sub-lattices (para- and ortho-position). 
Further derivatives with benzene-thiol end-groups and benzene or pyridine centers have been studied by \textcite{Liu2017}.
\textcite{Li2018} provide further support: 
Using electrochemical gating they demonstrate that
a meta-oriented diphenyl benzene structure has a much stronger
variation of the transmission with gate voltage as compared to the
para-coupled species. An on-off ratio of 200 has been achieved in this way. For a very similar experiment performed by \textcite{Huang2018} observation of an even larger on-off ratio of 500 has been reported. 

In addition to molecular rods, also more 
extended graphene-like 
structures have been investigated, 
{\it e.g.}, by \textcite{Sangtarash2016}. 
\begin{figure}[b]
\includegraphics[scale=0.18]{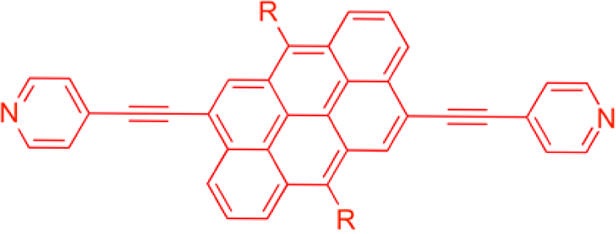}
\includegraphics[scale=0.18]{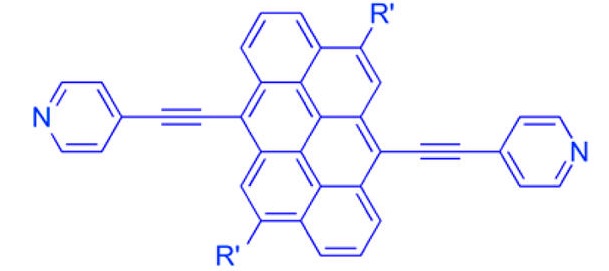}
	\caption{Molecules with Anthanthrene core studied by \textcite{Geng2015}. } 
\label{ft10}
\end{figure}
It was found that also in such cases shifting the contact positions
to neighboring sites can substantially influence the molecular transmission. 
Specifically, the typical transmission of the para-type coupling shown in Fig.\ref{ft10} 
was found experimentally to exceed the transmission of the meta-type contact by a factor of 30 
\cite{Geng2015}.  

Finally, we briefly address a recent development 
that points towards a possibility of 
mechanical control of destructive \qi. 
\textcite{Stefani2018} have observed a modulation 
of the conductance of a $\pi$-stacking molecular wire 
(a [2.2]paracyclophane-compound) 
within an MCBJ, by an order of magnitude upon pushing or pulling. 
The authors interpret their results as due to a sharp anti-resonance in the 
transmission function that can be relocated with respect to $\Fermi$ 
by mechanical manipulation. See also the earlier work by \textcite{Frisenda2016b}.

Mechanical control of destructive \qi\  is also thought to be responsible 
for the large conductance fluctuations seen in experiments on 
ferrocene-based molecular wires  
\cite{Camarasa2019}.
In this case the proposed effect derives from a rotational 
degree of freedom that can switch the junctions \qi\  from 
destructive to constructive and back. 

\subsection{Electrostatic effects and image charges}
The electrostatic environment for electrons on a molecule changes 
in various ways when the molecule is placed between two metal leads. 
For instance, if the Fermi energy of the metal lies outside the HOMO-LUMO gap,
charge is transfered between molecule and contact. 
Thus, a surface-dipole is generated, which can substantially alter the 
work function. 

A very dramatic demonstration of this effect is obtained when the surface coverage of adsorbents 
is asymmetric between the two leads. This asymmetry can be obtained when working in a polar solvent, into which the 
molecules of interest have been dissolved, as shown in the experiments by \textcite{Capozzi2015}. The experiments were 
performed using the STM-BJ technique, for Au tip and Au metal surface and a left-right {\em symmetric} oligomer of four 
units of thiophene-1,1-dioxide, having the same anchor groups at both ends. Despite the symmetry of this arrangement the 
\IV-curves for single-molecule junctions showed very strong asymmetry, with rectification ratio's, 
$R_{\rm rr}:= |I(+V)/I(-V)|$,  
above 100. The explanation offered for the observed asymmetry relies entirely on the geometric shape anisotropy of the tip and surface electrodes. The difference in size creates a difference in effective capacitance at the surface, and the largest voltage drop is found at the smallest capacitor, {\it i.c.} the tip electrode. The effect disappears in non-polar solvents.

Apart from a global adjustment of the electrochemical potential the presence of the metal electrodes affects the molecular levels in other ways. Any charge distribution on the molecule creates image charges in the metal electrodes, which result in an electrostatic energy shift of the molecular levels. In break junction experiments on Zn-porphyrin molecules, that show sharp molecular level resonances in the \IV-curves, \textcite{Perrin2013} showed that the position of the molecular resonances shifts strongly, over nearly 0.5 eV, as a function of the change of distance between the electrodes. The explanation offered for this shift is based on the image charge potentials, which are very sensitive to the distance of the molecule to the metal leads and to its orientation in the junction. 

Image charges are expected to shift energy levels, but can also change the symmetry of the molecular orbitals \cite{Kaasbjerg2011}. Such symmetry breaking was invoked to explain the observations of unexpectedly many charging levels in the experiments by \textcite{Kubatkin2003}. The experiments were performed by low-temperature deposition of OPE5 molecules into a junction between metal leads, which leads to weak metal-molecule coupling. The molecule in this configuration acts as a quantum dot, and by means of a back gate the charging state of the molecule could be varied. Surprisingly, up to 8 charging levels could be reached in a gate voltage window of -4 V to +4 V, which does not agree with homogeneous charging models for the molecule. The explanation comes from the distortion of the molecular levels by the image charges, by which local potential wells are formed at each end of the molecule near each of the two electrodes. This results in charging at the two ends of the molecules and keeps the charges at a 
distance from each 
other.

Together, these experiments show that electrostatic effects have a major influence on molecular junction properties, and lead to surprising effects. The effects are very sensitive to the shape and geometry of the metal-molecule junction but, qualitatively, the effects are known and well understood.

\subsection{Current-voltage characteristics}
Diode characteristics in molecular junctions formed the start of the field \cite{Aviram1974}, 
and are attractive because they represent the simplest 2-terminal functional property of molecules. A general discussion of mechanisms for non-linearities in current-voltage (\IV) traces has been given in Section~\ref{ss.comp.non-linear}. Here, we will consider the interpretation of experiments showing strongly non-linear \IV characteristics at high applied bias, {\it i.e.}, far from equilibrium. This includes asymmetric \IV characteristics for which the rectification ratio $R_{\rm rr} := |I(+V)/I(-V)| $ differs strongly from unity. It also includes non-monotonic \IV characteristics, where the current becomes smaller for increasing voltage, above a certain threshold value. The latter is known as negative differential resistance (NDR), and is also a sought-after property for use in devices such as oscillators and amplifiers. A recent review of diode characteristics in molecular junctions is  given by \textcite{Zhang2017}, and the theory has been presented in Section~\ref{ss.comp.non-linear} above. 

Experiments started with work on Langmuir-Blodgett films \cite{Geddes1990,Metzger2003} which demonstrated that the D-$\sigma$-A type of molecules proposed by Aviram and Ratner indeed show asymmetric \IV characteristics (here, D is a donor group, A is an acceptor group, and $\sigma$ represents a coupling by $\sigma$ bonds). However, they also showed that the properties of such junctions can be richer than anticipated, because the asymmetry sometimes had the opposite sign. 

With the advent of single-molecule techniques it soon became apparent that asymmetry is a rather common feature in molecular \IV characteristics, even for nominally symmetric molecular systems, of which an extreme example was already presented in the previous section  \cite{Capozzi2015}. Conversely, when a molecule has an asymmetric structure this property alone is not enough for producing a large rectification ratio. 

Simple one-level models are helpful for obtaining a first interpretation for many of these observations. For the \IV curves to be asymmetric the left-right mirror symmetry of the junction needs to be broken in at least one of many ways. 

As illustrated in Fig.~\ref{fig.diode+NDR}(a), the asymmetry may result from different barrier widths produced by asymmetric coupling of the molecule. In this case the voltage drop is largely concentrated at the wide barrier. 
The schematics in Figs.~\ref{fig.diode+NDR}(a) apply for molecules that have molecular orbitals which are delocalized over the full length of the molecular backbone. The externally applied electrical potential is assumed to produce voltage drops only over the linker groups to the molecular backbone, here represented as tunneling barriers, which will not be a valid assumption in general (see below). 

The mechanism of Aviram and Ratner requires two levels and is described by a D-$\sigma$-A type structure of the molecule. The mechanism proposed is based on incoherent hopping transfer between A and D sites, which only works when the coupling between the molecule and the leads is weak.  As a consequence, in this case the total current will be very small. Moreover, charging effects are likely to modify the outcome fundamentally. 

When the coupling to the leads is stronger, coherent transport also produces asymmetric \IV curves as a result of the shift of the relative positions of the two levels with respect to each other, as a function of the applied voltage \cite{Elbing2005},  Fig.~\ref{fig.diode+NDR}(b). At higher bias, when the levels cross, the same mechanism also produces NDR \cite{Perrin2015,Perrin2016}.

Fusing D and A moieties together directly may be viewed as a molecular representation of a p-n junction in semiconductors \cite{Ng2002}. However, the sign of the asymmetry of the \IV curves does not generally agree with this picture. The mechanism of the observed diode-like characteristics may be produced by voltage-induced breaking of the delocalization of the wavefunction across the molecule, see Section~\ref{ss.comp.non-linear} above and \textcite{Zhang2017}. 

\begin{figure}[!t]
\includegraphics[scale=0.3]{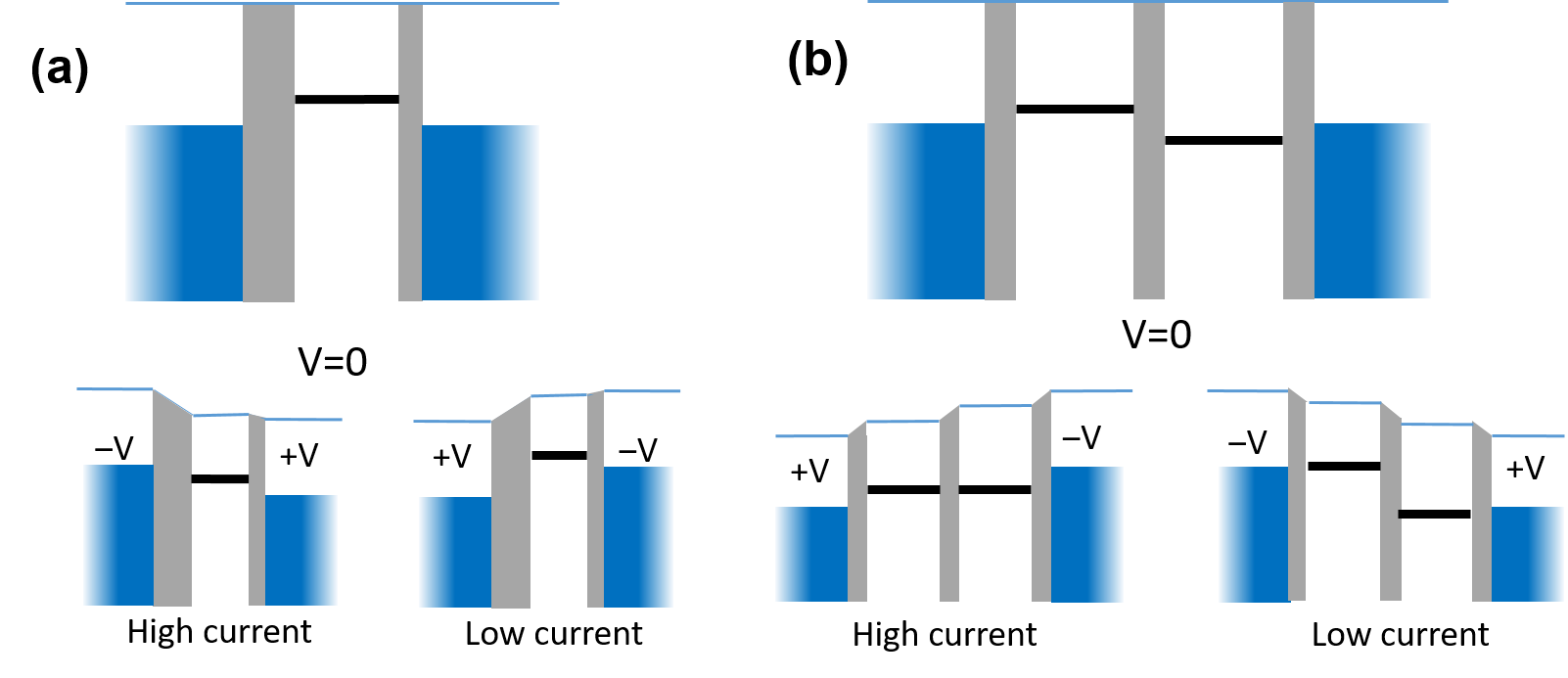}
	\caption{Schematics of molecular levels leading to asymmetric and non-monotonous \IV characteristics. In these diagrams the vertical scale represents energy, while the horizontal axis is a space coordinate. Filled states in leads are indicated by the shaded range (blue online). The lines at the top show the local electrostatic potential (vacuum level). The gray bars show effective energy barriers. The asymmetry can be due to differences in coupling to the two leads, represented by different barrier widths (a). When the molecule has two sites separated by an internal barrier (b), for coherent transport the current becomes high when the two levels are shifted into resonance by the applied potential. Level alignment was not considered in the original proposal by Aviram and Ratner. In that proposal the high current was taken to result from a cascade between the levels, involving relaxation by vibron excitation.}
\label{fig.diode+NDR}
\end{figure}

The widely observed asymmetries are often referred to as diode characteristics. However, an asymmetric \IV does not imply that the system is useful as a diode.
Large rectification ratio's $R_{\rm rr}$ are difficult to achieve at the single molecule level. Commercial diodes have $R_{\rm rr}$ of the order of $10^5$ to $10^8$.
The model calculations presented by \textcite{Armstrong2007} and \textcite{Garrigues2016} suggest that there is a maximum for $R_{\rm rr}$ for a molecular diode of about $10^{2}$ or  $10^{3}$, respectively.  However, the model assumptions underlying these estimates are not general enough to cover all possible molecular diode systems. Indeed, in recent experiments rectification ratios have been reported of up to $10^{3}$ \cite{Capozzi2015,Perrin2016}, up to $10^{4}$ \cite{Atesci2018}, and even $10^{5}$ \cite{Chen2017}. Explanations for the large rectification ratios invoke the role of two levels in the molecule, or additional electrostatic effects.

The full problem of non-equilibrium junctions and the evaluation of the \IV characteristics is quite involved, because local charge may change in response to the flowing current, the nature and shape of the molecular orbitals will be strongly affected by the field gradient, and the potential profile needs to be calculated self-consistently, setting apart further complications due to inelastic excitations. Few self-consistent calculations have been performed. However, such computations are of prime importance for the interpretation of systematic trends and for uncovering specific physical mechanisms \cite{Ruben2008, Perrin2013}. The work by  \textcite{Stokbro2003} and \textcite{Arnold2007} shows how complicated the problem may become. 
These calculations were done for a symmetric molecule, benzeneditiol, and similar computations for larger molecules are quite demanding. A self-consistent calculation for an asymmetric molecule, which incorporates a cobaltocene group, was done by \textcite{Liu2006}. 
The results show substantial asymmetry, but the rectification ratio remains modest, below 30. More important is that the self-consistent calculations show that the profile of the electrical potential neither agrees with the assumption that the potential drops over the connections of the molecule to the leads, nor that it drops over the molecule itself. The actual potential profile computed shows a combination of the two, and the shape of the profile changes upon reversing the polarity. The conclusion should be that simplified models may be very helpful as a guide, but are unlikely to be very accurate for describing properties of molecules far from equilibrium, such as needed for description of \IV characteristics.

\subsection{Thermal and thermo-electrical properties}
A difference in temperature $\Delta \temp$ between the two leads connecting a molecule in a junction induces a current across the junction, which is known as the Seebeck effect. Under open-circuit (or high external impedance) conditions a potential difference builds up which counter-acts this current. Upon reaching equilibrium the voltage induced by the temperature difference is given by the Seebeck coefficient $S$ (or the thermopower) of the junction, $\Delta V = S\, \Delta \temp$, with \cite{Houten1992},
\begin{equation}\label{Seebeck}
S = -\frac{\pi^2k_B^2\temp}{3\left| e \right|}\Eval{\frac{\partial \ln({\transmission}(E))}{\partial E}}{E=E_F}{}.
\end{equation}
Here, ${\transmission}(E)$ is the energy dependent transmission and $\temp$ is the average temperature. A picture of noninteracting particles has been assumed. 

Measurements of the thermopower provide additional information on the electronic properties of the junction \cite{Ludoph1999}. {\color{black} For single-molecule junctions, through} relation (\ref{Seebeck}) the sign of $S$ immediately translates into the nature of the nearest molecular orbital (HOMO or LUMO),\footnote{\color{black} A note of caution: when quantum interference between the molecular orbitals plays a role near the Fermi level this unambiguous interpretation may be lost, see Section~\ref{ss.QI-exp}. }
 and the combined knowledge of conductance and thermopower gives information on the distance of this level to the Fermi energy of the leads \cite{Paulsson2003,Reddy2007,Malen2009}.  {\color{black} Three methods for measuring thermopower have been reported: by measuring the thermally induced voltage \cite{Ludoph1999}, by measuring the thermally induced current \cite{Widawsky2013}, or by taking \IV curves with a temperature difference across the junction \cite{Rincon2016}.}

\textcite{Reddy2007} have reported the first single-molecule thermopower measurements on 1,4-benzenedithiol, 4,4'-dibenzenedithiol and 4,4''-tribenzenedithiol and they have demonstrated that these form p-type molecular junctions, {\it  i.e.}, the Fermi energy lies close to the HOMO. \textcite{Malen2009} studied three classes of molecules - phenylenediamines, phenylenedithiols and alkanedithiols of various length. The phenylene molecular wires show a thermopower that grows linearly with length of the molecular backbone, while the alkanes show the opposite trend. Electrical transport through alkanedithiols 
connected between gold leads had previously been argued to be influenced by gateway states at the connection to the leads \cite{Zhou2008,Zeng2002}, but it was difficult to separate the contributions of in-gap states from those of HOMO or LUMO tunneling. By the additional information obtained through thermopower \textcite{Malen2009} demonstrated the importance of gateway states. 

Generally, the position of the molecular levels, and therewith the transmission as a function of energy, is asymmetric with respect to the Fermi energy and, therefore,  breaks electron-hole symmetry. While this leads to a finite thermopower, the opposite effect can also be observed: imposing a current leads to asymmetric heating. The latter is more difficult to observe, and for this purpose \textcite{Lee2013} developed an advanced nanoscale-thermocouple integrated into a scanning tunneling probe. The probe measures the temperature of the tip apex contacting the molecule with a thermocouple mounted approximately 300 nm from the tip apex. Using this probe, the authors have studied heat dissipation in a system formed by attaching single molecules of 1,4-benzenediisonitrile and 1,4-benzenediamine between Au electrodes. They showed that, even for symmetric molecules, the heat dissipation takes place asymmetrically across the junction and depends on the applied bias polarity and the sign of the Seebeck coefficient 
$S$. For positive $S$ the electrons dissipate more heat at the junction from which they originate, while for negative $S$ 
more heat is dissipated at the receiving end. These techniques have more recently been extended to allow measurements of the thermal conductance of atomic and molecular junctions \cite{Cui2017}. 
For a comprehensive review on this topic we refer to \textcite{Cui2017a}.

{\color{black}
An interesting theoretical contribution with general validity for measurements of 
Seebeck coefficients has recently been made by \textcite{Rix2019}. 
It brings together a topic of the previous section, circulating currents,
and the Seebeck measurement. The authors observe that the measurement of $S$ as the ratio 
$\Delta V/\Delta T$ at zero charge current implies that the divergence of the associated 
current density vanishes, $\text{div}\, {\bf j}(\br) =0$, but not ${\bf j}(\br)$ itself. 
Circulating currents are possible under these conditions and will, in general, emerge. }
{\color{black}The nearly-open circuit conditions of thermopower experiments would suppress the regular
current and permit detection of the effects of the circulating currents. Evidently, such experiments will be 
very challenging.}

\subsection{IETS and sign inversion} \label{ss.IETS}  
Electronic transport through single molecules serves as an interesting playground for studying inelastic interactions of electrons with vibration modes (or `vibrons') in a single molecule. The inelastic signal offers a means of spectroscopy that aids in confirming the presence of the molecule under test. 

The strength of the electron-vibron coupling governs the nature of the electron transport. In the majority of the experiments reported in the literature the electron transit time  ({\it i.e.}, the inverse level broadening) in the molecule is short compared to the time needed for the ions to respond. The interaction can then be described by perturbation theory, at the level of the Born-Oppenheimer approximation. In the other limit, the transport is incoherent and leads, {\it e.g.}, to polaron formation on the molecule. Examples that classify into this limit are the experiments on long oligomers, beyond the transition from coherent to incoherent transport, described in Section~\ref{ss.length-dependence}, 
and the Franck-Condon blockade to be discussed below. Here, we will limit the discussion to weak electron-phonon coupling. 

The resolution of the experimental signal relies on a sharply defined Fermi level in the metal electrodes, which implies that clear signals are only found at low temperatures.
The first measurements demonstrating electron-vibron interaction on single molecules \cite{Stipe1998} were done for a single acetylene (C$_2$H$_2$) molecule on a Cu(100) surface, in a low-temperature UHV STM setup. A small but well-defined step upward, towards an increased conductance, was observed at a voltage equal to $\hbar\omega/e$, the energy of a vibrational excitation of  the molecule. This measurement was done in tunneling mode, at an electron transmission $\transmission \ll 1$ and the increase in conductance at energy $\hbar\omega$ was attributed to opening of an additional, inelastic transport channel. 

Measurements of vibration modes on a single molecule bound symmetrically between two leads were first done for a single H$_2$ molecule and Pt leads \cite{Smit2002,Djukic2005}. In this case a step downward, {\it i.e.} a decrease in the molecular conductance was recorded.\footnote{\color{black}\textcite{Kristensen2009} using first-principle calculations have shown that a step-up in conductance for such Pt-H$_2$-Pt systems should also be possible. This was attributed to the opening of an otherwise closed d-channel, along with the usual s-channel, when exciting a transverse hindered rotation mode for the H$_2$ molecule. Recently, such step-up has also been found in experiments \cite{Tewari2019}. As pointed out there, these steps in conductance are small and it is extremely difficult to distinguish them from the structures that could appear due to elastic scattering of electronic waves through defects in the Pt leads.} The critical difference between the two experiments is the transmission $\transmission$, which 
is close to unity for the latter experiment. A similar decrease in conductance at 
transmission close to unity was also recorded for mono-atomic chains of Au atoms \cite{Agrait2002}. 

The crossover in sign of the inelastic signals has been 
investigated near equilibrium perturbatively in the 
electron-phonon coupling energy $g$ using 
numerical \cite{Paulsson2008} and 
analytical approaches 
\cite{Egger2008,Entin2009}. 
The latter directly employ the Holstein model, 
but also the former approach effectively reduces to this model
when adopting the lowest-order expansion, 
\begin{equation}
\nonumber
 \hat H = (\varepsilon_0 + g\hat Q)\hat d^\dagger \hat d + 
 \hbar \omega_0 \hat a^\dagger \hat a 
 + \sum_{k}\left( V_k \hat c^\dagger_{k} \hat d + \text{h.c.}\right) 
 + \sum_{k} \epsilon_k \hat c^\dagger_k \hat c_k,
\end{equation}
where $\hat Q=\hat a^\dagger {+} \hat a$ is the displacement operator. 
It features a single fermionic level coupled to a 
vibration mode (bosonic) with frequency $\omega_0$;
for a perspective on strong-coupling phenomena in this model 
see \textcite{Thoss2018}. 

The structure of the general result for IETS corrections, as it emerges 
from the Holstein model, can be analyzed employing standard dimensional analysis. 
For simplicity, we consider the case of symmetric coupling $\Gamma=\Gamma_\mathcal{L}=\Gamma_\mathcal{R}$
and adopt the wide-band limit, in which $\Gamma$ is independent of energy. 
Focusing on zero-temperature, we have the parametric dependency,
\begin{equation}
 dI/dV = e^2/h\  \mff({\Vb}, \varepsilon_0, \Gamma, \omega_0,g),
\end{equation}
with $\mff$ being a dimensionless function of its arguments
that includes elastic and inelastic scattering processes.
We are interested in
the corrections to the $dI/dV$-curve induced by the coupling $g$.   
The natural dimensionless small parameter will be $g/\Gamma$: 
In the limit $g\ll\Gamma$ the dwell-time of the electron 
in the process of traversing the level, $\hbar/\Gamma$, 
is too small for the action, which is related to the electron-vibron coupling, 
to become effective. 
The leading IETS signal will be 
determined by 
{\color{black} the expression of second-order in this small 
parameter,}\footnote{The first order term in $g$ is proportional 
to the oscillator's displacement out of its equilibrium position.  
It is higher order in $\Vb$ and can be ignored 
when vibrational relaxation is fast enough.}
{\color{black} 
\begin{equation}
\label{e62} 
 \left. {\frac{dI^\text{\tiny IETS}}{dV}}\right|_{{e\Vb}{\gtrsim}\hbar\omega_0} \approx \frac{e^2}{h} 
\frac{g^2}{\Gamma^2} 
\mfftwo(\varepsilon_0/\Gamma,\omega_0/\Gamma). 
\end{equation}
}It has been written already for the case $e\Vb\gtrsim\hbar\omega_0$, 
since it is at these voltages where the sign of the IETS-correction manifests itself.
Also, we have accounted for $\mfftwo$ being dimensionless, 
so its three arguments combine into two dimensionless ratios. 

In general, $\mfftwo$ is a complicated function 
of its parameters \cite{Egger2008,Entin2009}. 
It takes a transparent form in the limit of a soft (very slow) vibration, 
{\it i.e.}, $\hbar\omega_0/\Gamma\to 0$ in \eqref{e62}, 
{\color{black}
\begin{equation}
\label{e63} 
\lim_{\hbar\omega_0/\Gamma\to 0} \left. {\frac{dI^\text{\tiny IETS}}{dV}}\right|_{{e\Vb}{\gtrsim}\hbar\omega_0} \approx 
\frac{e^2}{h} 
\frac{g^2}{\Gamma^2} 
\mfftwo(\varepsilon_0/\Gamma,0). 
\end{equation}
}
Intuitively, Eq. \eqref{e63} describes the jump of the 
conductance when the bias-voltage crosses the threshold energy 
$\hbar \omega_0$ under the assumption that the phonon-frequency is still 
small as compared to the level broadening.

After this step the model has in fact been simplified 
to such an extent that the sign of $\varepsilon_0$ 
no longer matters, because all reference scales except zero energy have dropped out. 
Consequently, the rhs of \eqref{e63} only depends on $(\varepsilon_0/\Gamma)^2$ and therefore 
can also be considered as a function of the zero-bias transmission, $\transmission$, 
which is an invertible function of the same argument. Summarizing, within the model
assumptions the IETS signal follows a `universal' function 
-- {\it i.e.}, independent of microscopic model parameters -- 
with the transmission $\transmission$ as the only remaining 
variable describing the molecular bridge at hand,
{\color{black}
\begin{equation}
\lim_{\hbar\omega_0/\Gamma\to 0}  \left. {\frac{dI^\text{\tiny IETS}}{dV}}\right|_{e\Vb{\gtrsim}\omega_0} \approx 
\frac{e^2}{h} 
\frac{g^2}{\Gamma^2} 
f(\transmission).
\end{equation}
}
For the case $\Gamma_{\mathcal L}/\Gamma_{\mathcal R}\eqqcolon  
\alpha$ with $\alpha \neq 1$ 
a more general result can be obtained along similar lines; we 
are then left with a two-parameter dependency $f(\transmission,\alpha)$. 

The function $f(\transmission,\alpha)$ has been calculated and 
discussed by \textcite{Paulsson2008}.
At high transmission, $\transmission{\simeq}1$,
the forward-scattering electronic states are nearly fully occupied, 
implying that electrons that undergo inelastic 
scattering by excitation of a vibron can only find unoccupied 
states by scattering backward. In the other limit, 
$\transmission\ll 1$, nearly all incoming states are scattered back 
elastically, such that after an inelastic scattering 
event with a vibron the electron only finds empty states 
in the forward scattering direction. Therefore, we expect 
a tendency that inelastic scattering leads 
to a decrease of the conductance 
at high transmission, $f<0$, 
and to an increase in conductance at low transmission,
$f>0$. The crossover occurs on a line in the 
$(\alpha,\transmission)$-plane where $f(\alpha,\transmission)=0$. 
The line was calculated by \textcite{Paulsson2008}, see also \textcite{Kim2013}; 
for soft vibrations and symmetric coupling, {\it i.e.} $\alpha{=}1$, 
the crossover takes place at $\transmission=0.5$.  

We emphasize that the model analysis is 
based on the wide-band limit and on assuming soft vibrations with a 
very fast relaxation mechanism. 
The more general problem also includes 
the back-action of the non-equilibrium vibron occupation, which results in a shift 
of the crossover point to higher transmission and, therefore, 
is considerably more complicated.  
A first attempt at solving the problem was made by \textcite{Urban2010}, 
which, however, neglected back-action related frequency renormalizations\footnote{We express our gratitude to Tom\'a{\v s} Novotn\'y for 
bringing these developments to our attention.}
\cite{Kaasbjerg2013}. 
More systematic 
treatments have been presented in subsequent work that have, in particular, 
also included this effect \cite{Novotny2011,Utsumi2013,Ueda2017}; 
{\color{black} for a numerical solution of the Holstein model including 
an analysis of IETS see \textcite{Schinabeck2016}}. 

Experimentally, a crossover between a step-up and a step-down at the bias voltage corresponding to the excitation of a molecular vibration was observed for single-molecule H$_2$O junctions between Pt leads \cite{Tal2008}, and for benzenedithiol between Au leads \cite{Kim2011}. The crossover point in the former experiment was found near $\transmission 	\approx 0.65$, but from shot noise measurements on the same junctions it was found that a second conduction channel contributes to the conductance. The dominant conductance channel has a transmission near 0.5 at the crossover point. The experiments by Kim \ea included an analysis of the asymmetry of the coupling to the two leads, $\alpha=\GR / \GL$, which could be obtained from the shape of the current-voltage characteristics. From a plot of the inelastic signal intensity, normalized to the conductance on the junction, a clear crossover was obtained near  $\transmission \approx 0.5$. In a follow-up piece of research \textcite{Karimi2016} demonstrated, using 
shot noise measurements, that the electron 
transport in BDT molecules is indeed due to a single conductance channel for $\transmission$ ranging from 0 to 0.6.

As was emphasized by  \textcite{Avriller2009,Schmidt2009,Haupt2009}, the conductance is just the first moment of the 
distribution of electron transfer probabilities. Inelastic signals are expected to show up in all moments of the 
distribution of electron transfer, and in particular in the second moment, which is known as shot noise. For symmetric 
junctions, at transmission $\transmission \simeq 1$, inelastic scattering is expected to produce increased noise 
signals. A crossover to a negative contribution takes place near $\transmission \approx 0.85$, and another crossover 
back to a positive noise contribution at $\transmission \approx 0.15$, again under the assumption of a large width of 
the electronic level. This inelastic noise signal was observed for short Au atomic chains \cite{Kumar2012}, and a cross 
over was seen from positive to negative inelastic noise contributions. However, the crossover point was found near 
$\transmission \approx 0.95$, higher than predicted for the simple 
model systems.   
The rather large value for the turn-around point was addressed in a subsequent theoretical study \cite{Avriller2012}, who adopted the approach of \textcite{Haupt2009,Haupt2010}
for an {\it ab-initio} treatment. 
This study did not observe any sign-change at all, so that the discrepancy between experiment and different theories appears to persist. {\color{black} More recently, by an hierarchical quantum master equation approach, \textcite{Schinabeck2019} obtained a shift in the transition between positive and negative inelastic noise contributions, that they attribute to the non-equilibrium occupation of vibrational modes. Unfortunately, the shift makes the discrepancy with experiment larger, not smaller.}

IETS is not governed by strict selection rules, such as apply for other forms of spectroscopy, including Raman scattering and IR absorption and emission. Nevertheless, symmetries of the molecular orbitals and symmetries of the vibration modes involved in the electron scattering lead to approximate selection rules, that are known as propensity rules \cite{Troisi2006b,Gagliardi2007}. For example, \textcite{Gagliardi2007} show that for all molecules bound to Au through a sulfur atom the IETS spectrum is dominated by the totally symmetric vibration modes. The amplitudes, on the other hand, are difficult to predict because they are extremely sensitive to interference between various contributions to the inelastic signal.
Another important difference with the other forms of spectroscopy is the intrinsic limitation in resolution due to the strong coupling of the vibration modes to those of the metal leads and due to the hybridization of the electronic levels of the molecule with the bulk states in the leads. Only in limiting cases, when the molecule is weakly coupled to both leads, a more quantitative comparison between theory and experiment becomes possible. This has been demonstrated recently by \textcite{Krane2018} in low-temperature STM experiments, where a monolayer of MoS$_2$ served to decouple the molecules from the substrate.

\textcite{Lykkebo2013} have predicted that overtones, {\it i.e.} multiples of the fundamental vibration frequencies, may in some case dominate the IETS spectrum. To our knowledge observation of overtones has not been reported for regular IETS spectra, but we will see that overtones dominate the spectra under conditions of Franck-Condon blockade, see Section~\ref{ss.Franck-Condon}.

\subsection{Coulomb blockade and Kondo effect}
The theory of Coulomb blockade and 
the Kondo effect has been reviewed briefly in Sec.~\ref{s.models}. 
Here, we show how molecular junctions allow testing
of long-standing predictions, and offer means of preparing new electron-correlated phases. 
While our review focuses on molecules contacted to leads, 
in this section we will also
mention a few STM experiments operated in the tunneling regime.
The development of our understanding of quantum transport through
open-shell molecules enjoyed cross-talk of both experimental
approaches \cite{Scott2010}.

\subsubsection{The single-impurity Anderson model in single-molecule junctions}
While the single-impurity Anderson model (SIAM) applies to various physical
systems\cite{Kouwenhoven2001,Scott2010}, 
molecular junctions occupy a prominent role,
because they allow for quantitative testing of 
theoretical predictions in a wide parameter window.

\textcite{Zhang2013} studied the 
$\mathrm C_{28}\mathrm H_{25}\mathrm O_2\mathrm N_4$ neutral radical 
adsorbed
on Au(111).
In the differential conductance a zero-bias resonance (ZBR)
appears. The latter is associated with the spin of the 
unpaired electron of the ra\-di\-cal. The strong temperature 
dependence of the ZBR, without saturation at low $\temp$,
points to the weak-coupling Kondo regime,
$\temp \gg \temp_\mathrm K$, where 
spin-flip scattering can be described by a perturbation theory
in the Kondo exchange coupling $J$, see Eq.~(\ref{e31})
\cite{Kondo1964, Appelbaum1966}.\footnote{The perturbation
approach breaks down when $\temp \approx \temp_\mathrm K$.}
To third order in $J$ the conductance at zero magnetic
field can be expressed by via,
\begin{equation}
\label{eq:pertJ3}
  G(eV, \temp) = G_\text{bg} + A\, f(eV/k_\mathrm B \temp),
\end{equation}
where $A$ is proportional to $J^3$ and the function $f$ is independent
of microscopic parameters. The background contribution $G_\text{bg}$ is
treated as a constant. The formula can be easily 
generalized to finite magnetic fields, where the only
additional parameter is the gyromagnetic constant $g$, which determines the
Zeeman energy, and consequently the splitting of the ZBR
into two symmetrically-positioned steps in the differential conductance \cite{Appelbaum1967}. 
Zhang \ea test
the accuracy of this theory by treating the temperature 
$\temp$ and $g$ as fitting parameters.  The $\temp$ obtained from the fits agrees with the experimental temperature to within an accuracy 
of 10\% (unless $\temp<6K$, where the perturbation theory is not
expected to be successful).  The value for $g = 1.93\pm 0.02$
is slightly reduced compared to the gas-phase,
reflecting the screening of the $g$-factor by the electron gas in the Au substrate.

Extending quantitative tests of the theory for single molecules suspended between two leads  down to
temperatures below $\temp_K$, (`strong coupling' Kondo regime)
poses a significant
experimental challenge, namely the stability of the system across temperature
differences spanning several orders of magnitude. 
This difficulty can be bypassed in molecular junctions where,
instead of varying the temperature, stretching of the molecular
junction drives changes of $\temp_\mathrm K$ by varying microscopic parameters.
\textcite{Zonda2019} have taken this approach in order to investigate the ZBR
in PTCDA suspended between Ag contacts.
The theoretical apparatus involves conductance calculations based
on the solution of the SIAM with the numerical renormalization group
\cite{Wilson1975,Bulla2008}.
The fitting procedure involved only two parameters, the couplings to the two leads 
$\Gamma_\mathcal L$ and $\Gamma_\mathcal R$, but successfully captures the
conductance spectra in the bias window $\pm 15$ mV, in both strong and
weak coupling regimes.

\subsubsection{Two-impurity Anderson model}
Following the theoretical overview in 
Sec.~\ref{sss.models.TIAM}, we present
a discussion of molecules with two relevant orbitals
in quarter and half filling.

{\it  Quarter filling: The SU(4) Kondo effect.}  
Beautiful realizations of the SU(4) Kondo effect
on a single entity were given
in carbon nanotubes \cite{Jarillo2005,Makarovski2007,Schmid2015}. 
In this case, the orbital degeneracy required for an SU(4)
Kondo effect arises from the valley degree
of freedom. However, the contacts and impurities 
induce mixing of the two orbitals and therefore disturbances of the SU(4) symmetry are significant. 
In molecular `quantum dots' such perturbations can be avoided 
by imposing  point symmetry of the compound system 
(molecule and contacts). Consider, {\it e.g.}, the case of transition-metal phthalocyanines (Pc)
adsorbed on a crystalline metallic surface. The gas-phase
Pc is symmetric around a fourfold axis, which gives rise
to doubly-degenerate levels. The fourfold axis
can be preserved when the molecule is placed on a cubic lattice. This is the case for 
nickel and copper phthalocyanines (NiPc and CuPc) adsorbed on Ag(100)
\cite{Korytar2011,Mugarza2012}. The degenerate ligand states
($e_\mathrm g$) contain a single electron
in the Coulomb blockade regime. The SU(4) character manifests itself in 
STM images of the Kondo resonance, which is uniformly
detected over the ligands of the Pc.

An SU(4) Kondo effect was also found for FePc adsorbed
on two distinct sites on Au(111) \cite{Minamitani2012}.
The gas phase FePc has a doubly
degenerate level, residing mainly on the d-shell.
The molecules are adsorbed on Au(111) on two sites, either 
the `on-top' sites, or the `bridge.'
The analysis of STM spectra with and without magnetic field
suggests that at the on-top adsorption site, the Kondo
effect is of the SU(4) type, and at the bridge site
it turns into an SU(2). This observation correlates
well with the strongly reduced symmetry of the bridge site.
The molecular SU(4) Kondo system allows to be 
assembled into a 2D array, as demonstrated by the same group
\cite{Tsukahara2011}. Such an SU(4) Kondo lattice
offers means to study the competition between orbital
and spin ordering \cite{Lobos2014,Fernandez2015}.

The above mentioned results
suggest that the delicate conditions required for the SU(4) Kondo effect
may be achieved in molecules more easily than in other systems.
The orbital mixing, {\it e.g.}, can vanish in carbon nanotubes due to symmetry
reasons. We remark that, additionally, the SU(4) Kondo
effect can be disturbed by certain interaction matrix elements, {\it i.e.} terms
in Eq.~(\ref{Eq:Hd}). The latter can have a lower symmetry in
molecules than in carbon nanotubes.

{\it  Molecules with two open shells.}  
Molecules with two open shells can be understood as having 
two unpaired spins. This allows realizing several
 important regimes of the TIAM at half filling.
\color{black}
These regimes are controlled by the sign and magnitude of the exchange coupling, 
see Sec.~\ref{sss.models.TIAM}. When the two respective orbitals belong
to the same d-shell,
the exchange is ferromagnetic (Coulomb exchange, Hund's rule) and strong
($\approx 1$eV), effectively locking both spins into a spin-one moment. When the overlap
between the two orbitals is small, Coulomb exchange is negligible and other
mechanisms take a leading role (superexchange, RKKY exchange). These mechanisms
are sensitive on the details of the system and the environment, allowing for
{\it in-situ} tuning, as shown below.
\normalcolor

\textsl{Underscreened Kondo effect.}
The signatures of an underscreened Kondo effect were observed for the
first time in a single entity, a C$_{60}$ molecule in a gold nanogap, by
\textcite{Roch2009}. The key in the identification of the underlying
spin model was the comparison of the temperature dependence of the conductance
with the universal dependencies obtained from the NRG calculations,
$G(\temp) = G_0f_S(\temp/\temp_\mathrm K)$, for a given impurity spin $S$. 
A second important hallmark of underscreened Kondo
physics is the splitting of the Kondo peak in a magnetic field.
Consistent with expectations for the underscreened Kondo effect, Roch \ea observed that the field required to split the peak
is very small. The interplay of the Kondo correlations of a spin 1
system and magnetic anisotropy was studied by
\textcite{Parks2010} in a Co complex. By mechanically distorting
the ligand field around the Co, the anisotropy energy can be
manipulated, providing means to switch on and off the underscreened
correlations without a magnetic field.

\textsl{Singlet-triplet transitions.}
Roch and coworkers also reported the observation of
a quantum phase transition
in an electromigrated C$_{60}$ junction \cite{Roch2008}.
As Roch \emph{et al.} argue, the transport measurements can be
interpreted by  the two-impurity Kondo model with only a single screening channel
at the accessible experimental temperatures. An important manifestation
of the quantum critical point is the appearance of a two-stage
Kondo screening process on the singlet side of the transition.
In the transport measurements, the second-stage Kondo effect is detected
by the formation of a dip in the middle of a broad first-stage
Kondo peak.

An interesting option to experimentally observe the
 singlet-triplet transition is to bring two open-shell
molecules into contact.
\textcite{Esat2016} have studied
a PTCDA-Au complex adsorbed on the surface of gold.
The LUMO of the complex is a delocalized $\pi$ orbital which
captures a single electron upon adsorption. When two
PTCDA-Au
complexes are brought close to each other both LUMO's overlap.
As Esat \ea argue, the direct exchange interaction is negligible,
and the physics is dictated by the charging energy of the LUMOs
along with the intermolecular hybridization. The pairs of PTCDA-Au complexes can
be found in various on-surface orientations, shown by the STM images.
The zero-bias features in the tunneling conductance
are either a single Kondo peak or a split resonance. 
The occurrence of two kinds of
zero-bias features was rationalized by NRG calculations on a TIAM,
showing that they correspond to (underscreened) triplet and singlet
phases of the two-molecule systems. Both phases are separated
by a quantum phase transition, driven by changing the orbital hybridization
$t$. Esat \ea employed {\it ab-initio} calculations of pairs of
gas-phase complexes in the orientations observed experimentally.
 The $t$ was estimated from the splitting
of the LUMO energies. The $t$ calculated in this way correlates well with
the observation of the split peak.

A prototypical two-spin molecule (a complex with two Ni$^{2+}$
 centers) was studied by \textcite{Zhang2015}
combining STM measurements on Cu(100) and first-principles
calculations.
The authors argued that the exchange interaction between the two spins
is much smaller than the Kondo temperature, so that both spins are
independently
Kondo-screened. The regime of independent Kondo screening becomes
interesting when larger spin arrays are arranged on the surfaces, {\it e.g.}, see
\textcite{Dilullo2012}.
Such a molecular Kondo lattice may exhibit a strongly-renormalized Fermi-liquid
phase known as the heavy-fermion phase, {\it e.g.}, see \textcite{Coleman2015}.

\subsubsection{The Kondo effect as evidence for open shell structure\label{sss.Open-shell}}

We pointed out in Sec.~\ref{sIV.A2} that above the Kondo temperature it can be tricky
 to differentiate between
 open and closed shell molecules solely based on transport spectroscopy.
 The reason is that in molecular junctions, typically,  the level spacing
 (e.g. the HOMO/LUMO gap) is comparable to the charging energy.
The appearance of a Kondo resonance pinned at zero bias
provides a direct proof of an open-shell nature of the contacted
molecule, as illustrated in Fig.~\ref{fig:kondo}. It is not surprising that the Kondo resonance (and its
sensitivity to microscopic details) allows to gain qualitative 
insights into the charge-transfer and level alignment of a molecule 
in contact with leads.
We review a few examples where such
understanding was achieved. In these works, the interpretation of 
the experimental results was often backed-up by {\it ab-initio}
DFT
calculations. Although state-of-the-art DFT incorrectly treats
the Coulomb blockade regime, it can provide qualitative insights into the
charge state and the character of the relevant frontier orbitals 
of the contacted molecule.

\begin{figure}
\includegraphics[width=.65\columnwidth]{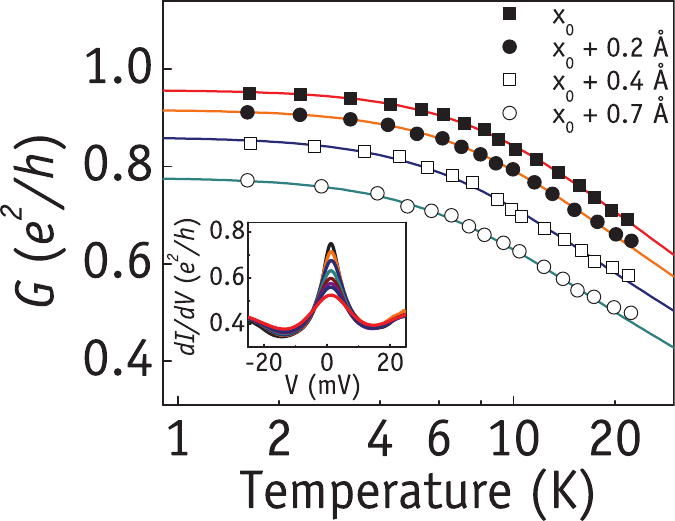}
\caption{\label{fig:kondo}{Temperature dependence of the 
linear conductance  of C$_{60}$ in a gold
junction for different electrode separations \cite{Parks2007}.
Solid lines are fits to the formula 
$G(T) = G_{0}\left[ 1+ \left(T/T_\mathrm K\right)^2
\left( 2^{1/s} -1 \right)\right]^{-s} + G_\text{b}$, where
$s=0.22$ and $G_\text{b}, G_{0}, T_\mathrm K$ are parameters.
The widely-used formula is a fit to the theoretical $G(T)$ from
a numerical renormalization group calculation \cite{Costi1994,Goldhaber1998}. 
The extracted $T_\mathrm K$ are (from the top trace to the bottom):
60.3 $\pm$2.4, 55.5 $\pm$ 0.9, 45.6 $\pm$ 1.9, 38.1 $\pm$1.2 K.
Inset: d$I/$d$V$ traces at $x_0 + 0.7$\AA.} Reproduced from \protect\textcite{Parks2007}. }
\end{figure}

 A first series of examples \cite{Parks2007,Temirov2008, Mugarza2011} comprises molecules that have a closed shell configuration when neutral, 
but a Kondo resonance appears when the molecule is contacted.
 \textcite{Temirov2008} studied PTCDA adsorbed on Ag. By contacting
 the molecule with an STM tip and gradually peeling off the molecule 
 from the substrate, Temirov \ea recorded the d$I/$d$V$ spectrum and
 observed the emergence of a zero-bias resonance ZBR on top of a weakly
 varying background. The resonance 
 width was approximately 15 meV. Standard tests of the Kondo effect are the
 magnetic field splitting and the temperature dependence; however, due to the
 rather large width of the ZBR such tests were not accessible.
 The identification of the ZBR with the Kondo resonance thus relied on the observed 
 pinning to the position at zero bias and on comparison with model and {\it ab-initio} 
calculations \cite{Greuling2011}.
 The findings can be rationalized by realizing, that
 the peeling represents an inverse adsorption process: adsorbed
 PTCDA accommodates two electrons in the LUMO and a gradual decoupling
 should lead to redox transitions. 
 A similar redox transition leading to the appearance of the Kondo effect was proposed by \textcite{Karan2018} to explain the d$I/$d$V$ changes in an iron porphyrin/Au junction.

The intricacies of the level alignment further deepen when
a molecule with two relevant orbitals is considered.
\textcite{Requist2014}
studied an NO molecule adsorbed on Au(111). The Kondo
resonance observed in STM was proposed to arise from a two-orbital system at quarter filling. 
DFT calculations offered the following scenario: the electron is distributed in both
 even and odd states ($2\pi^*_\mathrm e$, $2\pi^*_\mathrm o$), but the $2\pi^*_\mathrm e$
carries most of the electronic occupation.
When the DFT results were contrasted with an NRG calculation, correlation
effects invert the order of occupancies, so that the odd $2\pi_\mathrm o^*$ orbital
wins. We remark that in the case of NO adsorbents there is no
symmetry protection for the even and odd state, in contrast
to the SU(4) limit.
New computational approaches are being developed, that aim
at improved {\it ab-initio} descriptions of strongly-correlated phenomena in molecular junctions and adsorbents 
\cite{Jacob2015,Droghetti2017}.

 Molecular junctions offer new means of manipulating and controlling  
 the microscopics that enters the Kondo physics. For example, 
 Kondo resonances were switched on or off by
 changing the length of anchor groups \cite{Park2002},
 by current-induced dehydrogenation \cite{Zhao2005}
 or mechanical stretching \cite{Rakhmilevitch2014}.
 Pulses of electric current were used by \textcite{Miyamachi2012}
 to switch a molecule from a spinfull ($S=2$) to a
 spinless ($S=0$) state  on the spin-crossover compound 
 Fe(1,10-phenanthroline)$_2$ 
 (NCS)$_2$ on CuN/Cu(100). One of the keys to identifying
 the switching was the presence and absence of a Kondo
 peak in the respective spin states.
 
 We recall the exponential sensitivity of the Kondo temperature
 on the microscopic parameters, Eq.~(\ref{eq.tk}), which can be exploited
 as a sensitive probe of subtle interactions:
 \textcite{Jacobson2015} studied  CoH and CoH$_2$ complexes adsorbed on a spatially corrugated surface (a hexagonal
 boron nitride monolayer on Rh(111)) and systematically observed the Kondo
 ZBR only on adsorption sites of a specific type.

A robust ZBR indicates the conservation of localized radical (spin half) character
 of a molecule. Such observation was made by \textcite{Frisenda2015}
 when attaching a polychlorotriphenylmethyl radical to a pair
 of Au leads, and stretching the junction in a MCBJ setup by 0.5 nm. 
 The ZBR retained its width ($\approx 5$ mV), although
 the background conductance dropped ten times.

An important feature of the differential conductance spectrum
of open-shell molecules is the appearance of thresholds at finite
bias, which are due to inelastic spin excitations 
\cite{Hirjibehedin2007,Gaudenzi2017,Osorio2010,Fock2012}.
The finite-bias thresholds can be analyzed 
in external fields, supplied by the gate voltage and a
 magnetic field, which helps to determine the nature of the low-energy
excitations of the open-shell system.
{\color{black}
Unlike vibrational signals in IETS, the IETS features due to spin excitations are characterized by 
selection rules due to spin conservation (to the extent that spin-orbit 
interaction is weak). A remarkable consequence of the latter can be a suppression 
of zero-bias conductance, termed spin-blockade \cite{Romeike2007,Bruijckere2019}.
Orbital degrees of freedom lead to similar steps in the differential conductance \cite{Kuegel2018}.

}
Fundamental questions related to electronic correlations
were studied using single open-shell molecules coupled to
ferromagnetic \cite{Pasupathy2004, Fu2012} and
superconducting leads \cite{Franke2011, Hatter2015}. However,
a thorough discussion of these works is beyond the scope of this review.

\subsection{Franck-Condon blockade}\label{ss.Franck-Condon}
\newcommand{\braket}[2]{\langle #1 | #2 \rangle}
We consider a molecule, weakly coupled to the electrodes, 
with $N$ electrons, in its vibrational ground-state $\ket{N,0}$.
Furthermore, let the molecule be tuned
 near a charge-degeneracy point, so that 
 $\ket{N,0}$ and $\ket{N+1,0}$ are nearly degenerate.
If the electron-vibron coupling is strong, 
{\it i.e.}, the reorganization energy is large,
then the overlap $\braket{N+1,0}{N,0}$ is small.
Consequently, the low-bias  current is suppressed 
even though charge-fluctuations are not strongly hindered 
by Coulomb repulsion. 
At larger bias it is possible to excite the vibrational states, 
$\ket{N+1,n}$. According to Franck-Condon theory,
the transition probabilities are proportional to
the overlap squared, $|\braket{N,0}{N+1,n}|^2$ with 
\begin{equation}
\label{eq:fcb}
|\braket{N,0}{N+1,n}|^2 \propto e^{-\lambda}\lambda^n / n!,
\end{equation}
where $\lambda$ denotes the dimensionless electron-vibron
coupling. For strong coupling ($\lambda \gg 1$) the matrix element is exponentially
small at $n=0$, but reaches a maximum for finite $n=n_\text{max}$.
Thus, transport is recovered at finite bias when the electrons have excess energy $eV$ large enough for exciting 
$\sim n_{\rm max}$ vibrons, see Fig.~\ref{fig:fcpar}.
The unique properties of electronic transport in the Franck-Condon blockade (FCB) regime
were analyzed theoretically in \textcite{Koch2005,Wegewijs2005,Leijnse2008,Kaat2005} and other works.
\begin{figure}[!t]
\includegraphics[scale=0.22]{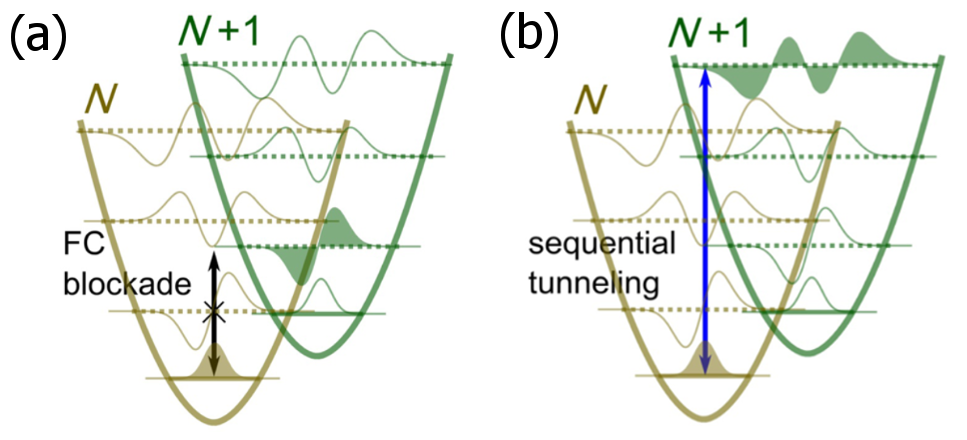}
\caption{\label{fig:fcpar}
Illustration of Franck-Condon blockade in a molecular junction assuming a parabolic 
potential surface with eigenenergies
$n\hbar\omega_0$ and vibronic (harmonic oscillator) wavefunctions.
(a) At low excitation levels transport is suppressed because 
the oscillator overlap between the $\ket{N,0}$ and $\ket{N+1,1}$ is small. 
(b) At higher bias,  higher vibrational levels $\ket{N+1,n}$, the overlap with 
$\ket{N,0}$ is enhanced and current starts to flow. 
Reprinted with permission from Lau {\it et al.}, Nano Lett. {\bf 16}, 170. Copyright (2016) American Chemical Society.
}
\end{figure}
Note, that FCB derives from the suppression of overlap matrix elements and hence 
can not be lifted by tuning the gate voltage (see Fig.~\ref{fig:franckcondon}), 
in a stark contrast to Coulomb blockade. 
This behavior has been taken as a hallmark in  the experimental discovery of FCB.
In the context of nanoscale devices, FCB was first observed in carbon
nanotubes \cite{Leturcq2009}, where, quite remarkably,
the blockade survives over several charge states $N$. 
A molecular FCB was first observed by \textcite{Burzuri2014} in a Fe$_4$ complex.

\begin{figure}[!b]
\includegraphics[width=0.95\columnwidth]{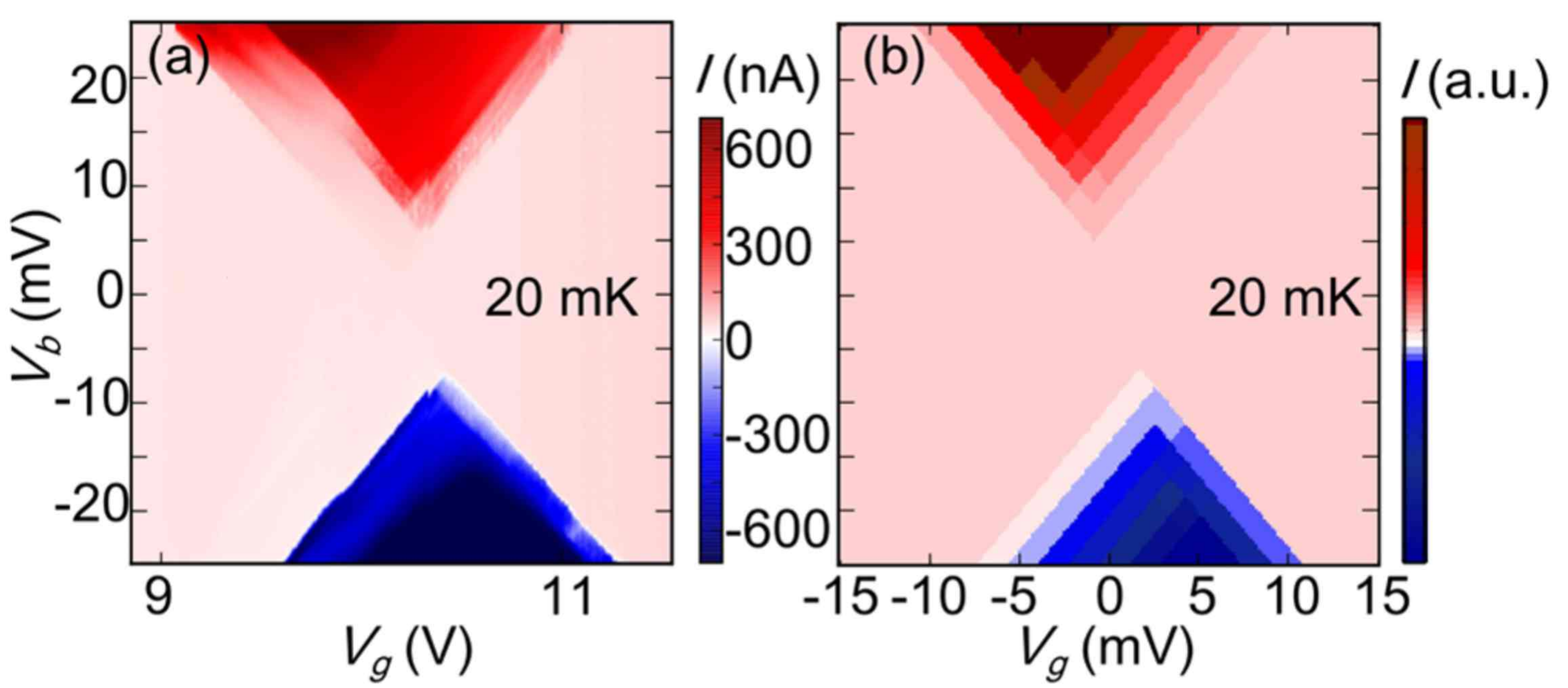}
\caption{\label{fig:franckcondon} (a) Current stability diagram of 
a functionalized C$_{60}$ bound to graphene nanoelectrodes 
at 20 mK. (b) Simulated current stability diagram with $\lambda = 3$,
$\hbar\omega_0 =1.7$meV. High current is shown in dark shades (red and blue online) 
and low current is light gray (pink online). The current is suppressed
for low bias voltage $|V_b| \lesssim 6$ mV because of
Franck-Condon blockade. The blockade can not be lifted
by changing the gate voltage $V_g$ For higher bias, the current is restored,
its detailed $V_{g,b}$--dependence bears strong imprint of
vibronic excitations. 
Reprinted with permission from Lau {\it et al.}, Nano Lett. {\bf 16}, 170. Copyright (2016) American Chemical Society.
}
\end{figure}

The FCB has some remarkable consequences for the charge-carrier
dynamics \cite{Koch2006}. Above a threshold bias of a few times $\hbar\omega_0$ 
the electron current is accompanied by excitations and de-excitations
of vibrons. The latter have their fingerprint in unusual
temporal fluctuations of the current, {\it i.e.}, in the current noise
(`avalanche transport'). 
For an extended analysis 
of noise and avalanches see \textcite{Schinabeck2014}.
%

The bistable behavior of the current in the FCB was recently 
observed in an experiment by  \textcite{Lau2015} for a functionalized C$_{60}$ molecule 
coupled to graphene electrodes.
The waiting times of the on and off phases were of the order 
of $10^{-1}$ s, extremely long for molecular vibration processes. 
The FCB with avalanche transport was also manifested by
 giant enhancement of noise above the regular shot noise level ($10^2-10^4$) in accord with the
 theoretical prediction \cite{Koch2005}.

\section{Case studies of quantitative comparison}\label{s.CaseStudies}

Having reviewed some of the most impressive phenomena observed in single-molecule junctions, for which we have a high level of qualitative understanding, we 
 now turn to a discussion of the quantitative level of agreement between computation and experiment. In many experiments the conductance is the sole quantity reported, and consequently this is also what computations often focus on. It will be useful to distinguish three conductance regimes, which leads us to organizing this section into the following thee subsections. 

\subsection{High zero-bias conductance}\label{s.High}
At high zero-bias conductance, with quantum conductance channels having transmission probabilities near unity, the experiments give access to many parameters other than just the conductance. The systems we consider here are small molecules, with strong hybridization of the molecular levels with the metallic states in the leads. As a consequence, the level broadening is much larger than any other energy scale in the problem, which makes the computations less sensitive to the unknown details of the metal-electrodes arrangement. 

A prime example is a hydrogen molecule, H$_2$, coupled to Pt leads at either side \cite{Smit2002}. Clean Pt leads can be obtained by means of the MCBJ technique at cryogenic vacuum conditions, and the molecules can be introduced into the gap by deposition from the vapor phase. The conductance histogram after deposition of H$_2$  shows many changes, indicating that many types of conducting bridges involving Pt-H bonds are being formed, but a prominent peak near a conductance of 1\,\go\  is the most conspicuous feature. The experiments mainly focus on analyzing the structure of the bridge associated with this histogram peak. 

\begin{figure}
\includegraphics[scale=0.3]{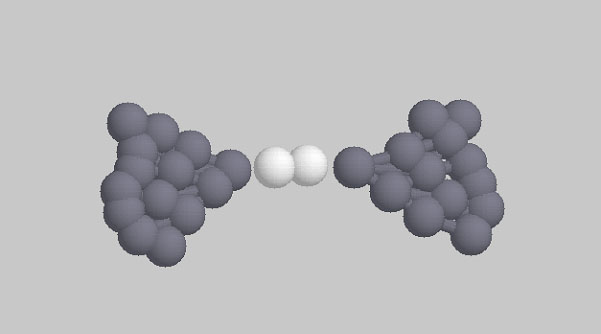}
\caption{Proposed Pt-H$_2$-Pt molecular configuration. The dark spheres represent Pt atoms of the metallic leads, for which the detailed arrangement is not known. The white spheres represent the H atoms that are aligned with the axis of the junction. }\label{fig.Pt-H-Pt}
\end{figure} 

The proposed configuration of the molecule in the contact is shown in Fig.~\ref{fig.Pt-H-Pt}. From a chemistry point of view, a more common arrangement of hydrogen between Pt atoms, with lower total energy, is one with the two hydrogen atoms aligned orthogonal to the junction axis, with one atom at either side. The latter configuration was proposed by \textcite{Garcia2004} as being associated with the conductance peak at 1\,\go\   in the histograms. In contrast, from their DFT based computations they found a conductance of only 0.2\,\go\   for the configuration illustrated in Fig.~\ref{fig.Pt-H-Pt}. However, at least three independent calculations for the latter configuration produce a conductance very close to 1\,\go\   \cite{Smit2002,Djukic2005,Garcia-Suarez2005,Thygesen2005}. By virtue of the many parameters that can be obtained from the experiments, including the vibration modes and their shift with isotope substitution, the contact stretching dependence of the vibration modes \cite{Djukic2005}, and the 
number of conductance channels derived from shot noise measurements \cite{Djukic2006}, the discrepancy between the calculations can be definitely decided in favor of the the latter group of computational results. Although the critical difference between these computations has not been identified, the example discussed here demonstrates the importance of benchmarking the computations against well-characterized experimental results. Barring the result by \textcite{Garcia2004} the quantitative agreement on the conductance between experiment and computations in this regime of transmission is near 10\%.

At first sight the high electron transmission of a closed-shell molecule such as H$_2$ is counter-intuitive. A very instructive discussion of the principles involved in arriving at this high conductance  is given in \textcite{Cuevas2003}. Although the HOMO-LUMO gap for the H$_2$ molecule is very large, the H$_2$ molecular orbitals hybridize strongly with the d-orbitals of Pt, which have a very large density of states, resulting in a high electron transmission. The other key experimental observation, namely that the conductance is carried by a just single channel, is rooted in the axial symmetry of the molecular bridge, illustrated in Fig.~\ref{fig.Pt-H-Pt}.

The approach of exploiting the chemical affinity of clean metal tips for direct binding to molecules, without the need for anchoring groups, has been extended to organic molecular systems, notably benzene \cite{Kiguchi2008} and the related oligoacenes \cite{Yelin2016}. The more elaborate molecular orbital structure of these aromatic molecules permits multiple conductance channels for a single molecule. With a total conductance for a Pt-benzene-Pt junction in the range from 0.1 to 1.3~\go. Analysis of shot noise shows that up to three channels participate in the electron transport. This agrees well with the accompanying DFT calculations, which offer the interpretation that the number of carbon atom bonds to each of the Pt tip atoms reduces from three to one in the process of gradual breaking of the junction \cite{Kiguchi2008}. 
Quantitative comparison between theory and experiment is influenced by the limited knowledge of the orientation of the molecule  in the experiment. Given the combination of channels obtained from shot noise measurements, and total conductance, the quantitative agreement is probably better than 30\%.

The molecular orientation and the effect of the number of carbon-tip bonds can be observed directly through low-temperature STM imaging at close tip-molecule distance. Approaching C$_{60}$ on Cu(111) by a Cu coated tip the work by \textcite{Schull2011} shows the effect of local bond formation with the tip, which clearly distinguishes sites on the molecule having stronger double C=C bond character (higher conductance) from those with predominantly single C-C bond character (lower conductance). In the case of C$_{60}$ the conductance is high because the LUMO  {\color{black} is triply degenerate and} nearly coincides with the Fermi energy of the metal. The DFT computations reproduce the different symmetries and bonding sites on a single C$_{60}$ molecule. However, differences between calculations and measured site-dependent conductance remain, which are partly attributed to a small tilt of the molecules that was not included in the calculations. 
Quantitatively, despite the high level of experimental information available, a gap of about a factor of 2 remains between experiments and calculations. 

In this regime of strongly coupled molecules and high conductance the quantitative agreement between the conductance in experiments and computations is generally good. The computational accuracy in reproducing the experimental results appear to be mostly limited by details of the experiments that remain unknown. One such effects comes from scattering of partial electron waves on defects in the metallic leads \cite{Ludoph1999a}, which can reduce the conductance typically by 10\%, and in exceptional cases even up to 50\%. Although more precise testing, for example by means of low-temperature STM experiments, would be desirable the present evidence suggests that LDA-based calculations in this regime capture the essential ingredients and that the computations have a high degree of predictive power. 

\subsection{Low zero-bias conductance}\label{s.Low}
Before turning to the most widely studied class of molecules, let us first consider the other extreme, the range of very low zero-bias conductance. As a practical criterion we take this to include those single-molecule junctions for which the zero-bias conductance, $G(0)$, lies below the measurement sensitivity of the experiments. This limit usually lies at $10^{-5}$~\go, but with proper electronics can be extended to $10^{-7}$~\go. For those junctions the molecules would be undetectable by the common histogram-based methods at low bias. Instead, the experimental evidence for the presence of the molecules comes from current-voltage (\IV) curves up to high voltage bias. The differential conductance shows a gap of low conductance until one observes a rise, often steep, which may lie at different values for the two bias polarities. If the junction remains stable one may observe  that the conductance reaches a peak value and comes down again at still higher bias. 

\begin{figure}
\includegraphics[scale=0.7]{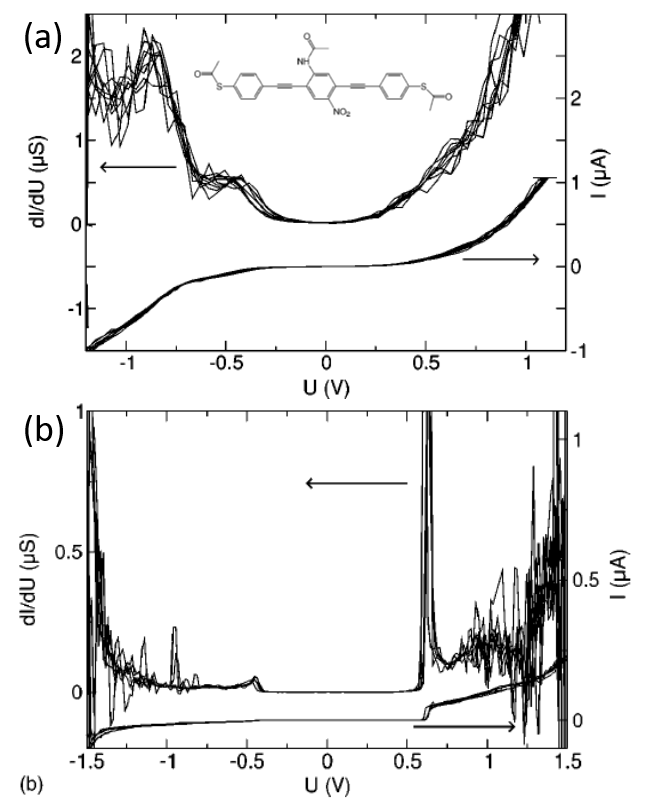}
\caption{Single molecule measurements on a thiol-coupled molecule (shown in the inset on top) having side groups that make it asymmetric. The current (right axis) and differential conductance (left axis) reflect the asymmetry of the molecule. The sign of the asymmetry is random, and changes between different measured junctions. The gap in the conductance observed at room temperature (a) become much more pronounced at 30~K (b). For each curve several consecutive sweeps are shown.  
Adapted from Reichert {\it et al.}, Appl. Phys. Lett. {\bf 82}, 4137 (2003) with the permission of AIP Publishing.
 }
\label{fig.Reichert-Weber}
\end{figure}

Figure~\ref{fig.Reichert-Weber} shows one of the first single-molecule measurements at low temperatures \cite{Reichert2003}, and gives a comparison with data for the same molecule obtained at room temperature.  The low-temperature data show a clear gap in the differential conductance until a sharp peak at about 0.5~V signals the line-up of the chemical potential in the leads with the nearest level on the molecule. The irreproducible fluctuations observed after the peaks are due to instability of the molecular junction under the influence of the high current flowing through the molecule. Comparing this data to the room-temperature curves in Figure~\ref{fig.Reichert-Weber}(a) reveals several important effects. First, the peaks in the differential conductance become strongly smeared. 
Second, the gap appears smaller, and the zero-bias conductance is larger, at room temperature. These two effects can be understood as resulting from the thermal smearing of the molecular level structure by rapid fluctuations between many configurations. 
This implies that the position of the molecular levels is very sensitive to details, as will be further demonstrated with experiments below. Obviously, 
one needs to be careful in interpreting conductance and gap structure from room temperature data.

The positions of the resonances observed in the differential conductance can be influenced by local electrostatic potentials. Low-temperature experiments using in-situ deposition of the metal electrodes \cite{Kubatkin2003}, experiments using electromigration \cite{Park2002,Liang2002,Osorio2007}, or three-terminal break junctions \cite{Perrin2013}, all demonstrate the sensitivity of the molecular level to the potential applied to a gate electrode. The advantage of a gate electrode is that one can observe this shift continuously as a function of a control parameter, {\it i.c.} the applied voltage. However, the actual potential at the molecule will be sensitive to the presence of ions and polar molecules (water), or even dielectrics. Moreover, the image charge distribution in the leads, and therefore the distance of the molecule to the leads and its orientation, also strongly affect the position of the resonances \cite{Perrin2013}. 

Computations for such molecular systems mainly aim at qualitative comparison and cannot easily be made quantitative for several reasons. First, the positions of the resonances are not reliably obtained from DFT type calculations, given the difficulties associated with the theory in predicting proper gap energies. Second, due to the exponentially small local density of states on the molecule near the Fermi energy, the positions of the resonances are extremely sensitive to the amount of charge transfer between the molecules and the leads.  Third, for weak coupling charging effects become important, which are not easily treated from first-principles in DFT. Finally, the calculations must be performed self-consistently under non-equilibrium conditions. While such calculations have been demonstrated for simple systems \cite{Stokbro2003,Arnold2007,Liu2006}, they are computationally demanding, as was discussed in Section~\ref{ss.comp.non-linear}.   

Apart from the position of the resonances, the shape of the resonances poses a problem. A straight-forward application of resonant tunneling models does not work: the position, height, and width  of the resonance cannot be simultaneously described by any choice for $\Gamma$; typically the observed resonance width is by far too large, as shown by the low-temperature MCBJ experiments by \textcite{Secker2011}. The conductance at the peak lies invariably much below \go, in contrast to the predictions from simple resonant tunneling models. Part of the explanation lies in the role of vibrations. When the resonance lies at low energy (tens of millivolts) clear vibrational side bands can be resolved, as was also shown systematically by \textcite{Osorio2007a} and \textcite{Lau2015}. When the resonance shifts to higher energy the vibrational side bands broaden and eventually cannot be resolved. Yet, they continue to determine the full width associated with the resonance. Thus, the width of the resonance is not set by 
temperature or by the strength of the coupling $\Gamma$ between the molecule and the leads, but is approximately set by the reorganization energy due to electron-vibration mode interaction \cite{Secker2011}.

Electronic resonances at still higher energy develop temporal fluctuations, with characteristic times that can be milliseconds \cite{Secker2011} or much longer \cite{Lortscher2006}. A typical \IV curve may show a smooth broadened step in current, but zooming in at the step one finds that the smooth transition actually is due to instrumental time-averaging of fluctuations of the conductance of the junction, roughly between the values of current before and after the step. Clearly the current induces fluctuations in the configuration of the molecule, as was also found for molecules with high zero-bias conductance, where they are triggered by the excitation of vibration modes \cite{Thijssen2006}. As pointed out by Secker {\it et al.}, given the many additional effects taking place at a resonance the similarity of the experimental curves with those obtained from resonant tunneling models must be regarded as being fortuitous. The broadening of the resonance in the differential conductance may also be taken as 
evidence that the lattice temperature of the molecule under high bias becomes very high.

\subsection{Intermediate zero-bias conductance}\label{s.Intermediate}
Having discussed the two extremes, high zero-bias conductance and low conductance, we find that the former type of junctions can be accurately described by computational models, while the latter can only be addressed qualitatively. Let us now review the large group of molecules with intermediate levels of zero-bias conductance, which comprises nearly all molecules studied by means of conductance histograms in break junctions. It goes beyond the scope of this review to present a complete overview of the many molecular systems in this class. Instead, we will focus on two widely studies molecular systems, benzenedithiol, a simple aromatic molecular system, and the non-aromatic alkanedithiols, that have been adopted as benchmark systems. We will propose alternative benchmark systems and conclude with a few general remarks.

\subsubsection{Benzenedithiol} \label{ss.BDT}
The first electronic transport measurements of a single molecule by \textcite{Reed1997} targeted a benzene-1,4-dithiol molecule between two gold electrodes. This system became one of the fruit-fly systems in this field of study. The choice appears to be a natural one, because it is one of the simplest systems that offer a fully conjugated path across the molecule. The benzene ring provides a delocalized $\pi$ electron path for conduction, and the sulfur group is assumed to bind covalently to the gold leads and to produce a small electronic barrier between the states of the metal leads and the $\pi$ electrons of the benzene ring. For the zero-bias conductance many different values have been reported, but typically it lies well above the experimental sensitivity threshold, while the differential conductance is strongly non-linear and shows peaks at high bias. In this sense the molecule has properties that are clearly intermediate between the two cases discussed in the sections above.

In some ways the choice of this fruit-fly system is also an unfortunate one, because there are few molecular systems for which the experimental results -- and there are many -- differ so widely. On the other hand, the challenge of understanding this simplest of all aromatic molecular junctions has stimulated many theoretical works and therewith helped to uncover many effects that need to be considered in evaluating molecular junctions. The transmission of the molecule is expected to be high for such a short molecule with a fully delocalized pi-orbital system. However,  the measured conductance is extremely sensitive to the details of the connection of the molecule to the Au metal leads, and the short molecule allows for many different arrangements. It turns out that there are many more factors influencing the observed conductance, that also have implications for the interpretation of experiments on other molecules. An overview of the experimental data is presented in Table~1 in the Supplemental Material, 
and representative conductance histograms from different experiments are compared in Fig.~\ref{fig.BDT}.

The original experiment by \textcite{Reed1997} was done in a break junction setup, but did not present any statistical analysis or histogram distributions. 
%
\begin{figure}[t]
\includegraphics[scale=0.20]{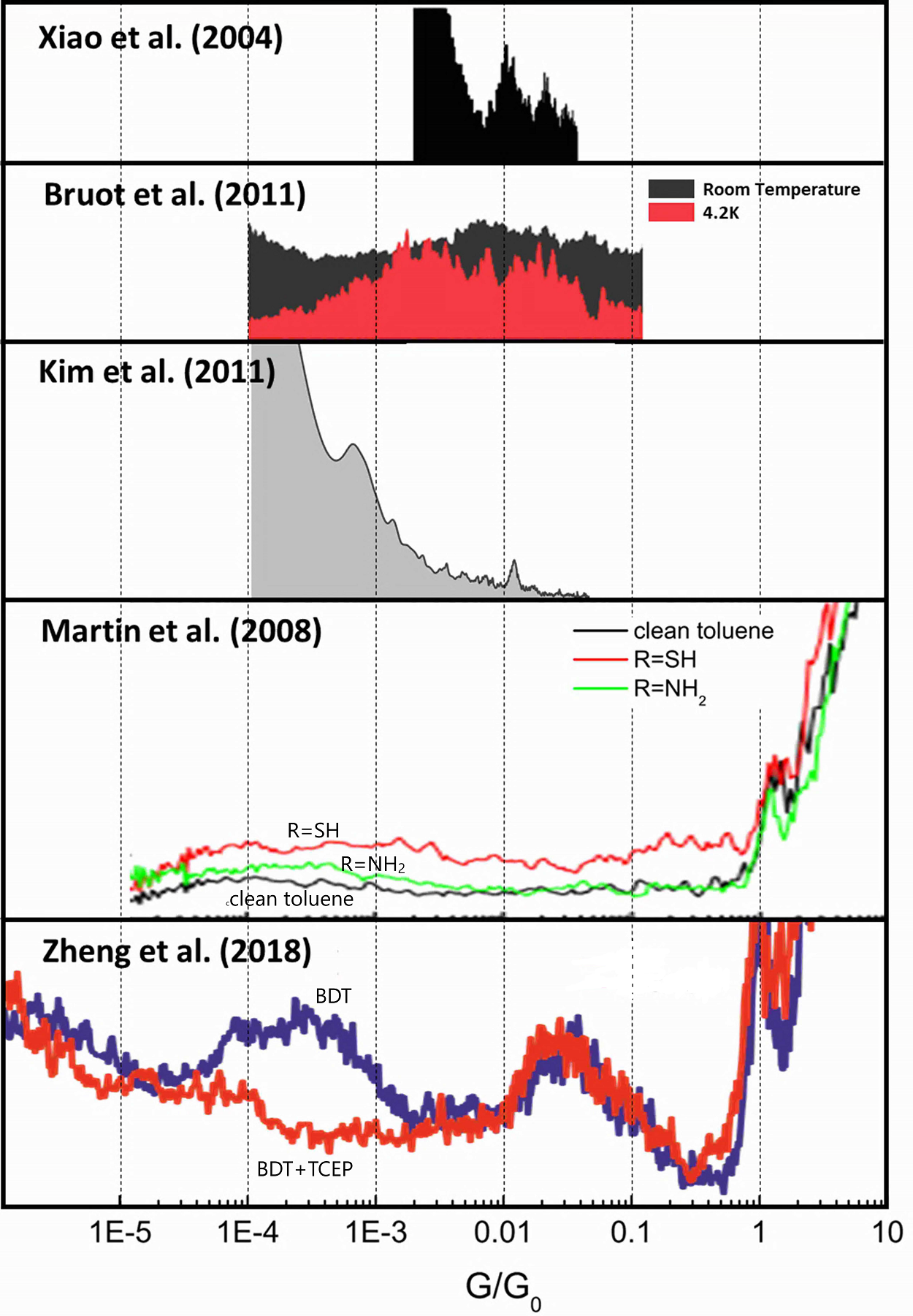}
\caption{A selection of conductance histogram data for Au-BDT-Au junctions, illustrating the wide variation and sensitivity to experimental conditions (color online) . The histograms have been rescaled for ease of comparison. The figure also illustrates differences in appearance of the data due to choice of the methods of constructing histograms, see also Section~\ref{ss:IID}. Three methods have been applied: (1) First binning data in linear conductance, and representing this in a histogram in linear conductance. (2) First binning data in linear conductance, and representing this in a histogram in the logarithm of conductance. (3) Binning the logarithm of conductance and  representing this in a histogram. The latter method is now the method of choice, but earlier data have used the other two. 
The data by Xiao \ea  (top panel) uses method (1) and has been mapped onto this logarithmic scale. The data by Kim \ea uses method (2). This enhances the data at low conductance, and the peaks are observed against a decaying background. The other three examples employ method (3). 
The graphs are adapted with permission from Xiao {\it et al.}, Nano Lett. {\bf 4}, 267.  $\copyright$ (2004) American Chemical Society;
Bruot  {\it et al.}, Nature Nano. {\bf 7}, 35. $\copyright$ Springer Nature (2011); 
Kim  {\it et al.}, Nano Lett. {\bf 11}, 3734. $\copyright$ (2011) American Chemical Society;
Martin  {\it et al.}, New J. Phys., {\bf 10}, 065008 (2008). $\copyright$ IOP Publishing and DPG. All rights reserved;
Zheng {\it et al.}, Chem. Sci. {\bf 9}, 5033 (2018), published by The Royal Society of Chemistry.
 } \label{fig.BDT}
\end{figure}
%
Therefore, one cannot judge whether the curves shown in the paper are representative for the Au-BDT-Au junctions. This situation improved after the introduction of statistical analysis based on conductance histograms \cite{Smit2002,Xu2003}.  \textcite{Xiao2004}  were the first to report such measurements for Au-BDT-Au, and found a series of peaks in the histogram at multiples of $1.1\cdot10^{-2}$~\go\   (Fig.~\ref{fig.BDT}, top). The measurements were done at room temperature at a bias of 0.2~V in a 0.1~M solution in NaClO$_4$. A systematic study of the bias dependence up to 0.6~V showed only a weak increase of the conductance with bias, by about a factor of two. 

This result brought the experimental values much closer to the theory, but also led to new confusion. The problem lies, at least partly, in the method of data selection. Most traces of conductance during breaking show only smooth exponential decay. For this reason Xiao \ea decided to select only those curves that have a distinct plateau region for inclusion in the histogram. The fact that this selection is done manually makes the criteria difficult to judge. 
The shape of the histogram has not been reproduced by other groups. In fact, a new study by \textcite{Ulrich2006}, by essentially the same technique, now using 1,2,4-trichlorobenzene as solvent, showed that there is no clear structure in the histograms when all curves are included. This conclusion did not change even after implementing an automated data selection procedure, or by taking the experiment to low temperatures (30~K).

Further work \cite{Fujii2008,Tsutsui2008,Kim2011,Martin2008} confirmed that the presence of the molecules in the junction leads to histogram counts spread over many orders of magnitude, ranging from below $10^{-5}$ to several times $10^{-1}$~\go. Only at high bias, above 1~V, Tsutsui \ea observed a broad peak at 0.1~\go.
Different experimental approaches have produced peaks in the histograms for BDT, without data selection, although the position of the peaks varies between the methods, suggesting that it is influenced by the choice of experimental method. Using acetyl-protected sulfur groups \textcite{Lortscher2007} studied the sulfur-coupled benzene molecule by MCBJ under UHV at room temperature, and found a single broad peak at $0.5\cdot10^{-4}$ \go. Note that the acetyl protection group ideally requires a de-protection agent to be removed in order to allow direct S-Au bond formation. Since the authors do not report using active de-protection there is a possibility that the nature of the bond to Au is different, possibly explaining the low conductance. The question of de-protection is of wider interest, and is related to the question whether the H atom of the thiol group splits off upon bonding to Au, as was discussed in Section~\ref{ss:VC}.\footnote{
For BDT this question may possibly be decided by analyzing the differential conductance.
\textcite{Stokbro2003} calculated \IV curves for BDT self-consistently, and the shapes for the \IV curves for the thiol bonded and the thiolate bonded molecules were found to be qualitatively different. The \IV curves for the thiolate bonded molecules show no gap, but a weak local minimum around zero bias at a high value of 0.46 \go. Instead, for the thiol bond the \IV curve shows a large gap. }

By their I(t) technique \textcite{Haiss2008} obtained for the same acetyl protected compound a conductance of $1.1\cdot 10^{-4}$ \go, agreeing within a factor of 2 with the value given by  L{\"o}rtscher \ea  Figure~\ref{fig.BDT} illustrates the wide variability in the appearance of conductance histograms for Au-BDT-Au single-molecule junctions. 

Recently, \textcite{Kaneko2019} demonstrated the role of different  molecule-metal anchoring configurations in the conductance for BDT. They used a self-breaking procedure, by which the contact is formed under the influence of spontaneous slow relaxation of the metal leads at room temperature at the last stage of breaking of a junction. This is likely to select only the more stable junction geometries, and they find a high conductance value at $2.4 \cdot 10^{-2}$\go, a medium value at $3.4 \cdot 10^{-3}$\go, and a low value at $3.9 \cdot 10^{-4}$\go. Through simultaneous measurements of surface-enhanced Raman signals of the same molecule, 
they propose at an interpretation for the three conductance states as due to binding to bridge, hollow, and top sites, respectively.

{\it Theory for BDT.} 
BDT also played an important role as a model system for benchmarking newly developed {\it ab-initio} based transport codes. For this purpose, BDT was suitable because on the one hand hand it is a molecule large enough for it to display typical features of molecular orbitals, while on the other hand it is small enough to keep the numerical effort manageable. 

First studies of the BDT-transmission relied on simplified approaches
such as, {\it e.g.}, H\"uckel models \cite{Emberly1998} and 
DFT-studies with electrodes treated as Jellium-models \cite{DiVentra2000}. 
{\color{black}These models were helpful in analyzing the possible role of multiple molecules in the junctions and variations in bonding configurations \cite{Emberly2001}.}
It was realized early on, however, that a proper description of transport 
of atomic scale systems, {\it i.e.} atomic wires and molecules, requires 
treating system and contacts on the same 
footing \cite{Yaliraki1999,Palacios2001,Brandbyge2002}, 
see  Section~\ref{s.Computational}. 
As a consequence, with time progressing, transport studies 
became more elaborate, and included an ever growing number of contact atoms.

To facilitate the extensive calculations, efficient
{\it ab-initio} codes have been developed. 
First consistent results for the BDT-transmission with Au contacts have been obtained  by \textcite{Xue2003};   
\textcite{Stokbro2003}; \textcite{Evers2004,Faleev2005,Ke2005,Thygesen2005a,Grigoriev2006,Kondo2006}; and by \textcite{Garcia-Suarez2007}.
In these studies minor differences in $\transmission(E)$ persist, {\it e.g.}, 
with respect to the energy-position of the d-band shoulder; 
they can plausibly be attributed to variations in the simulation details
concerning, {\it e.g.}, functional approximations and contact geometries.  
A systematic study of artifacts related to the finite size of the simulated electrodes 
has been performed by \textcite{Evers2006}. 

It turned out that improving the simulation technology
was associated, typically, with an increased zero-bias transmission. 
Perhaps somewhat surprisingly, this made the perceived discrepancy 
between DFT-based transport studies and experiments larger, not smaller.
Since simulations rely on an {\it ad-hoc} assumption concerning the junction's 
atomistic geometry, computational studies have been aimed at investigating the 
sensitivity of the transmission with respect to geometry 
changes.
As one would expect, at transmissions of order unity
very strong changes of the binding geometry are required to
reduce the transmission significantly. This is because 
large overlap-matrix elements have to be decimated. 
Hence, the transmission of BDT turned out to be rather robust against 
geometry changes, with observed variations of the order of percents \cite{Evers2004}.  \footnote{Note 
that a much larger sensitivity to the binding sites (on-top, bridge, or hollow) on a gold 
surface in the early work by \cite{Yaliraki1999}, with variations up to three orders of magnitude, is likely due to the 
approximations used for the coupling to the leads.}  

\textcite{Basch2005} 
obtained a large conductance of 0.5 \go\  for Au-S-benzene-S-Au binding to a flat Au surface. 
They also investigated the effect of atomic chain formation,
which happens when, upon stretching the junction,
Au-atoms are pulled out of the surface and form a connecting chain. 
In this case, the conductance becomes extremely sensitive to the alignment of the molecule with the chain axis, 
giving values ranging from nearly perfect transmission, down to $3\cdot10^{-4}$ \go. 

In similar computational studies, and assuming a thiolate bond,
\textcite{Sergueev2010}, \textcite{BorgesPontes2011} and \textcite{French2013} observe that the conductance can vary 
under stretching non-monotonically between 0.01 and 0.5 {\go}.
In particular, the conductance can increase with stretching (as in the experiment by \textcite{Bruot2012}). 

It is conceivable that at least in some experiments the formation of the molecular junction involves two or
more molecules. Investigating this possibility, \textcite{Strange2010}
find a rich landscape of possible configurations. 
If sufficient complexity is allowed for, the transmission can vary over orders of magnitude upon stretching. 

All in all, the computational studies offer a reasonable interpretation for the wide range of observed conductance values in the experiments.
Possibly these difficulties appear somewhat more pronounced for BDT than for other molecules because BDT is so short. 
At this point we conclude that, while at present a very detailed comparison between theory and experiment appears
to be premature, the computations nevertheless have uncovered many mechanisms that are likely to simultaneously act in producing
conductance values that vary by orders of magnitude, depending on the details of the experimental procedures.

\subsubsection{Alkanedithiols} \label{ss.ADT}
The second model system that has been widely studied is the series of alkanedithiols (ADT), again coupled between two gold electrodes.  We have discussed ADT in Section~\ref{ss.length-dependence}, where we were concerned with the systematic variation of the conductance with length. In contrast, here we will focus on the absolute numbers of the conductance, their variation between experiments, and quantitative comparison between experiment and theory. In other words, where the focus in Section~\ref{ss.length-dependence} was on the decay constant $\beta$, here we will be mostly concerned with the pre-factor $G_{\rm c}$.  

In contrast to BDT, for the alkanes the backbones of the molecules consist of non-conjugated bonds. Since all carbon-atoms are sp$^3$-hybridized, alkanes exhibit a band gap in the long wire limit and, thus, are insulating. Therefore, they form an instructive model system and complement nicely to the well studied metallic-atom chains \cite{Agrait2003}. Before the introduction of single-molecule techniques many electron transport experiments had already been reported for self assembled monolayers, see, {\it e.g.}, \cite{Salomon2003}, which we will not cover here. 
In view of the important role that ADT has played in the initial experimental investigations we will start with a few historical notes. 

{\it Early experiments.} 
\begin{figure}[t]
\centering\includegraphics[scale = 0.37]{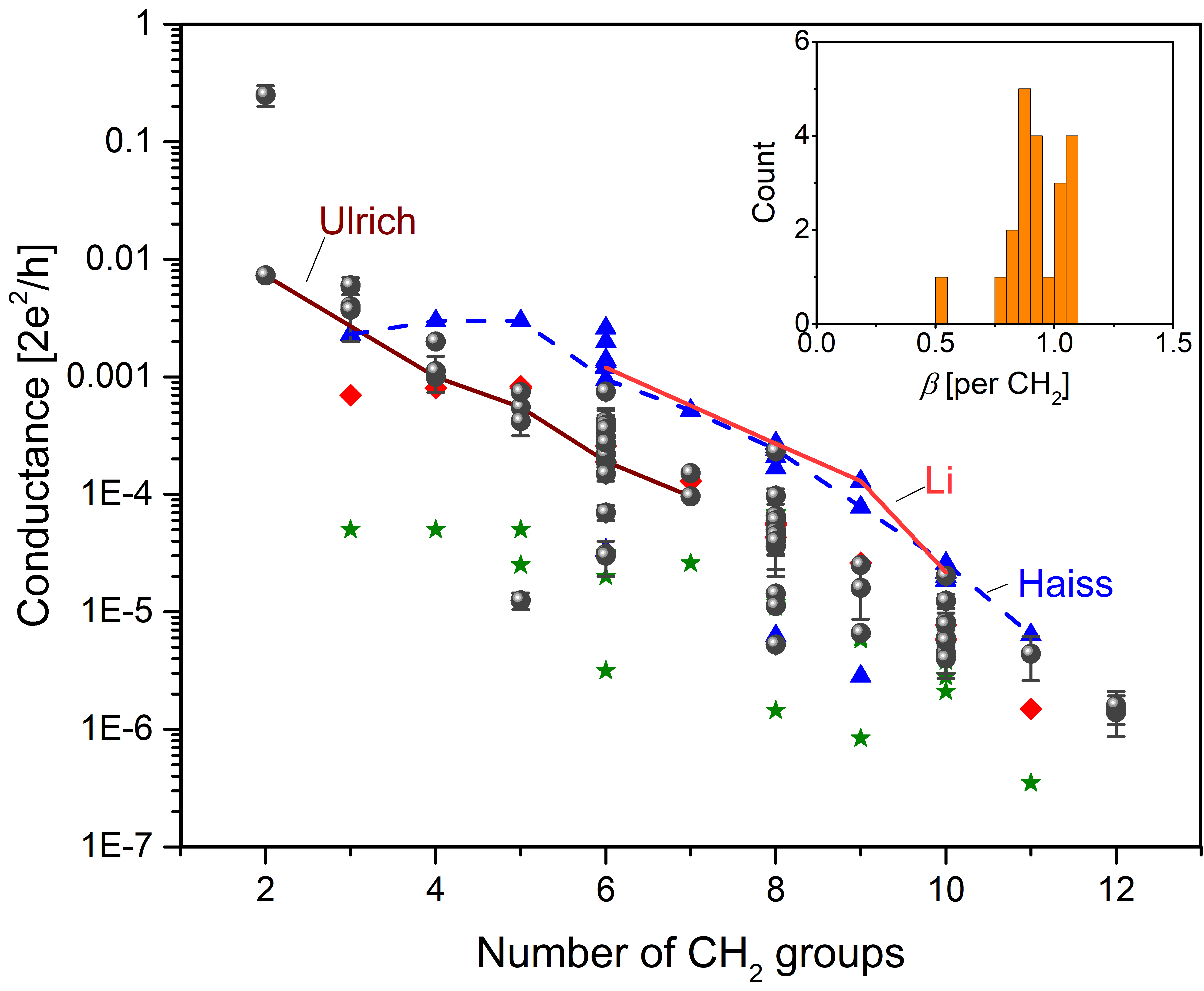}
\caption{Overview of measured conductance values for Au-ADT-Au junctions plotted on a semi-logarithmic scale as a function of the length of the backbone, in terms of number of CH$_2$ groups. 
Symbols are used for distinguishing low-conductance peaks (stars, green), medium conductance (diamonds, red), and high conductance peaks (triangles, blue), following the categories introduced by \textcite{Li2008}. Histograms that only show a single peak and for which the identification in terms of these three categories was not discussed are shown by gray bullets. The data includes published work from \protect\cite{Cui2001,Xu2003,Xu2003b,Haiss2004,He2006,Li2006,Gonzalez2006,Chen2006,Jang2006,Sek2006,Wierzbinski2006, Ulrich2006,Huang2007,Park2007,Nishikawa2007,Li2008,Hihath2008,Kiguchi2008a,Xia2008,Huisman2008,Song2009,Haiss2009,Song2010,Martin2010a,Kim2011a,Kim2011b,Arroyo2011, Hihath2015,Gil2018,Su2015}. The spread in the data is large, but the systematic trends in the separate works is shown by connecting the data points obtained from three works, namely \cite{Ulrich2006}, H-peak from \cite{Li2008}, and H-peak from \cite{Haiss2009}. This indicates that the decay constant $\beta$ is a more robust and reproducible property, as 
illustrated for the complete dataset by the histogram in the inset.  More complete details of the data collected here are given in the Supplemental Material.
\label{fig.ADT} }
\end{figure}

\textcite{Cui2001} were the first to contact and measure a single 1,8-octanedithiol (ADT8) molecule. They used a conducting-tip AFM, where the molecule of interest was inserted at very low concentration into a dense self-assembled monolayer (SAM) of the mono-thiol equivalent. Upon exposure of this molecule-covered Au(111) surface to a solution of gold nano-particles the dithiol molecules made a $-$SAu chemical bond with the nano-particles, and these were probed by the conducting AFM tip. They reported a conductance of $1.43\cdot 10^{-5}$\go~at 0.1~V bias in toluene solution. Later work from the same group for decanedithiol (ADT10) and dodecanedithiol (ADT12) showed that the conductance of the molecular junctions decreases exponentially with the length of the molecule, and they found a decay constant $\beta$ of $0.58\pm0.06$ per CH$_2$ group. This value is about a factor of two smaller than found by most groups,  which is possibly related to the effect of the dense SAM on the work function. 

This may also explain the discrepancy with the work by Nongjian Tao's group \cite{Xu2003,Xu2003b} that followed shortly after. They used their STM-BJ method to probe hexanedithiol (ADT6), ADT8 and ADT10 in solution. The conductances for the same wire lengths obtained in these works are more than an order of magnitude higher, and also the decay constant $\beta=1.04\pm0.05$ per CH$_2$  group is twice larger, but agrees well with values known from earlier work on self-assembled mono-layers. 

On the other hand, the results by Cui \ea received  support from \textcite{Haiss2004}, who used the $I(t)$ and $I(s)$ measurement techniques for probing a similar set of molecules  ADT6, ADT8 and nonanedithiol (ADT9). They found conductance values and a decay factor close to those of \textcite{Cui2001,Cui2002}.
The large discrepancy with the work from Tao's group was investigated by Haiss \ea by repeating experiments under similar conditions. Larger conductance jumps were indeed identified, and tentatively attributed to parallel conduction by gold atom chains. In our view, this is unlikely, since Au atomic chains are not stable under ambient conditions. We will return to this discrepancy below.  
In fact, there is another remarkable distinction between these experiments: where \textcite{Xu2003} find a distinctly non-linear dependence of the conductance on the applied bias voltage, in \cite{Haiss2004} the conductance scales linearly with bias up to 1~V. This is perhaps the most conspicuous indication that the two experiments are probing different objects.

{\it Overview of experimental results.} 
In the years following these reports, many groups have measured series of ADT molecules, mainly using different versions of break junction techniques. All results that we found are collected in Fig.~\ref{fig.ADT}, where the conductance is plotted on a semi-logarithmic scale against the length of the ADT molecules, expressed as the number of CH$_2$ groups.  Despite the simple character of the molecules we find that the conductance reported by different groups for each length of the molecule varies by nearly two orders of magnitude. On the other hand, the exponential decrease of conductance with length has been confirmed by many works (inset). Nearly all data are consistent with $\beta = 0.9 \pm 0.2$ per CH$_2$ group, apart from one exceptional result. This reproducibility, despite the large scatter in the absolute values of the conductance, is associated with the fact that the variation observed in the main panel of Fig.~\ref{fig.ADT} is much smaller if we only compare measurements done by the same technique 
under the same conditions. In order to illustrate this we have connected the points obtained from three such studies. The variation between studies illustrates the sensitivity of the conductance to details of the experimental techniques. Note that this poses difficulties when discussing comparison with computational results because it is not obvious how the experimental conditions affect the data, and how this translates into the geometry that theory should consider. 

This sensitivity to experimental conditions is highlighted by the fact that several groups have reported multiple conductance peaks (shown by different colors in Fig.~\ref{fig.ADT}), while other groups report only a single peak in the conductance histograms. Following \textcite{Li2008} we classify the peaks as low (L), medium (M), and high (H) conductance. In fact, each of these three types of peaks often comes as a series of about three peaks at integer multiples of a basic L, M, or H conductance. The appearance of such multiple peaks has been attributed to the formation of junctions with up to three molecules in parallel.  Although the details of the experiments differ, all three series of peaks appear in the works of \textcite{Li2008} and \textcite{Haiss2009} at similar values of conductance. The experiments are similar in that the tip of the STM is prevented from coming into metallic contact  with the metal surface. This is a common feature of the $I(s)$ technique employed by Haiss \ea, but Li \ea use a 
tip approach and retract procedure with a maximum setting of the conductance equal to $0.2$~\go. Perhaps this explains the close agreement between the two data sets, highlighted by the connected points in  Fig.~\ref{fig.ADT}. 

The multiple-peak structure for ADT has been reproduced by several other groups (the full data set is available as Supplemental Material). Employing amplifiers with a wide current sensing range, {\it e.g.}, by using a dual range current amplifier \cite{Li2008} or logarithmic amplifier, may be decisive for being able to observe multiple conductance peaks. Most standard MCBJ or STM-BJ experiments, that follow a procedure of indentation of the two metal electrodes into metallic contact, produce only a single peak in the conductance histogram. Comparing with the overview of data in Fig.~\ref{fig.ADT} suggests that this peak is most likely associated with the M-peak. 

The interpretation offered by \textcite{Li2008} for the three classes of peaks was based upon extensive DFT model calculations. The conductance, in this interpretation, is influenced by the bonding configuration of S on Au (on-top versus bridge site bonding) and by the presence of gauche conformations in the alkane chain. 
Although microscopic evidence for this interpretation cannot be obtained from experiment, the model has not been challenged to date. The calculation by Li \ea agrees with the observed value for the decay constant, $\beta\simeq 1$, and the absolute values of the conductance agree within about a factor 5, assuming the bonding motifs have been correctly attributed. The variation in bonding configurations and gauche conformations lead to variations in the conductance by about two orders of magnitude, which agrees with the spread in the experimental data, suggesting that the details of the experiment influence the averaging process leading to the appearance of peaks in the conductance histogram.

While most of the data reported in the literature agree with the exponential decrease of conductance with increasing length of the alkane chain, with a decay constant $\beta\simeq 1$, the data by Haiss \ea (blue connected points in Fig.~\ref{fig.ADT}) show an anomalous transition to a length-independent regime at $N<5$.  Possibly, this deviation is related to the large bias voltage of 0.6~V employed in this study. 


In hindsight, neither of the popular molecular systems BDT nor ADT is very suitable as benchmark system.  Below, we discuss a few systems that may be better suited for this purpose.

\subsubsection{Alternative benchmark systems}\label{sss.Alternative}

{\color{black} We present here molecular systems and measurement techniques that offer great perspectives as alternatives for quantitatively benchmarking computations against experiments.}
{
\color{black}
{\it Benzenediamine and alkanediamines.} 
Already early on, searches began for alternatives for the two systems discussed above. Venkataraman and coworkers proposed to replace thiol anchors by amine anchors  \cite{Venkataraman2006a,Quek2007}. They have been able to show that the peaks in the histograms are much more sharply defined, in particular for benzendiamine (BDA), as compared to benzenedithiol (BDT). While this result could not be reproduced in MCBJ experiments in vacuum \cite{Martin2008}, several other groups have found well-defined conductance peaks, reproducing the results by Venkataraman \ea to within a factor of 2, for various atmospheres and solvents. An overview of the reported data is given in Table~3 in the Supplemental Material.  

For alkanediamines (Au$-$NH$_2$$-$(CH$_2$)$_n$$-$NH$_2$$-$Au) the number of published results is more limited. While the original data by \textcite{Venkataraman2006a} were confirmed by \textcite{Chen2006}, in the latter work {\em two} sets of peaks were found, one agreeing with the first paper, and the second at about an order of magnitude lower conductance. {\color{black} They attribute the appearance of two peaks to different anchor group contact geometries.}

\textcite{Quek2007} present arguments, supported by extensive DFT computations, why amines may be more favorable than thiols for producing well-defined peaks in conductance histograms. The main observation was that the overall tendency for binding of amines to Au as compared to thiols is weaker. As a consequence, the bonding motifs for Au to amine do not vary much for different junction geometries. Furthermore, as a result of the isotropic nature of the Au 6s orbital small variations in bond angles and bond lengths would have a limited effect on the conductance. 
}

{\color{black} Despite their promising properties and  the fact that amines have been proposed as alternative benchmark systems as early as 2006, and most of the published data support this idea, this proposal has not been followed widely. 
As we have discussed in Section~\ref{sss.coupling}, the effectiveness of amine anchors may depend on the local environment of the molecules in the experiment.  As we will see next, the proposed properties for amines may not be uniqiue, and well-defined and reproducible conductance peaks can also be found for thiol-coupled molecules. 
}

{\it OPE3.}  
{\color{black} There is at least one example that thiol-coupled molecules may give sharply defined conductance peaks and very reproducible conductance
values across different platforms.
}
The molecule is the third member of the oligophenylene ethynylenes, OPE3, with S-Au anchoring. Other lengths of this oligomer have also been studied, but less frequently. The molecule is long enough for the the properties of the backbone to dominate the conductance over the properties of the anchor groups. Combining the work of fourteen experiments from seven independent research groups we find that the conductance of Au-S-OPE3-S-Au junctions all agree with the value $G=(2\pm1) \cdot 10^{-4}$\go \,
\cite{Xiao2005,Huber2008,Wu2008,Xing2010,Kaliginedi2012,Wen2013,Frisenda2013,Parker2014,Frisenda2015,Garcia2015,Frisenda2016a,Bopp2017}, 
as illustrated for a subset of the data from literature in Fig.~\ref{fig.OPE3}. 
\begin{figure}
\includegraphics[scale=0.24]{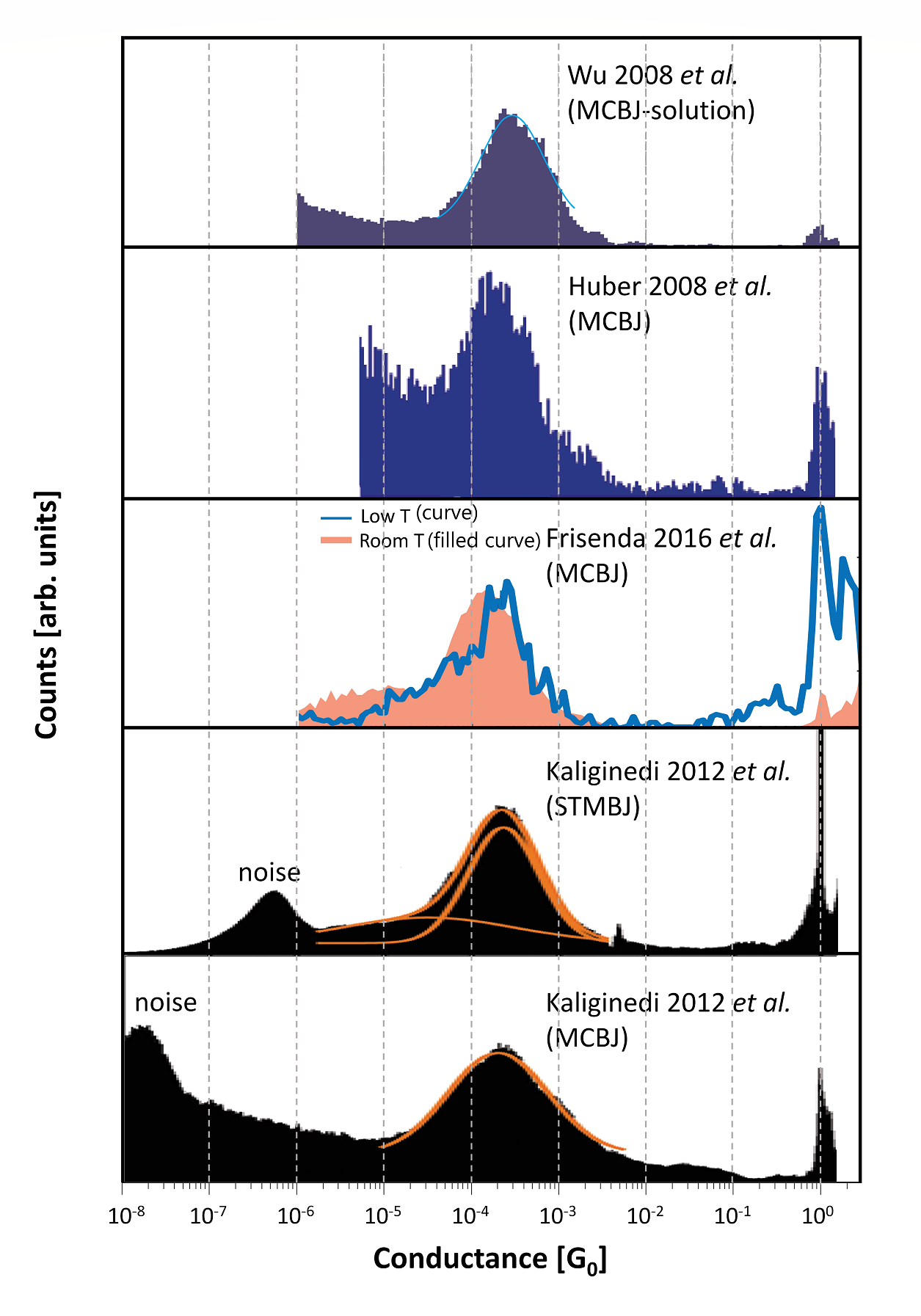}
\caption{ Comparison of conductance histograms for OPE3 obtained by different experimental methods, under different conditions, and by different experimental groups. In all cases the logarithm of the conductance was used for binning the data into a semi-log plot of counts as a function of the logarithm  of the conductance. The position of the peak in the conductance reproduces to within a factor of 2 in conductance, and also the width of the peak is fairly reproducible.
Adapted with permission from Huber {\it et al.}, J. Am. Chem. Soc. {\bf 130}, 1080. $\copyright$ (2008) American Chemical Society; 
Wu {\it et al.}, Nature Nano. {\bf 3}, 569. $\copyright$ Springer Nature (2008);
Kaliginedi {\it et al.}, J. Am. Chem. Soc. {\bf 134}, 5262. $\copyright$ (2012) American Chemical Society; 
Frisenda and van der Zant, Phys. Rev. Lett. {\bf 117}, 126804 (2016).
\label{fig.OPE3}}
\end{figure}
Compared to the two widely studied systems discussed above the spread in the conductance is remarkably small. As far as we are aware, the full reason for this reproducibility has not been elucidated. In particular, in view of the sensitivity of the alkanedithiols to top or bridge anchor sites on Au surfaces, the question arises why this does not lead to similar spread of reported values for OPE3. Whereas for ADT frequently multiple peaks have been reported, conductance histograms for OPE3 have only a single peak with a maximum in the range 1--3$\cdot 10^{-4}$~\go. 

Although many groups have presented computational results for OPE3, we could not find any systematic studies of the dependence of the conductance on anchoring site, {\it i.e.}, the choice between top, bridge, or hollow sites. Three calculations that assume hollow-site coupling of the molecule to a flat Au (111) surface consistently find a conductance for the molecular bridge that is two orders of magnitude higher than the experimental values: 0.01\go \cite{Wen2013}, 0.021\go \cite{Paulsson2006}, and 0.023\go \cite{Zheng2016}. Frisenda \ea have calculated the evolution of the conductance during stretching of a molecule bridging two pyramidal Au tips, and find a plateau at the final stages, where the binding is to a top site on both ends, of between 1 and $2\cdot 10^{-4}$\go, remarkably close to experiment \cite{Frisenda2015}. A partial study of the sensitivity to the binding site can be found in \cite{Kaliginedi2012}, showing a similar value of $4\cdot 10^{-4}$\go\  for top binding sites, which drops to $0.7 \cdot 10^{-4}$\go\  with one anchor moved to a hollow site.

From the experimental evidence we conclude that OPE poses an interesting candidate as a benchmark system, despite the fact that some of the reasons behind the reproducibility need to be elucidated.

{\it Low-temperature STM.} 
Low-temperature STM experiments offer the best perspectives for benchmarking computational methods. The presence, the identity, the position, and the orientation of the molecule can be obtained from the STM images. The cleanliness of tip and sample surface can be guaranteed, and the structure of at least one of the two electrodes, the surface, can be known in detail. Although the atomic arrangement of the metal tip is more difficult to characterize in detail, some information on the apex atom is available. While many experimental groups have studied the arrangement and structure of molecules deposited at surfaces, contacting of such molecules by the tip has been addressed by a much smaller community {\color{black}\cite{Joachim1995,Temirov2008,Lafferentz2009,Okuyama2018,Schmaus2011,Reecht2016}}. Some prominent examples of controlled lifting of a molecule or molecular chain can be found in the work of Tautz and Temirov and coworkers \cite{Temirov2008,Wagner2012}, or that of Grill and Schull and coworkers {\
color{black}\cite{Lafferentz2009,Koch2012,Reecht2015}}. However, the systems studied in these works introduce additional complications for quantitative comparison with computations, due to correlation effects in the former case (as discussed in Section~\ref{sss.Open-shell}), and the many molecular conformations of the long chain during lift off in the latter case. 

\begin{figure*}[t!]
\includegraphics[scale=0.6]{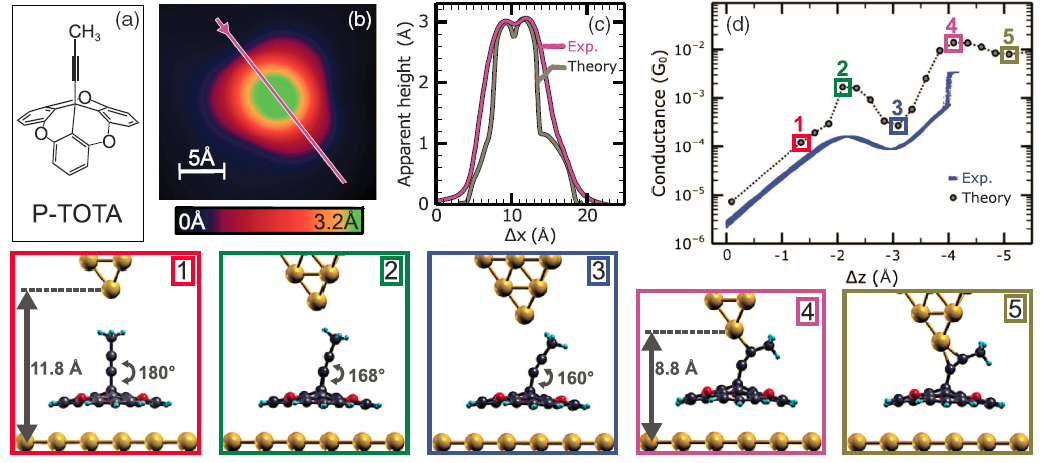}
\caption{ STM experiment on the tripod molecule propynyltrioxatriangulenium (P-TOTA), and comparison with computations. (a) Structure of the molecule. (b) Height profile of a single P-TOTA molecule as observed in an STM image recorded at 4.5K, for 100 mV bias, and 30 pA current set point (color online). (c) Height profile along the trajectory indicated in b and compared with computations. (d) Conductance-distance dependence for approach of the tip to the center of the molecule from above. The experimental curves are shown for 40 forward and backward traces. The configurations of the molecule in the computations for the positions of the tip at the numbered stages are shown in the lower panels.  Reproduced with permission from \protect\textcite{Jasper2017}.  \label{fig.tripod}}
\end{figure*}

A prime example of an experiment designed for close comparison with theory can be found in the recent paper from the Berndt group \cite{Jasper2017}, see Fig.~\ref{fig.tripod}. The molecule chosen for this study, propynyltrioxatriangulenium, has a tripod structure designed for resting on the Au(111) substrate, with a propylyl wire protruding straight up from its center. Approaching the center of the molecule with the tip from above the authors observe a variation of the conductance, which they interpret with the help of DFT/NEGF computations as follows: As the tip comes closer to the top methyl group the latter is repelled and the propylyl wire bends away to the side. 
Approaching further, the bent molecule exposes the carbon triple bond to the apex atom, to which a bond is formed, as is revealed by a jump of about an order of magnitude in the conductance. At this point the conductance is $3.5\cdot 10^{-3}$~\go\, and this conductance varies by less than a factor 1.4 between experiments with different tips and different molecules. This example illustrates the complementary type of information obtained from theory and experiment and a close agreement between the qualitative features. From this agreement we gain confidence in the information that DFT provides about the nature of the molecular orbitals involved, about the change in couplings as a consequence of mechanical deformation, and on the size of the molecular deformation that is induced by the tip. Quantitatively, the computational results for the conductance even agree to within a factor of $\sim 4$. The DFT calculations include long-range dispersion interactions, but have not been adjusted by scissor operators for 
the over-estimation of the HOMO-LUMO gap or for image charge shifts.

\subsubsection{Concluding remarks}
While our qualitative understanding of electron transport in molecular junctions has reached a high level, as we concluded in Section~\ref{s.Achievements}, often the quantitative agreement between experiments and theory is still not very firm. The overview above shows that many unknowns hamper a proper comparison. The reason why we stress this is that interesting physical effects may be overlooked if we cannot make a proper comparison. The problem resides, both, with experiment, where the choice of proper benchmark systems has not yet been made, and with theory, where the limitations of DFT become apparent in the treatment of the HOMO-LUMO gap, the effects of image charges in the leads, and in describing electron correlations.

As has been put forward by several groups, quantitative comparison becomes much better when considering ratio's of experimental data, {\it e.g.}, see the switching of conductance for two magnetic states of a molecule \cite{Schmaus2011}, or the comparison of different connection sites on a molecule, as in the `Magic Ratio's' discussed by \textcite{Geng2015} and further reviewed in \textcite{Ulcakar2019}. The results for ratio's compare quantitatively, because many of the aspects of the theory that are sensitive to matrix elements drop out.

Despite the remaining gap in quantitative agreement mentioned above, the role of DFT-based computations in guiding and interpreting the experiments should be strongly emphasized. Given the limited number of parameters that experiments typically give access to, 
computations complement the experiments and guide our understanding. 
In situations where the conductance of a single species 
is the only parameter available for comparison 
and the atomic structure is not fully known,  
we must remain cautious to avoid mis-interpretations.

\section{Selected open problems}\label{s.Clouds}

Successes and challenges of modeling and understanding the transport properties 
of molecular junctions have been addressed throughout this review. Here, we will address a few
specific examples of open questions that have not been fully resolved. 
These examples help to illustrate
 the prospects for further research and discoveries. 

\subsection{Experimental phenomena awaiting basic qualitative understanding} 
\label{ss.awaiting}
Arguably, the most exciting open problems concern strong experimental signatures 
that are awaiting a consistent qualitative explanation. 
As an intriguing example, 
we describe recent work by \textcite{Frisenda2016a} on an OPE3 molecule equipped with thiol linkers, coupled 
to Au leads. They found a sudden transition from a  
smooth conducting state into 
a peculiar insulating state, driven by subtle stretching of the molecular junction, see Fig. \ref{fig.Frisenda2016}.
\begin{figure}[b]
\includegraphics[scale=0.65]{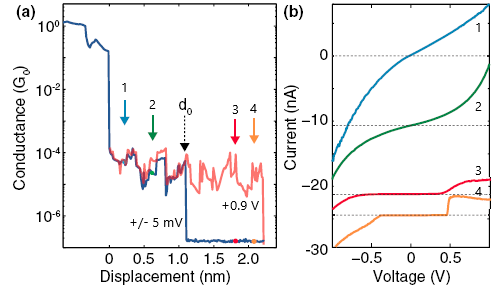}
\caption{Observations of anomalous behavior for an  OPE3 
molecule coupled by thiol bonds to Au in MCBJ experiments at 10 K. (a) Traces 
of conductance recorded while breaking the molecular junction, shown for low bias (obtained from the slope of $I(V)$ around $V=0$) and high bias voltage (obtained from the ratio $I/V$ at $V=0.9$ V).
While the two curves nearly coincide up to the point marked as $d_0$, at this point there is a sudden break in the curve 
for low bias voltage, while the high bias curve for the same junction continues without interruption. 
(b) Current-voltage curves shown for the four different electrode separations marked by the numbered arrows in (a).
A vertical offset has been applied for clarity.
Following a smooth evolution of $I(V)$ at lower electrode separation  
a sharp transition is seen to a gapped state. 
Reproduced with permission from \protect\textcite{Frisenda2016a}. \label{fig.Frisenda2016}}
\end{figure}
The authors favor an interpretation according to which the transition is one into
an emergent insulator regime dominated by strong correlation effects of the Coulomb blockade type. 
Clearly, this is a very conspicuous result, and the mechanism invoked calls for 
further investigations. On the experimental side, can we exclude alternative explanations?
Can we exclude that the state of binding of the molecule changes under the influence
of the combination of stretching and high voltage bias?
Can we find further evidence for such a sudden change of the character
of the electron transport, {\it e.g.}, by probing shot noise?

Other explanations are conceivable \cite{Frisenda2016a}, 
such as formation of a polaron or a change of the redox state, and these can be further explored 
by experimental and theoretical investigations. 

\subsection{Chirality induced spin selectivity}
An entire class of phenomena 
has been discovered experimentally by Ron Naaman, Dave Waldeck, and coworkers. 
Early measurements employed photo-emission of electrons from Au surfaces, where the electrons were transmitted 
through films of chiral molecules 
(double stranded DNA) adsorbed on the Au surfaces \cite{Ray2006,Goehler2011}. 
These works strongly suggested that the interaction of 
the spin of the transmitted electrons with the adsorbed layer
is very sensitive to the molecular handedness. 
By now, an impressive body of experimental work has been 
compiled revealing the presence of `chirality induced spin selectivity' ({\ciss})  under very diverse 
conditions, 
including single-molecule electron transport experiments \cite{Xie2011}; 
for a review, see \textcite{Naaman2015}. 
As was suggested most recently, {\ciss} may have very important 
technological applications as a novel method for enantiomeric separations, 
which is important, {\it e.g.}, in pharmaceutical production processes 
\cite{Banerjee-Ghosh2018}. 

Despite considerable investment into theoretical work,
as to the fundamental origin of {\ciss} there is no consensus, yet. 
In principle, there are two conceivable ways by which molecular
handedness could couple to the electron's spin producing 
a {\ciss} effect: 
(i) a mechanism mediated via spin-orbit coupling (\soc); 
(ii) a mechanism operating in the presence of a current flow that 
translates handedness into an induced 
magnetic flux. 
At present, all mechanisms invoked to explain 
strong {\ciss} seem to fail, underestimating the observed magnitude 
significantly \cite{Naaman2015}. 

Due to its prominent character and potential technological impact,
we offer a brief survey of the theoretical attempts to 
understand the {\ciss} phenomenon
in terms of technology and basic science.
Earlier theoretical treatises motivated by the 
transmission electron measurements \cite{Ray2006,Goehler2011}
have addressed {\ciss} in terms 
of scattering approaches \cite{Yeganeh2009,Medina2012,Eremko2013,Gerstin2013,Varela2014}.
Since our focus is on electron transport, we will refrain from following this direction, here. 

{\it  {\ciss} induced by {\soc} on the molecule.} 
A considerable number of theoretical works on {\ciss} report  
transmission calculations for various tight-binding models 
\cite{Guo2012,Guo2012b,Gutierrez2012,Guo2014,Guo2014b,Varela2016}.
\textcite{Gutierrez2012} consider a single-channel tight-binding 
model with nearest neighbor hopping and \soc.
The authors motivate their model parameters referring 
to DNA; 
the hopping parameter is reported to take values of 20--40 meV. 
Chirality enters the model indirectly via its feedback into the {\soc}. 
To find quantitative estimates of the latter, a heuristic argument is exploited 
that yields typical values for light atoms (C, B, N, O) 
of about 2 meV coupling strength. 
At present the accuracy of this estimate is not known. 
Indeed, scales of meV can be reached with light elements, 
{\it e.g.}, when promoting a carbon atom in graphene from sp$^2$- to sp$^3$-hybridization. 
%
%
%
{\color{black} However, as compared to this promotion, the chirality induced symmetry breaking 
should be weaker by a geometric factor. Therefore, 
the reliability of the heuristic estimate appears to be uncertain.

\textcite{Guo2012} have considered also double-stranded wires 
within a tight-binding description. These authors 
employ in their works \cite{Guo2012,Guo2012b,Guo2014,Guo2014b} 
model parameters similar to those of \textcite{Gutierrez2012}, 
so that the quantitative uncertainties carry over. 

{\it  Symmetry constraints on spin filtering in two terminal transport.} 
\textcite{Guo2012} and 
\textcite{Matityahu2016}
have emphasized the importance 
of two channels for the observation of {\ciss}. 
Indeed, it is well known that the {\soc} can be gauged out 
in single channel wires, so that spin filtering functionality based on {\soc}
is not expected \cite{Meyer2002}.
In non-interacting single-channel
wires, one can also  make an argument
based on time reversal symmetry according to 
which a single channel wire can never
act as a spin filter in a two-terminal measurement
\cite{Kiselev2005,Bardarson2008}.
Time reversal symmetry also has implications for the
conductance of interacting wires with several channels.
In this case 
an Onsager-type of symmetry relation holds
that inhibits spin filtering in two terminal measurements
\cite{Yang2019}.

{\it  Non-unitary effects.}  
A conceptually innovative work-around for the no-go theorems
has been put forward by \textcite{Guo2012,Guo2014} and further explored by
\textcite{Matityahu2016,Matityahu2017}. 
These authors investigate the effect of a third bath that 
the electrons traversing the chiral molecule may be coupled to.
The bath gives rise, in general, to non-unitary effects such as 
`dephasing' or `leakage'. Technically, this effect can be modeled
by a complex self-energy that can bring about 
spin-selective transport in the presence of spin-orbit interactions;
intuitively, evanescent waves associated with opposite 
spins have different decay lengths.
The overall magnitude of the filtering effect
thus brought about appears to be too small explain the main experimental features
observation.
}

{\it  {\ciss} induced by substrate mediated {\soc}.} 
Since straightforward {\soc} on the chiral molecules is suspected to be too weak in order to 
account for the experimentally observed magnitude of {\ciss}, 
it is natural to consider the role of the substrate. 
In most experimental situations, this substrate is taken to be a Au surface, 
where {\soc} can be considered sizable. 
Indeed, \textcite{Gerstin2013} have predicted that a considerable 
spin-polarization can be obtained by a mechanism that combines strong {\soc}
in the substrate with orbital angular momentum selectivity imposed by the 
chiral molecule.

{\it  Outlook.}  
At present, it appears that a large gap remains between the 
experimental observations and the quantitative estimates from theory. 
Currently, theoretical 
investigations are focusing on qualitative aspects, 
{\it e.g.}, on rigorous bounds set by 
symmetries, such as time-reversal \cite{Yang2019,Dalum2019}. 
Further experiments are required for guiding the theory and for limiting
the possible interpretations for this potentially very important phenomenon.

\subsection{Challenges to theory and modeling}
{\color{black} One out of several} important challenges for theory is achieving 
systematic, quantitative control of accuracy in electronic structure predictions
for molecule-metal interfaces. We illustrate this point with two specific examples.

{\it `Unphysical' values for fitted model parameters.} 
\textcite{Perrin2014} measured large negative differential resistance (NDR) effects
 in a thiolated arylethynylene with a
9,10-dihydroanthracene core, between Au electrodes.
Suggested by the molecular geometry illustrated in Fig.~\ref{fig.Perrin2014}, 
the differential conductance 
was modeled by a two-level model (TLM, see Sec.~\ref{sIV.A1}), 
where the two levels represent the 
left and right conjugated arms, separated by a non-conjugated linker in the middle. 
\begin{figure}[b]
\includegraphics[scale=0.17]{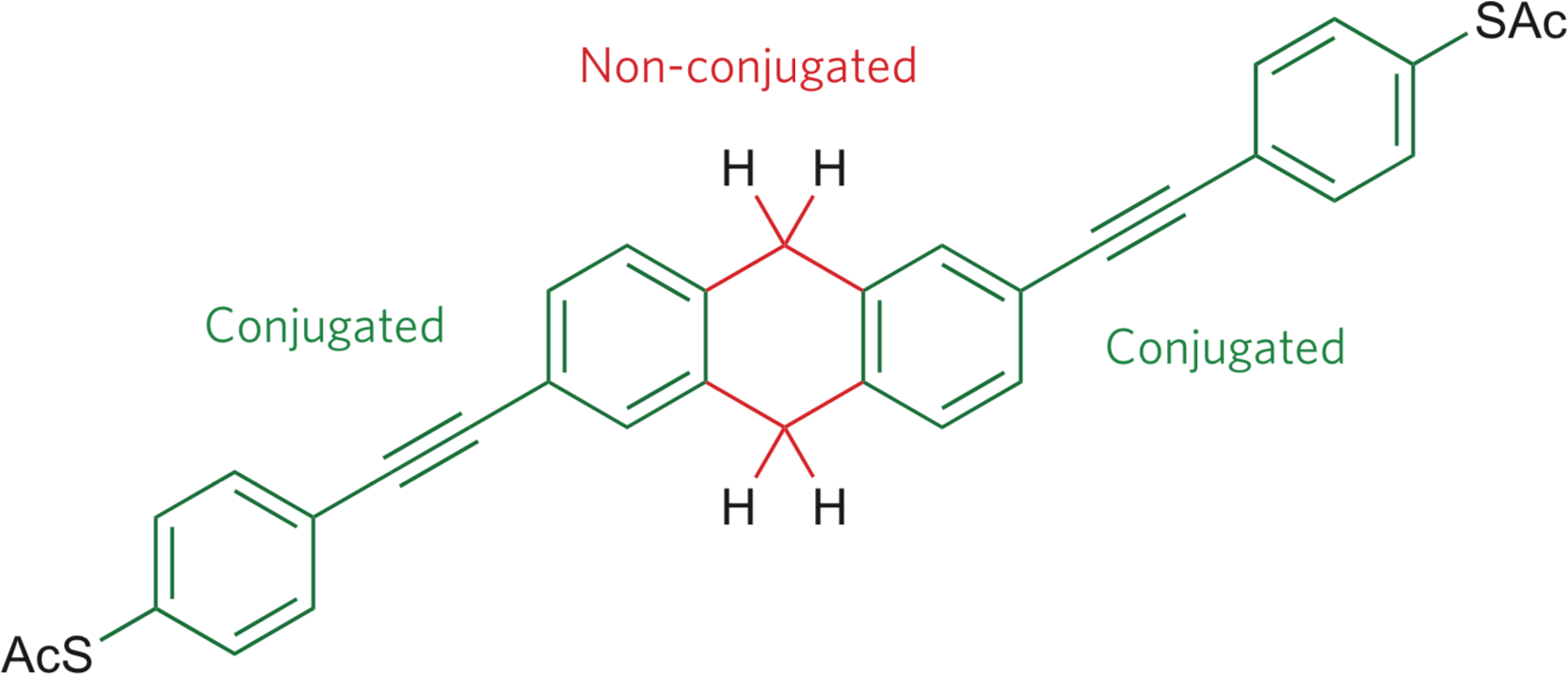}
\caption{A thiolated arylethynylene molecule 
with a 9,10-dihydroanthracene core (the three
rings in the center). 
The core divides the molecule into two conjugated parts.
\label{fig.Perrin2014}
}
\end{figure}
The experimental line-shape was seen 
to be reproduced amazingly well on a qualitative level including, in particular the strong NDR feature, 
even though interactions have not been accounted for. 
However, a quantitative agreement is achieved only after re-scaling 
the model-curve by a large factor of $7.2\cdot 10^{-5}$.

{\color{black} We suspect that the} large prefactor suggests the importance of correlation effects beyond 
the non-interacting TLM. The situation is not untypical in the sense 
that fitting formul{\ae} resulting from non-interacting 
 theories often lead to matching fits at the expense of 
choosing fitting parameters that are hard to justify physically without 
invoking very strong, interaction mediated renormalizations. 
In a way, we encounter a molecular analogue of the Fermi-liquid theory.
How to calculate the corresponding `Fermi-liquid corrections' 
quantitatively is 
an important open question in molecular scale electronics and 
neighboring fields.

{\it Quantitative DFT calculations.} 
The lack of control in {\it ab-initio} transport calculations is best illustrated
recalling the specific problem of alignment between molecular levels and the Fermi-energy associated
with metallic substrates. The problem is severe enough, leading sometimes even 
to qualitative discrepancies.

For example, \textcite{Wang2009} have studied the length dependence of the conductance of oligoyne
molecular wires ($N$=1,2 and 4 units).  The experimentally measured conductance drops by 
roughly a factor of two with increasing length. Inconsistent with this, DFT calculations employing
the LDA functional yield a weak increase of the conductance.
%
As the authors have argued, the discrepancy is due to an insufficient 
accuracy of the semi-local functional employed in the DFT study. 

One of the most important 
defects of such semi-local functionals is the 
neglect of the derivative discontinuity with the 
associated mis-treatment of the Coulomb-blockade.
Inaccuracies with respect to the Kohn-Sham energy-level alignment 
are an important consequence.
As we explained in \ref{sss.Beyond-GGA}, several authors have 
adopted an ad-hoc recipe (`scissors operator') as a post-DFT
repair treatment and thereby achieve more quantitative 
conductance values \cite{Quek2007,Mowbray2008,Quek2009}.
After following this procedure, 
a match of the trends between theory and 
experiment has also been restored in \textcite{Wang2009}.

A proper treatment of the Coulomb-blockade in intermediate-sized systems 
has been a huge challenge for electronic-structure calculations in 
quantum-chemistry and material sciences ever since such calculations have 
been attempted. While correlated methods are under continuous development, {\it e.g.}, 
based on many-body Green's functions, their application 
to quantum transport remains challenging, in particular, due to computational issues.

\section{Conclusions}\label{s.Conclusion}

The confrontation between theory and experiment that we have presented above leads to several important conclusions.

\subsection{Benchmark systems}
For a critical evaluation of comparison between theory and experiment 
it is convenient to distinguish three 
classes of molecular junctions, although admittedly 
these classes are not sharply defined. 
The first class, comprising small molecules with a conductance close to 1 \go, is well under control. 
For this class, experiments give access to many additional tools for characterization (shot noise, 
thermopower, inelastic electron tunneling spectroscopy, ...) that often allow for simple interpretations, 
and that assist the testing of theoretical models. 
It turns out that the model calculations in this limit are less sensitive to details of 
the electrode-molecule interface, and close 
quantitative agreement between theory and experiment has been obtained. 
Although sometimes there may be a factor $\sim 2$ between the experimental conductance and the value obtained in 
calculations, we feel that most of the discrepancy can be attributed to atomistic details in experiment 
that have not been included in the models.

The second class covers the other extreme, 
including molecular junctions having very low conductance, 
here defined as having a zero-bias conductance below the detection 
limit of the experiment. For this class, comparison between theory 
and experiment can only be made on a qualitative level. 
This is, first of all, because in this limit 
experiment and theory alike are very sensitive to details. 
Moreover, comparison can only be made for current levels at rather
high bias, {\it i.e.} at strongly out-of-equilibrium conditions, 
where the approximations underlying {\it ab-initio} transport calculations are least controlled. 
Nevertheless, many interesting phenomena have been 
uncovered for such junctions.

Finally, the third class includes molecules falling between these limits, which comprises the majority of 
molecular junctions studied to date. 
Contained in this set are two types of molecules 
that have played the roles of benchmark systems 
from early on,   
namely benzenedithiol (BDT) and the series of 
alkanedithiols (ADT). BDT has served an important role as workhorse in computational studies, where the results in essence 
agree between research groups, but it has been an unfortunate choice in terms of experiments. There is a large 
variability between various experiments, suggesting that the preferred binding configurations of the BDT 
molecule between metal leads depend on details of the experimental procedures. 
ADT has the advantage over BDT that it offers the possibility of studying a trend, {\it i.e.} the length dependence of the conductance. 
The systematics of this trend is well defined and reproducible in experiments and understood quantitatively. 
Contrasting to this, a large variability of the experimentally measured contact conductance is 
observed for ADT, which appears consistent with the findings for BDT.

As alternative benchmark system for comparison between experiment and theory we propose to give a more
prominent role to oligo(phenylene ethynylene)dithiol. It
appears that this molecule in the three-monomer long
version (OPE3) has been studied by many groups with
different methods. The conductance values and the
shape of the conductance histogram were found to be 
highly reproducible. At this moment it is still unclear why 
OPE3 would not suffer from the same problems as BDT or ADT. 
Two additional reasons for studying this molecular system
more closely are (i) the fact that the experimental and 
theoretical conductances still differ by two orders
of magnitude, and (ii) the observation of anomalous behavior
at low temperatures, as discussed in Section~\ref{ss.awaiting}. 
Other oligomers have also been studied \cite{Zheng2016,Kaliginedi2012,Xing2010,Lu2009}, 
but not yet by many groups and methods.
It is likely that many other suitable systems exist, that would provide 
closely reproducible conductance values. 

\subsection{Uncovering physical phenomena with robustness}
Despite many atomistic unknowns existing in single molecule junctions, 
a great number of phenomena have been uncovered and understood. 
Such phenomena are often associated with a sense of `robustness' with 
respect to the physical observable. 
Depending on the observable, the origin of this robustness varies 
and different mechanisms have been discussed in this 
review: For instance, 
in dimensionless ratios of observables, such as the magneto-resistance,
prefactors strongly fluctuating from junction to junction cancel out.
Similarly when analyzing trends, the relative conductance (taking {\it e.g.} 
the typical value) often is reproducible between different experimental settings, 
while the absolute values may not be. The experimental observation of destructive 
quantum interference (DQI) has strongly benefited from this kind of robustness. 
Indeed, the fact that simple counting rules have been found predicting 
DQI very reliably irrespective of chemical details of the anchor groups, 
can be understood as a manifestation of this robustness. 
Ultimately, it originates from the stability of the nodal structure 
of the frontier orbitals. 

We briefly discuss conditions of robustness with an eye on break-junction experiments. 
When a series of molecules is studied with a systematic variation in a single parameter ({\it 
e.g.} its length), one may expect that this single parameter dominates the trend in the conductance provided the 
experiment meets at least the following conditions: 
(a) the contact breaking procedure for all molecules under test must be performed following the same protocol, 
(b) the core of the molecule must dominantly determine the value of the conductance, 
and (c) the dominant anchoring configurations are not influenced by the parameter that is varied between the molecules.
For predictions, DFT studies can be helpful evaluating whether (b) and (c) are likely to be fulfilled.

A completely different, genuinely many-body mechanism of robustness 
is realized in the Kondo-effect: the Abrikosov-Suhl resonance is 
always situated at the Fermi-energy, so the existence of a zero-bias 
anomaly is guaranteed as long as there is a free spin. 
Since the free spin does not require fine tuning, 
the Kondo-effect is a generic encounter in open-shell molecules. 
As we have discussed in this review, metal-organic 
molecules turn out to be an ideal testbed for studying Kondo-physics, {\it e.g.}, 
its dependency on space and continuous parameter tunings,
or in the presence of competing spins and several channels.
Indeed, the elusive phenomenon of 
the underscreened Kondo effect was first observed in such systems. 

Robust are also many observed effects that relate to the coupling of electrons
to vibrations, such as the Franck-Condon blockade. 
For this reason, molecular junctions have offered a exquisite 
platform for investigating inelastic signals in the 
differential conductance and in noise, and the way 
these depend on the electron transmission 
probability of the junction.

\subsection{The important role of DFT-based computations}
Experiments provide access to only a limited number of 
observational parameters, in many cases only the 
value of the conductance averaged over many junction breaking 
cycles. This fact has led to some over-emphasis 
on the numerical value of the conductance. More important than the conductance 
value itself are the predicted qualitative features and trends. 
We recall that DFT serves as a guide to the experiments, 
where it may indicate which are the more energetically favorable bonding configurations, 
which molecular orbitals are likely to dominate the electron transport, 
and which symmetries apply to these electronic states. 
Moreover, trends in the experimental transport 
data are often closely reproduced, 
and are expected to be more reliably predicted, 
as advertised by {\it e.g.} the work by \textcite{Schmaus2011} and  
\textcite{Geng2015}.

However, there are cases, and we have discussed a few, 
where the discrepancy is larger and trends extracted from 
DFT-based transport calculations fail qualitatively. 
This indicates the presence of physics ignored in DFT 
- at least when employing conventional exchange-correlation functionals. 
Extensions may need to consider dynamics of image charges, 
strong electron-electron correlations, polaron formation, and other 
many body effects.

We emphasize that even in cases where the actual physics is not contained 
in DFT-based transport simulations, a well-informed setup of DFT-calculations 
can - and normally will - provide good guidance. By computing estimates for the level broadening $\Gamma$, 
for the molecular charging energy $U$, for the vibration frequencies and coupling matrices  
and more, it is possible to decide in which corner of the physical 
phase space the system finds itself. 
To provide an important example, we consider the Kondo effect. It is not 
captured by the available DFT-functionals. However, an open-shell calculation even 
with conventional functionals can indicate the existence of a free spin.
This observation then will be readily interpreted as a 
necessary (i.e. mean-field type) precursor for the Kondo-effect, 
which would be seen if more accurate tools would be available.

\subsection{Outlook}

The field of single-molecule transport can be seen as an 
active subfield of the much larger research 
area molecular-interface sciences.
As we have emphasized in the introduction,
only aspects of it could be covered in this review.
Similarly, also this outlook reports only selected emergent 
directions chosen from a great many promising activities ongoing worldwide. 

\subsubsection{Precision, Reproducibility, Control}
Quantitative electronic structure calculations 
of interfaces between molecules and condensed matter, 
such as 
metals or semi-conductors, will 
remain a challenge for the foreseeable future.
%
Moreover, despite shortcomings 
of density 
functional theory (DFT), trends can be predicted and designs can 
be optimized, even quantitatively. 
To facilitate such predictions with a perspective of high throughput, 
a workaround can be designed following a trend in 
materials sciences that replaces {\it ab-initio} based understanding 
by big data fitting: 
{\color{black} 
computations merely extrapolate between 
experimentally secured data `points'.
Such a data point consists of a structural information,
{\it e.g.}, type and length of molecular wire, anchor group, electrode 
material and distance, etc., together with a measure that decides about
`distances' in this configuration space. 
Further qualifying information includes 
parameters observed in transport experiments, {\it e.g.}, 
typical conductance, Seebeck-coefficient, forces, etc.
`Secured' implies that the data collected in the data point 
satisfies a set of quality criteria, in particular, 
reproducibility.
With increasing size of the secured data set, 
the number of fitting parameters that are being 
used in order to parameterize the extrapolating functionals
also grow, which may call for parametrization techniques, 
such as machine learning. 

Reproducibility can be established in a statistical sense as 
a property of a series of measurements.
The production of secured data sets with {\it individual} reproducibility 
requires not only a precise measurement of electronic structure features, 
but also needs to specify the corresponding atomistic geometry. Arguably,
the best prospects in this direction offer low-temperature scanning tunneling and atomic 
force measurements. Such studies will turn out to be crucial 
not only in the subfield of single-molecule transport but also 
from the broader perspective of molecular interfaces sciences.
}

\subsubsection{Towards novel phenomena - challenges for experiments}

We have covered in the review a number of theoretical developments 
that are well advanced, {\it e.g.}, the formation of polarons
in soft molecules or the Kondo-effect in systems with 
two coupled spin-degrees of freedom. In these situations the 
theory is awaiting experimental confirmation.
Further theoretical investigations are ongoing 
with good prospects  
for experimental testing. 

{\it Many-body quantum interference.}  
Quantum interference ({\qi }) in single-molecule transport 
has been a topic of intensive research
in the past decade with a very strong emphasis on single-particle
phenomena. Much less is known for situations where states interfere 
that consist of a few Slater determinants instead of just 
a single one. One would expect that many-body {\qi } should be 
a frequent encounter in molecular systems, but so far it has been 
identified only in exceptional cases \cite{Yu2017}. 

To give a perspective, we mention that 
many-body interference is a topic closely related to ongoing 
research directions in neighboring fields. 
For instance, in condensed matter physics the quantum-interference of many-body states
drives {\it many-body localization} 
\cite{Nandkishore2015,Bera2017}. 
The phenomenon is associated with a breakdown of diffusion in 
inhomogeneous wires that can persist even at elevated temperatures, 
provided that the coupling to the environment and vibrational modes
is sufficiently weak. To what extent single molecules could be a suitable 
testbed to study this fascinating phenomenon remains to be seen. 
{
\color{black}
{\it Light-matter interaction.} 
In many fields of science, light offers very powerful tools for investigation. 
For example, advanced spectroscopic tools have been developed that are capable of interrogating individual molecules 
embedded in an isolating matrix \cite{Orrit2014}.
For molecules connected to metallic leads fluorescence is quenched by electron-hole excitations in the leads. 
The metallic leads similarly hamper electroluminescence and other forms of light-emission and absorption. 
{\color{black} For example, quenching of optical excitations of the molecule plays an important role for achieving light-induced switching
in single-molecule junctions \cite{Dulic2003,Molen2008}. 
}

In contrast, there is one form of light-matter interaction that profits from the proximity of the metallic leads, 
which is Raman spectroscopy \cite{Ward2008,Natelson2013,Iwane2017}. In this case the light excites 
surface plasmon modes in the metallic tips, which lead to an enhanced electric field at the molecule. 
Other forms of light-matter interaction can be studied by separating the active center in the molecule from the 
metal by }{\color{black}  insulating layers, or insulating molecular wires. 
For this reason, light emitted from single-molecule
junctions has been seen first after separating the molecule from direct contact with the metallic leads \cite{Qiu2003,Marquardt2010,Reecht2014,Doppagne2018}.}

{\i Molecular wires.}  
Quite generally, long molecular wires open interesting 
perspectives for studying new effects 
related to strong electronic correlations and topology.  
For instance, polymethines can form topological insulators 
realizing, {\it e.g.}, the Su-Schrieffer-Heeger model \cite{Heeger1988}. 
Indeed, first indications of topological edge modes in single molecules   
{\color{black} have been reported only very recently 
by \textcite{Gunasekaran2018} and, in particular, 
by \textcite{Groning2018}. 
Much more is to be expected here to come, such as studies on the 
formation of topological defects (`solitons') 
and their propagation inside the wire.}
In wires that are closer to a metallic state, strong correlation 
effects are expected including spin liquids. 
As a relatively simple example illustrating the prospect, 
we mention the oligoacenes; {\color{black}they represent a research 
field in its own right, with a respectable body of literature, because
they might exhibit a correlated phase \cite{Shen2018}. 
Recently, it has been proposed that oligoacenes might
exhibit a band gap oscillating with increasing wire length \cite{Schmitteckert2017}; 
see \textcite{Setten2019} for a recent overview.
}

{\it Molecular Kondo chains.} 
Since single-molecule Kondo effects are abundant in the literature,
a natural next step would be to take a look at molecular Kondo-clusters
and networks interfacing a normal metal. While reports with such
systems exist, deviations from the single-molecule Kondo behavior
are rare \cite{Tsukahara2011,Dilullo2012,Fernandez2015}. The prospect, here, is to
engineer a strongly-correlated phase,
such as a heavy-fermion metal or a Kondo insulator.

{\it Molecules and superconductivity: towards Majorana modes.}  
When magnetic molecules weakly couple to superconductors, the remnant of the
molecule's magnetic moment can be seen as a pair of in-gap resonances 
\cite{Hatter2015,Island2017},
termed Yu-Shiba-Rusinov states. The versatility and control allowed by
the molecular design has facilitated the observation of a quantum 
phase-transition \cite{Farinacci2018}. The intense activity in this field is
motivated by theoretical predictions of Majorana modes in spin chains 
{\color{black} \cite{Choy2011,Nadj-Perge2013,Pientka2013}.}

\subsubsection{Towards time-dependent studies - Molecular Plasmonics}
The study of dynamical and light-induced 
phenomena in single-molecule transport is a long-standing
challenge in the field. A short review and 
perspective may be found in \cite{Thoss2018}. 
 
Indeed, single-molecule transport under illumination has 
been investigated already for quite some time \cite{Natelson2013}. 
However, a recent breakthrough has been reached by combining 
externally applied THz-pulses with low-temperature scanning 
tunneling microscopy \cite{Cocker2016}.  
In this way time-resolutions of 
100 fs have been achieved with perspectives 
to achieve even higher resolutions.

\subsubsection{Towards devices - {\ciss } and Molecular Nuclear Spintronics}

{\color{black}Many forms of switching in single-molecule junctions have been investigated,
and for a recent review we refer to \textcite{Ke2019}.}
The magnetic degrees of freedom of molecular matter are investigated in 
the fields of Molecular Magnets and Molecular Spintronics. 
Phenomena and activities in these fields are very rich 
and deserve reviews of their own. We have only 
touched upon the chirality induced spin selectivity; 
despite the effect not being well understood, 
applications are already under investigation, for instance for separating 
enantiomers in chemical synthesis \cite{Banerjee-Ghosh2018}.

We would like to mention yet another development
that is as speculative as it is stimulating: 
While the coherence times of electronic spins are believed to 
be too short in order to allow for quantum information processing, 
the lifetimes of nuclear spins located on isolated molecules 
are much longer. This observation has nourished hopes to use 
molecular-nuclear spins for applications in quantum computing  
\textcite{Moreno2018}.
Indeed, the experimental control of nuclear spin-states in isolated molecules 
has been demonstrated in an impressive sequence of experiments 
\cite{Thiele2014,Ganzhorn2016}.

\begin{acknowledgments}
Over the years we have profited from discussions with many scientists in the field, many more than we are able to list and acknowledge here. Specifically we want to express our gratitude to 
Alexey Bagrets,
Richard Berndt, 
Marius Buerkle, 
Andrea Donarini, 
David Egger, 
Karin Fink, 
Thomas Frederiksen, 
Per Hedeg{\aa}rd, 
Manabu Kiguchi, 
Denis Kochan, 
Leeor Kronik, 
Manohar Kumar, 
Stefan Kurth, 
Nicol{\'a}s Lorente, 
Magdalena Marganska-Lyzniak,  
Marcel Mayor, 
Sense Jan van der Molen, 
Ron Naaman, 
Jeff Neaton, 
Abraham Nitzan,
Tom{\'a}\v{s} Novotn{\'y},
Felix von Oppen, 
Jens Paaske, 
Fabian Pauly, 
Jascha Repp, 
Mario Ruben, 
Elke Scheer, 
Peter Schmitteckert, 
Gemma Solomon, 
Robert Stadler, 
Charles Stafford, 
Gianluca Stefanucci, 
Oren Tal, 
Jos Thijssen, 
Michael Thoss, 
{\color{black} H\'ector V\'azquez, }
Latha Venkataraman, 
Heiko Weber, 
Florian Weigend, 
Maarten Wegewijs, 
Wulf Wulfhekel, 
and 
Herre van der Zant.
{\color{black} It is a pleasure to thank Nicol{\'a}s Lorente, Jascha Repp, Michael Thoss and, in 
particular, Mark Hybertsen for careful reading of the manuscript and very helpful feedback.} 
This work was supported by the Netherlands Organization for Scientific Research (NWO/OCW),
as part of the Frontiers of Nanoscience program and program ENW.NVP.2018.001, by the Deutsche Forschungsgemeinschaft
under EV30/8-1 and EV30/11-1 and under SFB 1277 project B01 and project A03, and by the PRIMUS/Sci/09 programme of the Charles University.
\end{acknowledgments}

\bibliography{reviewRMP} 


\end{document}